\documentclass[11pt,epsfig]{article} 
\usepackage[utf8]{inputenc}
\usepackage[top=1in, left=0.95in, bottom=1.1in, right=0.95in]{geometry}
\usepackage[colorlinks=true,
linkcolor=blue,
urlcolor=red,
citecolor=red]{hyperref}
\usepackage[vcentermath,enableskew]{youngtab}
\usepackage{ytableau}
\usepackage{tikz}
\usepackage{lscape}
\usepackage{tabularx}
\usepackage[toc]{appendix}
\usepackage{longtable}
\usepackage{enumerate}
\usepackage{float}
\usepackage{subfig}
\usepackage{color}
\usepackage{xcolor}
\usepackage{multirow}
\usepackage{cite}

\let\counterwithin\relax
\usepackage{chngcntr}
\usepackage{amssymb, amsmath,mathrsfs}
\usepackage{multicol,multirow}
\usepackage{ulem}
\usepackage{graphics}
\usepackage{graphicx}
\usepackage{epsf}
\usepackage{epsfig}
\usepackage{float}
\usepackage{makecell}
\usepackage{multirow}
\usepackage{color}
\usepackage{xcolor}
\usepackage{simplewick}
\usepackage{amsmath}
\usepackage{amsfonts}
\usepackage{makeidx} 
\newcommand{\lra}[1]{\langle #1 \rangle}

\newcommand{\bin}[1]{(\overline{N} #1 N)}
\newcommand{\lrd}{\overleftrightarrow{D}}
\newcommand{\lrpartial}{{\negthickspace\stackrel{\leftrightarrow}{D}\negthickspace{}}}

\def\baselinestretch{1.2}

\begin{document}
\begin{center}

{\Large \textbf  {Complete CP-eigen Bases of Meson-Baryon Chiral Lagrangian \\ up to $p^5$-order}
}\\[10mm]


Chuan-Qiang Song$^{a,b,c}$\footnote{songchuanqiang21@mails.ucas.ac.cn},Hao Sun$^{a, b}$\footnote{sunhao@itp.ac.cn}, Jiang-Hao Yu$^{a, b, c, d}$\footnote{jhyu@itp.ac.cn}\\[10mm]

\noindent 
$^a${ \small School of Fundamental Physics and Mathematical Sciences, Hangzhou Institute for Advanced Study, UCAS, Hangzhou 310024, China}  \\
$^b${ \small School of Physical Sciences, University of Chinese Academy of Sciences,   Beijing 100049, P.R. China}   \\
$^c${ \small Institute of Theoretical Physics, Chinese Academy of Sciences,   Beijing 100190, P. R. China} \\
$^d${ \small International Centre for Theoretical Physics Asia-Pacific, Beijing/Hangzhou, China}\\[10mm]

\date{\today}   
          
\end{center}

\begin{abstract}
Chiral perturbation theory describes the low energy dynamics of mesons and baryons in terms of the nonlinear Goldstone boson and fermion degrees of freedom. Through the Young tensor technique, we construct the on-shell operator bases for the meson-baryon system up to $p^5$-order, using the chiral dimension power counting and heavy baryon expansion. For the Lorentz structure, additional treatments on off-shell external sources and operators with higher derivatives are necessarily considered, while for the internal structure, the invariant tensor basis is converted into the trace basis equivalently, and Cayley-Hamilton relations are utilized to classify different CP eigen-operators. Finally we present the complete operator set of $C$+$P$+, $C$+$P$-, $C$-$P$+, and $C$-$P$- eigen-operators at the $p^5$-order, and obtain the operator counting from the Hilbert series.  

\end{abstract}


\newpage
\setcounter{tocdepth}{3}
\setcounter{secnumdepth}{3}

\tableofcontents

\setcounter{footnote}{0}

\def\baselinestretch{1.5}
\counterwithin{equation}{section}

\newpage
\section{Introduction}
Chiral perturbation theory (ChPT)~\cite{Weinberg:1968de,Weinberg:1978kz,Gasser:1983yg,Gasser:1984gg,Gasser:1987rb} is an effective field theory (EFT) for quantum chromodynamics (QCD) in low energy region. For the $N_f$ flavor of light quarks, there are chiral symmetry $SU(N_f)_{L}\times SU(N_f)_{R}$ and its spontaneous symmetry breaking to $SU(N_f)_V$, while $N_f=3$. The $N_f^2-1$ Goldstone bosons correspond to the degrees of freedom for pseudoscalar mesons~\cite{Coleman:1969sm,Callan:1969sn,Weinberg:1978kz}: $\{\pi^-,\pi^+,\pi^0\}$ pseudoscalar meson triplet for $N_f=2$. $\{\pi^{\pm,0},\eta,K^{\pm,0},\bar{K}^0\}$ pseudoscalar meson octet $N_f=3$. In the ChPT, the perturbative chiral expansion is performed in terms of the ratio between the meson momentum transfer $p$ and the cutoff scale $\Lambda$: $p/\Lambda$. According to Weinberg's power counting rules, the chiral expansion corresponds to the loop expansion order by order. The chiral Lagrangian can be constructed systematically based on the nonlinear coset construction according to the chiral expansion. Considering the strong dynamics, the effective operators should be considered to be invariant under the parity ($P$), charge ($C$), and hermitian conjugation ($h.c.$) transformation. Nowadays, the complete CP-even chiral Lagrangian up to $p^8$-order~\cite{
Wess:1971yu,Witten:1983tw,Gasser:1983yg,Gasser:1984gg, 
Fearing:1994ga,Bijnens:1999sh,Bijnens:1999hw,Bijnens:2001bb,Ebertshauser:2001nj,Bijnens:2018lez,Hermansson-Truedsson:2020rtj,Bijnens:2023hyv} for pure mesonic system have been constructed for both normal and anomalous ones, and the complete operator set for $C$+$P$+, $C$+$P$-, $C$-$P$+, and $C$-$P$- eigen-operators are listed in Ref.~\cite{li2024complete}.


The chiral Lagrangian can be extended to describe the meson-baryon system by adding the baryon field into the building blocks~\cite{Gasser:1987rb}. In the meson-baryon ChPT, the baryons compose a doublet $\{p,n\}$ for the $SU(2)$ case and form an octet $\{\Sigma^{\pm,0},\Lambda,\Xi^{-,0},p,n\}$ for the $SU(3)$ case. Due to non-vanishing baryon mass $m_N \sim \Lambda$ in the chiral limit, the chiral expansion $p/\Lambda$ would be invalid when dealing with the power counting of the baryon operators. One of the solutions for such a problem was provided in the framework of heavy baryon chiral perturbation theory (HBChPT)~\cite{Krause:1990xc,Jenkins:1990jv,Ecker:1995rk}, which takes a non-relativistic projection of the baryon field. In the HBChPT, after taking additional expansion on $1/m_N \sim 1/\Lambda$, the correct chiral expansion $p/\Lambda$ is recovered. Moreover, the infrared regularization scheme~\cite{Ellis:1997kc,Becher:1999he} and the extended on-mass-shell renormalization scheme~\cite{Fuchs:2003qc} are also proposed to solve such problem while maintaining relativistic form. For $N_f=2$, both the relativistic form of the pion-nucleon operators and the heavy baryon projected chiral Lagrangian have been constructed up to the order $\mathcal{O}(p^4)$~\cite{Krause:1990xc,Ecker:1995rk,Fettes:1998ud,Fettes:2000gb}. And the $SU(3)$ meson-baryon chiral Lagrangian up to order $\mathcal{O}(p^4)$~\cite{Krause:1990xc,Frink:2004ic,Oller:2006yh,Frink:2006hx,Jiang:2016vax} have been obtained, although possible redundancies may still need to be examined.

Nowadays, higher order meson-baryon operators are necessarily needed in certain processes, such that the nucleon mass~\cite{McGovern:1998tm,McGovern:2006fm,Schindler:2006ha,Schindler:2007dr} has been calculated up to $\mathcal{O}(p^6)$ order and the counter-term from $\mathcal{O}(p^5)$ operators are needed. 
Moreover, the meson-baryon ChPT operators written in the literature are all $CP$-even, relevant to the strong dynamics. 
The other $CP$ operators are also important in the weak processes, such as the non-vanishing electric dipole moments (EDM)~\cite{An:2009zh,Engel:2013lsa,Bsaisou:2014oka,Dekens:2014jka}. And the $CP$-odd operators are also relevant for the $CP$ violating dark matter~\cite{Dekens:2022gha} and $P$ violating nuclear potential~\cite{Maekawa:2011vs}.

To construct the higher-order operators, the traditional method encounters several difficulties, especially for the meson-baryon ChPT which contains an increasing number of derivatives. In the chiral limit, the presence of heavy baryons necessitates consideration of more derivatives acting on the baryon, since derivatives of the baryon field scale as $\mathcal{O}(p^0)$ in the heavy baryon power counting scheme. As the number of derivatives increases, the Lorentz structure becomes very complicated, due to the following redundancies: the equations of motion (EOMs) of the fields, integration-by-part (IBP) techniques, and Fierz identities. Thus other techniques should be used to reduce the redundancies. Additionally, external sources should not be vanishing under the equations of motion. Furthermore, regarding to the trace basis of the internal structure, as the number of fields increases, more Cayley-Hamilton relations and repeated fields must be considered.

We employ the Young tensor technique~\cite{Li:2020gnx,Li:2020xlh,Li:2022tec} to construct the ChPT effective operators. This method proves to be efficient in resolving issues related to simplifying the complicated Lorentz structure through on-shell amplitude operator correspondence and streamlining the flavor structure via the invariant tensor basis. This technique has been applied in various effective field theories, including in the standard model effective field theory~\cite{Li:2020gnx,Li:2020xlh,Ren:2022tvi,Yang:2023ncf}, low energy effective field theory~\cite{Li:2020tsi}, Higgs effective field theory~\cite{Sun:2022snw,Sun:2022ssa,Sun:2022aag}, gravity effective field theory~\cite{Li:2023wdz}, and other EFTs~\cite{Li:2021tsq,Song:2023lxf,Song:2023jqm} involving new fields. Additionally, this technique has been applied to electroweak chiral Lagrangian~\cite{Sun:2022ssa,Sun:2022aag} and the ChPT~\cite{Low:2022iim,li2024complete}. In the meson-baryon ChPT, we construct the complete and independent meson-baryon chiral Lagrangian with explicit $C$ and $P$ eigenvalues, utilizing the chiral building blocks through the Young tensor technique up to $\mathcal{O}(p^5)$. The Adler zero condition~\cite{Low:2019ynd,Dai:2020cpk,Kampf:2021jvf,Low:2022iim,Sun:2022ssa,Sun:2022aag} imposed on the amplitudes can handle the operators involving Goldstone bosons. We manage the off-shell external sources, which are non-vanishing under their EOM, through the treatment in Ref.~\cite{Ren:2022tvi}.

Using the procedures documented in the Mathematica package ABC4EFT~\cite{Li:2020gnx,Li:2020xlh,Li:2022tec}, we obtain the  435 $C$+$P$+ operators, 387 $C$+$P$- operators, 352 $ C$-$P$+ operators, and 420 $C$-$P$- operators for the $SU(2)$ case at $\mathcal{O}(p^5)$ and 528 $C$+$P$+ operators, 475 $C$+$P$- operators, 413 $C$-$P$+ operators, and 450 $C$-$P$- operators for the $SU(3)$ case at $\mathcal{O}(p^4)$, although obtaining the $\mathcal{O}(p^5)$ results is straightforward. 
As a cross-check, we utilize the Hilbert series technique~\cite{Lehman:2015via,Henning:2015alf,Henning:2015daa,Marinissen:2020jmb,Graf:2020yxt,Bijnens:2022zqo} to obtain the numbers of the independent operators for the baryon ChPT both for the $SU(2)$ case and the $SU(3)$ case. Unlike the pure mesonic case, in the HBChPT~\cite{Krause:1990xc,Jenkins:1990jv,Ecker:1995rk} the types with more derivatives acting on baryons should be considered to be in the same order. Thus taking the heavy baryon expansion, the results from the Hilbert series can not be fully trusted, and they are considered to be cross-checked in the middle steps.


The paper is organized as follows. In Sec.~\ref{ChPT}, we review the chiral Lagrangian building blocks and their CP properties and the heavy baryon projection. In Sec.~\ref{invariant}, we briefly review the Young tensor method to construct the complete and independent effective operators, convert the flavor basis to the trace basis by the Cayley-Hamilton theorem, work out the number of operators for $CP$-eigen basis by the Hilbert series, and consider the operators with more derivatives acting on baryon in Sec.~\ref{more}, show the example of the whole procedure in Sec.~\ref{example}. In Sec.~\ref{conclusion}, the conclusion is drawn. We compared our $CP$-even operators up to $p^4$-order with the literature~\cite{Fettes:2000gb} in App.~\ref{comparison} and show the Cayley-Hamilton relations in App.~\ref{app: CH relations}. Finally, the complete $CP$-eigen bases are listed in the App.~\ref{app:SU2} and~\ref{app:SU3}. 

\section{Chiral Perturbation Theory}
\label{ChPT}

\subsection{Chiral Lagrangian and Building Blocks}

The ChPT is an effective field theory of QCD at the energy scale under $\Lambda_\chi \approx 1$ \text{GeV}. In this section, we will review its basic ingredients.

As a low-energy effective theory, the ChPT reveals the symmetry of the underlying theory, which is the chiral symmetry of the low-energy QCD. The QCD Lagrangian takes the form that
\begin{equation}
\label{eq: lQCD lagrangian}
    \mathcal{L}_0 = \Bar{q}_L iD\!\!\!\!\slash q_L + \Bar{q}_R iD\!\!\!\!\slash q_R -\frac{1}{4} G^a_{\mu\nu} G^a{}^{\mu\nu}\,,
\end{equation}
where $q_L$ and $q_R$ are light quarks,
\begin{align}
    q_L = (u_L\,,d_L\,,s_L)^T\,,\quad q_R = (u_R\,, d_R\,, s_R)^T\,.
\end{align}
The Lagrangian in Eq.~\eqref{eq: lQCD lagrangian} is invariant under a global symmetry $SU(3)_L\times SU(3)_R\times U(1)_V$, where the abelian group $U(1)_V$ represents baryon-number conservation, while the non-abelian group $\mathcal{G}=SU(3)_L \times SU(3)_V$ is called the chiral symmetry, under which the light quarks transform as
\begin{align}
    q_L &\rightarrow L q_L\,,\quad L\in SU(3)_L\,, \\
    q_R &\rightarrow R q_R\,,\quad R\in SU(3)_R\,.
\end{align}
According to the Nother's theorem, the chiral symmetry $\mathcal{G}$ corresponds to 16 conserving currents,
\begin{align}
    J_L^a{}^\mu = \Bar{q}_L \gamma^\mu \frac{\lambda^a}{2}q_L \,, \quad J_R^a{}^\mu = \Bar{q}_R \gamma^\mu \frac{\lambda^a}{2}q_R \,,
\end{align}
where $\lambda^a$ are the Gell-mann matrices. Alternatively, they can be combined as 
\begin{equation}
    J_V^a{}^\mu = J_R^a{}^\mu + J_L^a{}^\mu \,,\quad J_A^a{}^\mu = J_R^a{}^\mu - J_L^a{}^\mu\,,
\end{equation}
which transform under parity as
\begin{equation}
    J_V^a{}^\mu \rightarrow J_V^a{}^\mu \,,\quad J_A^a{}^\mu \rightarrow -J_A^a{}^\mu\,,
\end{equation}
called the vector currents and the axial currents respectively, corresponding to 16 conserving charges 
\begin{equation}
    Q_V^a = \int dx^3 J_V^a{}^0 \,,\quad Q_A^a = \int dx^3 J_A^a{}^0 \,,
\end{equation}
with opposite parity. However, there are no degenerate states of negative parity of the baryons with positive parity, which means the vacuum state is not annihilated by the axial charge $Q_A^a$, the chiral symmetry spontaneously breaks to its diagonal subgroup $\mathcal{H}=SU(3)_V$, whose generators are defined as $T^a = \lambda^a\otimes \mathbb{I}+ \mathbb{I}\otimes\lambda_a$. According to the Goldstone theorem, the broken symmetries $T^{\hat{a}}=\lambda^a\otimes \mathbb{I}- \mathbb{I}\otimes\lambda_a$ generate 8 massless Nambu-Goldstone bosons (NGBs) $\phi(p)^a$, sharing the same quantum number of the axial currents, and interpolate to them as
\begin{equation}
    \langle 0|J^a_A{}^\mu |\phi(p)^b\rangle = ip^\mu F_0 \delta^{ab}\,,
\end{equation}
where $F_0$ is the decay constant of the NGBs, which means these NGBs are pseudoscalar bosons with negative parity, corresponding to the pseudoscalar meson octets\footnote{The pseudoscalar meson octets are actually pseudo NGBs since they are not exactly massless.},
\begin{equation}
    \Pi(x)=\sum_{a=1}^8\phi^a(x)\lambda^a = \left(\begin{array}{ccc}
    \pi^0 + \frac{1}{\sqrt{3}}\eta & \sqrt{2}\pi^+ & \sqrt{2}K^+ \\
    \sqrt{2}\pi^- & -\pi^0+\frac{1}{\sqrt{3}}\eta & \sqrt{2}K^0 \\
    \sqrt{2}K^- & \sqrt{2}~\Bar{K}^0 & -\frac{2}{\sqrt{3}}\eta
    \end{array}\right)\,.
\end{equation}
For the $SU(2)$ case only the pions are included
\begin{equation}
    \Pi(x)=\sum_{I=1}^3\phi^I(x)\sigma^I = \left(\begin{array}{cc}
    \pi^0 & \sqrt{2}\pi^+ \\ \sqrt{2}\pi^-&-\pi^0
    \end{array}\right)\,.
\end{equation}

In addition to continuous chiral symmetry, discrete symmetries like the parity ($P$) and the charge conjugation ($C$) are also important in the ChPT. The ChPT Lagrangian can be divided into 4 disjoint sectors in terms of their eigenvalues under these two discrete symmetries,
\begin{equation}
\label{eq: ChPT lagrengiain}
    \mathcal{L}_{ChPT} = \mathcal{L}_{ChPT}^{C+P+} + \mathcal{L}_{ChPT}^{C+P-} + \mathcal{L}_{ChPT}^{C-P+} + \mathcal{L}_{ChPT}^{C-P-}\,.
\end{equation}
The $C$ and $P$ eigenvalues of an effective operator are related to the intrinsic $C$ and $P$ charges of the fields and their Lorentz and chiral structures, which will be discussed subsequently.

The building blocks of the ChPT can be described under the CCWZ~\cite{Coleman:1969sm,Callan:1969sn} coset construction, by which the NGBs after symmetry break $\mathcal{G}\rightarrow H$ are characterized non-linearly, and are collected in a unitary matrix
\begin{equation}
    u(x) = \exp\left(\frac{\phi(x)^a \lambda^a
    }{2F_0}\right)\,,
\end{equation}
which transforms under $SU(3)_V$ as $u(x)\rightarrow V u(x) V^{-1}$, where $V\in SU(3)_V$. Apart from the mesons, there exist external sources interacting with them, including the scalar source $s$, pseudoscalar source $p$, vector source $v_\mu$, and axial source $a_\mu$. To construct effective operators, it is convenient to choose these degrees of freedom as
\begin{align}
    u_\mu &= i(u^\dagger(\partial_\mu -ir_\mu)u-u(\partial_\mu -il_\mu)u^\dagger ) \,,\notag \\
    \Sigma_{\pm} &= u^\dagger \Sigma u^\dagger \pm u \Sigma^\dagger u \,, \notag \\
    f_{\pm}{}_{\mu\nu} &= u^\dagger f^R_{\mu\nu} u \pm u f^L_{\mu\nu} u^\dagger \,,\label{eq: building blocks}
\end{align}
where
\begin{align}
    \Sigma &= 2B(s+ip) \,,\notag \\
    f^R_{\mu\nu} &= \partial_\mu r_\nu - \partial_\nu r_\mu -i [r_\mu,r_\nu] \,,\quad r_\mu = v_\mu + a_\mu \,, \notag \\
    f^L_{\mu\nu} &= \partial_\mu l_\nu - \partial_\nu l_\mu -i [l_\mu,l_\nu] \,,\quad l_\mu = v_\mu - a_\mu \,, \label{eq: sources}
\end{align}
and $B$ is a constant related to the quark condensation. In particular, the trace of the meson $u_\mu$ and the field-strength tensor $f_{\pm}{}_{\mu\nu}$ is zero by the definition in Eq.~\eqref{eq: building blocks}, while the trace of $\Sigma_\pm$ is not. The choice of the degrees of freedom in Eq.~\eqref{eq: building blocks} is convenient since they transform under the group $\mathcal{H}=SU(3)_V$ universally,
\begin{equation}
    X \rightarrow g X g^{-1}\,, \quad g \in SU(3)_V\,,
\end{equation}
and $X = u_\mu\,,\chi_\pm \,, f_\pm$. Besides, their transformations under the $C$ and $P$ transformations are
\begin{align}
    C : \quad & X \rightarrow (-1)^X X^T \,, \\
    P : \quad & X \rightarrow \eta_X X \,,
\end{align}
where $X = u_\mu\,,\chi_\pm \,, f_\pm$ and $(-1)^X\,,\eta_X$ are intrinsic $C$ and $P$ charge respectively. These intrinsic changes are summarized in Tab.~\ref{tab: building blocks mesons}. And we give the conventions that $\Tilde{f}_\pm^{\mu\nu}=\epsilon^{\mu\nu\rho\lambda} f_{\pm\rho\lambda}$ and the Lorentz indices would be symmetric when the covariant derivatives act on $u_\mu$ 
\begin{equation}
    D^\mu D^\nu D^{...} u^{\lambda}=D^\mu D^\nu D^{...} u^{\lambda}+\text{full permutation of}\,\, {(\mu\nu\lambda...)}\,.
\end{equation}
\begin{table}
    \centering
    \begin{tabular}{|l|cccccc|}
    \hline
          $X$ & $u_\mu$ & $\Sigma_+$ & $\Sigma_-$ & $f_+$ & $f_-$ & $D_\mu$ \\
    \hline
     $d_\chi$ & $1$ & $2$ & $2$ & $2$ & $2$ & $1$ \\
$C$: $(-1)^X$ & $+$ & $+$ & $+$ & $-$ & $+$ & $+$ \\
$P$: $\eta_X$ & $-$ & $+$ & $-$ & $+$ & $-$ & $+$ \\
    \hline
    \end{tabular}
    \caption{The chiral dimension $d_\chi$ and the intrinsic $C$, $P$ charges of the bosonic degrees of freedom and the covariant derivatives $D_\mu$ on them.}
    \label{tab: building blocks mesons}
\end{table}

Within the CCWZ framework, the baryons $\Psi$ are included in the ChPT as a specific representation of the $SU(3)_V$. At this stage, the ChPT of the $SU(3)$ case and the one of the $SU(2)$ case are different, since the baryons in them are of different representations. If the ChPT considers all the 3 light quarks $u\,, d\,, s$, the light baryons composed of them form an octet
\begin{equation}
    B = \frac{B^a\lambda^a}{\sqrt{2}} = \left(\begin{array}{ccc}
\frac{1}{\sqrt{2}}\Sigma^0+\frac{1}{\sqrt{6}}\Lambda & \Sigma^+ & p \\
\Sigma^- & -\frac{1}{\sqrt{2}}\Sigma^0 + \frac{1}{\sqrt{6}}\Lambda & n \\
\Xi^- & \Xi^0 & -\frac{2}{\sqrt{6}}\Lambda 
    \end{array}\right) \rightarrow g B g^{-1}\,,\quad g \in SU(3)_V \,,
\end{equation}
while if the ChPT contains only the quarks $u\,, d$, the remaining baryons are proton and neutron, and they compose a doublet
\begin{equation}
    N = \left(\begin{array}{c}
p \\ n
    \end{array}\right) \rightarrow g N \,,\quad g \in SU(2)_V\,.
\end{equation}
The Dirac bilinears composed of baryons transform under $C$ and $P$ as 
\begin{align}
    C : \quad & (\Bar{\Psi}\Gamma \overleftrightarrow{D}^{\mu_1\mu_2\dots \mu_n}\Psi) \rightarrow (-1)^n(-1)^\Gamma (\Bar{\Psi}\Gamma \overleftrightarrow{D}^{\mu_1\mu_2\dots \mu_n}\Psi) \,, \label{eq: C transformation of bilinears}\\
    P : \quad & (\Bar{\Psi}\Gamma \overleftrightarrow{D}^{\mu_1\mu_2\dots \mu_n}\Psi) \rightarrow \eta_\Gamma (\Bar{\Psi}\Gamma \overleftrightarrow{D}^{\mu_1\mu_2\dots \mu_n}\Psi) \,,
\end{align}
where $\Psi = B\,, N$, $\overleftrightarrow{D}^{\mu_1\mu_2\dots \mu_n}$ is the short-hand form of $\overleftrightarrow{D}^{\mu_1}\overleftrightarrow{D}^{\mu_2}\dots\overleftrightarrow{D}^{\mu_n}$, and $\Gamma$ is some Dirac matrix. The derivative on the baryons $\overleftrightarrow{D}^\mu$ is defined as
\begin{equation}
    (\Bar{\Psi}\Gamma \overleftrightarrow{D}^{\mu}\Psi) = (\Bar{\Psi}\Gamma D^{\mu}\Psi) - ( D^{\mu}\Bar{\Psi}\Gamma\Psi)\,,
\end{equation}
thus contributes a factor $-1$ to the $C$ transformation of the bilineas in Eq.~\eqref{eq: C transformation of bilinears}. The factors $(-1)^\Gamma$ and $\eta_\Gamma$ are the intrinsic $C$ and $P$ charges of the Dirac matrices respectively, and they are listed in Tab.~\ref{tab: building blocks baryons}.

\begin{table}
    \centering
    \begin{tabular}{|l|ccccc||cc|}
    \hline
          $\Gamma$ & $1$ & $\gamma^5$ & $\gamma^\mu$ & $\gamma^5\gamma^\mu$ & $\sigma^{\mu\nu}$ & $\epsilon_{\mu\nu\rho\lambda}$ & $\overleftrightarrow{D}^\mu$\\
    \hline
          $d_\chi$ & $0$ & $1$ & $0$ & $0$ & $0$ & $0$ & $0$ \\
$C$: $(-1)^\Gamma$ & $+$ & $+$ & $-$ & $+$ & $-$ & $+$ & $-$\\
$P$: $\eta_\Gamma$ & $+$ & $-$ & $+$ & $-$ & $+$ & $-$ & $+$\\
\hline
    \end{tabular}
    \caption{The chiral dimension $d_\chi$ and the intrinsic charges of the $\gamma$-matrices, the totally antisymmetric tensor $\epsilon_{\mu\nu\rho\lambda}$, and the derivatives on the baryons $\overleftrightarrow{D}^\mu$.}
    \label{tab: building blocks baryons}
\end{table}

\subsection{Power Counting and Heavy Baryon Projection}

To organize effective operators in different order, a power-counting scheme is developed in the ChPT. Because the mesons participate in the interactions via the $u_\mu$, which is proportional to its momentum at lowe-energy, $u_\mu \propto p_\mu$, the Lagrangian can be regarded as a series expansion in terms of the small ratio $p/\Lambda$. We define the power of $p/\Lambda$ of an operator or a degree of freedom as its chiral dimension $d_\chi$, summarized as follows:
\begin{itemize}
    \item $d_\chi=1$ for $u_\mu$ and the derivative $D_\mu$ on it, while $d_\chi=0$ for $u$;
    \item According to the definitions of the external sources in Eq.\eqref{eq: sources} the chiral dimensions of both $\Sigma_\pm$ and $f_\pm$ are $d_\chi=2$.
\end{itemize}
The chiral dimensions of these bosonic degrees of freedom can be found in Tab.~\ref{tab: building blocks mesons}.

For the relativistic baryon field, the power counting is invalid due to its heavy mass $m_N \sim \Lambda$. We take the heavy baryon form~\cite{Krause:1990xc,Jenkins:1990jv} for the baryon field. According to the discussion above the baryon is so heavy that it behaves non-relativistic when interacting with other degrees of freedom, thus it is natural to expand the operators containing baryons in terms of the inverse of their mass $1/m$. For any time-like vector $v^\mu$ satisfying $v^2 = 1$, we can define the projection operators as $P_v^{\pm} = (1+v\!\!\!\slash)/2$, and the Dirac spinors can be expressed as
\begin{equation}
    \Psi(x) = e^{-imv\cdot x}\left[\underbrace{e^{imv\cdot x} P_v^+\Psi(x)}_{\equiv \mathcal{N}_v(x)} + \underbrace{e^{imv\cdot x}P_v^- \Psi(x)}_{\equiv \mathcal{H}_v(x)}\right]\,.
\end{equation}
Then a general meson-baryon interaction becomes
\begin{equation}
\mathcal{L}_{M B} = \bar{\mathcal{N}_v}A\mathcal{N}_v+\bar{\mathcal{H}_v}B\mathcal{N}_v+\bar{\mathcal{N}_v}\gamma^0B^\dagger\gamma^0\mathcal{H}_v -\bar{\mathcal{H}_v}C\mathcal{H}_v\,,
\end{equation}
where $A$, $B$, $C$ is composed of the mesonic fields. Integrating out the heavy component $\mathcal{H}_v$, and expanding $C^{-1}$ in inverse power of the nucleon mass, the Lagrangian can be expanded in terms of $\mathcal{N}_v$~\cite{Ecker:1995rk,Fettes:2000gb}, 
\begin{align}
    \mathcal{L}_{M B} \rightarrow &\bar{\mathcal{N}_v} (A+\gamma^0B^\dagger\gamma^0C^{-1}) \mathcal{N}_v \\
    &=\bar{\mathcal{N}_v}(A^{(1)}+A^{(2)}+\frac{1}{2m_N}\gamma^0B^{\dagger{(1)}}\gamma^0B^{(1)}+...)\mathcal{N}_v\,.
\end{align}

Thus the bilinears and derivatives of the baryon can be expanded to take the form
that
\begin{align}
\label{heavy baryon}
    \Bar{\Psi} 1 \Psi &= \Bar{\mathcal{N}_v} 1 \mathcal{N}_v +\frac{1}{m^2}\partial^2(\Bar{\mathcal{N}_v} 1\mathcal{N}_v)+...\,, \notag \\
    \Bar{\Psi} \gamma^5 \Psi &= \frac{1}{m_N}\partial_\mu(\Bar{\mathcal{N}_v} S^\mu \mathcal{N}_v)+... \,, \notag \\
    \Bar{\Psi} \gamma^\mu \Psi &= \Bar{\mathcal{N}_v} v^\mu \mathcal{N}_v +\frac{1}{2m_N}\Bar{\mathcal{N}_v}(\overleftrightarrow\partial^\mu-v^\mu v\!\!\!\slash) \mathcal{N}_v+\frac{1}{2m_N}\epsilon^{\mu\nu\rho\lambda}\partial_\nu(\Bar{\mathcal{N}_v}v_\rho S_\lambda \mathcal{N}_v)+...\,, \notag \\
    \Bar{\Psi} \gamma^5\gamma^\mu \Psi &= 2\Bar{\mathcal{N}_v} S^\mu \mathcal{N}_v -\frac{i}{2m_N}v^\mu\Bar{\mathcal{N}_v} S^\nu\overleftrightarrow\partial_\nu \mathcal{N}_v...\,, \notag \\
    \Bar{\Psi} \sigma^{\mu\nu} \Psi &= 2\epsilon^{\mu\nu\rho\lambda}\Bar{\mathcal{N}_v} v_\rho S_\lambda \mathcal{N}_v +\frac{1}{2m_N}(\epsilon^{\mu\rho\lambda\sigma}\Bar{\mathcal{N}_v} v_\sigma [S_\lambda,v^\nu]\overleftrightarrow\partial_\rho u\mathcal{N}_v-[v^\mu ,\partial^\nu](\Bar{\mathcal{N}_v} \mathcal{N}_v))...\,, \notag \\
    \Bar{\Psi} \gamma^5\sigma^{\mu\nu} \Psi &= 2\Bar{\mathcal{N}_v} [v^\mu ,S^\nu] \mathcal{N}_v +...\,, \notag \\
    \Bar{\Psi} \overleftrightarrow{D}^\mu \Psi &= \Bar{\mathcal{N}_v} v^\mu \mathcal{N}_v +\frac{1}{2m_N}\Bar{\mathcal{N}_v}\overleftrightarrow\partial^\mu \mathcal{N}_v+\frac{1}{2m_N}\epsilon^{\mu\nu\rho\lambda}\partial_\nu(\Bar{\mathcal{N}_v}v_\rho S_\lambda \mathcal{N}_v)...\,,  
\end{align}
and the $S^\mu$ in the Eq.~\eqref{heavy baryon} is the spin-operator of the baryon, and its definition and basic properties are
\begin{equation}
    S^\mu=\frac{i}{2}\gamma^5\sigma^{\mu\nu}v^\nu\,,\quad S\cdot v=0\,,\quad\{S^\mu,S^\nu\}=\frac{1}{2}\left(v^\mu v^\nu-g^{\mu\nu}\right)\,,\quad\left[S^\mu,S^\nu\right]=i\epsilon^{\mu\nu\rho\lambda}v_\rho S_\lambda\,.
\end{equation}
Furthermore, the heavy baryon projection do not change the part of the mesons and we choose the leading order(LO) for the heavy baryon projection. We take one of our operators for example
\begin{equation}
    (\Bar{N}\gamma^5\sigma^{\mu\nu} \overleftrightarrow{D}^\lambda[u_\mu,u_\nu] N)=(\Bar{N_v}[v^\mu,S^\nu] v^\lambda[u_\mu,u_\nu] N_v)+...\,.
\end{equation}
In Eq.~\eqref{heavy baryon}, the LO of $\Bar{\Psi} \gamma^5 \Psi$ vanishes thus it is attributed to a higher dimension.

In this work, we present the relativistic form of the effective operators and keep the $\gamma^5$ chiral dimension-1, which will be illustrated further when doing to comparison to the literature~\cite{Fettes:2000gb}. Therefore, the chiral dimension for baryon fields are summarized as
\begin{itemize}
    \item For the baryons, they are dimensionless and the derivatives on them are dimensionless due to their considerable baryon masses. Taking the proton as an example, the ratio $p/\Lambda$ is no longer small and does not contribute to $d_\chi$.     
    \item It should be emphasized that the chiral dimension of $\gamma^5$ is taken to be $1$, determined by the heavy baryon form. 
\end{itemize}

Once the degrees of freedom and their symmetries (both continuous and discrete) are identified, the Lagrangian of the ChPT can be constructed in terms of the power counting scheme. Then power counting $\mathcal{O}(p^\chi)$ for the operators is the chiral dimension $d_\chi$ of the operators. In this work, we construct the complete and independent operators up to $\mathcal{O}(p^5)$ in all 4 sectors in Eq.~\eqref{eq: ChPT lagrengiain} of the ChPT.

\newcommand{\calt}{\mathcal{T}}

\section{Construction of Invariants}
\label{invariant}

This section serves to present the methods to construct invariants used in this paper. It is a challenge to eliminate all the redundancies when writing down effective operators, which results in the operators in literature being neither complete nor independent. In this paper, we adopt 3 different algebra techniques, including the Young tensor method, the Cayley-Hamilton theorem, and the Hilbert series method to ensure the result here is minimal and complete.
\subsection{Young Tensor Technique}
\label{young}
The Young tensor technique~\cite{Li:2020gnx,Li:2020xlh,Li:2022tec} is a systematic method to construct independent group invariants explicitly. Next, we will outline the basic points of this method for the Lorentz and internal structures separately.

\subsubsection*{Lorentz Structure}

The Lie algebra of the Lorentz group $SO(3,1)$ is the complexification of the direct sum Lie algebra $SL(2,\mathbb{C})_l\oplus SL(2,\mathbb{C})_r$, thus all the irreducible representations of the Lorentz group is labeled by two half integers $(j_l,j_r)$.
where $j_l$ and $j_r$ labels the irreducible representations of $SL(2,\mathbb{C})$.

If we define the $\lambda_\alpha\in (\frac{1}{2},0)$ as a left-handed spinor and $\Tilde{\lambda}_{\dot{\alpha}}\in (0,\frac{1}{2})$ as a right-handed spinor, all other fields of other irreducible representation of the Lorentz group can be expressed as their product. In the ChPT the helicities of all the degrees of freedom are at most 1, and their representations and spinor forms are
\begin{align}
\label{define}
    \text{Scalar field: }&\phi\in (0,0) \sim 1\notag \\
    \text{Left-handed spinor field: } & \psi \in (\frac{1}{2},0) \sim \lambda_\alpha\notag \\
    \text{Right-handed spinor field: } & \psi^\dagger \in (0,\frac{1}{2}) \sim \Tilde{\lambda}_{\dot{\alpha}}\notag \\
    \text{Left-handed field strength tensor: } & F_L=\frac{F-i\tilde{F}}{2} \in (1,0) \sim \lambda_\alpha\lambda_\beta \notag \\
    \text{Right-handed field strength tensor: } & F_R=\frac{F+i\tilde{F}}{2} \in (0,1) \sim \Tilde{\lambda}_{\dot{\alpha}} \Tilde{\lambda}_{\dot{\beta}}\notag \\
    \text{Derivative: } & D \in (1,1) \sim \lambda_\alpha\Tilde{\lambda}_{\dot{\alpha}} \,,
\end{align}
thus an operator can be expressed as a product of spinors with all the left- and right-handed indices contracted by the asymmetric tensors $\epsilon^{\alpha\beta}\,,\epsilon^{\dot{\alpha}\dot{\beta}}$. Defining 
\begin{equation}
    \epsilon^{\alpha\beta} \lambda^i_\beta\lambda^j_\alpha = \langle ij \rangle\,,\quad \tilde{\lambda}^i_{\dot{\alpha}} \tilde{\lambda}^j_{\dot{\beta}}\epsilon^{\dot{\beta}\dot{\alpha}} = [ij]\,,
\end{equation}
a general operator takes the form that
\begin{equation}
    \mathcal{B} = \prod^n \langle ij\rangle \prod^{\tilde{n}} [kl]\,,
\end{equation}
where $n$ and $\tilde{n}$ are the half numbers of left- and right-handed spinor indices carried by the operator respectively. 

After such an operator-amplitude correspondence, the independent Lorentz structures can be constructed via the so-called primary Young diagrams, with the redundancies such as the EOM, IBP, the Fierz identities, and so on removed automatically. For an operator of $N$ fields whose helicities are $\{h_1,h_2,\dots,h_N\}$, and $k$ derivatives, the number of the left- and right-handed indices are determined,
\begin{equation}
\label{eq:nn}
    n=\frac{k}{2}-\sum_{h_i<0}h_i\,,\quad \tilde{n}=\frac{k}{2}+\sum_{h_i>0}h_i\,,
\end{equation}
and the corresponding primary Young diagram takes the form that
\begin{equation}
    \begin{tikzpicture}
\filldraw [draw = black, fill = cyan] (10pt,10pt) rectangle (22pt,22pt);
\filldraw [draw = black, fill = cyan] (10pt,22pt) rectangle (22pt,34pt);
\draw [densely dotted] (24pt,22pt) -- (32pt,22pt);
\filldraw [draw = black, fill = cyan] (10pt,46pt) rectangle (22pt,58pt);
\filldraw [draw = black, fill = cyan] (10pt,58pt) rectangle (22pt,70pt);
\draw [densely dotted] (24pt,58pt) -- (32pt,58pt);
\draw [densely dotted] (16pt,36pt) -- (16pt,46pt);
\filldraw [draw = black, fill = cyan] (34pt,10pt) rectangle (46pt,22pt);
\filldraw [draw = black, fill = cyan] (34pt,22pt) rectangle (46pt,34pt);
\filldraw [draw = black, fill = cyan] (34pt,46pt) rectangle (46pt,58pt);
\filldraw [draw = black, fill = cyan] (34pt,58pt) rectangle (46pt,70pt);
\draw [densely dotted] (40pt,36pt) -- (40pt,46pt);
\draw [|<-] (0pt,10pt)--(0pt,34pt);
\draw [|<-] (0pt,70pt)--(0pt,46pt);
\node (n2) at (-5pt,40pt) {\small $N-2$};

\draw [|<-] (10pt,80pt)--(22pt,80pt);
\draw [|<-] (46pt,80pt)--(34pt,80pt);
\node (nt) at (28pt,80pt) {\small $\tilde{n}$};


\draw (46pt,58pt) rectangle (58pt,70pt);
\draw (46pt,46pt) rectangle (58pt,58pt);
\draw [densely dotted] (60pt,58pt) -- (70pt,58pt);
\draw (70pt,58pt) rectangle (82pt,70pt);
\draw (70pt,46pt) rectangle (82pt,58pt);
\draw [|<-] (46pt,36pt)--(58pt,36pt);
\draw [|<-] (82pt,36pt)--(70pt,36pt);
\node (n) at (64pt,36pt) {\small $n$};
    \end{tikzpicture}\,,
\end{equation}
where the cyan boxes are from right-handed indices.
According to the group theory, the semi-standard Young tableaux (SSYT)~\cite{Henning:2019enq,Li:2020gnx,Li:2020xlh,Li:2022tec} associated with the primary Young diagram form a set of bases, called y-basis. Considering a specific operator, the numbers can be filled to the Young diagrams are also determined by $h_i$ and $k$,
\begin{equation}
\label{eq:numi}
    \# i = \sum_{h_i>0}h_i -2h_i + \frac{k}{2}\,,\quad i =1\,,2\,,\dots \,, N\,,    
\end{equation}

When applying this method to the ChPT, there are several remarks
\begin{itemize}
    \item The NGBs $u_\mu$ in the ChPT is of the representation $(\frac{1}{2},\frac{1}{2})$ of the Lorentz group meaning its amplitude satisfies Adler's zero condition, which constrains the y-basis. Such constraints can be extracted by taking the limit $\lim_{p\rightarrow p} \mathcal{B}(p)$, where $p$ is the momentum of the NGB, and are complemented as a system of linear equations of the original y-basis. The solution space of the system of linear equations contains the Lorentz structures satisfying Adler's zero condition.
    \item The Young tensor method adopts left- and right-handed spinors $\psi/\psi^\dagger$ as fundamental building blocks, but the operators composed of them are usually not $C\,, P$ eigenstates. thus we need to combine them to form Dirac spinors so that they have specific $C\,, P$ eigenvalues. A Dirac spinor $N$ is a direct sum of a left-handed spinor $N_L$ and a right-handed spinor $N_R$,  
    \begin{equation}
        N = \left(\begin{array}{c}
N_L \\ N_R
        \end{array}\right)\,.
    \end{equation}
    Considering a simple example of operator type $u\Sigma_+ N^2$, there are two Lorentz structures (the chiral symmetry has been discarded for simplicity for now)
    \begin{equation}
        \mathcal{B}_1 = (N^\dagger_L\bar{\sigma}^\mu N_L) u_\mu\Sigma_+ \,,\quad \mathcal{B}_2 = (N_R^\dagger \sigma^\mu N_R) u_\mu \Sigma_+\,.
    \end{equation}
    Under the parity transformation, they interchange with each other, since $P$ makes left- and right-handed spinors swap, and neither of them are $P$ eigenstates. Recalling 
    \begin{equation}
        \gamma^\mu = \left(\begin{array}{cc}
0 & \sigma^\mu \\ \overline{\sigma}^\mu & 0
        \end{array}\right)\,,
    \end{equation}
    there is
    \begin{equation}
        \mathcal{B}_1 = (\overline{N} \frac{1-\gamma^5}{2}\gamma^\mu N)u_\mu\Sigma_+\,,\quad \mathcal{B}_1 = (\overline{N} \frac{1+\gamma^5}{2}\gamma^\mu N)u_\mu\Sigma_+\,,
    \end{equation}
    thus they can be combined to give 
    \begin{equation}
        \mathcal{B}_1 + \mathcal{B}_2 =  (\overline{N}\gamma^\mu N)u_\mu\Sigma_+ \,,\quad -\mathcal{B}_1 + \mathcal{B}_2 =  (\overline{N}\gamma^5\gamma^\mu N)u_\mu\Sigma_+ \,,
    \end{equation}
    whose $P$ eigenvalues are $-1$ and $+1$ respectively since the intrinsic party charge of $u_\mu$ is $-1$.
\end{itemize}

\subsubsection*{Internal Flavor Structure}
The flavor group here refers to the special unitary group $SU(N)$, and the cases of $SU(2)$ and $SU(3)$ are needed in the ChPT.

\begin{table}
\ytableausetup
{boxsize=1em}
\ytableausetup
{aligntableaux=center}
    \centering
    \begin{tabular}{|c|c|c|c|c|c|c|c|c|c|}
    \hline
\multirow{2}{*}{$SU(2)$} & $\mathbf{1}$ & $\mathbf{2}$ & $\overline{\mathbf{2}}$ & $\mathbf{3}$ & \multirow{2}{*}{$SU(3)$} & $\mathbf{1}$ & $\mathbf{3}$ & $\overline{\mathbf{3}}$ & $\mathbf{8}$ \\
\cline{2-5} \cline{7-10}
                         & \ydiagram{1,1} & \ydiagram{1} & \ydiagram{1} & \ydiagram{2} & & \ydiagram{1,1,1} & \ydiagram{1} & \ydiagram{1,1} & \ydiagram{2,1} \\
    \hline
    \end{tabular}
    \caption{The corresponding Young diagrams of the first several irreducible representations of the $SU(2)$ and $SU(3)$ group.}
    \label{tab: young diagrams}
\end{table}

According to the group theory, every irreducible representation of the $SU(N)$ group corresponds to a standard Young diagram. The first several irreducible representations of $SU(2)$ and $SU(3)$ are presented in Tab.~\ref{tab: young diagrams}. The effective operators are gauge invariants, which means they are of the trivial representation $\mathbf{1}$, which corresponds to the Young diagram of the form
\begin{equation}
    SU(2)\sim \yng(2,2)\dots\yng(1,1)\,,\quad SU(3)\sim \yng(2,2,2)\dots\yng(1,1,1)\,.
\end{equation}
Thus the gauge invariants are the different Young diagrams of the shape above during the outer product of the Young diagrams in Tab.~\ref{tab: young diagrams}, which respects the Littlewood-Richardson rule. For example of 3 adjoint representations of $SU(3)$, the outer product can be performed as follows,

\begin{align}
    & \begin{ytableau}
i_1 & j_1 \\ k_1
    \end{ytableau} 
    \times \begin{ytableau}
*(green) i_2 & *(green) j_2 \\ *(green) k_2
    \end{ytableau}
    \times \begin{ytableau}
*(yellow) i_3 & *(yellow) j_3 \\ *(yellow) k_3
    \end{ytableau} = \left(\begin{ytableau}
i_1 & j_1 & *(green) i_2 \\
k_1 & *(green) j_2 \\
*(green) k_2 
    \end{ytableau} + 
\begin{ytableau}
i_1 & j_1 & *(green) i_2 \\
k_1 & *(green) k_2 \\
*(green) j_2 
    \end{ytableau}
    \right) \times 
    \begin{ytableau}
*(yellow) i_3 & *(yellow) j_3 \\ *(yellow) k_3
    \end{ytableau} \notag \\
    & = \begin{ytableau}
i_1 & j_1 & *(green) i_2 \\
k_1 & *(green) j_2 & *(yellow) i_3\\
*(green) k_2 & *(yellow) j_3 & *(yellow) k_3
    \end{ytableau} + 
\begin{ytableau}
i_1 & j_1 & *(green) i_2 \\
k_1 & *(green) k_2 & *(yellow) i_3\\
*(green) j_2 & *(yellow) j_3 & *(yellow) k_3
    \end{ytableau}\,,
\end{align} 
which implies there are 2 independent $SU(3)$ invariants composed of 3 adjoint representations called y-basis.
Their explicit form can be obtained by transferring every column in the trivial Young tableaux to an asymmetric tensor $\epsilon^{ijk}$, then multiplying them with the fields. Every field $\Phi$ of the $\mathbf{8}$ representation of $SU(3)$ can be expressed as
\begin{equation}
    \Phi_{ijk} = \Phi^a {\lambda^a}^l_i\epsilon_{ljk} \in 
    \begin{ytableau}
        j & i\\ k
    \end{ytableau}
    \,,
\end{equation}
then the two independent invariants take the form
\begin{equation}
    \left(\begin{ytableau}
i_1 & j_1 & *(green) i_2 \\
k_1 & *(green) j_2 & *(yellow) i_3\\
*(green) k_2 & *(yellow) j_3 & *(yellow) k_3
    \end{ytableau} \,,
    \begin{ytableau}
i_1 & j_1 & *(green) i_2 \\
k_1 & *(green) k_2 & *(yellow) i_3\\
*(green) j_2 & *(yellow) j_3 & *(yellow) k_3
    \end{ytableau}
    \right) 
     \rightarrow \left(d^{abc}\,,f^{abc}\right)\,.
\end{equation}
This result can be applied to the ChPT $ d_\chi =2$ operators of the $SU(3)$ case. 
Similar arguments for 4 $SU(3)$ adjoint representations gives 8 independent tensors, they are
\begin{align}
  & \calt_1 = d^{abe}d^{cde} \,,\quad \calt_2 = d^{abe}f^{cde} \,,\quad \calt_3 = f^{abe}f^{cde} \,,\quad \calt_3 = \delta^{ab}\delta^{cd}\,,\notag \\
  & \calt_5 = f^{abe}d^{cde} \,,\quad \calt_6 = \delta^{ac}\delta^{bd} \,,\quad \calt_7 = d^{ace}d^{bde} \,,\quad \calt_8 = d^{ace}f^{bde} \,, \label{eq: su3 tensor}
\end{align} 

While for the $SU(2)$ case, the situation is different since the nucleons are of $\mathbf{2}$ representation,
\begin{equation}
    N_i \sim \begin{ytableau}
        i 
    \end{ytableau}\,,
\end{equation}
thus the tensor product of two nucleons
\begin{align}
    \mathbf{2} \times \mathbf{2} &= \mathbf{3} + \mathbf{1} \notag \\
    \ydiagram{1} \times \ydiagram{1} &= \ydiagram{2} + \ydiagram{1,1} \,,
\end{align}
implies they can form an adjoint representation and a trivial representation, the latter of which means the nucleons contract with each other. Considering the case of 2 nucleons and other 2 bosonic fields, the y-basis are 
\begin{align}
    \begin{ytableau}
        i 
    \end{ytableau} \times 
    \begin{ytableau}
        j
    \end{ytableau} \times 
    \begin{ytableau}
        *(green) k & *(green) l 
    \end{ytableau} \times 
    \begin{ytableau}
        *(yellow) m & *(yellow) n
    \end{ytableau} &= \left(\begin{ytableau}
        i & j
    \end{ytableau} + \begin{ytableau}
        i \\ j
    \end{ytableau}\right) \times
    \begin{ytableau}
        *(green) k & *(green) l 
    \end{ytableau} \times 
    \begin{ytableau}
        *(yellow) m & *(yellow) n
    \end{ytableau} \notag \\
    &= \begin{ytableau}
        i & j & *(green) k \\
        *(green) l & *(yellow) m & *(yellow) n
    \end{ytableau} + 
    \begin{ytableau}
        i & *(green) k & *(green) l \\
        j & *(yellow) m & *(yellow) n
    \end{ytableau}\,.
\end{align}
Let the fields of $\mathbf{3}$ representation of the $SU(2)$ be $\Phi_{ij} = \Phi^I {\sigma^I}^l_i\epsilon_{lj}$, the two independent tensors are
\begin{align}
\mathcal{T}_1 = 
   \begin{ytableau}
        i & j & *(green) k \\
        *(green) l & *(yellow) m & *(yellow) n
    \end{ytableau} &= \epsilon^{il}\epsilon^{jm}\epsilon^{kn} {\sigma^I}^p_k\epsilon_{pl} {\sigma^J}^q_m \epsilon_{qn} = {\sigma^I}^i_k {\sigma^J}^k_j \propto \epsilon^{IJK}{\tau^K}^i_j \,, \notag \\
\mathcal{T}_2 = 
    \begin{ytableau}
        i & *(green) k & *(green) l \\
        j & *(yellow) m & *(yellow) n
    \end{ytableau} &= \epsilon^{ij}\epsilon^{km}\epsilon^{ln} {\sigma^I}^p_k\epsilon_{pl} {\sigma^J}^q_m \epsilon_{qn} = \epsilon^{ij} {\tau^I}^p_k\epsilon_{pq}{\tau^J}^q_m \epsilon^{km} \propto \epsilon^{ij} \delta^{IJ}\,.\label{eq: su2 tensors}
\end{align}

The effective operators are constructed in this paper up to $d_\chi=4$. According to the power-counting scheme discussed before there are most 4 bosonic degrees of freedom apart from the two baryons. In Tab.~\ref{tab: invariants} we present the numbers of the independent structures needed in the ChPT of $SU(2)$ and $SU(3)$ case respectively.

\begin{table}[]
    \centering
    \begin{tabular}{|c|c|c|c|c|c|}
\hline
$\mathcal{X}$ & $1$ & $\mathcal{A}$ & $\mathcal{A}\mathcal{B}$ & $\mathcal{A}\mathcal{B}\mathcal{C}$ & $\mathcal{A}\mathcal{B}\mathcal{C}\mathcal{D}$ \\
\hline
$SU(2):\quad \overline{N}N\mathcal{X} $ & 1 & 1 & 2 & 4 & 9 \\
\hline
$SU(3):\quad \overline{B}B\mathcal{X} $ & 1 & 2 & 8 & 32 & 145 \\
\hline
    \end{tabular}
    \caption{The number of independent chiral structures generated by the Young tensor method, where $\mathcal{A},\mathcal{B},\mathcal{C},\mathcal{D}$ are general bosonic degrees of freedom.}
    \label{tab: invariants}
\end{table}

If the effective operators contain repeated fields, there are extra constraints on the result in Tab.~\ref{tab: invariants}, which is realized in the Young tensor method by the projectors. Suppose an operator has $N$ repeated fields, which form a representation of the symmetric group $S_N$, the irreducible representations can be extracted via the projectors associated with different Young tableaux. For a Young tableau of $S_N$ group, its projector is defined as
\begin{equation}
    \mathcal{Y} = \frac{1}{N!} h \times v\,,
\end{equation}
where $h\,, v$ are the horizontal and vertical stabilizers. And more details can be found in the Ref.~\cite{Li:2020xlh,Li:2022tec,Fonseca:2019yya}. Considering the $SU(2)$ structures in Eq.~\eqref{eq: su2 tensors} as an example, if the two fields of the adjoint representation are repeated, their interchange $(IJ)\in S_2$ takes the matrix form that
\begin{equation}
    (IJ) = \left(\begin{array}{cc}
        -1 & 0 \\
         0 & 1
    \end{array}\right)\,,
\end{equation}
since $\epsilon^{IJK}$ is asymmetric and $\delta^{IJ}$ is symmetric, thus the symmetric representation and the asymmetric representation of $S_2$ can be obtained by the projectors
\begin{align}
    \mathcal{Y}(\young(IJ)) &= \frac{1}{2}(\mathbb{I}_{2\times 2} + (IJ) ) = 
    \left(\begin{array}{cc}
         0 & 0 \\
         0 & 1
    \end{array}\right)\,, \\
    \mathcal{Y}(\young(I,J)) &= \frac{1}{2}(\mathbb{I}_{2\times 2} - (IJ) ) = 
    \left(\begin{array}{cc}
         1 & 0 \\
         0 & 0
    \end{array}\right)\,,
\end{align}
as expected that $\calt_1$ is asymmetric and $\calt_2$ is symmetric. Actually, the application of the projectors can be regarded as the constraints on the original independent tensors in Eq.~\eqref{eq: su2 tensors}. For example, the symmetric projector on a tensor transforms them into the symmetric one that
\begin{equation}
    \mathcal{Y}(\young(IJ))\cdot\calt_i = \mathcal{Y}(\young(IJ))\cdot \mathcal{Y}(\young(IJ))\dot \calt_i = \mathcal{Y}(\young(IJ))\cdot \sum_{i=1}^2 \left[\mathcal{Y}(\young(IJ))\right]_{ij}\calt_j = \sum_{i=1}^2 \left[\mathcal{Y}(\young(IJ))\right]_{ij}\widetilde{\calt}_j\,,\quad i=1,2\,,
\end{equation}
where $\widetilde{\calt}_j = \mathcal{Y}(\young(IJ)) \calt_j$ is the symmetrized tensor. It gives a system of linear equations that 
\begin{equation}
    \left\{\begin{array}{l}
\widetilde{\calt}_1 = 0 \\
\widetilde{\calt}_2 = \widetilde{\calt}_2 
    \end{array}\right.\,,
\end{equation}
which is degenerate and there is only one independent equation $\widetilde{\calt}_1 = 0$. Thus after symmetrization, $\calt_1$ vanishes while $\calt_2$ survives.

For the $SU(3)$ tensors in Eq.~\eqref{eq: su3 tensor}, if the last two fields are repeated, the matrix representation of the interchange $(cd)\in S_2$ is hard to work out for the tensors in Eq.~\eqref{eq: su3 tensor} and should be obtained via the y-basis then converted to the tensor form for via a similar transformation, after which we can get the symmetric projector
\begin{equation}
    \mathcal{Y}(\young(cd)) = 
    \left(
\begin{array}{cccccccc}
 1 & 0 & 0 & 0 & 0 & 0 & 0 & 0 \\
 0 & 0 & 0 & 0 & 0 & 0 & 0 & 0 \\
 0 & 0 & 0 & 0 & 0 & 0 & 0 & 0 \\
 0 & 0 & 0 & 1 & 0 & 0 & 0 & 0 \\
 0 & 0 & 0 & 0 & 1 & 0 & 0 & 0 \\
 \frac{1}{2} & 0 & -\frac{1}{2} & -\frac{1}{6} & 0 & \frac{2}{3} & 1 & 0 \\
 -\frac{1}{3} & 0 & -\frac{1}{6} & \frac{1}{9} & 0 & \frac{2}{9} & \frac{1}{3} & 0 \\
 0 & 0 & 0 & 0 & \frac{1}{2} & 0 & 0 & 0 \\
\end{array}
\right)\,.
\end{equation}
it generates 4 independent equations on the tensors in Eq.~\eqref{eq: su3 tensor} that
\begin{align}
    \widetilde{\calt_2} &= 0\,, \notag \\ 
    \widetilde{\calt_3} &= 0\,, \notag \\
    \frac{1}{2}\widetilde{\calt_1} - \frac{1}{6}\widetilde{\calt_4}-\frac{1}{3}\widetilde{\calt_6}+\widetilde{\calt_7} &= 0 \,,\notag \\
    -\widetilde{\calt_8} +\frac{1}{2}\widetilde{\calt_5} &= 0 \,. \label{eq: symmetry constraint}
\end{align}

Up to 4 bosonic degrees of freedom, the numbers of symmetrized invariants can be generated via the Young tensor method and are summarized in Tab.~\ref{tab: symmetrized invariants}.  

\begin{table}[]
    \centering
    \begin{tabular}{|c|c|c|c|c|c|c|}
\hline
$\mathcal{X}$ & $\mathcal{A}^2$ & $\mathcal{A}^2\mathcal{B}$ & $\mathcal{A}^3$ & $\mathcal{A}^2\mathcal{B}\mathcal{C}$ & $\mathcal{A}^3\mathcal{B}$ & $\mathcal{A}^4$ \\
\hline
$SU(2):\quad \overline{N}N\mathcal{X} $ & 1 & 2 & 1 & 5 & 2 & 1 \\
\hline
$SU(3):\quad \overline{B}B\mathcal{X} $ & 4 & 16 & 4 & 73 & 25 & 7 \\
\hline
    \end{tabular}
    \caption{The number of symmetrized chiral structures generated by the Young tensor method, where $\mathcal{A},\mathcal{B},\mathcal{C}$ are general bosonic degrees of freedom.}
    \label{tab: symmetrized invariants}
\end{table}
\newpage
\subsubsection*{Adler zero condition and off-shell structure for external source}
\label{ex}
The amplitudes containing Goldstone bosons are constrained further by the Adler zero condition~\cite{Adler:1969gk}, which implies that any amplitudes tend to zero when a Goldstone momentum $p$ becomes soft,
\begin{equation}
    \lim_{p\rightarrow 0}\mathcal{M}(p) = 0\,.
\end{equation}
In the Young tensor method, the Adler zero condition is expressed by a system of linear equations~\cite{Sun:2022snw,Sun:2022ssa,Low:2022iim}.

For example, we consider an operator type with $N$ fields, of which one is Goldtone bosons. The Lorentz basis is of dimension-$d_N$, $\{\mathcal{B}_i|i=1,2,\dots,d_N\}$, and a general operator should be expressed in the form
\begin{equation}
    \mathcal{M}(p) = \sum_{i=1}^{d_N} c_i \mathcal{B}_i(p)\,,
\end{equation}
where $p$ is the Goldstone boson's momentum. The Adler zero condition means this general amplitude becomes zero after the limit that
\begin{equation}
\label{eq: adz1}
    \lim_{p\rightarrow 0} \mathcal{M}(p) = \sum_{i=1}^{d_N}c_i \left(\lim_{p\rightarrow 0}\mathcal{B}_i\right) = \sum_{i=1}^{d_N}c_i \tilde{\mathcal{B}}_i = 0\,,
\end{equation}
where $\tilde{\mathcal{B}}_i = \lim_{p\rightarrow 0}\mathcal{B}_i$ are the basic amplitudes after the same limit, which are generally no more independent and can be expanded on the original basis $\mathcal{B}_i$,
\begin{equation}
\label{eq: adz2}
    \tilde{\mathcal{B}}_i = \sum_{j=1}^{d_N} K_{ij}\mathcal{B}_j\,.
\end{equation}
with the coefficients $K_{ij}$. Substitute Eq.~\eqref{eq: adz2} in the Eq.~\eqref{eq: adz1} and interchange the summation order, the Adler zero condition becomes
\begin{equation}
    \sum_{j=1}^{d_N} \left(\sum_{i=1}^{d_N} c_i K_{ij}\right)\mathcal{B}_j = 0\,,
\end{equation}
which means all the coefficients vanish,
\begin{equation}
    \sum_{i=1}^{d_N}c_i K_{ij} = 0\,, \quad j=1,2,\dots\,, d_N\,.
\end{equation}
Thus we obtain a system of linear equations about the expansion coefficients $c_i$, whose solutions contain all the amplitude satisfying the Adler zero condition, and the independent ones compose the wanted Lorentz basis.

As for the external sources $\Sigma_\pm$ and $f_\pm$, 
since they are not dynamic degrees of freedom of the ChPT, the operators containing their EOM terms can not be eliminated via field redefinition. In the Young tensor method, such a condition implies a relaxation of the amplitude's on-shell condition,
\begin{align}
    D^2 \Sigma_\pm{}_i \neq 0 & \rightarrow \langle  ii\rangle [ii] \neq 0\,, \label{eq: offshell1}\\
    D_\mu f_{\pm L}^{\mu\nu}{}_{i} \neq 0 & \rightarrow \langle ii\rangle |i] \neq 0 \,,  \label{eq: offshell2} \\
    D_\mu f_{\pm R}^{\mu\nu}{}_{i} \neq 0 & \rightarrow [ ii] |i\rangle \neq 0 \label{eq: offshell3}\,.
\end{align}
However, the above conditions of $f_\pm{}_L$ and $f_\pm{}_R$ are not independent, since of the Bianchi identity,
\begin{equation} \label{eq: Bian_FL_FR}
    D_\mu \tilde{f}_\pm^{\mu\nu} = 0\,,
\end{equation}
or
\begin{equation}
    D_\mu f_{\pm L}^{\mu\nu} - D_\mu f_{\pm R}^{\mu\nu} = 0\,.
\end{equation}

Thus to amount for the EOMs of the external sources, the conditions from Eq.~\eqref{eq: offshell1} and Eq.~\eqref{eq: offshell3} should be applied to the amplitudes\footnote{The condition in Eq.~\eqref{eq: offshell2} is not included since of the Bianchi identity.}. In Ref.~\cite{Ren:2022tvi} the authors have extended the Young tensor method to reach this goal, the main idea is to distinguish the momentum and field indices in the amplitudes and construct all the spinor contractions, which is redundant and can be reduced to minimal set by the relations such as IBP, the Fierz identities and so on. In this paper, we adopt the same management and leave the details in Ref.~\cite{Ren:2022tvi}.

\subsection{Trace Basis and Cayley-Hamilton Theorem}

The independent invariant tensor for the flavor structure has been obtained through the Young tensor method, it is convenient to translate the invariant tensor basis to the trace basis that the $CP$ properties will be more obvious. When constructing the group invariants composed of the adjoint representations solely, which is the $SU(3)$ case in the ChPT, all the invariants are the matrix traces and are constrained by 
Cayley-Hamilton theorem. Thus, we use the Cayley-Hamilton theorem to reduce redundancies in the trace basis and convert the invariant tensor basis to the trace basis.

Actually, the Cayley-Hamilton theorem is a general theorem shared by any squared matrice, whose states that every $n\times n$ matrix $\mathcal{A}$ must satisfy its characteristic equation,
\begin{equation}
  \left.p(\lambda)\right|_{\lambda=\mathcal{A}} = \left.det(\lambda \mathbb{I}_{n\times n}-\mathcal{A})\right|_{\lambda=\mathcal{A}} = 0\,,
\end{equation}
where $\mathbb{I}_{n\times n}$ is $n\times n$ identity matrix. For any $2\times 2$ matrix $\mathcal{A}$, the Cayley-Hamilton theorem gives 
\begin{equation}
\label{eq: ch-2}
  \mathcal{A}^2 = \lra{\mathcal{A}} \mathcal{A} + \frac{1}{2}(\lra{\mathcal{A}^2}-\lra{\mathcal{A}}^2)\mathbb{I}_{2\times 2}\,,
\end{equation}
while for the $3\times 3$ case, the Cayley-Hamilton theorem states that
\begin{equation}
  \label{eq: ch-3}
  \mathcal{A}^3 = \mathcal{A}^2\lra{\mathcal{A}}-\frac{1}{2}\mathcal{A}(\lra{\mathcal{A}}^2-\lra{\mathcal{A}^2})+\frac{1}{6}(\lra{\mathcal{A}}^3-3\lra{\mathcal{A}^2}\lra{\mathcal{A}}+2\lra{\mathcal{A}^3})\mathbb{I}_{3\times 3}\,.
\end{equation}
In this section we focus on the $SU(3)$ invariants, which means the involving $3\times 3$ matrices (fields) are traceless since they are of the $SU(3)$ adjoint representation, the equation in Eq.~\eqref{eq: ch-3} becomes
\begin{equation}
  \mathcal{A}^3 = \frac{1}{2}\mathcal{A}\lra{\mathcal{A}^2} + \frac{1}{3}\lra{\mathcal{A}^3}\,.
\end{equation}
The Cayley-Hamilton theorem is also helpful in the ChPT of the $SU(2)$ case, which will be commented on at the end of this section.
Let $\mathcal{A}=\mathcal{A}+\mathcal{B}+\mathcal{C}$ and extract different terms proportional to $\mathcal{A}^{n_1}\mathcal{B}^{n_2}\mathcal{\mathcal{C}}^{n_3}\,,(n_1+n_2+n_3=3)$, Eq.~\eqref{eq: ch-3} generates various relations, which we call Cayley-Hamilton relation (CH relations), for examples, 
the CH relations proportional to $\mathcal{A}\mathcal{B}\mathcal{C}\,,\mathcal{A}\mathcal{B}^2$ and $\mathcal{A}^3$ are 
\begin{align}
  \mathcal{A}\mathcal{B}\mathcal{C} + \mathcal{A}\mathcal{C}\mathcal{B} + \mathcal{B}\mathcal{A}\mathcal{C} + \mathcal{C}\mathcal{A}\mathcal{B} + \mathcal{B}\mathcal{C}\mathcal{A} + \mathcal{C}\mathcal{B}\mathcal{A} &= \mathcal{A}\lra{\mathcal{B}\mathcal{C}} + \mathcal{B}\lra{\mathcal{A}\mathcal{C}} + \mathcal{C}\lra{\mathcal{A}\mathcal{B}} + \lra{\mathcal{A}\mathcal{B}\mathcal{C}} + \lra{\mathcal{A}\mathcal{C}\mathcal{B}}\,, \label{eq: CH relation 1}\\
  \mathcal{A}\mathcal{B}\mathcal{B} + \mathcal{B}\mathcal{A}\mathcal{B} + \mathcal{B}\mathcal{B}\mathcal{A} &= \frac{1}{2} \mathcal{A} \lra{\mathcal{B}\mathcal{B}} + \mathcal{B}\lra{\mathcal{A}\mathcal{B}} + \lra{\mathcal{A}\mathcal{B}\mathcal{B}} \,, \\
  \mathcal{A}^3 &= \frac{1}{2}\mathcal{A}\lra{\mathcal{A}^2} + \frac{1}{3}\lra{\mathcal{A}^3} \,,
\end{align}
respectively, where the cyclic property of the matrix trace is used, and the last relation proportional to $\mathcal{A}^3$ is coincident with the original equation in Eq.~\eqref{eq: ch-3}, as expected.
Such CH relations generate constraints when there are more than 3 fields in the invariants and reduce all the traces to a minimal set, called trace basis.

Considering the simplest case of 4 distinct fields $\mathcal{A},\mathcal{B},\mathcal{C},\mathcal{D}$, there are totally 9 traces composed by these fields,
\begin{align}
  & \lra{\mathcal{A}\mathcal{D}}\lra{\mathcal{B}\mathcal{C}}\,, \lra{\mathcal{A}\mathcal{C}}\lra{\mathcal{B}\mathcal{D}}\,, \lra{\mathcal{A}\mathcal{B}}\lra{\mathcal{C}\mathcal{D}}\,, \notag \\
  & \lra{\mathcal{A}\mathcal{B}\mathcal{C}\mathcal{D}}\,, \lra{\mathcal{A}\mathcal{C}\mathcal{B}\mathcal{D}}\,, \lra{\mathcal{A}\mathcal{B}\mathcal{D}\mathcal{C}}\,, \lra{\mathcal{A}\mathcal{C}\mathcal{D}\mathcal{B}}\,, \lra{\mathcal{A}\mathcal{D}\mathcal{B}\mathcal{C}}\,, \lra{\mathcal{A}\mathcal{D}\mathcal{C}\mathcal{B}}\,,
\end{align}
which are related by the CH relation in Eq.~\eqref{eq: CH relation 1}. Taking the cyclic property of the traces into account, there is only one independent constraint,
\begin{align}
  & -\lra{\mathcal{A}\mathcal{D}}\lra{\mathcal{B}\mathcal{C}} - \lra{\mathcal{A}\mathcal{C}}\lra{\mathcal{B}\mathcal{D}} - \lra{\mathcal{A}\mathcal{B}}\lra{\mathcal{C}\mathcal{D}} \notag \\
  & + \lra{\mathcal{A}\mathcal{B}\mathcal{C}\mathcal{D}} + \lra{\mathcal{A}\mathcal{C}\mathcal{B}\mathcal{D}} + \lra{\mathcal{A}\mathcal{B}\mathcal{D}\mathcal{C}} + \lra{\mathcal{A}\mathcal{C}\mathcal{D}\mathcal{B}} + \lra{\mathcal{A}\mathcal{D}\mathcal{B}\mathcal{C}} + \lra{\mathcal{A}\mathcal{D}\mathcal{C}\mathcal{B}} \notag \\
  & = 0\,, \label{eq: CH relation of 4 fields}
\end{align}
which means there are only 8 independent traces and they can be chosen as
\begin{align}
  & T_1 = \lra{\mathcal{A}\mathcal{C}}\lra{\mathcal{B}\mathcal{D}}\,,\quad T_2 = \lra{\mathcal{A}\mathcal{B}}\lra{\mathcal{C}\mathcal{D}} \,,\notag \\
  & T_3 = \lra{\mathcal{A}\mathcal{B}\mathcal{C}\mathcal{D}} \,,\quad T_4 = \lra{\mathcal{A}\mathcal{B}\mathcal{D}\mathcal{C}} \,,\quad T_5 = \lra{\mathcal{A}\mathcal{C}\mathcal{B}\mathcal{D}} \notag \\
  & T_6 = \lra{\mathcal{A}\mathcal{C}\mathcal{D}\mathcal{B}} \,,\quad T_7 = \lra{\mathcal{A}\mathcal{D}\mathcal{B}\mathcal{C}} \,,\quad T_8 = \lra{\mathcal{A}\mathcal{D}\mathcal{C}\mathcal{B}} \,. \label{eq: trace basis}
\end{align} 
These traces can be related to the tensors in Eq.~\eqref{eq: su3 tensor} obtained by the Young tensor method, which can be realized by expressing the fields in terms of $SU(3)$ generators $\lambda^a$, 
\begin{equation}
  \mathcal{A} = \mathcal{A}_a \lambda^a\,,\quad \mathcal{B} = \mathcal{B}_b \lambda^b\,,\quad \mathcal{C}= \mathcal{C}_c\lambda^c \,,\quad \mathcal{D} = \mathcal{D}_d \lambda^d \,,
\end{equation}
Defining
\begin{equation}
  \frac{T_i}{\mathcal{A}_a \mathcal{B}_b \mathcal{C}_c \mathcal{D}_d} = \mathcal{A}_a^{-1}\mathcal{B}_b^{-1}\mathcal{C}_c^{-1}\mathcal{D}_d^{-1} T_i\,, 
\end{equation}
the trace basis takes the form
\begin{align}
  \frac{T_1}{\mathcal{A}_a\mathcal{B}_b\mathcal{C}_c\mathcal{D}_d} &= \lra{\lambda^a\lambda^c}\lra{\lambda^b\lambda^d} = 4\delta^{ac}\delta^{bd}\,, \\
  \frac{T_2}{\mathcal{A}_a\mathcal{B}_b\mathcal{C}_c\mathcal{D}_d} &= \lra{\lambda^a\lambda^b}\lra{\lambda^c\lambda^d} = 4\delta^{ab}\delta^{cd}\,, \\
  \frac{T_3}{\mathcal{A}_a\mathcal{B}_b\mathcal{C}_c\mathcal{D}_d} &= \lra{\lambda^a\lambda^b\lambda^c\lambda^d} = \frac{4}{3}\delta^{ab}\delta^{cd} + 2 d^{abe}d^{cde} - 2 f^{abe}f^{cde} + 2i d^{abe}f^{cde} + 2i f^{abe}d^{cde} \,, \\
  \frac{T_4}{\mathcal{A}_a\mathcal{B}_b\mathcal{C}_c\mathcal{D}_d} &= \lra{\lambda^a\lambda^b\lambda^d\lambda^c} = \frac{4}{3}\delta^{ab}\delta^{cd} + 2 d^{abe}d^{cde} + 2 f^{abe}f^{cde} - 2i d^{abe}f^{cde} + 2i f^{abe}d^{cde} \,, \\
  \frac{T_5}{\mathcal{A}_a\mathcal{B}_b\mathcal{C}_c\mathcal{D}_d} &= \lra{\lambda^a\lambda^c\lambda^b\lambda^d} = = \frac{4}{3}\lra{\lambda^a\lambda^c}\lra{\lambda^b\lambda^d} + 4 d^{ace}d^{bde} + 4i d^{ace}f^{bde} + \lra{\lambda^a\lambda^b\lambda^d\lambda^c}  \\
  \frac{T_6}{\mathcal{A}_a\mathcal{B}_b\mathcal{C}_c\mathcal{D}_d} &= \lra{\lambda^b\lambda^a\lambda^c\lambda^d} = \frac{4}{3}\delta^{ab}\delta^{cd} + 2 d^{abe}d^{cde} + 2 f^{abe}f^{cde} + 2i d^{abe}f^{cde} - 2i f^{abe}d^{cde} \,, \\
  \frac{T_7}{\mathcal{A}_a\mathcal{B}_b\mathcal{C}_c\mathcal{D}_d} &= \lra{\lambda^a\lambda^d\lambda^b\lambda^c} = \lra{\{\lambda^a,\lambda^c\}\lambda^d\lambda^b} - \lra{\lambda^b\lambda^a\lambda^c\lambda^d} \notag \\
  & = \frac{1}{2}\lra{\{\lambda^a,\lambda^c\}\{\lambda^b,\lambda^d\}} + \frac{1}{2}\lra{\{\lambda^a,\lambda^d\} [\lambda^d,\lambda^b]} - \lra{\lambda^a\lambda^b\lambda^c\lambda^c} \notag \\
  & = \frac{4}{3}\lra{\lambda^a\lambda^c}\lra{\lambda^b\lambda^d} + 4 d^{ace}d^{bde} - 4i d^{ace}f^{bde} + \lra{\lambda^b\lambda^a\lambda^c\lambda^d} \,, \\
  \frac{T_8}{\mathcal{A}_a\mathcal{B}_b\mathcal{C}_c\mathcal{D}_d} &= \lra{\lambda^b\lambda^a\lambda^d\lambda^c} = \frac{4}{3}\delta^{ab}\delta^{cd} + 2 d^{abe}d^{cde} - 2 f^{abe}f^{cde} - 2i d^{abe}f^{cde} - 2i f^{abe}d^{cde} \,, 
\end{align} 
which implies the trace basis and the m-basis can be converted to each other non-degenerately. 

At this stage, it can be seen that the CH relation and the trace basis are inspiring when constructing the effective operators of definite $CP$ eigenvalues. For example, the CH relation in Eq.~\eqref{eq: CH relation 1} implies that
\begin{align}
  \delta^{ac}\delta^{cd}+\delta^{ad}\delta^{bc} &= \frac{1}{{4\mathcal{A}_a\mathcal{B}_b\mathcal{C}_c\mathcal{D}_d}} \left[\lra{\mathcal{A}\mathcal{C}}\lra{\mathcal{B}\mathcal{D}} + \lra{\mathcal{A}\mathcal{D}}\lra{\mathcal{B}\mathcal{C}}\right] \notag \\
  &= \frac{1}{{4\mathcal{A}_a\mathcal{B}_b\mathcal{C}_c\mathcal{D}_d}} (T_2 + T_3 + T_4 + T_5 + T_6 + T_7 + T_8) \notag \\
  &= f(\calt_{1-5}\,,\calt_{7,8})\,,
\end{align}
which means $\delta^{ac}\delta^{cd}+\delta^{ad}\delta^{bc}$ is not independent and can be expanded by the m-bases $\calt_{1-5}\,,\calt_{7,8}$.
Such a relation is important after the identification that
\begin{equation}
  \mathcal{A} \rightarrow \overline{B}\,,\quad \mathcal{B} \rightarrow B \,,\quad \mathcal{C}\rightarrow u_\mu \,,\quad \mathcal{D} \rightarrow \Sigma_+\,,
\end{equation}
which means we are constructing the effective operators of type $u\Sigma_+N^2$, then the redundancy of the $\delta^{ac}\delta^{cd}+\delta^{ad}\delta^{bc}$ implies that
the $C$-even operator 
\begin{equation}
  \lra{\overline{B}\gamma^5\gamma^\mu u_\mu}\lra{B\Sigma_+} + \lra{\overline{B}\gamma^5\gamma^\mu \Sigma_+}\lra{B u_\mu} = \lra{\overline{B}\gamma^5\gamma^\mu u_\mu}\lra{B\Sigma_+} + h.c. \,,
\end{equation}
is not independent while the $C$-odd operator 
\begin{equation}
  \lra{\overline{B}\gamma^5\gamma^\mu u_\mu}\lra{B\Sigma_+} - \lra{\overline{B}\gamma^5\gamma^\mu \Sigma_+}\lra{B u_\mu} = \lra{\overline{B}\gamma^5\gamma^\mu u_\mu}\lra{B\Sigma_+} - h.c.\,,
\end{equation}
is indeed independent. Moreover, we can find all the combinations of the trace bases with specific $C$ eigenvalues. Let $\mathcal{A}\rightarrow \overline{B}, \mathcal{B}\rightarrow B$ and $\mathcal{C},\mathcal{D}$ unchanged represent any non-baryon building blocks, a general operator composed of them $\lra{\overline{N},N,\mathcal{C},\mathcal{D}}$
transforms under $C$ as
\begin{equation}
  \lra{\overline{N},N,\mathcal{C},\mathcal{D}} \rightarrow p_C \lra{\overline{N},N,\mathcal{D},\mathcal{C}}\,,
\end{equation}
which means that $\mathcal{C},\mathcal{D}$ interchanged and an overall factor $p_C$ arises, which is defined as $p_C = (-1)^\Gamma (-1)^\mathcal{C} (-1)^\mathcal{D}$, where $(-1)^\Gamma$ is the charge from Dirac bilinears and $(-1)^\mathcal{C}\,,(-1)^\mathcal{D}$ are the intrinsic charges of the building blocks $\mathcal{C},\mathcal{D}$.
Then all the traces in Eq.~\eqref{eq: trace basis} can be combined to form two disjoint sectors, one has $C$ eigenvalue $p_C$ and the other has $C$ eigenvalue $-p_C$,
\begin{align}
  p_C: & \notag \\
         & \lra{\mathcal{A}\mathcal{B}\mathcal{C}\mathcal{D}} + \lra{\mathcal{A}\mathcal{B}\mathcal{D}\mathcal{C}} \notag \\
         & \lra{\mathcal{A}\mathcal{C}\mathcal{D}\mathcal{B}} + \lra{\mathcal{A}\mathcal{D}\mathcal{C}\mathcal{B}} \notag \\
         & \lra{\mathcal{A}\mathcal{C}\mathcal{B}\mathcal{D}} \notag \\
         & \lra{\mathcal{A}\mathcal{D}\mathcal{B}\mathcal{C}} \notag \\
         & \lra{\mathcal{A}\mathcal{B}}\lra{\mathcal{C}\mathcal{D}} \notag \\
  -p_C: & \notag \\
         & \lra{\mathcal{A}\mathcal{B}\mathcal{C}\mathcal{D}} - \lra{\mathcal{A}\mathcal{B}\mathcal{D}\mathcal{C}} \notag \\
         & \lra{\mathcal{A}\mathcal{C}\mathcal{D}\mathcal{B}} - \lra{\mathcal{A}\mathcal{D}\mathcal{C}\mathcal{B}} \notag \\
         & \lra{\mathcal{A}\mathcal{C}}\lra{\mathcal{B}\mathcal{D}} - \lra{\mathcal{A}\mathcal{D}}\lra{\mathcal{B}\mathcal{C}} \,,
\end{align}
It should be stressed that there are 5 traces in the $p_C$ sector and 3 traces in the $-p_C$ sector, and these two numbers are independent of the choice of explicit basis in Eq.~\eqref{eq: trace basis}.

The situation becomes complicated for five fields, $\mathcal{A}\,,\mathcal{B}\,,\mathcal{C}\,,\mathcal{D}\,,E$ since there is more than one CH relation. Actually finding and selecting all the independent CH relation is a challenge, but the Young tensor method can give helpful hints. 
Five fields can form two kinds of traces, one is the single trace of length 5 and the other is the product of a length-2 trace and a length-3 trace, thus the total number of complete but non-independent traces is 
\begin{equation}
  C_5^3 \times 2! + 4! = 44\,,
\end{equation} 
As we already know, the Young tensor method, which can generate all the independent tensors directly via the outer product of the Young diagrams, reveals that there are only 32 independent structures, thus 12 independent CH relations are needed.
Applying Eq.~\eqref{eq: CH relation 1} to the length-5 traces, it is not hard to find the 12 CH relations are 
\begin{align}
  1) \quad & \langle \mathcal{A}\mathcal{B}\mathcal{E}\mathcal{C}\mathcal{D}\rangle +\langle \mathcal{A}\mathcal{B}\mathcal{E}\mathcal{D}\mathcal{C}\rangle +\langle \mathcal{A}\mathcal{C}\mathcal{B}\mathcal{E}\mathcal{D}\rangle +\langle \mathcal{A}\mathcal{C}\mathcal{D}\mathcal{B}\mathcal{E}\rangle +\langle \mathcal{A}\mathcal{D}\mathcal{B}\mathcal{E}\mathcal{C}\rangle +\langle \mathcal{A}\mathcal{D}\mathcal{C}\mathcal{B}\mathcal{E}\rangle  \notag \\
         = & \uline{\langle \mathcal{A}\mathcal{B}\mathcal{E}\rangle\langle \mathcal{C}\mathcal{D}\rangle} +\langle \mathcal{A}\mathcal{C}\mathcal{D}\rangle\langle \mathcal{B}\mathcal{E}\rangle +\uline{\langle \mathcal{A}\mathcal{C}\rangle\langle \mathcal{B}\mathcal{E}\mathcal{D}\rangle} +\uline{\langle \mathcal{A}\mathcal{D}\mathcal{C}\rangle\langle \mathcal{B}\mathcal{E}\rangle} +\uline{\langle \mathcal{A}\mathcal{D}\rangle\langle \mathcal{B}\mathcal{E}\mathcal{C}\rangle} \\
  2) \quad & \langle \mathcal{A}\mathcal{C}\mathcal{D}\mathcal{E}\mathcal{B}\rangle +\langle \mathcal{A}\mathcal{C}\mathcal{E}\mathcal{B}\mathcal{D}\rangle +\langle \mathcal{A}\mathcal{D}\mathcal{C}\mathcal{E}\mathcal{B}\rangle +\langle \mathcal{A}\mathcal{D}\mathcal{E}\mathcal{B}\mathcal{C}\rangle +\langle \mathcal{A}\mathcal{E}\mathcal{B}\mathcal{C}\mathcal{D}\rangle +\langle \mathcal{A}\mathcal{E}\mathcal{B}\mathcal{D}\mathcal{C}\rangle  \notag \\
         = & \langle \mathcal{A}\mathcal{C}\mathcal{D}\rangle\langle \mathcal{B}\mathcal{E}\rangle +\uline{\langle \mathcal{A}\mathcal{C}\rangle\langle \mathcal{B}\mathcal{D}\mathcal{E}\rangle} +\uline{\langle \mathcal{A}\mathcal{D}\mathcal{C}\rangle\langle \mathcal{B}\mathcal{E}\rangle} +\uline{\langle \mathcal{A}\mathcal{D}\rangle\langle \mathcal{B}\mathcal{C}\mathcal{E}\rangle} +\langle \mathcal{A}\mathcal{E}\mathcal{B}\rangle\langle \mathcal{C}\mathcal{D}\rangle \\
  3) \quad & \langle \mathcal{A}\mathcal{B}\mathcal{C}\mathcal{D}\mathcal{E}\rangle +\langle \mathcal{A}\mathcal{B}\mathcal{D}\mathcal{E}\mathcal{C}\rangle +\langle \mathcal{A}\mathcal{C}\mathcal{B}\mathcal{D}\mathcal{E}\rangle +\langle \mathcal{A}\mathcal{C}\mathcal{D}\mathcal{E}\mathcal{B}\rangle +\langle \mathcal{A}\mathcal{D}\mathcal{E}\mathcal{B}\mathcal{C}\rangle +\langle \mathcal{A}\mathcal{D}\mathcal{E}\mathcal{C}\mathcal{B}\rangle  \notag \\
         = & \langle \mathcal{A}\mathcal{B}\mathcal{C}\rangle\langle \mathcal{D}\mathcal{E}\rangle +\langle \mathcal{A}\mathcal{B}\rangle\langle \mathcal{C}\mathcal{D}\mathcal{E}\rangle +\langle \mathcal{A}\mathcal{C}\mathcal{B}\rangle\langle \mathcal{D}\mathcal{E}\rangle +\uline{\langle \mathcal{A}\mathcal{C}\rangle\langle \mathcal{B}\mathcal{D}\mathcal{E}\rangle} +\uline{\langle \mathcal{A}\mathcal{D}\mathcal{E}\rangle\langle \mathcal{B}\mathcal{C}\rangle} \\
  4) \quad & \langle \mathcal{A}\mathcal{B}\mathcal{C}\mathcal{E}\mathcal{D}\rangle +\langle \mathcal{A}\mathcal{B}\mathcal{E}\mathcal{D}\mathcal{C}\rangle +\langle \mathcal{A}\mathcal{C}\mathcal{B}\mathcal{E}\mathcal{D}\rangle +\langle \mathcal{A}\mathcal{C}\mathcal{E}\mathcal{D}\mathcal{B}\rangle +\langle \mathcal{A}\mathcal{E}\mathcal{D}\mathcal{B}\mathcal{C}\rangle +\langle \mathcal{A}\mathcal{E}\mathcal{D}\mathcal{C}\mathcal{B}\rangle  \notag \\
         = & \langle \mathcal{A}\mathcal{B}\mathcal{C}\rangle\langle \mathcal{D}\mathcal{E}\rangle +\langle \mathcal{A}\mathcal{B}\rangle\langle \mathcal{C}\mathcal{E}\mathcal{D}\rangle +\langle \mathcal{A}\mathcal{C}\mathcal{B}\rangle\langle \mathcal{D}\mathcal{E}\rangle +\uline{\langle \mathcal{A}\mathcal{C}\rangle\langle \mathcal{B}\mathcal{E}\mathcal{D}\rangle} +\uline{\langle \mathcal{A}\mathcal{E}\mathcal{D}\rangle\langle \mathcal{B}\mathcal{C}\rangle} \\
  5) \quad & \langle \mathcal{A}\mathcal{B}\mathcal{C}\mathcal{E}\mathcal{D}\rangle +\langle \mathcal{A}\mathcal{B}\mathcal{D}\mathcal{C}\mathcal{E}\rangle +\langle \mathcal{A}\mathcal{C}\mathcal{E}\mathcal{B}\mathcal{D}\rangle +\langle \mathcal{A}\mathcal{C}\mathcal{E}\mathcal{D}\mathcal{B}\rangle +\langle \mathcal{A}\mathcal{D}\mathcal{B}\mathcal{C}\mathcal{E}\rangle +\langle \mathcal{A}\mathcal{D}\mathcal{C}\mathcal{E}\mathcal{B}\rangle  \notag \\
         = & \langle \mathcal{A}\mathcal{B}\mathcal{D}\rangle\langle \mathcal{C}\mathcal{E}\rangle +\langle \mathcal{A}\mathcal{B}\rangle\langle \mathcal{C}\mathcal{E}\mathcal{D}\rangle +\uline{\langle \mathcal{A}\mathcal{C}\mathcal{E}\rangle\langle \mathcal{B}\mathcal{D}\rangle} +\langle \mathcal{A}\mathcal{D}\mathcal{B}\rangle\langle \mathcal{C}\mathcal{E}\rangle +\uline{\langle \mathcal{A}\mathcal{D}\rangle\langle \mathcal{B}\mathcal{C}\mathcal{E}\rangle} \\
  6) \quad & \langle \mathcal{A}\mathcal{B}\mathcal{D}\mathcal{E}\mathcal{C}\rangle +\langle \mathcal{A}\mathcal{B}\mathcal{E}\mathcal{C}\mathcal{D}\rangle +\langle \mathcal{A}\mathcal{D}\mathcal{B}\mathcal{E}\mathcal{C}\rangle +\langle \mathcal{A}\mathcal{D}\mathcal{E}\mathcal{C}\mathcal{B}\rangle +\langle \mathcal{A}\mathcal{E}\mathcal{C}\mathcal{B}\mathcal{D}\rangle +\langle \mathcal{A}\mathcal{E}\mathcal{C}\mathcal{D}\mathcal{B}\rangle  \notag \\
         = & \langle \mathcal{A}\mathcal{B}\mathcal{D}\rangle\langle \mathcal{C}\mathcal{E}\rangle +\langle \mathcal{A}\mathcal{B}\rangle\langle \mathcal{C}\mathcal{D}\mathcal{E}\rangle +\langle \mathcal{A}\mathcal{D}\mathcal{B}\rangle\langle \mathcal{C}\mathcal{E}\rangle +\uline{\langle \mathcal{A}\mathcal{D}\rangle\langle \mathcal{B}\mathcal{E}\mathcal{C}\rangle} +\uline{\langle \mathcal{A}\mathcal{E}\mathcal{C}\rangle\langle \mathcal{B}\mathcal{D}\rangle}
\end{align}
\begin{align}
  7) \quad & \langle \mathcal{A}\mathcal{B}\mathcal{C}\mathcal{D}\mathcal{E}\rangle +\langle \mathcal{A}\mathcal{B}\mathcal{E}\mathcal{C}\mathcal{D}\rangle +\langle \mathcal{A}\mathcal{C}\mathcal{D}\mathcal{B}\mathcal{E}\rangle +\langle \mathcal{A}\mathcal{C}\mathcal{D}\mathcal{E}\mathcal{B}\rangle +\langle \mathcal{A}\mathcal{E}\mathcal{B}\mathcal{C}\mathcal{D}\rangle +\langle \mathcal{A}\mathcal{E}\mathcal{C}\mathcal{D}\mathcal{B}\rangle  \notag \\
         = & \uline{\langle \mathcal{A}\mathcal{B}\mathcal{E}\rangle\langle \mathcal{C}\mathcal{D}\rangle} +\langle \mathcal{A}\mathcal{B}\rangle\langle \mathcal{C}\mathcal{D}\mathcal{E}\rangle +\uline{\langle \mathcal{A}\mathcal{C}\mathcal{D}\rangle\langle \mathcal{B}\mathcal{E}\rangle} +\langle \mathcal{A}\mathcal{E}\mathcal{B}\rangle\langle \mathcal{C}\mathcal{D}\rangle +\uline{\langle \mathcal{A}\mathcal{E}\rangle\langle \mathcal{B}\mathcal{C}\mathcal{D}\rangle} \\
  8) \quad & \langle \mathcal{A}\mathcal{B}\mathcal{D}\mathcal{C}\mathcal{E}\rangle +\langle \mathcal{A}\mathcal{B}\mathcal{E}\mathcal{D}\mathcal{C}\rangle +\langle \mathcal{A}\mathcal{D}\mathcal{C}\mathcal{B}\mathcal{E}\rangle +\langle \mathcal{A}\mathcal{D}\mathcal{C}\mathcal{E}\mathcal{B}\rangle +\langle \mathcal{A}\mathcal{E}\mathcal{B}\mathcal{D}\mathcal{C}\rangle +\langle \mathcal{A}\mathcal{E}\mathcal{D}\mathcal{C}\mathcal{B}\rangle  \notag \\
         = & \uline{\langle \mathcal{A}\mathcal{B}\mathcal{E}\rangle\langle \mathcal{C}\mathcal{D}\rangle} +\langle \mathcal{A}\mathcal{B}\rangle\langle \mathcal{C}\mathcal{E}\mathcal{D}\rangle +\uline{\langle \mathcal{A}\mathcal{D}\mathcal{C}\rangle\langle \mathcal{B}\mathcal{E}\rangle} +\langle \mathcal{A}\mathcal{E}\mathcal{B}\rangle\langle \mathcal{C}\mathcal{D}\rangle +\uline{\langle \mathcal{A}\mathcal{E}\rangle\langle \mathcal{B}\mathcal{D}\mathcal{C}\rangle} \\
  9) \quad & \langle \mathcal{A}\mathcal{B}\mathcal{C}\mathcal{D}\mathcal{E}\rangle +\langle \mathcal{A}\mathcal{B}\mathcal{C}\mathcal{E}\mathcal{D}\rangle +\langle \mathcal{A}\mathcal{D}\mathcal{B}\mathcal{C}\mathcal{E}\rangle +\langle \mathcal{A}\mathcal{D}\mathcal{E}\mathcal{B}\mathcal{C}\rangle +\langle \mathcal{A}\mathcal{E}\mathcal{B}\mathcal{C}\mathcal{D}\rangle +\langle \mathcal{A}\mathcal{E}\mathcal{D}\mathcal{B}\mathcal{C}\rangle  \notag \\
         = & \langle \mathcal{A}\mathcal{B}\mathcal{C}\rangle\langle \mathcal{D}\mathcal{E}\rangle +\uline{\langle \mathcal{A}\mathcal{D}\mathcal{E}\rangle\langle \mathcal{B}\mathcal{C}\rangle} +\uline{\langle \mathcal{A}\mathcal{D}\rangle\langle \mathcal{B}\mathcal{C}\mathcal{E}\rangle} +\uline{\langle \mathcal{A}\mathcal{E}\mathcal{D}\rangle\langle \mathcal{B}\mathcal{C}\rangle} +\uline{\langle \mathcal{A}\mathcal{E}\rangle\langle \mathcal{B}\mathcal{C}\mathcal{D}\rangle} \\
  10) \quad & \langle \mathcal{A}\mathcal{C}\mathcal{B}\mathcal{D}\mathcal{E}\rangle +\langle \mathcal{A}\mathcal{C}\mathcal{B}\mathcal{E}\mathcal{D}\rangle +\langle \mathcal{A}\mathcal{D}\mathcal{C}\mathcal{B}\mathcal{E}\rangle +\langle \mathcal{A}\mathcal{D}\mathcal{E}\mathcal{C}\mathcal{B}\rangle +\langle \mathcal{A}\mathcal{E}\mathcal{C}\mathcal{B}\mathcal{D}\rangle +\langle \mathcal{A}\mathcal{E}\mathcal{D}\mathcal{C}\mathcal{B}\rangle  \notag \\
          = & \langle \mathcal{A}\mathcal{C}\mathcal{B}\rangle\langle \mathcal{D}\mathcal{E}\rangle +\uline{\langle \mathcal{A}\mathcal{D}\mathcal{E}\rangle\langle \mathcal{B}\mathcal{C}\rangle} +\uline{\langle \mathcal{A}\mathcal{D}\rangle\langle \mathcal{B}\mathcal{E}\mathcal{C}\rangle} +\uline{\langle \mathcal{A}\mathcal{E}\mathcal{D}\rangle\langle \mathcal{B}\mathcal{C}\rangle} +\uline{\langle \mathcal{A}\mathcal{E}\rangle\langle \mathcal{B}\mathcal{D}\mathcal{C}\rangle} \\
  11) \quad & \langle \mathcal{A}\mathcal{B}\mathcal{D}\mathcal{C}\mathcal{E}\rangle +\langle \mathcal{A}\mathcal{B}\mathcal{D}\mathcal{E}\mathcal{C}\rangle +\langle \mathcal{A}\mathcal{C}\mathcal{B}\mathcal{D}\mathcal{E}\rangle +\langle \mathcal{A}\mathcal{C}\mathcal{E}\mathcal{B}\mathcal{D}\rangle +\langle \mathcal{A}\mathcal{E}\mathcal{B}\mathcal{D}\mathcal{C}\rangle +\langle \mathcal{A}\mathcal{E}\mathcal{C}\mathcal{B}\mathcal{D}\rangle  \notag \\
          = & \langle \mathcal{A}\mathcal{B}\mathcal{D}\rangle\langle \mathcal{C}\mathcal{E}\rangle +\uline{\langle \mathcal{A}\mathcal{C}\mathcal{E}\rangle\langle \mathcal{B}\mathcal{D}\rangle} +\uline{\langle \mathcal{A}\mathcal{C}\rangle\langle \mathcal{B}\mathcal{D}\mathcal{E}\rangle} +\uline{\langle \mathcal{A}\mathcal{E}\mathcal{C}\rangle\langle \mathcal{B}\mathcal{D}\rangle} +\uline{\langle \mathcal{A}\mathcal{E}\rangle\langle \mathcal{B}\mathcal{D}\mathcal{C}\rangle} \\
  12) \quad & \langle \mathcal{A}\mathcal{C}\mathcal{D}\mathcal{B}\mathcal{E}\rangle +\langle \mathcal{A}\mathcal{C}\mathcal{E}\mathcal{D}\mathcal{B}\rangle +\langle \mathcal{A}\mathcal{D}\mathcal{B}\mathcal{C}\mathcal{E}\rangle +\langle \mathcal{A}\mathcal{D}\mathcal{B}\mathcal{E}\mathcal{C}\rangle +\langle \mathcal{A}\mathcal{E}\mathcal{C}\mathcal{D}\mathcal{B}\rangle +\langle \mathcal{A}\mathcal{E}\mathcal{D}\mathcal{B}\mathcal{C}\rangle  \notag \\
          = & \uline{\langle \mathcal{A}\mathcal{C}\mathcal{E}\rangle\langle \mathcal{B}\mathcal{D}\rangle} +\uline{\langle \mathcal{A}\mathcal{C}\rangle\langle \mathcal{B}\mathcal{E}\mathcal{D}\rangle} +\langle \mathcal{A}\mathcal{D}\mathcal{B}\rangle\langle \mathcal{C}\mathcal{E}\rangle +\uline{\langle \mathcal{A}\mathcal{E}\mathcal{C}\rangle\langle \mathcal{B}\mathcal{D}\rangle} +\uline{\langle \mathcal{A}\mathcal{E}\rangle\langle \mathcal{B}\mathcal{C}\mathcal{D}\rangle}
\end{align}
where the underlined traces are the ones eliminated. Thus the remaining 32 independent traces are 
\begin{align}
  & \langle \mathcal{A}\mathcal{B}\mathcal{C}\mathcal{D}\mathcal{E} \rangle \,,\quad \langle \mathcal{A}\mathcal{B}\mathcal{C}\mathcal{E}\mathcal{D} \rangle \,,\quad \langle \mathcal{A}\mathcal{B}\mathcal{D}\mathcal{C}\mathcal{E} \rangle \,,\quad \langle \mathcal{A}\mathcal{B}\mathcal{D}\mathcal{E}\mathcal{C} \rangle \,,\quad \langle \mathcal{A}\mathcal{B}\mathcal{E}\mathcal{C}\mathcal{D} \rangle \,,\quad \langle \mathcal{A}\mathcal{B}\mathcal{E}\mathcal{D}\mathcal{C} \rangle \,,\notag \\
  & \langle \mathcal{A}\mathcal{C}\mathcal{B}\mathcal{D}\mathcal{E} \rangle \,,\quad \langle \mathcal{A}\mathcal{C}\mathcal{B}\mathcal{E}\mathcal{D} \rangle \,,\quad \langle \mathcal{A}\mathcal{C}\mathcal{D}\mathcal{B}\mathcal{E} \rangle \,,\quad \langle \mathcal{A}\mathcal{C}\mathcal{D}\mathcal{E}\mathcal{B} \rangle \,,\quad \langle \mathcal{A}\mathcal{C}\mathcal{E}\mathcal{B}\mathcal{D} \rangle \,,\quad \langle \mathcal{A}\mathcal{C}\mathcal{E}\mathcal{D}\mathcal{B} \rangle \,,\notag \\
  & \langle \mathcal{A}\mathcal{D}\mathcal{B}\mathcal{C}\mathcal{E} \rangle \,,\quad \langle \mathcal{A}\mathcal{D}\mathcal{B}\mathcal{E}\mathcal{C} \rangle \,,\quad \langle \mathcal{A}\mathcal{D}\mathcal{C}\mathcal{B}\mathcal{E} \rangle \,,\quad \langle \mathcal{A}\mathcal{D}\mathcal{C}\mathcal{E}\mathcal{B} \rangle \,,\quad \langle \mathcal{A}\mathcal{D}\mathcal{E}\mathcal{B}\mathcal{C} \rangle \,,\quad \langle \mathcal{A}\mathcal{D}\mathcal{E}\mathcal{C}\mathcal{B} \rangle \,,\notag \\
  & \langle \mathcal{A}\mathcal{E}\mathcal{B}\mathcal{C}\mathcal{D} \rangle \,,\quad \langle \mathcal{A}\mathcal{E}\mathcal{B}\mathcal{D}\mathcal{C} \rangle \,,\quad \langle \mathcal{A}\mathcal{E}\mathcal{C}\mathcal{B}\mathcal{D} \rangle \,,\quad \langle \mathcal{A}\mathcal{E}\mathcal{C}\mathcal{D}\mathcal{B} \rangle \,,\quad \langle \mathcal{A}\mathcal{E}\mathcal{D}\mathcal{B}\mathcal{C} \rangle \,,\quad \langle \mathcal{A}\mathcal{E}\mathcal{D}\mathcal{C}\mathcal{B} \rangle \,,\notag \\
  & \langle \mathcal{A}\mathcal{B}\mathcal{C} \rangle\langle \mathcal{D}\mathcal{E} \rangle \,,\quad \langle \mathcal{A}\mathcal{B}\mathcal{D} \rangle\langle \mathcal{C}\mathcal{E} \rangle \,,\quad \langle \mathcal{A}\mathcal{C}\mathcal{D} \rangle\langle \mathcal{B}\mathcal{E} \rangle \,,\quad \langle \mathcal{A}\mathcal{C}\mathcal{B} \rangle\langle \mathcal{D}\mathcal{E} \rangle \,, \notag \\
  & \langle \mathcal{A}\mathcal{D}\mathcal{B} \rangle\langle \mathcal{C}\mathcal{E} \rangle \,,\quad \langle \mathcal{A}\mathcal{E}\mathcal{B} \rangle\langle \mathcal{C}\mathcal{D} \rangle \,,\quad \langle \mathcal{A}\mathcal{B} \rangle\langle \mathcal{C}\mathcal{D}\mathcal{E} \rangle \,,\quad \langle \mathcal{A}\mathcal{C} \rangle\langle \mathcal{C}\mathcal{E}\mathcal{D} \rangle \,,
\end{align}
and the combinations of specific $C$ eigenvalues are
\begin{align}
\label{cp+}
  p_C : & \notag \\
        & \langle \mathcal{A}\mathcal{B}\mathcal{C}\mathcal{D}\mathcal{E} \rangle + \langle \mathcal{A}\mathcal{B}\mathcal{E}\mathcal{D}\mathcal{C} \rangle \,,\quad \langle \mathcal{A}\mathcal{B}\mathcal{C}\mathcal{E}\mathcal{D} \rangle + \langle \mathcal{A}\mathcal{B}\mathcal{D}\mathcal{E}\mathcal{C} \rangle \,,\quad \langle \mathcal{A}\mathcal{B}\mathcal{D}\mathcal{C}\mathcal{E} \rangle + \langle \mathcal{A}\mathcal{B}\mathcal{E}\mathcal{C}\mathcal{D} \rangle \,,\notag \\
        & \langle \mathcal{A}\mathcal{C}\mathcal{B}\mathcal{D}\mathcal{E} \rangle + \langle \mathcal{A}\mathcal{C}\mathcal{B}\mathcal{E}\mathcal{D} \rangle \,,\quad \langle \mathcal{A}\mathcal{D}\mathcal{B}\mathcal{C}\mathcal{E} \rangle + \langle \mathcal{A}\mathcal{C}\mathcal{B}\mathcal{E}\mathcal{C} \rangle \,,\quad \langle \mathcal{A}\mathcal{E}\mathcal{B}\mathcal{C}\mathcal{D} \rangle + \langle \mathcal{A}\mathcal{E}\mathcal{B}\mathcal{D}\mathcal{C} \rangle \,,\notag \\
        & \langle \mathcal{A}\mathcal{C}\mathcal{D}\mathcal{B}\mathcal{E} \rangle + \langle \mathcal{A}\mathcal{D}\mathcal{C}\mathcal{B}\mathcal{E} \rangle \,,\quad \langle \mathcal{A}\mathcal{C}\mathcal{E}\mathcal{B}\mathcal{D} \rangle + \langle \mathcal{A}\mathcal{E}\mathcal{C}\mathcal{B}\mathcal{D} \rangle \,,\quad \langle \mathcal{A}\mathcal{D}\mathcal{E}\mathcal{B}\mathcal{C} \rangle + \langle \mathcal{A}\mathcal{E}\mathcal{D}\mathcal{B}\mathcal{C} \rangle \,,\notag \\
        & \langle \mathcal{A}\mathcal{C}\mathcal{D}\mathcal{E}\mathcal{B} \rangle + \langle \mathcal{A}\mathcal{E}\mathcal{D}\mathcal{C}\mathcal{B} \rangle \,,\quad \langle \mathcal{A}\mathcal{C}\mathcal{E}\mathcal{D}\mathcal{B} \rangle + \langle \mathcal{A}\mathcal{D}\mathcal{E}\mathcal{C}\mathcal{B} \rangle \,,\quad \langle \mathcal{A}\mathcal{D}\mathcal{C}\mathcal{E}\mathcal{B} \rangle + \langle \mathcal{A}\mathcal{E}\mathcal{C}\mathcal{D}\mathcal{B} \rangle \,,\notag \\
        & \langle \mathcal{A}\mathcal{B}\mathcal{C} \rangle \langle \mathcal{D}\mathcal{E} \rangle \,,\quad \langle \mathcal{A}\mathcal{B}\mathcal{D} \rangle\langle \mathcal{D}\mathcal{E} \rangle \,,\quad \langle \mathcal{A}\mathcal{C}\mathcal{B}\rangle\langle \mathcal{D}\mathcal{E} \rangle \,, \quad \langle \mathcal{A}\mathcal{D}\mathcal{B}\rangle\langle \mathcal{C}\mathcal{E} \rangle \,,\notag \\
        & \langle \mathcal{A}\mathcal{E}\mathcal{B} \rangle \langle \mathcal{C}\mathcal{D} \rangle \,,\quad \langle \mathcal{A}\mathcal{B} \rangle\langle \mathcal{C}\mathcal{D}\mathcal{E} \rangle + \langle \mathcal{A}\mathcal{B} \rangle\langle \mathcal{C}\mathcal{E}\mathcal{D} \rangle \,, \\
  -p_C: & \notag \\
  \label{cp-}
        & \langle \mathcal{A}\mathcal{B}\mathcal{C}\mathcal{D}\mathcal{E} \rangle - \langle \mathcal{A}\mathcal{B}\mathcal{E}\mathcal{D}\mathcal{C} \rangle \,,\quad \langle \mathcal{A}\mathcal{B}\mathcal{C}\mathcal{E}\mathcal{D} \rangle - \langle \mathcal{A}\mathcal{B}\mathcal{D}\mathcal{E}\mathcal{C} \rangle \,,\quad \langle \mathcal{A}\mathcal{B}\mathcal{D}\mathcal{C}\mathcal{E} \rangle - \langle \mathcal{A}\mathcal{B}\mathcal{E}\mathcal{C}\mathcal{D} \rangle \,,\notag \\
        & \langle \mathcal{A}\mathcal{C}\mathcal{B}\mathcal{D}\mathcal{E} \rangle - \langle \mathcal{A}\mathcal{C}\mathcal{B}\mathcal{E}\mathcal{D} \rangle \,,\quad \langle \mathcal{A}\mathcal{D}\mathcal{B}\mathcal{C}\mathcal{E} \rangle - \langle \mathcal{A}\mathcal{C}\mathcal{B}\mathcal{E}\mathcal{C} \rangle \,,\quad \langle \mathcal{A}\mathcal{E}\mathcal{B}\mathcal{C}\mathcal{D} \rangle - \langle \mathcal{A}\mathcal{E}\mathcal{B}\mathcal{D}\mathcal{C} \rangle \,,\notag \\
        & \langle \mathcal{A}\mathcal{C}\mathcal{D}\mathcal{B}\mathcal{E} \rangle - \langle \mathcal{A}\mathcal{D}\mathcal{C}\mathcal{B}\mathcal{E} \rangle \,,\quad \langle \mathcal{A}\mathcal{C}\mathcal{E}\mathcal{B}\mathcal{D} \rangle - \langle \mathcal{A}\mathcal{E}\mathcal{C}\mathcal{B}\mathcal{D} \rangle \,,\quad \langle \mathcal{A}\mathcal{D}\mathcal{E}\mathcal{B}\mathcal{C} \rangle - \langle \mathcal{A}\mathcal{E}\mathcal{D}\mathcal{B}\mathcal{C} \rangle \,,\notag \\
        & \langle \mathcal{A}\mathcal{C}\mathcal{D}\mathcal{E}\mathcal{B} \rangle - \langle \mathcal{A}\mathcal{E}\mathcal{D}\mathcal{C}\mathcal{B} \rangle \,,\quad \langle \mathcal{A}\mathcal{C}\mathcal{E}\mathcal{D}\mathcal{B} \rangle - \langle \mathcal{A}\mathcal{D}\mathcal{E}\mathcal{C}\mathcal{B} \rangle \,,\quad \langle \mathcal{A}\mathcal{D}\mathcal{C}\mathcal{E}\mathcal{B} \rangle - \langle \mathcal{A}\mathcal{E}\mathcal{C}\mathcal{D}\mathcal{B} \rangle \,,\notag \\
        & \langle \mathcal{A}\mathcal{B} \rangle\langle \mathcal{C}\mathcal{D}\mathcal{E} \rangle - \langle \mathcal{A}\mathcal{B} \rangle\langle \mathcal{C}\mathcal{E}\mathcal{D} \rangle \,,\quad \langle \mathcal{A}\mathcal{C}\mathcal{D} \rangle\langle \mathcal{B}\mathcal{E} \rangle - \langle \mathcal{B}\mathcal{D}\mathcal{C} \rangle\langle \mathcal{A}\mathcal{E} \rangle \,,
\end{align}
where $p_C = (-1)^\Gamma (-1)^\mathcal{C} (-1)^\mathcal{D} (-1)^\mathcal{E}$. There are 18 combinations of $C$ eigenvalue $p_C$, and 14 combinations of $C$ eigenvalue $-p_C$.

As for 6 fields, the Young tenor implies there are 145 independent invariants, but there are total
\begin{equation}
    5! + C_6^4 \times 3! + \frac{C_6^4 \times C_4^2 }{3!} + C_6^3 \times 2! = 265\,,
\end{equation}
thus 120 independent CH relations are needed, which have been found and are presented in the App.~\ref{app: CH relations}.

\subsubsection*{Repeated fields}

Apart from the general case, the Cayley-Hamilton relation can be used to find out the trace basis when there are repeated fields. Considering the case of 4 fields of the $SU(3)$ adjoint representation, $\mathcal{A},\mathcal{B},\mathcal{C},\mathcal{C}$, where the last two are repeated fields, there are totally 5 distinct traces,
\begin{equation}
    \lra{\mathcal{A}\mathcal{B}}\lra{\mathcal{C}\mathcal{C}}\,,\quad \lra{\mathcal{A}\mathcal{C}}\lra{\mathcal{B}\mathcal{C}}\,,\quad \lra{\mathcal{A}\mathcal{B}\mathcal{C}\mathcal{C}}\,,\quad \lra{\mathcal{A}\mathcal{C}\mathcal{B}\mathcal{C}}\,,\quad \lra{\mathcal{A}\mathcal{C}\mathcal{C}\mathcal{B}} = 0 \,,
\end{equation}
and a unique CH relation
\begin{equation}
\label{eq: CH relation of repeated 4 fields}
    -\lra{\mathcal{A}\mathcal{B}}\lra{\mathcal{C}\mathcal{C}} - 2 \lra{\mathcal{A}\mathcal{C}}\lra{\mathcal{B}\mathcal{C}} + 2\lra{\mathcal{A}\mathcal{B}\mathcal{C}\mathcal{C}} + 2\lra{\mathcal{A}\mathcal{C}\mathcal{B}\mathcal{C}} + 2\lra{\mathcal{A}\mathcal{C}\mathcal{C}\mathcal{B}} = 0 \,,
\end{equation}
which can be obtained from Eq.~\eqref{eq: CH relation of 4 fields} by letting $\mathcal{D}\rightarrow \mathcal{C}$. Thus there are only 4 independent traces, coincident with the Young tensor result in Tab.~\ref{tab: symmetrized invariants}. Actually, the CH relation in Eq.~\eqref{eq: CH relation of repeated 4 fields} is the same as the constraints of the symmetrized tensors obtained by the projectors in Eq.~\eqref{eq: symmetry constraint}. To see that express the fields in terms of the $SU(3)$ generators then substitute them back to Eq.~\eqref{eq: CH relation of repeated 4 fields}. Because
\begin{align}
    \frac{\lra{\mathcal{A}\mathcal{B}}\lra{\mathcal{C}\mathcal{C}}}{\mathcal{A}_a \mathcal{B}_a \mathcal{C}_c \mathcal{C}_d} &= \frac{1}{2}(\lra{\lambda^a\lambda^b}\lra{\lambda^c\lambda^d}+\lra{\lambda^a\lambda^b}\lra{\lambda^d\lambda^c}) = 4\widetilde{\delta^{ab}\delta^{cd}} \,,\notag \\
    \frac{\lra{\mathcal{A}\mathcal{C}}\lra{\mathcal{B}\mathcal{C}}}{\mathcal{A}_a \mathcal{B}_a \mathcal{C}_c \mathcal{C}_d} &= \frac{1}{2}(\lra{\lambda^a\lambda^c}\lra{\lambda^b\lambda^d}+\lra{\lambda^a\lambda^d}\lra{\lambda^c\lambda^b}) = 4 \times \frac{1}{2}(\delta^{ac}\delta^{bd}+\delta^{ad}\delta^{bc}) = 4\widetilde{\delta^{ac}\delta^{bd}} \,,\notag \\
    \frac{\lra{\mathcal{A}\mathcal{B}\mathcal{C}\mathcal{C}}}{\mathcal{A}_a \mathcal{B}_a \mathcal{C}_c \mathcal{C}_d} &= \frac{1}{2}(\lra{\lambda^a\lambda^b\lambda^c\lambda^d}+\lra{\lambda^a\lambda^b\lambda^d\lambda^c}) = \frac{4}{3}\widetilde{\delta^{ab}\delta^{cd}}+2\widetilde{d^{abe}d^{cde}} + 2i\widetilde{f^{abe}d^{cde}}\,, \notag \\
    \frac{\lra{\mathcal{A}\mathcal{C}\mathcal{B}\mathcal{C}}}{\mathcal{A}_a \mathcal{B}_a \mathcal{C}_c \mathcal{C}_d} &= \frac{1}{2}(\lra{\lambda^a\lambda^c\lambda^b\lambda^d}+\lra{\lambda^a\lambda^d\lambda^b\lambda^c}) = \frac{4}{3}\widetilde{\delta^{ac}\delta^{bd}}+2\widetilde{d^{ace}d^{bde}} - 2\widetilde{f^{ace}f^{bde}} + 2i \widetilde{f^{ace}d^{bde}} + 2i \widetilde{d^{ace}f^{bde}} \,,\notag \\
    \frac{\lra{\mathcal{A}\mathcal{C}\mathcal{C}\mathcal{B}}}{\mathcal{A}_a \mathcal{B}_a \mathcal{C}_c \mathcal{C}_d} &= \frac{1}{2}(\lra{\lambda^a\lambda^c\lambda^d\lambda^b}+\lra{\lambda^a\lambda^d\lambda^c\lambda^b}) = \frac{4}{3}\widetilde{\delta^{ac}\delta^{bd}}+2\widetilde{d^{ace}d^{bde}} + 2\widetilde{f^{ace}f^{bde}} + 2i \widetilde{f^{ace}d^{bde}} - 2i \widetilde{d^{ace}f^{bde}} \,,
\end{align}
with tilde means the symmetrization of the $c,d$ indices,
Eq.~\eqref{eq: CH relation of repeated 4 fields} becomes
\begin{align}
    & -\frac{1}{3}\widetilde{\delta^{ac}\delta^{bd}}-\frac{1}{2}\widetilde{\delta^{ab}\delta^{cd}} + \frac{1}{2}\widetilde{d^{abe}d^{cde}} + \widetilde{d^{ace}d^{bde}} \,,\notag \\
    & + i(\frac{1}{2} \widetilde{f^{abe}d^{cde}} + \widetilde{f^{ace}d^{bde}}) = 0\,,
\end{align}
which means its real and imaginary parts vanish simultaneously,
\begin{align}
    -\frac{1}{3}\widetilde{\calt_6}-\frac{1}{6}\widetilde{\calt_4}+\frac{1}{2}\widetilde{\calt_1}+\widetilde{\calt_7} &= 0 \notag \\
    \frac{1}{2} \widetilde{\calt_5} - \widetilde{\calt_8} &= 0\,, \label{eq: constraints on traces}
\end{align}
where we have made the replacement by the tensors in Eq.~\eqref{eq: su3 tensor}, and utilized the relation
\begin{align}
    \widetilde{\calt_8} &= \widetilde{d^{ace}f^{bde}} = \frac{1}{4i\mathcal{A}_a \mathcal{B}_a \mathcal{C}_c \mathcal{C}_d} \lra{[\mathcal{B},\mathcal{C}]\{\mathcal{A},\mathcal{C}\}} = -\frac{1}{4i\mathcal{A}_a \mathcal{B}_a \mathcal{C}_c \mathcal{C}_d} \lra{\{\mathcal{B},\mathcal{C}\}[\mathcal{A},\mathcal{C}]}\notag \\
    & = -\widetilde{f^{ace}d^{bde}}\,.
\end{align}
Noticing that the equations in Eq.~\eqref{eq: constraints on traces} are the same as the last two in Eq.~\eqref{eq: symmetry constraint}, we can conclude that the CH relations of repeated fields are related to the constraints on the symmetrized tensors obtained by the projectors. We obtain the independent CH relations with repeated fields up to 6 fields (two baryons and 4 bosonic fields) and present them in App.~\ref{app: CH relations}.

\subsubsection*{The Cayley-Hamilton relations of $SU(2)$ case}

Although the nucleons are of the $SU(2)$ fundamental representations, they can form a bilinear object by outer product,
\begin{equation}
    \left\{\begin{array}{l}
N \rightarrow g N \\
\overline{N} \rightarrow \overline{N} g^{-1}
    \end{array}\right. \rightarrow N\overline{N} \rightarrow gN\overline{N}g^{-1} \,,
\end{equation}
which, in particular, is not traceless, $\lra{N\overline{N}} = (\overline{N}N)\neq 0$. Thus the invariants composed of two nucleons and some other bosonic fields are equivalent to the invariants composed of a bilinear object $N\overline{N}$ and the bosonic fields. Considering the invariants composed by $\mathcal{A} = N\overline{N}\,,\mathcal{B}\,,\mathcal{C}$, of which there are two independent tensors already known from the Young tensor method in Eq.~\eqref{eq: su2 tensors}, let $\mathcal{A}= \mathcal{A}+\mathcal{B}$ and extract the terms proportional to $\mathcal{A}\mathcal{B}$, the Cayley-Hamilton theorem of $SU(2)$ group in Eq.~\eqref{eq: ch-2} states
\begin{equation}
    \mathcal{A}\mathcal{B} + \mathcal{B}\mathcal{A} = \lra{\mathcal{A}}\mathcal{B} + \lra{\mathcal{B}}\mathcal{A} + \lra{\mathcal{A}\mathcal{B}} - \lra{\mathcal{A}}\lra{\mathcal{B}}\,,
\end{equation}
and is further reduced to 
\begin{equation}
    \mathcal{A}\mathcal{B} + \mathcal{B}\mathcal{A} = \lra{\mathcal{A}}\mathcal{B} + \lra{\mathcal{A}\mathcal{B}} \,,
\end{equation}
due to the traceless $\mathcal{B}$. Multiplying by $\mathcal{C}$ and taking trace, the unique CH relation reads as
\begin{equation}
    \lra{\mathcal{A}\mathcal{B}\mathcal{C}} + \lra{\mathcal{A}\mathcal{C}\mathcal{B}} = \lra{\mathcal{A}}\lra{\mathcal{B}\mathcal{C}}\,,
\end{equation}
If we choose the two independent traces to be
\begin{equation}
    \lra{\mathcal{A}[\mathcal{B},\mathcal{C}]}\,,\quad \lra{\mathcal{A}}\lra{\mathcal{B}\mathcal{C}}\,,
\end{equation}
they can be corresponded to the tensors in Eq.~\eqref{eq: su2 tensors}.

\subsection{Hilbert Series for $CP$ Eigenstates}

The Hilbert series is a generating function of the group invariants~\cite{Lehman:2015coa,Lehman:2015via,Henning:2015alf,Henning:2015daa,Henning:2017fpj,Bijnens:2022zqo,Marinissen:2020jmb}, which takes the form that 
\begin{equation}
    HS = 1 + n_1 Q + n_2 Q^2 + n_2 Q^3 +\dots\,,
\end{equation}
where $Q$ is a field of some representation of a group, and $n_i$ is the number of the independent invariants composed of $i$ $Q$'s. Thus it can present the number of independent invariants, which is helpful when constructing explicit operators.

Since the Lorentz group $SO(3,1)\cong SU(2)_l\times SU(2)_r$ and the chiral symmetry $SU(N)_V$, where $N=2,3$ is compact, the Hilbert series can be generated via the orthogonality of the group characters. Suppose there is only one field $Q$ of some representation $r$ of a compact group $G$, the character is $\chi_r(z)$, where the $z$ is the variable of the maximal ring of $G$, then the number of the independent invariants composed of $m$ $Q$ is the multiplicity of the trivial representation $1$ while doing the tensor product decomposition of $\underbrace{r \times \dots \times r}_{m}$, which can be obtained by the character orthogonality relation
\begin{equation}
\label{eq: integral}
    n_m = \int dG \chi^*_1(z)\text{Sym}\left(\underbrace{\chi_r(z)\chi_r(z)\dots \chi_r(z)}_{m}\right)\,,
\end{equation}
where $dG$ is the invariant measure of $G$.
The character of the trivial representation $\chi_1=1$, and Sym implies the symmetric product of the characters, which can be generated by the Plethystic Exponential (PE) 
\begin{equation}
    \text{PE}(Q,\chi_r(z)) = \exp\left\{\sum_{k\geq 1}\frac{Q^k\chi_r(z^k)}{k}\right\}\,,
\end{equation}
via
\begin{equation}
    \text{Sym}\left(\underbrace{\chi_r(z)\chi_r(z)\dots \chi_r(z)}_{m}\right) = \left.\frac{d^m}{dQ^m}\text{PE}(Q,\chi_r(z))\right|_{Q=0} \,.
\end{equation}
The integral in Eq.~\eqref{eq: integral} can be generalized to multi-field and multi-group case directly,
\begin{align}
    n_{m_1,m_2,\dots,m_f} &= \int dG_1(z_1) \int dG_2(z_2) \dots \int dG_g(z_g) \notag \\
    & \times \prod_{i=1}^f\left(\left.\frac{d^{m_i}}{dQ^{m_i}}\text{PE}(Q_i,\prod_{j=1}^g\chi_{r_{ij}}(z_j))\right|_{Q_i=0}\right)\,,
\end{align}
where $n_{m_1,m_2,\dots,n_f}$ is the number of independent invariants composed of $m_1$ $Q_1$'s, $m_2$ $Q_2$'s, $\dots$, and $m_f$ $Q_f$'s, $ r_{ij} $ is the representation under group $G_j$ of the field $Q_i$.
We have set $\chi_1(z_j)^*=1$ for $j=1,2,\dots,g$, and the complete Hilbert series takes the form that
\begin{align}
    HS &= 1+\sum_{m_1,\dots m_f}  n_{m_1,m_2,\dots,m_f} ({Q_1}^{m_1}{Q_2}^{m_2}\dots {Q_f}^{m_f}) \notag \\
    &= \int \prod_{j=1}^g dG_j(z_j) \prod_{i=1}^f\text{PE}(Q_i,\prod_{j=1}^g \chi_{r_{ij}}(z_j)) \,.
\end{align}

When applying to the physical model, it should be noticed there are some remarks:
\begin{itemize}
    \item The fermions respect the Fermion-Dirac statistic, which means the product of the characters in Eq.~\eqref{eq: integral} is not symmetric but asymmetric. Such products can be generated by the fermion PE, noted as $\text{PE}_f$,
    \begin{equation}
        \text{PE}_f(Q,\chi_r(z)) = \exp\left\{\sum_{k\geq 1}\frac{(-1)^{k-1}Q^k\chi_r(z^k)}{k}\right\}\,,
    \end{equation}
    \item There are two redundancies, equation of motion (EOM) and integration-by-part (IBP) when counting the Lorentz group invariants, both of which are about the derivative $D$. In Ref~\cite{Henning:2015alf}. the authors extended the Lorentz group to the conformal group, where the Lorentz group irreducible representations are enlarged to infinite representation by the derivative. As a consequence, the Lorentz group characters are modified by an overall factor $P(D,z)$ associated with the total derivatives contributions, and should be removed before the integration. On the other hand, the EOM redundancy is removed by the subtraction between different conformal group representations.
    \item In the ChPT, the external sources are of no EOM, so their Lorentz group characters should only be extended to the conformal group characters, but not be modified by subtractions.
\end{itemize}
After these modifications, the integration applied to the ChPT directly is
\begin{equation}
\label{eq: HS1}
    HS = \int dSO(3,1)(z) \int dSU(N)(x) \frac{1}{P(D, z)} \prod_i \text{PE}_f(\psi_i,\chi_{i}(z,x)) \prod_j \text{PE}(\phi_j,\chi_j(z,x))\,,
\end{equation}
where $dSO(3,1)(z)$ and $dSU(N)(x)$ are the invariant measure of the Lorentz group and the chiral group, with $z,x$ their maximal ring variable respectively, $P(D,z)$ is complemented for the total derivative terms, and the characters of the fields take the factor form
\begin{equation}
    \chi_i (z,x) = \chi(z)_i\times \chi_i(x)\,,
\end{equation}
where $\chi(z)_i$ is the character of the Lorentz group and $\chi(x)_i$ is the one of the chiral group.

The Hilbert series can be extended to contain discrete symmetries such as $C$ and $P$, which means the group $SO(3,1)\times SU(N)$ is enlarged by two discrete groups $\{I, C\}$ and $\{I,P\}$ isomorphic to the $\mathbb{Z}_2$ group~\cite{Graf:2020yxt, Sun:2022aag}. As a consequence, the group manifold is divided into disjoint unions
\begin{align}
   & SO(3,1)\times SU(N) \rtimes\{I,C\} \rtimes\{I,P\} \notag \\
   & \cong SO(3,1)^+ \times SU(N)^+ \bigsqcup SO(3,1)^+ \times SU(N)^- \bigsqcup SO(3,1)^- \times SU(N)^+ \bigsqcup SO(3,1)^- \times SU(N)^-\,,
\end{align}
and the integral in Eq.~\eqref{eq: HS1} should be performed individually,
\begin{equation}
    HS \rightarrow \left\{\begin{array}{l}
HS^{++} \\
HS^{-+} \\
HS^{+-} \\
HS^{--} 
    \end{array}\right. \,,
\end{equation}
of which the characters and the integral variables need to be modified in different branches, details can be found in Ref.~\cite{Graf:2020yxt, Sun:2022aag}. 
The Hilbert series of the operators with explicit $C$ and $P$ eigenvalues can be obtained via their combinations
\begin{align}
    HS^{total} &= HS^{C+P+} \,, \\
    HS^{C+} &= \frac{1}{2}(HS^{++}+HS^{-+}) \,, \\ 
    HS^{P+} &= \frac{1}{2}(HS^{++}+HS^{+-}) \,, \\ 
    HS^{C+P+} &= \frac{1}{4}(HS^{++}+HS^{-+}+HS^{--}+HS^{+-})\,.
\end{align}

For the first time, the number of the operators for each type with all the $C\,, P$ eigenvalues up to order $\mathcal{O}(p^5)$ for the $SU(2)$ case and up to order $\mathcal{O}(p^4)$ for the $SU(3)$ case with no heavy baryon projection has been obtained in this paper, which will be presented as follows. 

For the $SU(2)$ case, the Hilbert series of 4 different $CP$ sectors are given as follows, 
\begin{align}
    HS^{C+P+}=&f_+N^2+2u^2N^2+\Sigma_-N^2+\langle\Sigma_-\rangle N^2+\langle\Sigma_-\rangle N^2+Df_-N^2+Df_+N^2+2uf_-N^2+2uf_+N^2\nonumber\\
    &+2Du^2N^2+3u^3N^2+u\Sigma_-N^2+u\Sigma_+N^2+u\langle\Sigma_+\rangle N^2+3f_-^2N^2+D^2f_+N^2+3f_-f_+N^2\nonumber\\
    &+3f_+^2N^2+5Duf_-N^2+5Duf_+N^2+3D^2u^2N^2+3u^2f_-N^2+7u^2f_+N^2+4Du^3N^2\nonumber\\
    &+4u^4N^2+D^2\Sigma_-N^2+f_-\Sigma_-N^2+f_+\Sigma_-N^2+2Du\Sigma_-N^2+3u^2\Sigma_-N^2+\Sigma_-^2N^2\nonumber\\
    &+D^2\Sigma_+N^2+f_-\Sigma_+N^2+f_+\Sigma_+N^2+2Du\Sigma_+N^2+3u^2\Sigma_+N^2+\Sigma_-\Sigma_+N^2+\Sigma_+^2N^2\nonumber\\
    &+D^2\langle\Sigma_-\rangle N^2+f_+\langle\Sigma-\rangle N^2+Du\langle\Sigma_-\rangle N^2+2u^2\langle\Sigma_-\rangle+\Sigma_-\langle\Sigma_-\rangle N^2+\Sigma_+\langle\Sigma_-\rangle N^2\nonumber\\
    &+{\langle\Sigma-\rangle}^2 N^2+D^2\langle\Sigma_+\rangle N^2+f_+\langle\Sigma_+\rangle N^2+Du\langle\Sigma_+\rangle N^2+2u^2\langle\Sigma_+\rangle N^2+\Sigma_-\langle\Sigma_+\rangle N^2\nonumber\\
    &+\Sigma_+\langle\Sigma_+\rangle N^2+\langle\Sigma_-\rangle\langle\Sigma_+\rangle N^2+{\langle\Sigma_+\rangle}^2 N^2+D^3 f_- N^2+D^3 f_+ N^2+3 D^3 N^2 u^2+8 D^2 f_- N^2 u\nonumber\\
    &+8 D^2 f_+ N^2 u+9 D^2 N^2 u^3+3 D^2 N^2 \Sigma_- u+3 D^2 N^2 \Sigma_+ u+D^2 N^2 u \langle\Sigma_-\rangle+2 D^2 N^2 u \langle\Sigma_+\rangle\nonumber\\
    &+5 D f_-^2 N^2+8 D f_- f_+ N^2+3 D f_- N^2 \Sigma_-+3 D f_- N^2 \Sigma_++20 D f_- N^2 u^2+D f_- N^2 \langle\Sigma_-\rangle\nonumber\\
    &+2 D f_- N^2 \langle\Sigma_+\rangle+5 D f_+^2 N^2+3 D f_+ N^2 \Sigma_-+3 D f_+ N^2 \Sigma_++20 D f_+ N^2 u^2+D f_+ N^2 \langle\Sigma_-\rangle\nonumber\\
    &+2 D f_+ N^2 \langle\Sigma_+\rangle+D N^2 \Sigma_-^2+D N^2 \Sigma_- \Sigma_++D N^2 \Sigma_- \langle\Sigma_+\rangle+D N^2 \Sigma_+^2+D N^2 \Sigma_+ \langle\Sigma_-\rangle\nonumber\\
    &+10 D N^2 u^4+7 D N^2 \Sigma_- u^2+5 D N^2 \Sigma_+ u^2+3 D N^2 u^2 \langle\Sigma_-\rangle+3 D N^2 u^2 \langle\Sigma_+\rangle+D N^2 \langle\Sigma_-\rangle \langle\Sigma_+\rangle\nonumber\\
    &+7 f_-^2 N^2 u+12 f_- f_+ N^2 u+9 f_- N^2 u^3+4 f_- N^2 \Sigma_- u+4 f_- N^2 \Sigma_+ u+2 f_- N^2 u \langle\Sigma_-\rangle\nonumber\\
    &+2 f_- N^2 u \langle\Sigma_+\rangle+7 f_+^2 N^2 u+9 f_+ N^2 u^3+4 f_+ N^2 \Sigma_- u+4 f_+ N^2 \Sigma_+ u+2 f_+ N^2 u \langle\Sigma_-\rangle\nonumber\\
    &+2 f_+ N^2 u \langle\Sigma_+\rangle+5 N^2 u^5+3 N^2 \Sigma_- u^3+3 N^2 \Sigma_+ u^3+3 N^2 u^3 \langle\Sigma_+\rangle+2 N^2 \Sigma_-^2 u+N^2 \Sigma_- \Sigma_+ u\nonumber\\
    &+N^2 \Sigma_- u \langle\Sigma_-\rangle+N^2 \Sigma_- u \langle\Sigma_+\rangle+2 N^2 \Sigma_+^2 u+N^2 \Sigma_+ u \langle\Sigma_-\rangle+N^2 \Sigma_+ u \langle\Sigma_+\rangle+N^2 u \langle\Sigma_-\rangle^2\nonumber\\
    &+N^2 u \langle\Sigma_+\rangle^2\,,
    \end{align}
    \begin{align}
    HS^{C+P-}=&f_-N^2+2uf_-N^2+2uf_+N^2+u\Sigma_-N^2+u\Sigma_+N^2+u\langle\Sigma_-\rangle N^2++3f_-^2N^2+D^2f_+N^2\nonumber\\
    &+3f_+^2N^2+5Duf_-N^2+5Duf_+N^2+3D^2u^2N^2+3u^2f_-N^2+7u^2f_+N^2+4Du^3N^2\nonumber\\
    &+4u^4N^2+D^2\Sigma_-N^2+f_-\Sigma_-N^2+f_+\Sigma_-N^2+2Du\Sigma_-N^2+3u^2\Sigma_-N^2+\Sigma_-^2N^2+\Sigma_+^2N^2\nonumber\\
    &+D^2\Sigma_+N^2+f_-\Sigma_+N^2+f_+\Sigma_+N^2+2Du\Sigma_+N^2+3u^2\Sigma_+N^2+\Sigma_-\Sigma_+N^2+3f_-f_+N^2\nonumber\\
    &+D^2\langle\Sigma_-\rangle N^2+f_+\langle\Sigma-\rangle N^2+Du\langle\Sigma_-\rangle N^2+2u^2\langle\Sigma_-\rangle+\Sigma_-\langle\Sigma_-\rangle N^2+\Sigma_+\langle\Sigma_-\rangle N^2\nonumber\\
    &+{\langle\Sigma-\rangle}^2 N^2+D^2\langle\Sigma_+\rangle N^2+f_+\langle\Sigma_+\rangle N^2+Du\langle\Sigma_+\rangle N^2+2u^2\langle\Sigma_+\rangle N^2+\Sigma_-\langle\Sigma_+\rangle N^2\nonumber\\
    &+\Sigma_+\langle\Sigma_+\rangle N^2+\langle\Sigma_-\rangle\langle\Sigma_+\rangle N^2+{\langle\Sigma_+\rangle}^2 N^2+8 D^2 f_- N^2 u+8 D^2 f_+ N^2 u+5 D^2 N^2 u^3\nonumber\\
    &+3 D^2 N^2 \Sigma_- u+3 D^2 N^2 \Sigma_+ u+2 D^2 N^2 u \langle\Sigma_-\rangle+D^2 N^2 u \langle\Sigma_+\rangle+3 D f_-^2 N^2+8 D f_- f_+ N^2\nonumber\\
    &+3 D f_- N^2 \Sigma_-+3 D f_- N^2 \Sigma_++16 D f_- N^2 u^2+2 D f_- N^2 \langle\Sigma_-\rangle+D f_- N^2 \langle\Sigma_+\rangle+3 D f_+^2 N^2\nonumber\\
    &+3 D f_+ N^2 \Sigma_-+3 D f_+ N^2 \Sigma_++16 D f_+ N^2 u^2+2 D f_+ N^2 \langle\Sigma_-\rangle+D f_+ N^2 \langle\Sigma_+\rangle+D N^2 \Sigma_- \Sigma_+\nonumber\\
    &+D N^2 \Sigma_- \langle\Sigma_-\rangle+D N^2 \Sigma_+ \langle\Sigma_+\rangle+6 D N^2 u^4+5 D N^2 \Sigma_- u^2+7 D N^2 \Sigma_+ u^2+3 D N^2 u^2 \langle\Sigma_-\rangle\nonumber\\
    &+3 D N^2 u^2 \langle\Sigma_+\rangle+5 f_-^2 N^2 u+12 f_- f_+ N^2 u+9 f_- N^2 u^3+4 f_- N^2 \Sigma_- u+4 f_- N^2 \Sigma_+ u\nonumber\\
    &+2 f_- N^2 u \langle\Sigma_-\rangle+2 f_- N^2 u \langle\Sigma_+\rangle+5 f_+^2 N^2 u+9 f_+ N^2 u^3+4 f_+ N^2 \Sigma_- u+4 f_+ N^2 \Sigma_+ u\nonumber\\
    &+2 f_+ N^2 u \langle\Sigma_-\rangle+2 f_+ N^2 u \langle\Sigma_+\rangle+3 N^2 \Sigma_- u^3+3 N^2 \Sigma_+ u^3+3 N^2 u^3 \langle\Sigma_-\rangle+3 N^2 \Sigma_- \Sigma_+ u\nonumber\\
    &+N^2 \Sigma_- u \langle\Sigma_-\rangle+N^2 \Sigma_- u \langle\Sigma_+\rangle+N^2 \Sigma_+ u \langle\Sigma_-\rangle+N^2 \Sigma_+ u \langle\Sigma_+\rangle+N^2 u \langle\Sigma_-\rangle \langle\Sigma_+\rangle\,,
    \end{align}
    \begin{align}
    HS^{C-P+}=&f_+N^2+2u^2N^2+\Sigma_-N^2+\langle\Sigma_-\rangle N^2+\langle\Sigma_-\rangle N^2+2uf_-N^2+2uf_+N^2+u\Sigma_+N^2+u\Sigma_+N^2\nonumber\\
    &+u\langle\Sigma_-\rangle N^2+D^2f_-N^2+3f_-f_+N^2+5Duf_-N^2+5Duf_+N^2+7u^2f_-N^2+3u^2f_+N^2\nonumber\\
    &+3Du^3N^2+f_-\Sigma_-N^2+f_+\Sigma_-N^2+2Du\Sigma_-N^2+u^2\Sigma_-N^2+f_-\Sigma_+N^2+f_+\Sigma_+N^2\nonumber\\
    &+2Du\Sigma_+N^2+u^2\Sigma_+N^2+\Sigma_-\Sigma_+N^2+f_-\langle \Sigma_-\rangle N^2+Du\langle\Sigma_-\rangle N^2+Du\langle\Sigma_-\rangle N^2+f_-\langle \Sigma_+\rangle N^2\nonumber\\
    &8 D^2 f_- N^2 u+8 D^2 f_+ N^2 u+5 D^2 N^2 u^3+3 D^2 N^2 \Sigma_- u+3 D^2 N^2 \Sigma_+ u+2 D^2 N^2 u \langle\Sigma_-\rangle\nonumber\\
    &+D^2 N^2 u \langle\Sigma_+\rangle+3 D f_-^2 N^2+8 D f_- f_+ N^2+3 D f_- N^2 \Sigma_-+3 D f_- N^2 \Sigma_++16 D f_- N^2 u^2\nonumber\\
    &+2 D f_- N^2 \langle\Sigma_-\rangle+D f_- N^2 \langle\Sigma_+\rangle+3 D f_+^2 N^2+3 D f_+ N^2 \Sigma_-+3 D f_+ N^2 \Sigma_++16 D f_+ N^2 u^2\nonumber\\
    &+2 D f_+ N^2 \langle\Sigma_-\rangle+D f_+ N^2 \langle\Sigma_+\rangle+D N^2 \Sigma_- \Sigma_++D N^2 \Sigma_- \langle\Sigma_-\rangle+D N^2 \Sigma_+ \langle\Sigma_+\rangle+6 D N^2 u^4\nonumber\\
    &+5 D N^2 \Sigma_- u^2+7 D N^2 \Sigma_+ u^2+3 D N^2 u^2 \langle\Sigma_-\rangle+3 D N^2 u^2 \langle\Sigma_+\rangle+5 f_-^2 N^2 u+12 f_- f_+ N^2 u\nonumber\\
    &+9 f_- N^2 u^3+4 f_- N^2 \Sigma_- u+4 f_- N^2 \Sigma_+ u+2 f_- N^2 u \langle\Sigma_-\rangle+2 f_- N^2 u \langle\Sigma_+\rangle+5 f_+^2 N^2 u\nonumber\\
    &+9 f_+ N^2 u^3+4 f_+ N^2 \Sigma_- u+4 f_+ N^2 \Sigma_+ u+2 f_+ N^2 u \langle\Sigma_-\rangle+2 f_+ N^2 u \langle\Sigma_+\rangle+3 N^2 \Sigma_- u^3\nonumber\\
    &+3 N^2 \Sigma_+ u^3+3 N^2 u^3 \langle\Sigma_-\rangle+3 N^2 \Sigma_- \Sigma_+ u+N^2 \Sigma_- u \langle\Sigma_-\rangle+N^2 \Sigma_- u \langle\Sigma_+\rangle+N^2 \Sigma_+ u \langle\Sigma_-\rangle\nonumber\\
    &+N^2 \Sigma_+ u \langle\Sigma_+\rangle+N^2 u \langle\Sigma_-\rangle \langle\Sigma_+\rangle\,,
    \end{align}
    \begin{align}
    HS^{C-P-}=&f_-N^2+Df_-N^2+Df_+N^2+2uf_-N^2+2uf_+N^2+2Du^2N^2+3u^3N^2+u\Sigma_-N^2+u\Sigma_+N^2\nonumber\\
    &+u\langle\Sigma_+\rangle N^2+D^2f_-N^2+3f_-f_+N^2+5Duf_-N^2+5Duf_+N^2+7u^2f_-N^2+3u^2f_+N^2\nonumber\\
    &+3Du^3N^2+f_-\Sigma_-N^2+f_+\Sigma_-N^2+2Du\Sigma_-N^2+u^2\Sigma_-N^2+f_-\Sigma_+N^2+f_+\Sigma_+N^2\nonumber\\
    &+2Du\Sigma_+N^2+u^2\Sigma_+N^2+\Sigma_-\Sigma_+N^2+f_-\langle \Sigma_-\rangle N^2+Du\langle\Sigma_-\rangle N^2+Du\langle\Sigma_-\rangle N^2+f_-\langle \Sigma_+\rangle N^2\nonumber\\
    &+D^3 f_- N^2+D^3 f_+ N^2+3 D^3 N^2 u^2+8 D^2 f_- N^2 u+8 D^2 f_+ N^2 u+9 D^2 N^2 u^3+3 D^2 N^2 \Sigma_- u\nonumber\\
    &+3 D^2 N^2 \Sigma_+ u+D^2 N^2 u \langle\Sigma_-\rangle+2 D^2 N^2 u \langle\Sigma_+\rangle+5 D f_-^2 N^2+8 D f_- f_+ N^2+3 D f_- N^2 \Sigma_-\nonumber\\
    &+3 D f_- N^2 \Sigma_++20 D f_- N^2 u^2+D f_- N^2 \langle\Sigma_-\rangle+2 D f_- N^2 \langle\Sigma_+\rangle+5 D f_+^2 N^2+3 D f_+ N^2 \Sigma_-\nonumber\\
    &+3 D f_+ N^2 \Sigma_++20 D f_+ N^2 u^2+D f_+ N^2 \langle\Sigma_-\rangle+2 D f_+ N^2 \langle\Sigma_+\rangle+D N^2 \Sigma_-^2+D N^2 \Sigma_- \Sigma_+\nonumber\\
    &+D N^2 \Sigma_- \langle\Sigma_+\rangle+D N^2 \Sigma_+^2+D N^2 \Sigma_+ \langle\Sigma_-\rangle+10 D N^2 u^4+7 D N^2 \Sigma_- u^2+5 D N^2 \Sigma_+ u^2\nonumber\\
    &+3 D N^2 u^2 \langle\Sigma_-\rangle+3 D N^2 u^2 \langle\Sigma_+\rangle+D N^2 \langle\Sigma_-\rangle \langle\Sigma_+\rangle+7 f_-^2 N^2 u+12 f_- f_+ N^2 u+9 f_- N^2 u^3\nonumber\\
    &+4 f_- N^2 \Sigma_- u+4 f_- N^2 \Sigma_+ u+2 f_- N^2 u \langle\Sigma_-\rangle+2 f_- N^2 u \langle\Sigma_+\rangle+7 f_+^2 N^2 u+9 f_+ N^2 u^3\nonumber\\
    &+4 f_+ N^2 \Sigma_- u+4 f_+ N^2 \Sigma_+ u+2 f_+ N^2 u \langle\Sigma_-\rangle+2 f_+ N^2 u \langle\Sigma_+\rangle+5 N^2 u^5+3 N^2 \Sigma_- u^3\nonumber\\
    &+3 N^2 \Sigma_+ u^3+3 N^2 u^3 \langle\Sigma_+\rangle+2 N^2 \Sigma_-^2 u+N^2 \Sigma_- \Sigma_+ u+N^2 \Sigma_- u \langle\Sigma_-\rangle+N^2 \Sigma_- u \langle\Sigma_+\rangle\nonumber\\
    &+2 N^2 \Sigma_+^2 u+N^2 \Sigma_+ u \langle\Sigma_-\rangle+N^2 \Sigma_+ u \langle\Sigma_+\rangle+N^2 u \langle\Sigma_-\rangle^2+N^2 u \langle\Sigma_+\rangle^2\,,
\end{align}
while for the $SU(3)$ case, these Hilbert series are
\begin{align}
    HS^{C+P+}=&f_+B^2+2u^2B^2+\Sigma_-B^2+\langle\Sigma_-\rangle B^2+\langle\Sigma_-\rangle B^2+2Df_-B^2+2Df_+B^2+8uf_-B^2+8uf_+B^2\nonumber\\
    &+7Du^2B^2+16u^3B^2+3u\Sigma_-B^2+5u\Sigma_+B^2+2u\langle\Sigma_+\rangle B^2+11f_-^2B^2+2D^2f_+B^2+11f_-f_+B^2\nonumber\\
    &+11f_+^2B^2+19Duf_-B^2+21Duf_+B^2+11D^2u^2B^2+33u^2f_-B^2+47u^2f_+B^2+31Du^3B^2\nonumber\\
    &+41u^4B^2+2D^2\Sigma_-B^2+3f_-\Sigma_-B^2+5f_+\Sigma_-B^2+8Du\Sigma_-B^2+20u^2\Sigma_-B^2+4\Sigma_-^2B^2\nonumber\\
    &+2D^2\Sigma_+B^2+3f_-\Sigma_+B^2+5f_+\Sigma_+B^2+8Du\Sigma_+B^2+20u^2\Sigma_+B^2+5\Sigma_-\Sigma_+B^2+4\Sigma_+^2B^2\nonumber\\
    &+D^2\langle\Sigma_-\rangle B^2+2f_+\langle\Sigma-\rangle B^2+2Du\langle\Sigma_-\rangle B^2+7u^2\langle\Sigma_-\rangle+2\Sigma_-\langle\Sigma_-\rangle B^2+2\Sigma_+\langle\Sigma_-\rangle B^2\nonumber\\
    &+{\langle\Sigma-\rangle}^2 B^2+D^2\langle\Sigma_+\rangle B^2+2f_+\langle\Sigma_+\rangle B^2+2Du\langle\Sigma_+\rangle B^2+7u^2\langle\Sigma_+\rangle B^2+2\Sigma_-\langle\Sigma_+\rangle B^2\nonumber\\
    &+2\Sigma_+\langle\Sigma_+\rangle B^2+\langle\Sigma_-\rangle\langle\Sigma_+\rangle B^2+{\langle\Sigma_+\rangle}^2 B^2\,,
    \end{align}
    \begin{align}
    HS^{C+P-}=&f_-B^2+8uf_-B^2+8uf_+B^2+5u\Sigma_-B^2+3u\Sigma_+B^2+2u\langle\Sigma_-\rangle B^2+Du^2B^2+6u^3B^2+11f_-^2B^2\nonumber\\
    &+2D^2f_+B^2+11f_-f_+B^2+11f_+^2B^2+19Duf_-B^2+21Duf_+B^2+11D^2u^2B^2+33u^2f_-B^2\nonumber\\
    &+47u^2f_+B^2+31Du^3B^2+41u^4B^2+2D^2\Sigma_-B^2+3f_-\Sigma_-B^2+5f_+\Sigma_-B^2+8Du\Sigma_-B^2\nonumber\\
    &+20u^2\Sigma_-B^2+4\Sigma_-^2B^2+2D^2\Sigma_+B^2+3f_-\Sigma_+B^2+5f_+\Sigma_+B^2+8Du\Sigma_+B^2+20u^2\Sigma_+B^2\nonumber\\
    &+5\Sigma_-\Sigma_+B^2+4\Sigma_+^2B^2+D^2\langle\Sigma_-\rangle B^2+2f_+\langle\Sigma-\rangle B^2+2Du\langle\Sigma_-\rangle B^2+7u^2\langle\Sigma_-\rangle+2\Sigma_-\langle\Sigma_-\rangle B^2\nonumber\\
    &+2\Sigma_+\langle\Sigma_-\rangle B^2+{\langle\Sigma-\rangle}^2 B^2+D^2\langle\Sigma_+\rangle B^2+2f_+\langle\Sigma_+\rangle B^2+2Du\langle\Sigma_+\rangle B^2+7u^2\langle\Sigma_+\rangle B^2\nonumber\\
    &+2\Sigma_-\langle\Sigma_+\rangle B^2+2\Sigma_+\langle\Sigma_+\rangle B^2+\langle\Sigma_-\rangle\langle\Sigma_+\rangle B^2+{\langle\Sigma_+\rangle}^2 B^2\,,
    \end{align}
    \begin{align}
    HS^{C-P+}=&f_+B^2+2u^2B^2+\Sigma_-B^2+\langle\Sigma_-\rangle B^2+\langle\Sigma_-\rangle B^2++8uf_-B^2+8uf_+B^2+5u\Sigma_-B^2+3u\Sigma_+B^2\nonumber\\
    &+2u\langle\Sigma_-\rangle B^2+Du^2B^2+6u^3B^2+2D^2f_-B^2+f_-^2B^2+f_+^2B^2+13f_-f_+B^2+21Duf_-B^2\nonumber\\
    &+19Duf_+B^2+D^2u^2B^2+47u^2f_-B^2+33u^2f_+B^2+27Du^3B^2+20u^4B^2+5f_-\Sigma_-B^2\nonumber\\
    &+3f_+\Sigma_-B^2+8Du\Sigma_-B^2+12u^2\Sigma_-B^2+5f_-\Sigma_+B^2+3f_+\Sigma_+B^2+8Du\Sigma_+B^2\nonumber\\
    &+12u^2\Sigma_+B^2+3\Sigma_-\Sigma_+B^2+2f_-\langle \Sigma_-\rangle B^2+2Du\langle\Sigma_-\rangle B^2+2Du\langle\Sigma_+\rangle B^2+2f_-\langle \Sigma_+\rangle B^2\nonumber\\&
    +2f_-\langle \Sigma_-\rangle B^2+u^2\langle\Sigma_-\rangle N^2+u^2\langle\Sigma_+\rangle N^2\,,
    \end{align}
    \begin{align}
    HS^{C-P-}=&f_-B^2+2Df_-B^2+2Df_+B^2+8uf_-B^2+8uf_+B^2+7Du^2B^2+16u^3B^2+3u\Sigma_-B^2+5u\Sigma_+B^2\nonumber\\
    &+2u\langle\Sigma_+\rangle B^2+2D^2f_-B^2+f_-^2B^2+f_+^2B^2+13f_-f_+B^2+21Duf_-B^2+19Duf_+B^2+D^2u^2B^2\nonumber\\
    &+47u^2f_-B^2+33u^2f_+B^2+27Du^3B^2+20u^4B^2+5f_-\Sigma_-B^2+3f_+\Sigma_-B^2+8Du\Sigma_-B^2\nonumber\\
    &+12u^2\Sigma_-B^2+5f_-\Sigma_+B^2+3f_+\Sigma_+B^2+8Du\Sigma_+B^2+12u^2\Sigma_+B^2+3\Sigma_-\Sigma_+B^2+2f_-\langle \Sigma_-\rangle B^2\nonumber\\
    &+2Du\langle\Sigma_-\rangle B^2+2Du\langle\Sigma_+\rangle B^2+2f_-\langle \Sigma_+\rangle B^2+2f_-\langle \Sigma_-\rangle B^2+u^2\langle\Sigma_-\rangle N^2+u^2\langle\Sigma_+\rangle N^2\,.
\end{align}
\subsection{Operators with More Covariant Derivatives}
\label{more}

The baryon mass does not vanish in the chiral limit, and the heavy baryon formulation can be treated to solve the problem. In the HBChPT~\cite{Krause:1990xc,Jenkins:1990jv}, the baryon sector is projected to the non-relativistic formulation. And the derivatives acting on baryon fields count to $\mathcal{O}(p^0)$ which means the operators in the type with more derivatives may count in the same order as the operators in the type of less derivatives after the heavy baryon projection, for example 
\begin{align}
    (\bar N \gamma^5\gamma^\mu u^\nu\bar N)\langle u_\mu u_\nu\rangle&=(\bar{N_v}S^\mu u^\nu N_v)\langle u_\mu u_\nu \rangle+...\,,\\
    (\bar N \gamma^5\gamma^\mu\lrd^\nu\lrd^\rho u_\nu\bar N)\langle u_\mu u_\rho\rangle&=(\bar{N_v}S^\mu v^\nu v^\rho u_\nu N_v)\langle u_\mu u_\rho \rangle+...\,,
\end{align}
the symbol $...$ stands for higher order in the heavy baryon projection, both of the operators count to $\mathcal{O}(p^3)$, but there are different derivatives in the two operators.

It seems that the number of $\lrd^\mu$ acting on baryon is arbitrary at the fixed order. In fact, there exists the maximum number of these derivatives for each type, following the cases below:
\begin{itemize}
    \item If one operator contains two derivatives contracted applying on the baryons, it should be the same type without the two derivatives. For example \begin{equation}
       (\bar N\lrd^\mu\lrd_\mu\lrd^\nu u_\rho N)\langle u_\nu u^\rho \rangle\rightarrow (\bar N\lrd^\nu u_\rho N)\langle u_\nu u^\rho \rangle\,.
    \end{equation} 
    \item The covariant derivatives $\lrd^\mu \lrd^\nu...$ acting on the baryon field are totally symmetric for the Lorentz indices. The Lorentz structure for the generic operator is
\begin{equation}
    \varepsilon_{\sigma\gamma\alpha\beta}(\Bar{N}\lrd^\mu \lrd^\nu \lrd^\rho \lrd^\lambda\lrd^\sigma... N)(f_{\mu\nu}...)(u_\rho ...)(D_\lambda\Sigma...)\,. 
\end{equation}
Since the Lorentz indices for $f_{\mu\nu}$ and $\varepsilon_{\mu\nu\rho\lambda}$ are antisymmetric, the derivatives contracting with them can not apply on the baryons, which also provides limits on the number of derivatives on the baryons.
    \item In the heavy baryon projection, due to Eq.~\eqref{heavy baryon}, the $\gamma^\mu$ should be equal to $\lrd^\mu$
     and we maintain the $\gamma^\mu$ that the Lorentz indices of the $\gamma^\mu \lrd^\nu$ is considered to be symmetric in our notation. As for the $\bar N\gamma^5 \sigma^{\mu\nu}N$ structure
\begin{equation}
    (\Bar{N}\gamma^5\sigma^{\mu\nu}N)=(\Bar{N}\gamma^5\gamma^\mu\lrd^\nu N-\Bar{N}\gamma^5\gamma^\nu\lrd^\mu N )+...\,,
\end{equation}
this means that the three operators are not independent. We should choose two operators for these and Lorentz indices are antisymmetric in $\sigma^{\mu\nu}$, it will be convenient to treat the $\gamma^5\gamma^\mu \lrd^\nu$ to be symmetric in our notation.
\end{itemize}
Following above, the maximum number of the covariant derivatives acting on the baryons for each type has been listed in Tab.~\ref{ number},


We take one of the $SU(2)$ $\mathcal{O}(p^5)$ type: $fu^3N^2$ for example. The maximum number of Lorentz indices in this type will be
\begin{equation}
    (\Bar{N}\Gamma^{\mu\nu\rho\lambda\sigma}N)(u_\mu u_\nu u_\rho)(f_{\lambda\sigma})\,.
\end{equation}
Due to the antisymmetric Lorentz indices of $f_{\mu\nu}$, the number of the covariant derivatives acting on the baryon field can not be 5. Here, we do not consider the $\varepsilon_{\mu\nu\rho\lambda}$ because of its antisymmetric Lorentz indices which will reduce the number of the covariant derivatives acting on baryons. And the number will not be 4 since the remaining Lorentz indices $\gamma^\mu$ or $\gamma^5\gamma^\mu$ indicates the symmetry that will take the operator to be zero in our notation. So in our notation, the maximum number of the covariant derivatives for type $fu^3N^2$ is 3. Then we can construct the type $Dfu^3N^2$, $D^2fu^3N^2$, $D^3fu^3N^2$ and select the order $\mathcal{O}(p^5)$ operators.

Now we can compare our result with the one in the literature~\cite{Fettes:2000gb}, shown in the App.~\ref{comparison}. Moreover, we give the whole operators lists in the App.~\ref{app:SU2} and~\ref{app:SU3} and show the numbers of operators for each $CP$ eigenstates order by order in Tab.~\ref{op number}.
\begin{table}
    \centering
    \resizebox{\linewidth}{!}{
    \begin{tabular}{|c|c|c|c|c|c|c|c|c|c|}
    \hline
       & $u^2N^2$ & $\Sigma N^2$&$fN^2$&$Du^2N^2$&$u^3N^2$&$u\Sigma N^2$&$ufN^2$&$DfN^2$&$D^2u^2N^2$ \\
        \hline
      number  &1&0&0&2&2&0&1&0&3\\
         \hline
       &$Du^3N^2$&$u^4N^2$ &$Du\Sigma N^2$&$u^2\Sigma N^2$&$DufN^2$&$u^2fN^2$&$f^2N^2$&$\Sigma^2N^2$&$\Sigma fN^2$\\
       \hline
       number &3&3&1&1&2&2&1&0&0\\
         \hline
       & $D^2fN^2$&$D^2\Sigma N^2$& $Df\Sigma N^2$&$D\Sigma^2N^2$&$Df^2N^2$&$D^2fuN^2$&$D^2\Sigma uN^2$&$D^3u^2N^2$&$f^2uN^2$\\
        \hline
       number &0&0&1&0&1&2&2&4&1\\
         \hline
       &$\Sigma^2uN^2$&$f\Sigma uN^2$&$Dfu^2N^2$&$D\Sigma u^2N^2$&$D^2u^3N^2$&$fu^3N^2$&$\Sigma u^3N^2$&$Du^4N^2$&$u^5N^2$\\
       \hline
       number &0&1&3&2&4&3&2&4&4\\
         \hline
    \end{tabular}
    }
    \caption{The maximum number of the additional covariant derivatives acting on the baryons of the types up to $\mathcal{O}(p^5)$.}
    \label{ number}
\end{table}

\begin{table}
    \centering
    \begin{tabular}{|c|c|c|c|c|c|}
    \hline
         &order&$C+P+$&$C+P-$&$C-P+$&$C-P-$  \\
         \hline
         \multirow{4}{*}{$SU(2)$}&$\mathcal{O}(p^2)$&6&4&1&2\\
         &$\mathcal{O}(p^3)$&21&14&9&17\\
         &$\mathcal{O}(p^4)$&95&85&67&73\\
         &$\mathcal{O}(p^5)$&435&387&352&420\\
         \hline
         \multirow{3}{*}{$SU(3)$}&$\mathcal{O}(p^2)$&16&8&3&7\\
         &$\mathcal{O}(p^3)$&78&66&48&70\\
         &$\mathcal{O}(p^4)$&528&475&413&450\\
         \hline
    \end{tabular}
    \caption{The number of operators for each $CP$ eigenstates.}
    \label{op number}
\end{table}

\subsection{Chiral Dim-5 Operator Example}
\label{example}

We illustrate the procedure above by the non-trivial type $Df_+u^2N^2$ which contains the Adler zero condition, the repeated fields, and the EOM. The $f_\pm^{\mu\nu}$ and $\Tilde{f}_\pm^{\mu\nu}$ in the operators can be converted to $F_R^{\mu\nu}$ and $F_L^{\mu\nu}$ form
\begin{equation}
    \begin{aligned}
        &f_{\pm\mu\nu} = F_{\pm L\mu\nu}+F_{\pm R\mu\nu}\,, \\
        &\Tilde{f}_{\pm\mu\nu} = F_{\pm R\mu\nu}-F_{\pm L\mu\nu}\,. \\
    \end{aligned}
\end{equation}
There are four class: $\psi_R\psi_R^\dagger F_{+R}\phi^2 D^3$, $\psi_R\psi_R^\dagger F_{+L}\phi^2 D^3$, $\psi_L\psi_L^\dagger F_{+R}\phi^2 D^3$ and $\psi_L\psi_L^\dagger F_{+L}\phi^2 D^3$ for the type $Df_+u^2N^2$ in the procedure. We first consider the class $\psi_R\psi_R^\dagger F_{+R}\phi^2 D^3$, there are 5 fields and 3 derivatives in the class $\psi_R\psi_R^\dagger F_{+R}\phi^2 D^3$ and their helicities are $\{-1/2,1/2,1,0,0\}$, while the last two Goldstone Bosons are repeated fields. According to Eq.~\eqref{eq:nn}, we obtain that $n=2,\,\tilde n=3$ and the corresponding primary Young diagram is
\begin{equation}
    \ydiagram{5,5,3}\,.
\end{equation}
The numbers of all indices that are used to fill the diagram can be obtained by Eq.~\eqref{eq:numi} that
\begin{equation}
    \# 1=4,\,\quad\#2=2,\,\quad\#3=1,\,\quad\#4=3,\,\quad\#5=3.
\end{equation}
Then we can obtain 11 SSYTs,
\begin{align}
        &\begin{ytableau}
            1&1&1&1&2\\
            2&3&4&4&4\\
            5&5&5
        \end{ytableau}\,,\quad\begin{ytableau}
            1&1&1&1&2\\
            2&3&4&4&5\\
            4&5&5
        \end{ytableau}\,,\quad\begin{ytableau}
            1&1&1&1&2\\
            2&3&4&5&5\\
            4&4&5
        \end{ytableau}\,,\quad\begin{ytableau}
            1&1&1&1&2\\
            2&4&4&4&5\\
            3&5&5
        \end{ytableau}\,,\quad\begin{ytableau}
            1&1&1&1&3\\
            2&2&4&4&5\\
            4&5&5
        \end{ytableau}\,,\quad\begin{ytableau}
            1&1&1&1&3\\
            2&2&4&4&4\\
            5&5&5
        \end{ytableau}\,,\notag\\
       & \begin{ytableau}
            1&1&1&1&3\\
            2&2&4&5&5\\
            4&4&5
        \end{ytableau}\,,\quad\begin{ytableau}
            1&1&1&1&4\\
            2&2&3&4&5\\
            4&5&5
        \end{ytableau}\,,\quad\begin{ytableau}
            1&1&1&1&4\\
            2&2&3&5&5\\
            4&4&5
        \end{ytableau}\,,\quad\begin{ytableau}
            1&1&1&1&4\\
            2&2&4&5&5\\
            3&4&5
        \end{ytableau}\,,\quad\begin{ytableau}
            1&1&1&1&4\\
            2&2&4&4&5\\
            3&5&5
        \end{ytableau}\,,
\end{align}
and the corresponding Lorentz y-basis are
\begin{align}
    \mathcal{M}_1&=\langle14\rangle\langle24\rangle[23][24][34]\,,\notag\\
    \mathcal{M}_2&=\langle14\rangle\langle25\rangle[23][24][35]\,,\notag\\
    \mathcal{M}_3&=\langle15\rangle\langle25\rangle[23][25][35]\,,\notag\\
    \mathcal{M}_4&=\langle14\rangle\langle25\rangle{[23]}^2[45]\,,\notag\\
    \mathcal{M}_5&=\langle14\rangle\langle35\rangle[23][34][35]\,,\notag\\
    \mathcal{M}_6&=\langle14\rangle\langle34\rangle[23]{[34]}^2\,,\notag\\
    \mathcal{M}_7&=\langle15\rangle\langle35\rangle[23]{[35]}^2\,,\notag\\
    \mathcal{M}_8&=\langle14\rangle\langle45\rangle[24][34][35]\,,\notag\\
    \mathcal{M}_9&=\langle14\rangle\langle45\rangle[24]{[35]}^2\,,\notag\\
    \mathcal{M}_{10}&=\langle15\rangle\langle45\rangle[23][35][45]\,,\notag\\
    \mathcal{M}_{11}&=\langle14\rangle\langle45\rangle[23][24][45]\,.
\end{align}

Furthermore, the Adler zero condition should be considered that the amplitudes should be zero whenever particle 4 or 5 becomes soft. Thus there are 7 bases that satisfy the Adler zero condition. According to Eq.~\eqref{define}, we can translate this Lorentz y-basis to the form of operators and choose the monomials as the Lorentz m-basis and more details can be found in the Ref.~\cite{Li:2020gnx,Li:2020xlh,Li:2022tec}. The Adler zero condition and translating can be summed up as a $7\times11$ transformation matrix 
\begin{equation}
    \left(\begin{array}{ccccccccccc}
         -\frac{1}{4}&\frac{1}{4}&-\frac{1}{4}&\frac{1}{4}&-\frac{1}{4}&\frac{1}{4}&0&\frac{1}{4}&-\frac{1}{4}&\frac{1}{4}&-\frac{1}{4}  \\
         0&\frac{1}{4}&0&-\frac{1}{4}&0&\frac{1}{4}&\frac{1}{4}&0&0&&-\frac{1}{4}  \\
         0&-\frac{1}{4}&0&0&\frac{1}{4}&-\frac{1}{4}&-\frac{1}{4}&0&0&-\frac{1}{4}&\frac{1}{4}  \\
         0&-\frac{1}{4}&0&0&\frac{1}{4}&-\frac{1}{4}&0&0&0&-\frac{1}{4}&\frac{1}{4}  \\
         0&0&0&0&0&0&0&0&0&\frac{1}{4}&0  \\
         0&\frac{1}{4}&0&0&-\frac{1}{4}&\frac{1}{4}&0&0&0&0&\frac{1}{4}  \\
         \frac{1}{4}&-\frac{1}{4}&-\frac{1}{4}&0&0&\frac{1}{4}&0&-\frac{1}{4}&\frac{1}{4}&-\frac{1}{4}&\frac{1}{4}  
    \end{array}\right)\,,
\end{equation}
and then the Lorentz m-basis can be obtained
\begin{align}
    \mathcal{B}_{1R}&=\psi_{R1}\Bar{\sigma}^\mu D^\lambda\psi_{R2}^\dagger F_{R3}^{\mu\nu}D_\nu \phi_4 D_\lambda\phi_5\,,\notag\\
    \mathcal{B}_{2R}&=\psi_{R1}\Bar{\sigma}^\mu D^\lambda\psi_{R2}^\dagger F_{R3}^{\mu\nu}D_\lambda \phi_4 D_\nu\phi_5\,,\notag\\
    \mathcal{B}_{3R}&=\psi_{R1}\Bar{\sigma}^\mu\psi_{R2}^\dagger F_{R3}^{\mu\nu}D_\nu D^\lambda\phi_4 D_\lambda\phi_5\,,\notag\\
    \mathcal{B}_{4R}&=\psi_{R1}\Bar{\sigma}^\mu\psi_{R2}^\dagger F_{R3}^{\mu\nu}D^\lambda \phi_4 D_\nu D_\lambda\phi_5\,,\notag\\
    \mathcal{B}_{5R}&=\psi_{R1}\Bar{\sigma}^\mu \psi_{R2}^\dagger D^\lambda F_{R3}^{\mu\nu}D_\nu \phi_4 D_\lambda\phi_5\,,\notag\\
    \mathcal{B}_{6R}&=\psi_{R1}\Bar{\sigma}^\mu \psi_{R2}^\dagger F_{R3}^{\nu\lambda}D_\nu D_\mu\phi_4 D_\lambda\phi_5\,,\notag\\
    \mathcal{B}_{7R}&=\psi_{R1}\Bar{\sigma}^\mu \psi_{R2}^\dagger F_{R3}^{\nu\lambda}D_\lambda\phi_4 D_\nu D_\mu\phi_5\,.
\end{align}
Moreover, there are 5 fields and 3 derivates in the class $\psi_R\psi_R^\dagger F_{+L}\phi^2 D^3$, and their helicities are $\{-1/2,1/2,-1,0,0\}$. Thus the Lorentz m-basis of $\psi_R\psi_R^\dagger F_{+L}\phi^2 D^3$ are similar to the class $\psi_R\psi_R^\dagger F_{+R}\phi^2 D^3$
\begin{align}
    \mathcal{B}_{1L}&=\psi_{R1}\Bar{\sigma}^\mu D^\lambda\psi_{R2}^\dagger F_{L3}^{\mu\nu}D_\nu \phi_4 D_\lambda\phi_5\,,\notag\\
    \mathcal{B}_{2L}&=\psi_{R1}\Bar{\sigma}^\mu D^\lambda\psi_{R2}^\dagger F_{L3}^{\mu\nu}D_\lambda \phi_4 D_\nu\phi_5\,,\notag\\
    \mathcal{B}_{3L}&=\psi_{R1}\Bar{\sigma}^\mu\psi_{R2}^\dagger F_{L3}^{\mu\nu}D_\nu D^\lambda\phi_4 D_\lambda\phi_5\,,\notag\\
    \mathcal{B}_{4L}&=\psi_{R1}\Bar{\sigma}^\mu\psi_{R2}^\dagger F_{L3}^{\mu\nu}D^\lambda \phi_4 D_\nu D_\lambda\phi_5\,,\notag\\
    \mathcal{B}_{5L}&=\psi_{R1}\Bar{\sigma}^\mu \psi_{R2}^\dagger D^\lambda F_{L3}^{\mu\nu}D_\nu \phi_4 D_\lambda\phi_5\,,\notag\\
    \mathcal{B}_{6L}&=\psi_{R1}\Bar{\sigma}^\mu \psi_{R2}^\dagger F_{L3}^{\nu\lambda}D_\nu D_\mu\phi_4 D_\lambda\phi_5\,,\notag\\
    \mathcal{B}_{7L}&=\psi_{R1}\Bar{\sigma}^\mu \psi_{R2}^\dagger F_{L3}^{\nu\lambda}D_\lambda\phi_4 D_\nu D_\mu\phi_5\,.
\end{align}

In addition, the EOMs for external source $F_R$ should not be considered while the $F_L$ is considered due to the discussion in Sec.~\ref{young}, and more details can be found in Ref.~\cite{Ren:2022tvi}. For the Goldstone field $\phi$, there is no such constraint. Therefore, the Lorentz m-basis for the $\psi_R\psi_R^\dagger F_{+L}\phi^2 D^3$ will not be changed and the $\psi_R\psi_R^\dagger F_{+R}\phi^2 D^3$ Lorentz m-basis becomes
\begin{align}
    \mathcal{B}_{1R}&=\psi_{R1}\Bar{\sigma}^\mu D^\lambda\psi_{R2}^\dagger F_{R3}^{\mu\nu}D_\nu \phi_4 D_\lambda\phi_5\,,\notag\\
    \mathcal{B}_{2R}&=\psi_{R1}\Bar{\sigma}^\mu D^\lambda\psi_{R2}^\dagger F_{R3}^{\mu\nu}D_\lambda \phi_4 D_\nu\phi_5\,,\notag\\
    \mathcal{B}_{3R}&=\psi_{R1}\Bar{\sigma}^\mu\psi_{R2}^\dagger F_{R3}^{\mu\nu}D_\nu D^\lambda\phi_4 D_\lambda\phi_5\,,\notag\\
    \mathcal{B}_{4R}&=\psi_{R1}\Bar{\sigma}^\mu\psi_{R2}^\dagger F_{R3}^{\mu\nu}D^\lambda \phi_4 D_\nu D_\lambda\phi_5\,,\notag\\
    \mathcal{B}_{5R}&=\psi_{R1}\Bar{\sigma}^\mu \psi_{R2}^\dagger D^\lambda F_{R3}^{\mu\nu}D_\nu \phi_4 D_\lambda\phi_5\,,\notag\\
    \mathcal{B}_{6R}&=\psi_{R1}\Bar{\sigma}^\mu \psi_{R2}^\dagger F_{R3}^{\nu\lambda}D_\nu D_\mu\phi_4 D_\lambda\phi_5\,,\notag\\
    \mathcal{B}_{7R}&=\psi_{R1}\Bar{\sigma}^\mu \psi_{R2}^\dagger F_{R3}^{\nu\lambda}D_\lambda\phi_4 D_\nu D_\mu\phi_5\,,\notag\\
    \mathcal{B}_{8R}&=\psi_{R1}\Bar{\sigma}^\mu \psi_{R2}^\dagger D^\lambda F_{R3}^{\mu\nu}D_\lambda \phi_4 D_\nu\phi_5\,,\notag\\
    \mathcal{B}_{9R}&=\psi_{R1}\Bar{\sigma}^\mu \psi_{R2}^\dagger D_\nu F_{R3}^{\mu\nu}D_\lambda \phi_4 D^\lambda\phi_5\,,\notag\\
    \mathcal{B}_{10R}&=\psi_{R1}\Bar{\sigma}^\mu \psi_{R2}^\dagger D_\nu F_{R3}^{\nu\lambda} D_\mu\phi_4 D_\lambda\phi_5\,,\notag\\
    \mathcal{B}_{11R}&=\psi_{R1}\Bar{\sigma}^\mu \psi_{R2}^\dagger D_\nu F_{R3}^{\nu\lambda} D_\lambda\phi_4 D_\mu\phi_5\,.
\end{align}
The number of operators can be cross-checked by the Hilbert series.
The other two class $\psi_L\psi_L^\dagger F_{+R}\phi^2 D^3$ and $\psi_L\psi_L^\dagger F_{+L}\phi^2 D^3$ would be trivial to obtain by replacing $\psi_R\psi_R^\dagger$ with $\psi_L\psi_L^\dagger$.

As for the internal flavor y-basis, the $SU(2)$ tensors can be obtained by similar SSYT techniques in the $SU(2)$ case. For this type, there are 4 flavor SSYTs
\begin{align}
    \begin{ytableau}
        i_1&i_2&i_3&j_3\\
        i_4&j_4&i_5&j_5
    \end{ytableau}\,,\quad\begin{ytableau}
        i_1&i_2&i_3&i_4\\
        j_3&j_4&i_5&j_5
    \end{ytableau}\,,\quad\begin{ytableau}
        i_1&i_3&j_3&i_4\\
        i_2&j_4&i_5&j_5
    \end{ytableau}\,,\quad\begin{ytableau}
        i_1&i_2&j_3&i_4\\
        i_3&j_4&i_5&j_5
    \end{ytableau}\,.
\end{align}
We can convert them to obtain the flavor m-basis
\begin{align}
    \mathcal{T}_1^m&=\delta^{i_4}_{i_5} \epsilon^{IJK}\,,\notag\\
    \mathcal{T}_2^m&=\delta^{IJ}\delta^{KL}{(\tau^L)}^{i_4}_{i_5}\,,\notag\\
    \mathcal{T}_3^m&=\delta^{IK}\delta^{JL}{(\tau^L)}^{i_4}_{i_5}\,,\notag\\
    \mathcal{T}_4^m&=\delta^{IL}\delta^{JK}{(\tau^L)}^{i_4}_{i_5}\,.\\
\end{align}
And for the $SU(3)$ case, there are 32 flavor SSYTs and we can also obtain the flavor m-basis through SSYT techniques
\begin{equation}
\begin{aligned}
&\delta^{DE} d^{ABC},\, d^{ACF} d^{BFG} d^{DEG},\, d^{ACF} d^{BEG} d^{DFG},\, f^{ACG} d^{BFG} d^{DEF},\,
\\
&d^{ACF} d^{BFG} f^{DEG},\, d^{ACF} f^{BEG} d^{DFG},\, f^{ACG} d^{BEF} d^{DFG},\, d^{ACF} d^{BEG} f^{DFG},\,
\\
&f^{ACF} d^{BFG} f^{DEG},\, f^{ACF} f^{BEG} d^{DFG},\, d^{ACF} f^{BFG} f^{DEG},\, f^{ACF} f^{BEG} f^{DEG},\,
\\
&f^{ACG} d^{BEF} f^{DFG},\, f^{ACF} f^{BFG} f^{DEG},\, f^{ACF} f^{BEG} f^{DFG},\, \delta^{AC} d^{BDE},\,
\\
&\delta^{BE} f^{ACD},\, d^{ABF} d^{CFG} d^{DEG},\, d^{ABF} d^{CFG} f^{DEG},\, d^{ABF} f^{CFG} f^{DEG},\,
\\
&f^{ABF} f^{CFG} f^{DEG},\, d^{ABF} d^{CEG} d^{DFG},\, f^{ABG} d^{CFG} d^{DFG},\, f^{ABG} d^{CEF} d^{DFG},\,
\\
&d^{ABF} f^{CEG} f^{DFG},\, f^{ABG} f^{CFG} d^{DEF},\, f^{ABF} f^{CEG} f^{DFG},\, \delta^{AC} f^{BDE},\, 
\\
&d^{ABF} f^{CEG} d^{DFG},\, f^{ABF} f^{CEG} d^{DFG},\, d^{ADF} d^{BFG} d^{CEG},\, f^{ADG} d^{BFG} d^{CEF},\,
\\
&f^{ADG} d^{BFG} d^{DEF},\,
\end{aligned}
\end{equation}
in which the $f^{ABC}$ and the $d^{ABC}$ are defined in the $SU(3)$ group 
\begin{equation} \label{eq: SU3_Structure_Constant}
\begin{aligned}
    f^{A B C}=\frac{1}{4i}Tr([\lambda^A,\lambda^B]\lambda^C) \,,\quad
    d^{A B C}=\frac{1}{4}Tr(\{\lambda^A,\lambda^B\}\lambda^C) \,.
\end{aligned}
\end{equation}
The tensor product of the Lorentz m-basis and the flavor m-basis gives $2\times(7+11)\times4=144$ in the $SU(2)$ case and $2\times(7+11)\times32=1152$ in the $SU(3)$ case for the whole type $Df_+u^2N^2$, but they are not all physical when considering the repeated fields.

The Goldstone Bosons $\phi_4$ and $\phi_5$ are the repeated field in this type, thus only the operators symmetric under the permutations of them are physical. In the Lorentz space, the generator of $S_2$ in the m-basis for the  $\psi_R\psi_R^\dagger F_{+R}\phi^2 D^3$ is
\begin{equation}
\mathcal{D}_{\mathcal{B}L}=\left(
\begin{array}{ccccccccccc}
 0 & 0 & 0 & 1 & 0 & 0 & 0 & 0 & 0 & 0 & 0 \\
 0 & 0 & -1 & 0 & 0 & 0 & 0 & 0 & 0 & 0 & 0 \\
 0 & -1 & 0 & 0 & 0 & 0 & 0 & 0 & 0 & 0 & 0 \\
 1 & 0 & 0 & 0 & 0 & 0 & 0 & 0 & 0 & 0 & 0 \\
 0 & 0 & 0 & 0 & 0 & 0 & 1 & 0 & 0 & 0 & 0 \\
 0 & 0 & 0 & 0 & 0 & 1 & 0 & 0 & 0 & 0 & 0 \\
 0 & 0 & 0 & 0 & 1 & 0 & 0 & 0 & 0 & 0 & 0 \\
 0 & 0 & 0 & 0 & 0 & 0 & 0 & 0 & 1 & 0 & 0 \\
 0 & 0 & 0 & 0 & 0 & 0 & 0 & 1 & 0 & 0 & 0 \\
 0 & 0 & 0 & 0 & 0 & 0 & 0 & 0 & 0 & 0 & 1 \\
 0 & 0 & 0 & 0 & 0 & 0 & 0 & 0 & 0 & 1 & 0 \\
\end{array}
\right).    
\end{equation}
And similarly, the generator in the Lorentz space for the $\psi_R\psi_R^\dagger F_{+L}\phi^2 D^3$ is
\begin{equation}
\mathcal{D}_{\mathcal{B}R}=\left(\begin{array}{ccccccc}
     0&0&0&1&0&0&0  \\
     0&0&-1&0&0&0&0  \\
     0&-1&0&0&0&0&0  \\
     1&0&0&0&0&0&0  \\
     0&1&1&0&1&0&0  \\
     0&0&0&0&0&0&1  \\
     0&0&0&0&0&1&0  \\
\end{array}\right)\,.    
\end{equation}
For $SU(2)$ case, generators of $S_2$ in the m-basis is
\begin{equation}
\mathcal{D}_{SU(2)}=\left(\begin{array}{ccccccccc}
     -1&0&0&0\\
     0&1&0&0 \\
     0&0&0&1 \\
     0&0&1&0 \\
\end{array}\right)    
\end{equation}
And for the $SU(3)$ case, the generator is
\begin{equation}
    \mathcal{D}_{SU(3)}=\left(
\begin{matrix}
\begin{smallmatrix}
 1 & 0 & 0 & 0 & 0 & 0 & 0 & 0 & 0 & 0 & 0 & 0 & 0 & 0 & 0 & 0 & 0 & 0 & 0 & 0 & 0 & 0 & 0 & 0 & 0 & 0 & 0 & 0 & 0 & 0 & 0 & 0 \\
 0 & 1 & 0 & 0 & 0 & 0 & 0 & 0 & 0 & 0 & 0 & 0 & 0 & 0 & 0 & 0 & 0 & 0 & 0 & 0 & 0 & 0 & 0 & 0 & 0 & 0 & 0 & 0 & 0 & 0 & 0 & 0 \\
 \frac{2}{3} & 0 & -1 & 0 & 0 & 0 & 0 & 0 & 0 & 0 & -\frac{1}{3} & \frac{2}{3} & 0 & 0 & 0 & 0 & 0 & 0 & 0 & 0 & 0 & 0 & 0 & 0 & 0 & 0 & 0 & 0 & 0 & 0 & 0 & 0 \\
 0 & 0 & 0 & 1 & 0 & 0 & 0 & 0 & 0 & 0 & 0 & 0 & 0 & 0 & 0 & 0 & 0 & 0 & 0 & 0 & 0 & 0 & 0 & 0 & 0 & 0 & 0 & 0 & 0 & 0 & 0 & 0 \\
 0 & 0 & 0 & 0 & -1 & 0 & 0 & 0 & 0 & 0 & 0 & 0 & 0 & 0 & 0 & 0 & 0 & 0 & 0 & 0 & 0 & 0 & 0 & 0 & 0 & 0 & 0 & 0 & 0 & 0 & 0 & 0 \\
 0 & 0 & 0 & 0 & 1 & 0 & 0 & 1 & 0 & 0 & 0 & 0 & 0 & 0 & 0 & 0 & 0 & 0 & 0 & 0 & 0 & 0 & 0 & 0 & 0 & 0 & 0 & 0 & 0 & 0 & 0 & 0 \\
 0 & 0 & 0 & -1 & 0 & 0 & 0 & 0 & 0 & 0 & 0 & 0 & 0 & \frac{2}{3} & -\frac{1}{3} & 0 & \frac{2}{3} & 0 & 0 & 0 & 0 & 0 & 0 & 0 & 0 & 0 & 0 & 0 & 0 & 0 & 0 & 0 \\
 0 & 0 & 0 & 0 & 1 & 1 & 0 & 0 & 0 & 0 & 0 & 0 & 0 & 0 & 0 & 0 & 0 & 0 & 0 & 0 & 0 & 0 & 0 & 0 & 0 & 0 & 0 & 0 & 0 & 0 & 0 & 0 \\
 0 & 0 & 0 & 0 & 0 & 0 & 0 & 0 & -1 & 0 & 0 & 0 & 0 & 0 & 0 & 0 & 0 & 0 & 0 & 0 & 0 & 0 & 0 & 0 & 0 & 0 & 0 & 0 & 0 & 0 & 0 & 0 \\
 0 & 0 & 0 & 0 & 0 & 0 & 0 & 0 & 1 & 0 & 0 & 0 & -1 & 0 & 0 & 0 & 0 & 0 & 0 & 0 & 0 & 0 & 0 & 0 & 0 & 0 & 0 & 0 & 0 & 0 & 0 & 0 \\
 0 & 0 & 0 & 0 & 0 & 0 & 0 & 0 & 0 & 0 & -1 & 0 & 0 & 0 & 0 & 0 & 0 & 0 & 0 & 0 & 0 & 0 & 0 & 0 & 0 & 0 & 0 & 0 & 0 & 0 & 0 & 0 \\
 0 & 0 & 0 & 0 & 0 & 0 & 0 & 0 & 0 & 0 & -1 & 1 & 0 & 0 & 0 & 0 & 0 & 0 & 0 & 0 & 0 & 0 & 0 & 0 & 0 & 0 & 0 & 0 & 0 & 0 & 0 & 0 \\
 0 & 0 & 0 & 0 & 0 & 0 & 0 & 0 & -1 & -1 & 0 & 0 & 0 & 0 & 0 & 0 & 0 & 0 & 0 & 0 & 0 & 0 & 0 & 0 & 0 & 0 & 0 & 0 & 0 & 0 & 0 & 0 \\
 0 & 0 & 0 & 0 & 0 & 0 & 0 & 0 & 0 & 0 & 0 & 0 & 0 & -1 & 0 & 0 & 0 & 0 & 0 & 0 & 0 & 0 & 0 & 0 & 0 & 0 & 0 & 0 & 0 & 0 & 0 & 0 \\
 0 & 0 & 0 & 0 & 0 & 0 & 0 & 0 & 0 & 0 & 0 & 0 & 0 & -1 & 1 & 0 & 0 & 0 & 0 & 0 & 0 & 0 & 0 & 0 & 0 & 0 & 0 & 0 & 0 & 0 & 0 & 0 \\
 0 & 0 & 0 & 0 & 0 & 0 & 0 & 0 & 0 & 0 & 0 & 0 & 0 & 0 & 0 & 1 & 0 & 0 & 0 & 0 & 0 & 0 & 0 & 0 & 0 & 0 & 0 & 0 & 0 & 0 & 0 & 0 \\
 0 & 0 & 0 & \frac{3}{2} & 0 & 0 & \frac{3}{2} & 0 & 0 & 0 & 0 & 0 & 0 & \frac{1}{2} & \frac{1}{2} & 0 & 0 & 0 & 0 & 0 & 0 & 0 & 0 & 0 & 0 & 0 & 0 & 0 & 0 & 0 & 0 & 0 \\
 0 & 0 & 0 & 0 & 0 & 0 & 0 & 0 & 0 & 0 & 0 & 0 & 0 & 0 & 0 & 0 & 0 & 1 & 0 & 0 & 0 & 0 & 0 & 0 & 0 & 0 & 0 & 0 & 0 & 0 & 0 & 0 \\
 0 & 0 & 0 & 0 & 0 & 0 & 0 & 0 & 0 & 0 & 0 & 0 & 0 & 0 & 0 & 0 & 0 & 0 & -1 & 0 & 0 & 0 & 0 & 0 & 0 & 0 & 0 & 0 & 0 & 0 & 0 & 0 \\
 0 & 0 & 0 & 0 & 0 & 0 & 0 & 0 & 0 & 0 & 0 & 0 & 0 & 0 & 0 & 0 & 0 & 0 & 0 & -1 & 0 & 0 & 0 & 0 & 0 & 0 & 0 & 0 & 0 & 0 & 0 & 0 \\
 0 & 0 & 0 & 0 & 0 & 0 & 0 & 0 & 0 & 0 & 0 & 0 & 0 & 0 & 0 & 0 & 0 & 0 & 0 & 0 & -1 & 0 & 0 & 0 & 0 & 0 & 0 & 0 & 0 & 0 & 0 & 0 \\
 \frac{2}{3} & 0 & 0 & 0 & 0 & 0 & 0 & 0 & 0 & 0 & 0 & 0 & 0 & 0 & 0 & 0 & 0 & 0 & 0 & -\frac{1}{3} & 0 & -1 & 0 & 0 & \frac{2}{3} & 0 & 0 & 0 & 0 & 0 & 0 & 0 \\
 0 & 0 & 0 & 0 & 0 & 0 & 0 & 0 & 0 & 0 & 0 & 0 & 0 & 0 & 0 & 0 & 0 & 0 & 0 & 0 & 0 & 0 & 1 & 0 & 0 & 0 & 0 & 0 & 0 & 0 & 0 & 0 \\
 0 & 0 & 0 & 1 & 0 & 0 & -1 & 0 & 0 & 0 & 0 & 0 & 0 & -1 & 1 & 0 & -\frac{2}{3} & 0 & 0 & 0 & -\frac{1}{3} & 0 & 0 & -1 & 0 & 0 & \frac{2}{3} & 0 & 0 & 0 & 0 & 0 \\
 0 & 0 & 0 & 0 & 0 & 0 & 0 & 0 & 0 & 0 & 0 & 0 & 0 & 0 & 0 & 0 & 0 & 0 & 0 & -1 & 0 & 0 & 0 & 0 & 1 & 0 & 0 & 0 & 0 & 0 & 0 & 0 \\
 0 & 0 & 0 & 0 & 0 & 0 & 0 & 0 & 0 & 0 & 0 & 0 & 0 & 0 & 0 & 0 & 0 & 0 & 0 & 0 & 0 & 0 & 0 & 0 & 0 & 1 & 0 & 0 & 0 & 0 & 0 & 0 \\
 0 & 0 & 0 & 0 & 0 & 0 & 0 & 0 & 0 & 0 & 0 & 0 & 0 & 0 & 0 & 0 & 0 & 0 & 0 & 0 & -1 & 0 & 0 & 0 & 0 & 0 & 1 & 0 & 0 & 0 & 0 & 0 \\
 0 & 0 & 0 & 0 & 0 & 0 & 0 & 0 & 0 & 0 & 0 & 0 & 0 & 0 & 0 & 0 & 0 & 0 & 0 & 0 & 0 & 0 & 0 & 0 & 0 & 0 & 0 & -1 & 0 & 0 & 0 & 0 \\
 0 & 0 & 0 & 1 & -1 & -1 & 0 & -1 & 0 & 0 & 0 & 0 & 0 & 0 & 0 & 0 & 0 & 0 & 0 & 0 & 0 & 0 & 1 & 0 & 0 & 0 & 0 & 0 & -1 & 0 & 0 & 0 \\
 0 & 0 & 0 & 0 & 0 & 0 & 0 & 0 & 0 & 0 & 0 & 0 & 0 & 0 & 0 & 0 & 0 & 0 & 0 & 0 & 0 & 0 & 0 & 0 & 0 & 1 & 0 & 0 & 0 & -1 & 0 & 0 \\
 0 & 0 & -\frac{5}{4} & 0 & 0 & 0 & 0 & 0 & \frac{11}{12} & \frac{1}{12} & -\frac{1}{12} & \frac{5}{12} & \frac{1}{12} & 0 & 0 & 0 & 0 & 0 & 0 & \frac{13}{12} & 0 & \frac{5}{4} & 0 & 0 & -\frac{5}{12} & \frac{5}{12} & 0 & 0 & 0 & -\frac{5}{6} & 1 & 0 \\
 0 & 0 & 0 & -\frac{3}{4} & -\frac{3}{4} & -\frac{1}{2} & \frac{1}{4} & 0 & 0 & 0 & 0 & 0 & 0 & \frac{1}{2} & -\frac{3}{4} & 0 & \frac{2}{3} & 0 & -\frac{5}{4} & 0 & \frac{1}{3} & 0 & \frac{1}{4} & \frac{5}{4} & 0 & 0 & -\frac{5}{12} & \frac{1}{3} & -\frac{1}{2} & 0 & 0 & 1 \\
 \end{smallmatrix}
\end{matrix}
\right)
\end{equation}\,,
thus the generator of the overall operator space is the tensor product of them
\begin{equation}
    \mathcal{D}=\mathcal{D}_{SU(2)/SU(3)}\otimes\mathcal{D}_{\mathcal{B}(L/R)}\,.
\end{equation}

The idempotent element of the subspace is symmetric under the permutations of the $\phi_4$ and the $\phi_5$, which is symbolized by the young diagram \ydiagram{2}
\begin{equation}
    \mathcal{Y}(\young(45))=I+\mathcal{D}\,,
\end{equation}
in which $I$ is the identity matrix. We take the $\mathcal{D}=\mathcal{D}_{SU(2)}\otimes \mathcal{D}_{\mathcal{B}L}$ for example, 
\begin{equation}
\mathcal{Y}(\young(45))=
\left(\begin{matrix}
\begin{smallmatrix}
 1 & 0 & 0 & -1 & 0 & 0 & 0 & 0 & 0 & 0 & 0 & 0 & 0 & 0 & 0 & 0 & 0 & 0 & 0 & 0 & 0 & 0 & 0 & 0 & 0 & 0 & 0 & 0 & 0 & 0 & 0 & 0 & 0 & 0 & 0 & 0 & 0 & 0 & 0 & 0 & 0 & 0 & 0 & 0 \\
 0 & 1 & 1 & 0 & 0 & 0 & 0 & 0 & 0 & 0 & 0 & 0 & 0 & 0 & 0 & 0 & 0 & 0 & 0 & 0 & 0 & 0 & 0 & 0 & 0 & 0 & 0 & 0 & 0 & 0 & 0 & 0 & 0 & 0 & 0 & 0 & 0 & 0 & 0 & 0 & 0 & 0 & 0 & 0 \\
 0 & 1 & 1 & 0 & 0 & 0 & 0 & 0 & 0 & 0 & 0 & 0 & 0 & 0 & 0 & 0 & 0 & 0 & 0 & 0 & 0 & 0 & 0 & 0 & 0 & 0 & 0 & 0 & 0 & 0 & 0 & 0 & 0 & 0 & 0 & 0 & 0 & 0 & 0 & 0 & 0 & 0 & 0 & 0 \\
 -1 & 0 & 0 & 1 & 0 & 0 & 0 & 0 & 0 & 0 & 0 & 0 & 0 & 0 & 0 & 0 & 0 & 0 & 0 & 0 & 0 & 0 & 0 & 0 & 0 & 0 & 0 & 0 & 0 & 0 & 0 & 0 & 0 & 0 & 0 & 0 & 0 & 0 & 0 & 0 & 0 & 0 & 0 & 0 \\
 0 & 0 & 0 & 0 & 1 & 0 & -1 & 0 & 0 & 0 & 0 & 0 & 0 & 0 & 0 & 0 & 0 & 0 & 0 & 0 & 0 & 0 & 0 & 0 & 0 & 0 & 0 & 0 & 0 & 0 & 0 & 0 & 0 & 0 & 0 & 0 & 0 & 0 & 0 & 0 & 0 & 0 & 0 & 0 \\
 0 & 0 & 0 & 0 & 0 & 0 & 0 & 0 & 0 & 0 & 0 & 0 & 0 & 0 & 0 & 0 & 0 & 0 & 0 & 0 & 0 & 0 & 0 & 0 & 0 & 0 & 0 & 0 & 0 & 0 & 0 & 0 & 0 & 0 & 0 & 0 & 0 & 0 & 0 & 0 & 0 & 0 & 0 & 0 \\
 0 & 0 & 0 & 0 & -1 & 0 & 1 & 0 & 0 & 0 & 0 & 0 & 0 & 0 & 0 & 0 & 0 & 0 & 0 & 0 & 0 & 0 & 0 & 0 & 0 & 0 & 0 & 0 & 0 & 0 & 0 & 0 & 0 & 0 & 0 & 0 & 0 & 0 & 0 & 0 & 0 & 0 & 0 & 0 \\
 0 & 0 & 0 & 0 & 0 & 0 & 0 & 1 & -1 & 0 & 0 & 0 & 0 & 0 & 0 & 0 & 0 & 0 & 0 & 0 & 0 & 0 & 0 & 0 & 0 & 0 & 0 & 0 & 0 & 0 & 0 & 0 & 0 & 0 & 0 & 0 & 0 & 0 & 0 & 0 & 0 & 0 & 0 & 0 \\
 0 & 0 & 0 & 0 & 0 & 0 & 0 & -1 & 1 & 0 & 0 & 0 & 0 & 0 & 0 & 0 & 0 & 0 & 0 & 0 & 0 & 0 & 0 & 0 & 0 & 0 & 0 & 0 & 0 & 0 & 0 & 0 & 0 & 0 & 0 & 0 & 0 & 0 & 0 & 0 & 0 & 0 & 0 & 0 \\
 0 & 0 & 0 & 0 & 0 & 0 & 0 & 0 & 0 & 1 & -1 & 0 & 0 & 0 & 0 & 0 & 0 & 0 & 0 & 0 & 0 & 0 & 0 & 0 & 0 & 0 & 0 & 0 & 0 & 0 & 0 & 0 & 0 & 0 & 0 & 0 & 0 & 0 & 0 & 0 & 0 & 0 & 0 & 0 \\
 0 & 0 & 0 & 0 & 0 & 0 & 0 & 0 & 0 & -1 & 1 & 0 & 0 & 0 & 0 & 0 & 0 & 0 & 0 & 0 & 0 & 0 & 0 & 0 & 0 & 0 & 0 & 0 & 0 & 0 & 0 & 0 & 0 & 0 & 0 & 0 & 0 & 0 & 0 & 0 & 0 & 0 & 0 & 0 \\
 0 & 0 & 0 & 0 & 0 & 0 & 0 & 0 & 0 & 0 & 0 & 1 & 0 & 0 & 0 & 0 & 0 & 0 & 0 & 0 & 0 & 0 & 0 & 0 & 0 & 1 & 0 & 0 & 0 & 0 & 0 & 0 & 0 & 0 & 0 & 0 & 0 & 0 & 0 & 0 & 0 & 0 & 0 & 0 \\
 0 & 0 & 0 & 0 & 0 & 0 & 0 & 0 & 0 & 0 & 0 & 0 & 1 & 0 & 0 & 0 & 0 & 0 & 0 & 0 & 0 & 0 & 0 & 0 & -1 & 0 & 0 & 0 & 0 & 0 & 0 & 0 & 0 & 0 & 0 & 0 & 0 & 0 & 0 & 0 & 0 & 0 & 0 & 0 \\
 0 & 0 & 0 & 0 & 0 & 0 & 0 & 0 & 0 & 0 & 0 & 0 & 0 & 1 & 0 & 0 & 0 & 0 & 0 & 0 & 0 & 0 & 0 & -1 & 0 & 0 & 0 & 0 & 0 & 0 & 0 & 0 & 0 & 0 & 0 & 0 & 0 & 0 & 0 & 0 & 0 & 0 & 0 & 0 \\
 0 & 0 & 0 & 0 & 0 & 0 & 0 & 0 & 0 & 0 & 0 & 0 & 0 & 0 & 1 & 0 & 0 & 0 & 0 & 0 & 0 & 0 & 1 & 0 & 0 & 0 & 0 & 0 & 0 & 0 & 0 & 0 & 0 & 0 & 0 & 0 & 0 & 0 & 0 & 0 & 0 & 0 & 0 & 0 \\
 0 & 0 & 0 & 0 & 0 & 0 & 0 & 0 & 0 & 0 & 0 & 0 & 0 & 0 & 0 & 1 & 0 & 0 & 0 & 0 & 0 & 0 & 0 & 0 & 0 & 0 & 0 & 0 & 1 & 0 & 0 & 0 & 0 & 0 & 0 & 0 & 0 & 0 & 0 & 0 & 0 & 0 & 0 & 0 \\
 0 & 0 & 0 & 0 & 0 & 0 & 0 & 0 & 0 & 0 & 0 & 0 & 0 & 0 & 0 & 0 & 1 & 0 & 0 & 0 & 0 & 0 & 0 & 0 & 0 & 0 & 0 & 1 & 0 & 0 & 0 & 0 & 0 & 0 & 0 & 0 & 0 & 0 & 0 & 0 & 0 & 0 & 0 & 0 \\
 0 & 0 & 0 & 0 & 0 & 0 & 0 & 0 & 0 & 0 & 0 & 0 & 0 & 0 & 0 & 0 & 0 & 1 & 0 & 0 & 0 & 0 & 0 & 0 & 0 & 0 & 1 & 0 & 0 & 0 & 0 & 0 & 0 & 0 & 0 & 0 & 0 & 0 & 0 & 0 & 0 & 0 & 0 & 0 \\
 0 & 0 & 0 & 0 & 0 & 0 & 0 & 0 & 0 & 0 & 0 & 0 & 0 & 0 & 0 & 0 & 0 & 0 & 1 & 0 & 0 & 0 & 0 & 0 & 0 & 0 & 0 & 0 & 0 & 0 & 1 & 0 & 0 & 0 & 0 & 0 & 0 & 0 & 0 & 0 & 0 & 0 & 0 & 0 \\
 0 & 0 & 0 & 0 & 0 & 0 & 0 & 0 & 0 & 0 & 0 & 0 & 0 & 0 & 0 & 0 & 0 & 0 & 0 & 1 & 0 & 0 & 0 & 0 & 0 & 0 & 0 & 0 & 0 & 1 & 0 & 0 & 0 & 0 & 0 & 0 & 0 & 0 & 0 & 0 & 0 & 0 & 0 & 0 \\
 0 & 0 & 0 & 0 & 0 & 0 & 0 & 0 & 0 & 0 & 0 & 0 & 0 & 0 & 0 & 0 & 0 & 0 & 0 & 0 & 1 & 0 & 0 & 0 & 0 & 0 & 0 & 0 & 0 & 0 & 0 & 0 & 1 & 0 & 0 & 0 & 0 & 0 & 0 & 0 & 0 & 0 & 0 & 0 \\
 0 & 0 & 0 & 0 & 0 & 0 & 0 & 0 & 0 & 0 & 0 & 0 & 0 & 0 & 0 & 0 & 0 & 0 & 0 & 0 & 0 & 1 & 0 & 0 & 0 & 0 & 0 & 0 & 0 & 0 & 0 & 1 & 0 & 0 & 0 & 0 & 0 & 0 & 0 & 0 & 0 & 0 & 0 & 0 \\
 0 & 0 & 0 & 0 & 0 & 0 & 0 & 0 & 0 & 0 & 0 & 0 & 0 & 0 & 1 & 0 & 0 & 0 & 0 & 0 & 0 & 0 & 1 & 0 & 0 & 0 & 0 & 0 & 0 & 0 & 0 & 0 & 0 & 0 & 0 & 0 & 0 & 0 & 0 & 0 & 0 & 0 & 0 & 0 \\
 0 & 0 & 0 & 0 & 0 & 0 & 0 & 0 & 0 & 0 & 0 & 0 & 0 & -1 & 0 & 0 & 0 & 0 & 0 & 0 & 0 & 0 & 0 & 1 & 0 & 0 & 0 & 0 & 0 & 0 & 0 & 0 & 0 & 0 & 0 & 0 & 0 & 0 & 0 & 0 & 0 & 0 & 0 & 0 \\
 0 & 0 & 0 & 0 & 0 & 0 & 0 & 0 & 0 & 0 & 0 & 0 & -1 & 0 & 0 & 0 & 0 & 0 & 0 & 0 & 0 & 0 & 0 & 0 & 1 & 0 & 0 & 0 & 0 & 0 & 0 & 0 & 0 & 0 & 0 & 0 & 0 & 0 & 0 & 0 & 0 & 0 & 0 & 0 \\
 0 & 0 & 0 & 0 & 0 & 0 & 0 & 0 & 0 & 0 & 0 & 1 & 0 & 0 & 0 & 0 & 0 & 0 & 0 & 0 & 0 & 0 & 0 & 0 & 0 & 1 & 0 & 0 & 0 & 0 & 0 & 0 & 0 & 0 & 0 & 0 & 0 & 0 & 0 & 0 & 0 & 0 & 0 & 0 \\
 0 & 0 & 0 & 0 & 0 & 0 & 0 & 0 & 0 & 0 & 0 & 0 & 0 & 0 & 0 & 0 & 0 & 1 & 0 & 0 & 0 & 0 & 0 & 0 & 0 & 0 & 1 & 0 & 0 & 0 & 0 & 0 & 0 & 0 & 0 & 0 & 0 & 0 & 0 & 0 & 0 & 0 & 0 & 0 \\
 0 & 0 & 0 & 0 & 0 & 0 & 0 & 0 & 0 & 0 & 0 & 0 & 0 & 0 & 0 & 0 & 1 & 0 & 0 & 0 & 0 & 0 & 0 & 0 & 0 & 0 & 0 & 1 & 0 & 0 & 0 & 0 & 0 & 0 & 0 & 0 & 0 & 0 & 0 & 0 & 0 & 0 & 0 & 0 \\
 0 & 0 & 0 & 0 & 0 & 0 & 0 & 0 & 0 & 0 & 0 & 0 & 0 & 0 & 0 & 1 & 0 & 0 & 0 & 0 & 0 & 0 & 0 & 0 & 0 & 0 & 0 & 0 & 1 & 0 & 0 & 0 & 0 & 0 & 0 & 0 & 0 & 0 & 0 & 0 & 0 & 0 & 0 & 0 \\
 0 & 0 & 0 & 0 & 0 & 0 & 0 & 0 & 0 & 0 & 0 & 0 & 0 & 0 & 0 & 0 & 0 & 0 & 0 & 1 & 0 & 0 & 0 & 0 & 0 & 0 & 0 & 0 & 0 & 1 & 0 & 0 & 0 & 0 & 0 & 0 & 0 & 0 & 0 & 0 & 0 & 0 & 0 & 0 \\
 0 & 0 & 0 & 0 & 0 & 0 & 0 & 0 & 0 & 0 & 0 & 0 & 0 & 0 & 0 & 0 & 0 & 0 & 1 & 0 & 0 & 0 & 0 & 0 & 0 & 0 & 0 & 0 & 0 & 0 & 1 & 0 & 0 & 0 & 0 & 0 & 0 & 0 & 0 & 0 & 0 & 0 & 0 & 0 \\
 0 & 0 & 0 & 0 & 0 & 0 & 0 & 0 & 0 & 0 & 0 & 0 & 0 & 0 & 0 & 0 & 0 & 0 & 0 & 0 & 0 & 1 & 0 & 0 & 0 & 0 & 0 & 0 & 0 & 0 & 0 & 1 & 0 & 0 & 0 & 0 & 0 & 0 & 0 & 0 & 0 & 0 & 0 & 0 \\
 0 & 0 & 0 & 0 & 0 & 0 & 0 & 0 & 0 & 0 & 0 & 0 & 0 & 0 & 0 & 0 & 0 & 0 & 0 & 0 & 1 & 0 & 0 & 0 & 0 & 0 & 0 & 0 & 0 & 0 & 0 & 0 & 1 & 0 & 0 & 0 & 0 & 0 & 0 & 0 & 0 & 0 & 0 & 0 \\
 0 & 0 & 0 & 0 & 0 & 0 & 0 & 0 & 0 & 0 & 0 & 0 & 0 & 0 & 0 & 0 & 0 & 0 & 0 & 0 & 0 & 0 & 0 & 0 & 0 & 0 & 0 & 0 & 0 & 0 & 0 & 0 & 0 & 1 & 0 & 0 & 1 & 0 & 0 & 0 & 0 & 0 & 0 & 0 \\
 0 & 0 & 0 & 0 & 0 & 0 & 0 & 0 & 0 & 0 & 0 & 0 & 0 & 0 & 0 & 0 & 0 & 0 & 0 & 0 & 0 & 0 & 0 & 0 & 0 & 0 & 0 & 0 & 0 & 0 & 0 & 0 & 0 & 0 & 1 & -1 & 0 & 0 & 0 & 0 & 0 & 0 & 0 & 0 \\
 0 & 0 & 0 & 0 & 0 & 0 & 0 & 0 & 0 & 0 & 0 & 0 & 0 & 0 & 0 & 0 & 0 & 0 & 0 & 0 & 0 & 0 & 0 & 0 & 0 & 0 & 0 & 0 & 0 & 0 & 0 & 0 & 0 & 0 & -1 & 1 & 0 & 0 & 0 & 0 & 0 & 0 & 0 & 0 \\
 0 & 0 & 0 & 0 & 0 & 0 & 0 & 0 & 0 & 0 & 0 & 0 & 0 & 0 & 0 & 0 & 0 & 0 & 0 & 0 & 0 & 0 & 0 & 0 & 0 & 0 & 0 & 0 & 0 & 0 & 0 & 0 & 0 & 1 & 0 & 0 & 1 & 0 & 0 & 0 & 0 & 0 & 0 & 0 \\
 0 & 0 & 0 & 0 & 0 & 0 & 0 & 0 & 0 & 0 & 0 & 0 & 0 & 0 & 0 & 0 & 0 & 0 & 0 & 0 & 0 & 0 & 0 & 0 & 0 & 0 & 0 & 0 & 0 & 0 & 0 & 0 & 0 & 0 & 0 & 0 & 0 & 1 & 0 & 1 & 0 & 0 & 0 & 0 \\
 0 & 0 & 0 & 0 & 0 & 0 & 0 & 0 & 0 & 0 & 0 & 0 & 0 & 0 & 0 & 0 & 0 & 0 & 0 & 0 & 0 & 0 & 0 & 0 & 0 & 0 & 0 & 0 & 0 & 0 & 0 & 0 & 0 & 0 & 0 & 0 & 0 & 0 & 2 & 0 & 0 & 0 & 0 & 0 \\
 0 & 0 & 0 & 0 & 0 & 0 & 0 & 0 & 0 & 0 & 0 & 0 & 0 & 0 & 0 & 0 & 0 & 0 & 0 & 0 & 0 & 0 & 0 & 0 & 0 & 0 & 0 & 0 & 0 & 0 & 0 & 0 & 0 & 0 & 0 & 0 & 0 & 1 & 0 & 1 & 0 & 0 & 0 & 0 \\
 0 & 0 & 0 & 0 & 0 & 0 & 0 & 0 & 0 & 0 & 0 & 0 & 0 & 0 & 0 & 0 & 0 & 0 & 0 & 0 & 0 & 0 & 0 & 0 & 0 & 0 & 0 & 0 & 0 & 0 & 0 & 0 & 0 & 0 & 0 & 0 & 0 & 0 & 0 & 0 & 1 & 1 & 0 & 0 \\
 0 & 0 & 0 & 0 & 0 & 0 & 0 & 0 & 0 & 0 & 0 & 0 & 0 & 0 & 0 & 0 & 0 & 0 & 0 & 0 & 0 & 0 & 0 & 0 & 0 & 0 & 0 & 0 & 0 & 0 & 0 & 0 & 0 & 0 & 0 & 0 & 0 & 0 & 0 & 0 & 1 & 1 & 0 & 0 \\
 0 & 0 & 0 & 0 & 0 & 0 & 0 & 0 & 0 & 0 & 0 & 0 & 0 & 0 & 0 & 0 & 0 & 0 & 0 & 0 & 0 & 0 & 0 & 0 & 0 & 0 & 0 & 0 & 0 & 0 & 0 & 0 & 0 & 0 & 0 & 0 & 0 & 0 & 0 & 0 & 0 & 0 & 1 & 1 \\
 0 & 0 & 0 & 0 & 0 & 0 & 0 & 0 & 0 & 0 & 0 & 0 & 0 & 0 & 0 & 0 & 0 & 0 & 0 & 0 & 0 & 0 & 0 & 0 & 0 & 0 & 0 & 0 & 0 & 0 & 0 & 0 & 0 & 0 & 0 & 0 & 0 & 0 & 0 & 0 & 0 & 0 & 1 & 1 \\
\end{smallmatrix}
\end{matrix}\right)\,.
\end{equation}
The rank of the matrix is 22 and we can obtain ranks of the matrix for the other three class that are 14, 22 and 14 in the $SU(2)$ case. Similarly, the matrix $\mathcal{Y}(\young(45))=I+\mathcal{D}_{SU(3)}\otimes \mathcal{D}_{\mathcal{B}L/R}$ for the $SU(3)$ case can be also obtained, they are $352\times352$ and $224\times224$ matrix. The ranks of the matrix for the four class are 176, 112, 176 and 112.
Thus there are 72 independent operators for $SU(2)$ case and 576 independent operators for $SU(3)$ case in the type $Df_+u^2N^2$.

Then, we can get the effective operators for each class in the type $Df_+u^2N^2$ and we need to translate the effective operators to the $C$ and $P$ eigenstates. 
Recalling
\begin{align}
        N = \left(\begin{array}{c}
N_L \\ N_R
        \end{array}\right)\,,\quad F_{L/R}^{\mu\nu}=\frac{1}{2}(f^{\mu\nu}\mp i\frac{1}{2}\Tilde{f}^{\mu\nu})\,,\notag\\
        \gamma^\mu = \left(\begin{array}{cc}
0 & \sigma^\mu \\ \bar{\sigma}^\mu & 0
        \end{array}\right)\,,\quad\sigma^{\mu\nu}=\frac{i}{2}=\left(\begin{array}{cc}
 \sigma^{\mu\nu}&0 \\0& \bar{\sigma}^{\mu\nu} 
        \end{array}\right)\,.
\end{align}
For $SU(2)$ case, the 72 independent effective operators are
\begin{align}
    \mathcal{O}_1={\tau^K}^i_j({N_R}_i \Bar{\sigma}^\mu {N_R^\dagger}^j)F_{+R\mu\nu}^I(D^\nu D^\lambda u^I) (D_\lambda u^K)\,,\quad\mathcal{O}_2={\tau^K}^i_j({N_R}_i \Bar{\sigma}^\mu {N_R^\dagger}^j)F_{+R\mu\nu}^I(D^\nu D^\lambda u^K) (D_\lambda u^I)\,,\notag\\
    \mathcal{O}_3={\tau^K}^i_j({N_R}_i \Bar{\sigma}^\mu {N_R^\dagger}^j)F_{+R\mu\nu}^K(D^\nu D^\lambda u^I) (D_\lambda u^I)\,,\quad\mathcal{O}_4=\epsilon_{IJK}({N_R}_i \Bar{\sigma}^\mu {N_R^\dagger}^i)F_{+R\mu\nu}^I(D^\nu D^\lambda u^J) (D_\lambda u^K)\,,\notag\\
    \mathcal{O}_5={\tau^K}^i_j({N_R}_i \Bar{\sigma}^\mu {N_R^\dagger}^j)F_{+R\nu\lambda}^I(D^\nu D_\mu u^I) (D^\lambda u^K)\,,\quad\mathcal{O}_6={\tau^K}^i_j({N_R}_i \Bar{\sigma}^\mu {N_R^\dagger}^j)F_{+R\nu\lambda}^I(D^\nu D_\mu u^K) (D^\lambda u^I)\,,\notag\\
    \mathcal{O}_7={\tau^K}^i_j({N_R}_i \Bar{\sigma}^\mu {N_R^\dagger}^j)F_{+R\nu\lambda}^K(D^\nu D_\mu u^I) (D^\lambda u^I)\,,\quad\mathcal{O}_8=\epsilon_{IJK}({N_R}_i \Bar{\sigma}^\mu {N_R^\dagger}^i)F_{+R\nu\lambda}^I(D^\nu D_\mu u^J) (D^\lambda u^K)\,,\notag\\
    \mathcal{O}_9={\tau^K}^i_j({N_R}_i \Bar{\sigma}^\mu {N_R^\dagger}^j)D_\nu F_{+R\mu\lambda}^I(D^\nu u^I) (D^\lambda u^K)\,,\quad\mathcal{O}_{10}={\tau^K}^i_j({N_R}_i \Bar{\sigma}^\mu {N_R^\dagger}^j)D_\nu F_{+R\mu\lambda}^I(D^\nu u^K) (D^\lambda u^I)\,,\notag\\
    \mathcal{O}_{11}={\tau^K}^i_j({N_R}_i \Bar{\sigma}^\mu {N_R^\dagger}^j)D_\nu F_{+R\mu\lambda}^K(D^\nu u^I) (D^\lambda u^I)\,,\quad\mathcal{O}_{12}=\epsilon_{IJK}({N_R}_i \Bar{\sigma}^\mu {N_R^\dagger}^i)D_\nu F_{+R\mu\lambda}^I(D^\nu u^I) (D^\lambda u^K)\,,\notag\\
    \mathcal{O}_{13}={\tau^K}^i_j({N_R}_i \Bar{\sigma}^\mu {N_R^\dagger}^j)D^\nu F_{+R\nu\lambda}^I(D_\mu u^I) (D^\lambda u^K)\,,\quad\mathcal{O}_{14}={\tau^K}^i_j({N_R}_i \Bar{\sigma}^\mu {N_R^\dagger}^j)D^\nu F_{+R\nu\lambda}^I(D_\mu u^K) (D^\lambda u^I)\,,\notag\\
    \mathcal{O}_{15}={\tau^K}^i_j({N_R}_i \Bar{\sigma}^\mu {N_R^\dagger}^j)D^\nu F_{+R\nu\lambda}^K(D_\mu u^I) (D^\lambda u^I)\,,\quad\mathcal{O}_{16}=\epsilon_{IJK}({N_R}_i \Bar{\sigma}^\mu {N_R^\dagger}^i)D^\nu F_{+R\nu\lambda}^I(D_\mu u^J) (D^\lambda u^K)\,,\notag\\
    \mathcal{O}_{17}={\tau^K}^i_j({N_R}_i \Bar{\sigma}^\mu {N_R^\dagger}^j)D^\nu F_{+R\mu\nu}^I(D_\lambda u^I) (D^\lambda u^K)\,,\quad\mathcal{O}_{18}={\tau^K}^i_j({N_R}_i \Bar{\sigma}^\mu {N_R^\dagger}^j)D^\nu F_{+R\mu\nu}^K(D_\lambda u^I) (D^\lambda u^I)\,,\notag\\
    \mathcal{O}_{19}={\tau^K}^i_j(D^\lambda{N_R}_i \Bar{\sigma}^\mu {N_R^\dagger}^j) F_{+R\mu\nu}^I(D_\lambda u^I) (D^\nu u^K)\,,\quad\mathcal{O}_{20}={\tau^K}^i_j(D^\lambda{N_R}_i \Bar{\sigma}^\mu {N_R^\dagger}^j) F_{+R\mu\nu}^I(D_\lambda u^K) (D^\nu u^I)\,,\notag\\        
    \mathcal{O}_{21}={\tau^K}^i_j(D^\lambda{N_R}_i \Bar{\sigma}^\mu {N_R^\dagger}^j) F_{+R\mu\nu}^K(D_\lambda u^I) (D^\nu u^I)\,,\quad\mathcal{O}_{22}=\epsilon_{IJK}(D^\lambda{N_R}_i \Bar{\sigma}^\mu {N_R^\dagger}^i) F_{+R\mu\nu}^I(D_\lambda u^J) (D^\nu u^K)\,,\notag\\
    \mathcal{O}_{23}={\tau^K}^i_j({N_R}_i \Bar{\sigma}^\mu {N_R^\dagger}^j)F_{+L\mu\nu}^I(D^\nu D^\lambda u^I) (D_\lambda u^K)\,,\quad\mathcal{O}_{24}={\tau^K}^i_j({N_R}_i \Bar{\sigma}^\mu {N_R^\dagger}^j)F_{+L\mu\nu}^I(D^\nu D^\lambda u^K) (D_\lambda u^I)\,,\notag\\
    \mathcal{O}_{25}={\tau^K}^i_j({N_R}_i \Bar{\sigma}^\mu {N_R^\dagger}^j)F_{+L\mu\nu}^I(D^\nu D^\lambda u^I) (D_\lambda u^K)\,,\quad\mathcal{O}_{26}=\epsilon_{IJK}({N_R}_i \Bar{\sigma}^\mu {N_R^\dagger}^i)F_{+L\mu\nu}^I(D^\nu D^\lambda u^J) (D_\lambda u^K)\,,\notag\\
    \mathcal{O}_{27}={\tau^K}^i_j({N_R}_i \Bar{\sigma}^\mu {N_R^\dagger}^j)F_{+L\nu\lambda}^I(D^\nu D_\mu u^I) (D^\lambda u^K)\,,\quad\mathcal{O}_{28}={\tau^K}^i_j({N_R}_i \Bar{\sigma}^\mu {N_R^\dagger}^j)F_{+L\nu\lambda}^I(D^\nu D_\mu u^K) (D^\lambda u^I)\,,\notag\\
    \mathcal{O}_{29}={\tau^K}^i_j({N_R}_i \Bar{\sigma}^\mu {N_R^\dagger}^j)F_{+L\nu\lambda}^K(D^\nu D_\mu u^I) (D^\lambda u^I)\,,\quad\mathcal{O}_{30}=\epsilon_{IJK}({N_R}_i \Bar{\sigma}^\mu {N_R^\dagger}^i)F_{+L\nu\lambda}^I(D^\nu D_\mu u^J) (D^\lambda u^K)\,,\notag\\
    \mathcal{O}_{31}={\tau^K}^i_j({N_R}_i \Bar{\sigma}^\mu {N_R^\dagger}^j)D_\nu F_{+L\mu\lambda}^I(D^\nu u^I) (D^\lambda u^K)\,,\quad\mathcal{O}_{32}={\tau^K}^i_j({N_R}_i \Bar{\sigma}^\mu {N_R^\dagger}^j)D_\nu F_{+L\mu\lambda}^I(D^\nu u^K) (D^\lambda u^I)\,,\notag\\
    \mathcal{O}_{33}={\tau^K}^i_j(D^\lambda{N_R}_i \Bar{\sigma}^\mu {N_R^\dagger}^j) F_{+L\mu\nu}^I(D_\lambda u^I) (D^\nu u^K)\,,\quad\mathcal{O}_{34}={\tau^K}^i_j(D^\lambda{N_R}_i \Bar{\sigma}^\mu {N_R^\dagger}^j) F_{+L\mu\nu}^I(D_\lambda u^K) (D^\nu u^I)\,,\notag\\
    \mathcal{O}_{35}={\tau^K}^i_j(D^\lambda{N_R}_i \Bar{\sigma}^\mu {N_R^\dagger}^j) F_{+L\mu\nu}^K(D_\lambda u^I) (D^\nu u^I)\,,\quad\mathcal{O}_{36}=\epsilon_{IJK}(D^\lambda{N_R}_i \Bar{\sigma}^\mu {N_R^\dagger}^i) F_{+L\mu\nu}^I(D_\lambda u^J) (D^\nu u^K)\,,\notag\\
    \mathcal{O}_{37}={\tau^K}^i_j({N_L}_i \sigma^\mu {N_L^\dagger}^j)F_{+R\mu\nu}^I(D^\nu D^\lambda u^I) (D_\lambda u^K)\,,\quad\mathcal{O}_{38}={\tau^K}^i_j({N_L}_i \sigma^\mu {N_L^\dagger}^j)F_{+R\mu\nu}^I(D^\nu D^\lambda u^K) (D_\lambda u^I)\,,\notag\\
    \mathcal{O}_{39}={\tau^K}^i_j({N_L}_i \sigma^\mu {N_L^\dagger}^j)F_{+R\mu\nu}^K(D^\nu D^\lambda u^I) (D_\lambda u^I)\,,\quad\mathcal{O}_{40}=\epsilon_{IJK}({N_L}_i \sigma^\mu {N_L^\dagger}^i)F_{+R\mu\nu}^I(D^\nu D^\lambda u^J) (D_\lambda u^K)\,,\notag\\
    \mathcal{O}_{41}={\tau^K}^i_j({N_L}_i \sigma^\mu {N_L^\dagger}^j)F_{+R\nu\lambda}^I(D^\nu D_\mu u^I) (D^\lambda u^K)\,,\quad\mathcal{O}_{42}={\tau^K}^i_j({N_L}_i \sigma^\mu {N_L^\dagger}^j)F_{+R\nu\lambda}^I(D^\nu D_\mu u^K) (D^\lambda u^I)\,,\notag\\
    \mathcal{O}_{43}={\tau^K}^i_j({N_L}_i \sigma^\mu {N_L^\dagger}^j)F_{+R\nu\lambda}^K(D^\nu D_\mu u^I) (D^\lambda u^I)\,,\quad\mathcal{O}_{44}=\epsilon_{IJK}({N_L}_i \sigma^\mu {N_L^\dagger}^i)F_{+R\nu\lambda}^I(D^\nu D_\mu u^J) (D^\lambda u^K)\,,\notag\\
    \mathcal{O}_{45}={\tau^K}^i_j({N_L}_i \sigma^\mu {N_L^\dagger}^j)D_\nu F_{+R\mu\lambda}^I(D^\nu u^I) (D^\lambda u^K)\,,\quad\mathcal{O}_{46}={\tau^K}^i_j({N_L}_i \sigma^\mu {N_L^\dagger}^j)D_\nu F_{+R\mu\lambda}^I(D^\nu u^K) (D^\lambda u^I)\,,\notag\\
    \mathcal{O}_{47}={\tau^K}^i_j({N_L}_i \sigma^\mu {N_L^\dagger}^j)D_\nu F_{+R\mu\lambda}^K(D^\nu u^I) (D^\lambda u^I)\,,\quad\mathcal{O}_{48}=\epsilon_{IJK}({N_L}_i \sigma^\mu {N_L^\dagger}^i)D_\nu F_{+R\mu\lambda}^I(D^\nu u^I) (D^\lambda u^K)\,,\notag\\
    \mathcal{O}_{49}={\tau^K}^i_j({N_L}_i \sigma^\mu {N_L^\dagger}^j)D^\nu F_{+R\nu\lambda}^I(D_\mu u^I) (D^\lambda u^K)\,,\quad\mathcal{O}_{50}={\tau^K}^i_j({N_L}_i \sigma^\mu {N_L^\dagger}^j)D^\nu F_{+R\nu\lambda}^I(D_\mu u^K) (D^\lambda u^I)\,,\notag
    \end{align}
    \begin{align}
    \mathcal{O}_{51}={\tau^K}^i_j({N_L}_i \sigma^\mu {N_L^\dagger}^j)D^\nu F_{+R\nu\lambda}^K(D_\mu u^I) (D^\lambda u^I)\,,\quad\mathcal{O}_{52}=\epsilon_{IJK}({N_L}_i \sigma^\mu {N_L^\dagger}^i)D^\nu F_{+R\nu\lambda}^I(D_\mu u^J) (D^\lambda u^K)\,,\notag\\
    \mathcal{O}_{53}={\tau^K}^i_j({N_L}_i \sigma^\mu {N_L^\dagger}^j)D^\nu F_{+R\mu\nu}^I(D_\lambda u^I) (D^\lambda u^K)\,,\quad\mathcal{O}_{54}={\tau^K}^i_j({N_L}_i \sigma^\mu {N_L^\dagger}^j)D^\nu F_{+R\mu\nu}^K(D_\lambda u^I) (D^\lambda u^I)\,,\notag\\
    \mathcal{O}_{55}={\tau^K}^i_j(D^\lambda{N_L}_i \sigma^\mu {N_L^\dagger}^j) F_{+R\mu\nu}^I(D_\lambda u^I) (D^\nu u^K)\,,\quad\mathcal{O}_{56}={\tau^K}^i_j(D^\lambda{N_L}_i \sigma^\mu {N_L^\dagger}^j) F_{+R\mu\nu}^I(D_\lambda u^K) (D^\nu u^I)\,,\notag\\
    \mathcal{O}_{57}={\tau^K}^i_j(D^\lambda{N_L}_i \sigma^\mu {N_L^\dagger}^j) F_{+R\mu\nu}^K(D_\lambda u^I) (D^\nu u^I)\,,\quad\mathcal{O}_{58}=\epsilon_{IJK}(D^\lambda{N_L}_i \sigma^\mu {N_L^\dagger}^i) F_{+R\mu\nu}^I(D_\lambda u^J) (D^\nu u^K)\,,\notag\\
    \mathcal{O}_{59}={\tau^K}^i_j({N_L}_i \sigma^\mu {N_L^\dagger}^j)F_{+L\mu\nu}^I(D^\nu D^\lambda u^I) (D_\lambda u^K)\,,\quad\mathcal{O}_{60}={\tau^K}^i_j({N_L}_i \sigma^\mu {N_L^\dagger}^j)F_{+L\mu\nu}^I(D^\nu D^\lambda u^K) (D_\lambda u^I)\,,\notag\\
    \mathcal{O}_{61}={\tau^K}^i_j({N_L}_i \sigma^\mu {N_L^\dagger}^j)F_{+L\mu\nu}^I(D^\nu D^\lambda u^I) (D_\lambda u^K)\,,\quad\mathcal{O}_{62}=\epsilon_{IJK}({N_L}_i \sigma^\mu {N_L^\dagger}^i)F_{+L\mu\nu}^I(D^\nu D^\lambda u^J) (D_\lambda u^K)\,,\notag\\
    \mathcal{O}_{63}={\tau^K}^i_j({N_L}_i \sigma^\mu {N_L^\dagger}^j)F_{+L\nu\lambda}^I(D^\nu D_\mu u^I) (D^\lambda u^K)\,,\quad\mathcal{O}_{64}={\tau^K}^i_j({N_L}_i \sigma^\mu {N_L^\dagger}^j)F_{+L\nu\lambda}^I(D^\nu D_\mu u^K) (D^\lambda u^I)\,,\notag\\
    \mathcal{O}_{65}={\tau^K}^i_j({N_L}_i \sigma^\mu {N_L^\dagger}^j)F_{+L\nu\lambda}^K(D^\nu D_\mu u^I) (D^\lambda u^I)\,,\quad\mathcal{O}_{66}=\epsilon_{IJK}({N_L}_i \sigma^\mu {N_L^\dagger}^i)F_{+L\nu\lambda}^I(D^\nu D_\mu u^J) (D^\lambda u^K)\,,\notag\\
    \mathcal{O}_{67}={\tau^K}^i_j({N_L}_i \sigma^\mu {N_L^\dagger}^j)D_\nu F_{+L\mu\lambda}^I(D^\nu u^I) (D^\lambda u^K)\,,\quad\mathcal{O}_{68}={\tau^K}^i_j({N_L}_i \sigma^\mu {N_L^\dagger}^j)D_\nu F_{+L\mu\lambda}^I(D^\nu u^K) (D^\lambda u^I)\,,\notag\\
    \mathcal{O}_{69}={\tau^K}^i_j(D^\lambda{N_L}_i \sigma^\mu {N_L^\dagger}^j) F_{+L\mu\nu}^I(D_\lambda u^I) (D^\nu u^K)\,,\quad\mathcal{O}_{70}={\tau^K}^i_j(D^\lambda{N_L}_i \sigma^\mu {N_L^\dagger}^j) F_{+L\mu\nu}^I(D_\lambda u^K) (D^\nu u^I)\,,\notag\\
    \mathcal{O}_{71}={\tau^K}^i_j(D^\lambda{N_L}_i \sigma^\mu {N_L^\dagger}^j) F_{+L\mu\nu}^K(D_\lambda u^I) (D^\nu u^I)\,,\quad\mathcal{O}_{72}=\epsilon_{IJK}(D^\lambda{N_L}_i \sigma^\mu {N_L^\dagger}^i) F_{+L\mu\nu}^I(D_\lambda u^J) (D^\nu u^K)\,.
\end{align}

The operators $\mathcal{O}_{1-8}$, $\mathcal{O}_{23-30}$, $\mathcal{O}_{37-44}$ and $\mathcal{O}_{59-66}$ can be trivially combined into explicit $CP$ eigenstates. For example
\begin{align}
    (\Bar{N}\gamma^\mu u_\lambda N)\langle f_+^{\mu\nu}D^\lambda u_\nu\rangle=\mathcal{O}_1+\mathcal{O}_{23}+\mathcal{O}_{37}+\mathcal{O}_{59}\,,\notag\\
    (\Bar{N}\gamma^5\gamma^\mu u_\lambda N)\langle f_+^{\mu\nu}D^\lambda u_\nu\rangle=\mathcal{O}_1+\mathcal{O}_{23}-\mathcal{O}_{37}-\mathcal{O}_{59}\,,\notag\\
    (\Bar{N}\gamma^\mu u_\lambda N)\langle \Tilde{f}_+^{\mu\nu}D^\lambda u_\nu\rangle=\mathcal{O}_1-\mathcal{O}_{23}+\mathcal{O}_{37}-\mathcal{O}_{59}\,,\notag\\
    (\Bar{N}\gamma^5\gamma^\mu u_\lambda N)\langle \Tilde{f}_+^{\mu\nu}D^\lambda u_\nu\rangle=\mathcal{O}_1-\mathcal{O}_{23}-\mathcal{O}_{37}+\mathcal{O}_{59}\,.
\end{align}
And the other explicit eigenstates operator can be obtained by combining $\mathcal{O}_{19-22}$, $\mathcal{O}_{33-36}$, $\mathcal{O}_{55-58}$ and $\mathcal{O}_{69-72}$. Furthermore, since the number of operators involving in the $F_{+L\mu\nu}$ and $F_{+R\mu\nu}$ are different, combination of the operators $\mathcal{O}_{9-18}$, $\mathcal{O}_{9-18}$, $\mathcal{O}_{31,32}$, $\mathcal{O}_{45-54}$, $\mathcal{O}_{67,68}$ will be a little different. The $D^\nu \Tilde{f}_{\nu\lambda}$ should be vanish due to Eq.~\eqref{eq: Bian_FL_FR}. Therefore, the explicit $CP$ eigenvalues for the whole 72 independent operators have been listed in Tab.~\ref{CP eigenvalue}.
\begin{table}
    \centering
    \resizebox{\linewidth}{!}{
    \begin{tabular}{|c|c|c|c|}
    \hline
         $C+P+$&$C+P-$&$C-P+$&$C-P-$  \\
         \hline
         $(\Bar{N}\gamma^\mu u_\lambda N)\langle f_+^{\mu\nu} D^\lambda u_\nu\rangle$&$(\Bar{N}\gamma^5\gamma^\mu N)\langle f_+^{\mu\nu}[u_\lambda ,D^\lambda u_\nu]\rangle$&$(\Bar{N}\gamma^\mu N)\langle f_+^{\mu\nu}[u_\lambda ,D^\lambda u_\nu]\rangle$&$(\Bar{N}\gamma^5\gamma^\mu u_\lambda N)\langle f_+^{\mu\nu} D^\lambda u_\nu\rangle$\\
         $(\Bar{N}\gamma^\mu f_+^{\mu\nu} N)\langle u_\lambda D^\lambda u_\nu\rangle$&$(\Bar{N}\gamma^\mu u_\lambda N)\langle \Tilde{f}_+^{\mu\nu} D^\lambda u_\nu\rangle$&$(\Bar{N}\gamma^5\gamma^\mu u_\lambda N)\langle \Tilde{f}_+^{\mu\nu} D^\lambda u_\nu\rangle$&$(\Bar{N}\gamma^5\gamma^\mu f_+^{\mu\nu} N)\langle u_\lambda D^\lambda u_\nu\rangle$\\
         $(\Bar{N}\gamma^\mu  D^\lambda u_\nu N)\langle f_+^{\mu\nu}u_\lambda\rangle$&$(\Bar{N}\gamma^\mu \Tilde{f}_+^{\mu\nu} N)\langle u_\lambda D^\lambda u_\nu\rangle$&$(\Bar{N}\gamma^5\gamma^\mu \Tilde{f}_+^{\mu\nu} N)\langle u_\lambda D^\lambda u_\nu\rangle$&$(\Bar{N}\gamma^5\gamma^\mu  D^\lambda u_\nu N)\langle f_+^{\mu\nu}u_\lambda\rangle$\\
         $(\Bar{N}\gamma^5\gamma^\mu N)\langle \Tilde{f}_+^{\mu\nu}[u_\lambda ,D^\lambda u_\nu]\rangle$&$(\Bar{N}\gamma^\mu  D^\lambda u_\nu N)\langle \Tilde{f}_+^{\mu\nu}u_\lambda\rangle$&$(\Bar{N}\gamma^5\gamma^\mu  D^\lambda u_\nu N)\langle \Tilde{f}_+^{\mu\nu}u_\lambda\rangle$&$(\Bar{N}\gamma^\mu N)\langle \Tilde{f}_+^{\mu\nu}[u_\lambda ,D^\lambda u_\nu]\rangle$\\
         $(\Bar{N}\gamma^\mu u_\lambda N)\langle f_+^{\nu\lambda} D_\mu u_\nu\rangle$&$(\Bar{N}\gamma^5\gamma^\mu N)\langle f_+^{\nu\lambda}[u_\lambda ,D_\mu u_\nu]\rangle$&$(\Bar{N}\gamma^\mu N)\langle f_+^{\nu\lambda}[u_\lambda ,D_\mu u_\nu]\rangle$&$(\Bar{N}\gamma^5\gamma^\mu u_\lambda N)\langle f_+^{\nu\lambda} D_\mu u_\nu\rangle$\\
         $(\Bar{N}\gamma^\mu f_+^{\nu\lambda} N)\langle u_\lambda D_\mu u_\nu\rangle$&$(\Bar{N}\gamma^\mu u_\lambda N)\langle \Tilde{f}_+^{\nu\lambda} D_\mu u_\nu\rangle$&$(\Bar{N}\gamma^5\gamma^\mu u_\lambda N)\langle \Tilde{f}_+^{\nu\lambda} D_\mu u_\nu\rangle$&$(\Bar{N}\gamma^5\gamma^\mu f_+^{\nu\lambda} N)\langle u_\lambda D_\mu u_\nu\rangle$\\
         $(\Bar{N}\gamma^\mu  D_\mu u_\nu N)\langle f_+^{\nu\lambda}u_\lambda\rangle$&$(\Bar{N}\gamma^\mu \Tilde{f}_+^{\nu\lambda} N)\langle u_\lambda D_\mu u_\nu\rangle$&$(\Bar{N}\gamma^5\gamma^\mu \Tilde{f}_+^{\nu\lambda} N)\langle u_\lambda D_\mu u_\nu\rangle$&$(\Bar{N}\gamma^5\gamma^\mu  D_\mu u_\nu N)\langle f_+^{\nu\lambda}u_\lambda\rangle$\\
         $(\Bar{N}\gamma^5\gamma^\mu N)\langle \Tilde{f}_+^{\nu\lambda}[u_\lambda ,D_\mu u_\nu]\rangle$&$(\Bar{N}\gamma^\mu  D_\mu u_\nu N)\langle \Tilde{f}_+^{\nu\lambda}u_\lambda\rangle$&$(\Bar{N}\gamma^5\gamma^\mu  D_\mu u_\nu N)\langle \Tilde{f}_+^{\nu\lambda}u_\lambda\rangle$&$(\Bar{N}\gamma^\mu N)\langle \Tilde{f}_+^{\nu\lambda}[u_\lambda ,D_\mu u_\nu]\rangle$\\ 
         $(\Bar{N}\gamma^\mu u_\lambda N)\langle D^\lambda f_+^{\mu\nu} u_\nu\rangle$&$(\Bar{N}\gamma^\mu u_\lambda N)\langle D^\lambda \Tilde{f}_+^{\mu\nu} u_\nu\rangle$&$(\Bar{N}\gamma^\mu N)\langle D^\lambda f_+^{\mu\nu} [u_\nu,u_\lambda]\rangle$&$(\Bar{N}\gamma^5\gamma^\mu u_\lambda N)\langle D^\lambda f_+^{\mu\nu} u_\nu\rangle$\\
         $(\Bar{N}\gamma^\mu D^\lambda f_+^{\mu\nu} N)\langle u_\lambda u_\nu\rangle$&$(\Bar{N}\gamma^\mu D^\lambda \Tilde{f}_+^{\mu\nu} N)\langle u_\lambda u_\nu\rangle$&$(\Bar{N}\gamma^\mu  N)\langle D_\nu f_+^{\nu\lambda}[u_\mu,u_\lambda]\rangle$&$(\Bar{N}\gamma^5\gamma^\mu D^\lambda f_+^{\mu\nu} N)\langle u_\lambda u_\nu\rangle$\\
         $(\Bar{N}\gamma^\mu u_\nu N)\langle D^\lambda f_+^{\mu\nu} u_\lambda\rangle$&$(\Bar{N}\gamma^5\gamma^\mu N)\langle D^\lambda f_+^{\mu\nu} [u_\nu,u_\lambda]\rangle$&$(\Bar{N}\gamma^5\gamma^\mu u_\lambda N)\langle D^\lambda \Tilde{f}_+^{\mu\nu} u_\nu\rangle$&$(\Bar{N}\gamma^5\gamma^\mu u_\nu N)\langle D^\lambda f_+^{\mu\nu} u_\lambda\rangle$\\
         $(\Bar{N}\gamma^\mu D_\nu f_+^{\nu\lambda} N)\langle u_\lambda u_\mu\rangle$&$(\Bar{N}\gamma^5\gamma^\mu  N)\langle D_\nu f_+^{\nu\lambda}[u_\mu,u_\lambda]\rangle$&$(\Bar{N}\gamma^5\gamma^\mu D^\lambda \Tilde{f}_+^{\mu\nu} N)\langle u_\lambda u_\nu\rangle$&$(\Bar{N}\gamma^5\gamma^\mu D_\nu f_+^{\nu\lambda} N)\langle u_\lambda u_\mu\rangle$\\
         $(\Bar{N}\gamma^\mu u_\lambda N)\langle D_\nu f_+^{\nu\lambda} u_\mu\rangle$&&&$(\Bar{N}\gamma^5\gamma^\mu u_\lambda N)\langle D_\nu f_+^{\nu\lambda} u_\mu\rangle$\\
         $(\Bar{N}\gamma^\mu u_\mu N)\langle u_\lambda D_\nu f_+^{\nu\lambda}\rangle$&&&$(\Bar{N}\gamma^5\gamma^\mu u_\mu N)\langle u_\lambda D_\nu f_+^{\nu\lambda}\rangle$\\
         $(\Bar{N}\gamma^\mu D_\nu f_+^{\nu\mu} N)\langle u_\lambda u^\lambda\rangle$&&&$(\Bar{N}\gamma^5\gamma^\mu D_\nu f_+^{\nu\mu} N)\langle u_\lambda u^\lambda\rangle$\\
         $(\Bar{N}\gamma^\mu u_\lambda N)\langle u^\lambda D_\nu f_+^{\nu\mu}\rangle$&&&$(\Bar{N}\gamma^5\gamma^\mu u_\lambda N)\langle u^\lambda D_\nu f_+^{\nu\mu}\rangle$\\
         \hline
         $(\Bar{N}\gamma^\mu \lrd^\lambda N)\langle f_+^{\mu\nu}[u_\lambda, u_\nu]\rangle$&$(\Bar{N}\gamma^\mu \lrd^\lambda N)\langle \Tilde{f}_+^{\mu\nu}[u_\lambda, u_\nu]\rangle$&$(\Bar{N}\gamma^\mu \lrd^\lambda f_+^{\mu\nu} N)\langle u_\lambda u_\nu\rangle$&$(\Bar{N}\gamma^\mu \lrd^\lambda \Tilde{f}_+^{\mu\nu} N)\langle u_\lambda u_\nu\rangle$\\
         $(\Bar{N}\gamma^5\gamma^\mu \lrd^\lambda \Tilde{f}_+^{\mu\nu} N)\langle u_\lambda u_\nu\rangle$&$(\Bar{N}\gamma^5\gamma^\mu \lrd^\lambda f_+^{\mu\nu} N)\langle u_\lambda u_\nu\rangle$&$(\Bar{N}\gamma^\mu \lrd^\lambda u_\lambda N)\langle f_+^{\mu\nu}u_\nu\rangle$&$(\Bar{N}\gamma^\mu \lrd^\lambda u_\lambda N)\langle \Tilde{f}_+^{\mu\nu}u_\nu\rangle$\\
         $(\Bar{N}\gamma^5\gamma^\mu \lrd^\lambda u_\lambda N)\langle \Tilde{f}_+^{\mu\nu}u_\nu\rangle$&$(\Bar{N}\gamma^5\gamma^\mu \lrd^\lambda u_\lambda N)\langle f_+^{\mu\nu}u_\nu\rangle$&$(\Bar{N}\gamma^\mu \lrd^\lambda u_\nu N)\langle f_+^{\mu\nu} u_\lambda \rangle$&$(\Bar{N}\gamma^\mu \lrd^\lambda u_\nu N)\langle \Tilde{f}_+^{\mu\nu} u_\lambda \rangle$\\
         $(\Bar{N}\gamma^5\gamma^\mu \lrd^\lambda u_\nu N)\langle \Tilde{f}_+^{\mu\nu} u_\lambda \rangle$&$(\Bar{N}\gamma^5\gamma^\mu \lrd^\lambda u_\nu N)\langle f_+^{\mu\nu} u_\lambda \rangle$&$(\Bar{N}\gamma^5\gamma^\mu \lrd^\lambda N)\langle \Tilde{f}_+^{\mu\nu}[u_\lambda, u_\nu]\rangle$&$(\Bar{N}\gamma^5\gamma^\mu \lrd^\lambda N)\langle f_+^{\mu\nu}[u_\lambda, u_\nu]\rangle$\\
         \hline
    \end{tabular}
    }
    \caption{Explicit CP eigenvalues for the 72 independent operators.}
    \label{CP eigenvalue}
\end{table}

In addition, due to the heavy baryon mass in the chiral limit, the derivatives acting on the baryon count to be $\mathcal{O}{(p^0)}$, thus there are some operators in the type $Df_+u^2N^2$ that contribute to $\mathcal{O}{(p^4)}$, and we distinguish this situation from the previous ones in the Tab.~\ref{CP eigenvalue}, and we will put this 16 operators into the type $fu^2N^2$ in our result. And some of the operators in the type with more covariant derivatives can contribute to $\mathcal{O}(p^5)$. Since the operators are classified by the power counting, the types $D^2fu^2N^2$, $D^3fu^2N^2$ and $D^4fu^2N^2$ should also be considered through the same procedure, and we select the operators that contribute to order $\mathcal{O}(p^5)$.

As for the $SU(3)$ case, it is similar to the $SU(2)$ case since they share the same Lorentz basis. The flavor $SU(3)$ m-basis is a combination of $d^{ABC}$ and $f^{ABC}$. According to the Cayley-Hamilton theorem, our independent flavor m-basis can be converted to the trace basis. There will be 16 independent flavor m-basis with the repeated fields in the type $Df_+u^2N^2$, and we convert them to be of the specific $C$ eigenvalues 
\begin{align}
\label{cp2+}
  p_C : & \notag \\
        & \langle \mathcal{A}\mathcal{C}\mathcal{B}\mathcal{D}\mathcal{D} \rangle \,,\quad  \langle \mathcal{A}\mathcal{B}\mathcal{D}\mathcal{C}\mathcal{D} \rangle \,,\quad \langle \mathcal{A}\mathcal{D}\mathcal{D}\mathcal{B}\mathcal{C} \rangle \,,\quad \langle \mathcal{A}\mathcal{D}\mathcal{C}\mathcal{D}\mathcal{B} \rangle \,,\notag \\
        & \langle \mathcal{A}\mathcal{B}\mathcal{C}\mathcal{D}\mathcal{D} \rangle + \langle \mathcal{A}\mathcal{B}\mathcal{D}\mathcal{D}\mathcal{C} \rangle \,,\quad \langle \mathcal{A}\mathcal{D}\mathcal{B}\mathcal{C}\mathcal{D} \rangle + \langle \mathcal{A}\mathcal{D}\mathcal{B}\mathcal{D}\mathcal{C} \rangle \,,\notag \\
        & \langle \mathcal{A}\mathcal{C}\mathcal{D}\mathcal{B}\mathcal{D} \rangle + \langle \mathcal{A}\mathcal{D}\mathcal{C}\mathcal{B}\mathcal{D} \rangle \,,\quad\langle \mathcal{A}\mathcal{C}\mathcal{D}\mathcal{D}\mathcal{B} \rangle + \langle \mathcal{A}\mathcal{D}\mathcal{D}\mathcal{C}\mathcal{B} \rangle \,,\notag \\
        & \langle \mathcal{A}\mathcal{B}\mathcal{C} \rangle \langle \mathcal{D}\mathcal{D} \rangle \,,\quad \langle \mathcal{A}\mathcal{B}\mathcal{D} \rangle\langle \mathcal{C}\mathcal{D} \rangle \,,\quad \langle \mathcal{A}\mathcal{B} \rangle\langle \mathcal{C}\mathcal{D}\mathcal{D} \rangle + \langle \mathcal{A}\mathcal{B} \rangle\langle \mathcal{C}\mathcal{D}\mathcal{D} \rangle \,, \\
  -p_C: & \notag \\
  \label{cp2-}
        & \langle \mathcal{A}\mathcal{B}\mathcal{C}\mathcal{D}\mathcal{D} \rangle - \langle \mathcal{A}\mathcal{B}\mathcal{D}\mathcal{D}\mathcal{C} \rangle \,,\quad\langle \mathcal{A}\mathcal{D}\mathcal{B}\mathcal{C}\mathcal{D} \rangle - \langle \mathcal{A}\mathcal{D}\mathcal{B}\mathcal{D}\mathcal{C} \rangle  \,,\notag \\
        & \langle \mathcal{A}\mathcal{C}\mathcal{D}\mathcal{B}\mathcal{D} \rangle + \langle \mathcal{A}\mathcal{D}\mathcal{C}\mathcal{B}\mathcal{D} \rangle  \,,\quad \langle \mathcal{A}\mathcal{C}\mathcal{D}\mathcal{D}\mathcal{B} \rangle + \langle \mathcal{A}\mathcal{D}\mathcal{D}\mathcal{C}\mathcal{B} \rangle  \,,\notag \\
        & \langle \mathcal{A}\mathcal{C}\mathcal{D} \rangle\langle \mathcal{B}\mathcal{D} \rangle - \langle \mathcal{B}\mathcal{D}\mathcal{C} \rangle\langle \mathcal{A}\mathcal{D} \rangle \,.
\end{align}
Thus combining the Lorentz m-basis and the trace basis, through the Eq.~\eqref{cp+}, ~\eqref{cp-}, ~\eqref{cp2+}, ~\eqref{cp2-} and Tab.~\ref{tab: building blocks mesons}, ~\ref{tab: building blocks baryons}, we can obtain the explicit $CP$ eigen operator basis. 

\section{Conclusions}
\label{conclusion}

We constructed the complete and independent meson-baryon $CP$-eigen bases of the ChPT up to the $p^5$-order for the first time. For the $SU(2)$ case, we obtained totally 557 $C$+$P$+ operators, 490 $C$+$P$- operators, 429 $C$-$P$+ operators, 512 $C$-$P$- operators from $p^2$-order to $p^5$-order and the numbers are 435, 387, 352, 420 at the $p^5$-order. And we compared our $CP$-even operators with the literature~\cite{Fettes:2000gb} up to $p^4$ order in the App.~\ref{comparison}. As for the $SU(3)$ case, the totally number of operators $C$-$P$+, $C$-$P$+, $C$-$P$+ and $C$-$P$+ are 622, 549, 464, 527 from $p^2$-order to $p^4$-order, in which there are 528 $C$-$P$+ operators, 475 $C$-$P$+ operators, 413 $C$-$P$+ operators and 450 $C$-$P$ operators at $p^4$-order. 

The young tensor technique is utilized to construct the operator bases with certain modifications and improvements: the off-shell external sources, the trace basis, and the $CP$ eigenbasis, etc. There are some advantages in this work compared with other literature:
\begin{itemize}
    \item Compared with the traditional operator construction on the baryon-meson system~\cite{Krause:1990xc,Ecker:1995rk,Fettes:2000gb,Frink:2004ic,Oller:2006yh,Frink:2006hx,Jiang:2016vax}, the on-shell amplitudes from the SSYTs of the primary Young diagrams provide the complete and independent Lorentz structure from the group theoretic principles automatically. 
    \item We obtain the $\mathcal{O}(p^5)$ operator sets of the meson-baryon ChPT for the first time, and the different $CP$ eigenstates have also been classified in our work. The Hilbert series of the meson-baryon operators for each $CP$ eigenstate in the $SU(2)$ and $SU(3)$ case are also obtained.
    \item We show the equivalence between the trace basis and the invariant tensor basis, and provide a systematic procedure to convert from one to another via the Cayley-Hamilton theorem.
    \item The operator type obtained by the Hilbert series is inappropriate for the HBChPT. We have also obtained these operators in the HBChPT by adding additional covariant derivatives for each type.
\end{itemize} 

The higher dimensional operators obtained in this work could have many applications. The $C$+$P$+ $\mathcal{O}(p^5)$ operators can be used in the $\mathcal{O}(p^5)$ nucleon mass calculation. Other applications include the following: the other $CP$-eigen bases can be applied to the $CP$ violating process when the meson-baryon interactions are considered such as the non-vanishing EDM~\cite{Engel:2013lsa,Bsaisou:2014oka,Dekens:2014jka}, the $CP$ violating dark matter~\cite{Dekens:2022gha}, and the $P$ violating nuclear potential~\cite{Maekawa:2011vs}, etc.

\section*{Acknowledgments}

We would like to express our gratitude to Xuan-He Li for the valuable assistance provided and for the insightful discussions throughout the course of this work. This work is supported by the National Science Foundation of China under Grants No. 12347105,
No. 12375099 and No. 12047503, and the National Key Research and Development Program of China
Grant No. 2020YFC2201501, No. 2021YFA0718304.

\bibliography{ref}

\appendix

\section{Comparison with Literature up to $p^4$ order for $SU(2)$ }
\label{comparison}
In this appendix, we present the comparison with the $SU(2)$ operators up to $\mathcal{O}(p^4)$ obtained in this paper and the previous literature~\cite{Fettes:2000gb}. And the operators in this section follow the conventions discussed in Sec.~\ref{ChPT}.

When comparing operators in the following tables, the following remarks should be taken:
\begin{itemize}
    \item The definition $\Sigma_\pm$, $f_\pm^{\mu\nu}$ and $D^\mu u^\nu$ in this paper are the same as $\chi_\pm$, $F_\pm^{\mu\nu}$ and $h^{\mu\nu}$ in Ref.~\cite{Fettes:2000gb}.
    \item Moreover, the trace of $f_\pm^{\mu\nu}$ is considered to be zero that only the trace $\langle F_-\rangle$ is zero in Ref.~\cite{Fettes:2000gb}.
    \item Special attention is required for the literature that $\Tilde{\chi}=\chi-\frac{1}{2}\langle \chi\rangle$, thus we neglect the translation between the $\Tilde{\chi}$ and $\chi$. 
    \item  In our notation
\begin{equation}  
\Tilde{f}_\pm^{\mu\nu}=\epsilon^{\mu\nu\rho\lambda}f_{\pm\rho\lambda}\,,\quad \Bar{N}\gamma^5 N=\frac{1}{m_N}D^\mu(\bar {N_v}S_\mu N_v)\,.
\end{equation}
\end{itemize}

The first column of the Tab.~\ref{com p2}, ~\ref{com p3} and ~\ref{com p4} provides all types of operators and the second column provides the label $i$ for the operator $\mathcal{O}^{++}_i$ defined in Tab.~\ref{Dim-2}, ~\ref{Dim-3} and ~\ref{Dim-4} one by one. Moreover, the non-relativistic (NR)-form of our operators has been obtained in the third column of the table through the heavy baryon projection. Furthermore, the relationship between our results and the operators in Ref.~\cite{Fettes:2000gb} are listed in the fourth column and the nucleon bilinear should be combined with these operators to form the whole chiral Lagrangian 
\begin{equation}
    \mathcal{L}_{\pi N}=\sum_{i\,,n} \Bar{\Psi}O_{i}^{(n)}\Psi\,,
\end{equation}
in which the $n$ means the order of the operators and $i$ is the mark for operators.

\setlength\LTleft{\fill}
\subsection*{$\mathcal{O}(p^2)$}
\begin{center}
\begin{longtable}{|c|c|c|c|}
\hline
Type & $\mathcal{O}^{++(2)}_i$ & NR-form & Ref.~\cite{Fettes:2000gb} \\
\hline
\multirow{3}{*}{$u^2N^2$}
& 1 &$(\Bar{N}_vN_v)\langle u\cdot u\rangle$ 
& $O^{(2)}_3=\frac{1}{2}\left\langle u\cdot u\right\rangle $ \\ 
& 2 &$(\Bar{N}_v[S_\mu,S_\nu]N_v)\langle u^\mu,u^\nu\rangle$ 
&$O^{(2)}_4=\frac{1}{2}\left[S^{\mu},S^{\nu}\right]  \left[  u_{\mu},u_{\nu}\right]  $ \\ 
& 3 &$(\Bar{N}_v v_\mu v_\nu N_v)\langle u^\mu u^\nu\rangle$ 
&  $O^{(2)}_2=\frac{1}{2}\left\langle \left(  v\cdot u\right)^{2}\right\rangle $ \\ 
\hline
\multirow{2}{*}{$\Sigma N^2$}
& 4 &$(\Bar{N}_v \Sigma_+N_v)$ 
&$O^{(2)}_5=\widetilde{\chi}_{+}$ \\ 
& 5 & $(\Bar{N}_vN_v)\langle \Sigma_+\rangle$
& $O^{(2)}_1=\left\langle \chi_{+}\right\rangle $ \\ 
\hline
\multirow{1}{*}{$fN^2$} 
& 6 &$(\Bar{N}_v[S_\mu,S_\nu]f_{+\mu\nu}N_v)$ 
&$O^{(2)}_6=-\frac{i}{4m}\left[  S^{\mu},S^{\nu}\right]  F_{\mu\nu}^{+}$ \\ 
\hline
\caption{Comparison between our results and Ref.~\cite{Fettes:2000gb} at $p^2$ order. The second column only provides the label $i$ for the operator $\mathcal{O}^{++}_i$ defined in Tab.~\ref{Dim-2}. }
\label{com p2}
\end{longtable}
\end{center}
\begin{landscape}
\subsection*{$\mathcal{O}(p^3)$}
\begin{center}
    \centering
    \begin{longtable}{|c|c|c|c|}
\hline
Type & $\mathcal{O}^{++(3)}_i$ & NR-form &Ref.~\cite{Fettes:2000gb} \\
\hline

\hline
\multirow{3}{*}{$Du^2N^2$}
& 1 &$(\Bar{N}_v v^\mu[D^\nu u_\mu,u_\nu]N_v)$ & $O^{(3)}_1+O^{(3)}_2=i[u_{\mu},[v\cdot D,u^{\mu}]]+i[u_{\mu},[D^{\mu},v\cdot u]]$	\\	
& 2 & $(\Bar{N}_v[S_\mu,S_\nu]u^\lambda N_v)\langle u^\mu D_\lambda u^\nu\rangle$& $O^{(3)}_{14}-O^{(3)}_{15}-i[S^{\mu},S^{\nu}]\langle\lbrack v\cdot D,u_{\mu}u_{\nu}\rangle-(-i)[S^{\mu},S^{\nu}]\langle u_{\mu}[D_{\nu},v\cdot u]\rangle$	\\	
& 3 &$(\Bar{N}_vv^\mu v^\nu v^\rho [D_\mu u_\nu,u_\rho]N_v)$ & $O^{(3)}_3=i[v\cdot u,[v\cdot D,v\cdot u]]$ \\	
\hline
\multirow{5}{*}{$u^3N^2$}
& 4 &$\epsilon^{\mu\nu\rho\lambda}(\Bar{N}_vv_\mu N_v)\langle[u_\nu,u_\rho]u_\lambda\rangle$ & $O^{(3)}_4=i\epsilon^{\mu\nu\alpha\beta}v_{\beta}\langle u_{\mu}u_{\nu}u_{\alpha}\rangle$ \\	
& 5 & $(\Bar{N}_v S_\mu u_\nu N_v)\langle u^\mu u^\nu\rangle$& $O^{(3)}_{11}S^{\mu}u^{\nu}\langle u_{\mu}u_{\nu}\rangle $ \\	
& 6 & $(\Bar{N}_v S_\mu u^\mu N_v)\langle u_\nu u^\nu\rangle$&$O^{(3)}_{10}=S\cdot u\langle u^{2}\rangle $ \\	
& 7 &$(\Bar{N}_v [S_\mu ,v_\nu] v_\rho u^\mu N_v)\langle u^\nu u^\rho\rangle$ &$O^{(3)}_{12}-O^{(3)}_{13}=S\cdot u\langle(v\cdot u)^{2}\rangle-\langle S\cdot u\,\,\,v\cdot u\rangle v\cdot u$ \\	
& 8 &$(\Bar{N}_v S_\mu v_\nu v_\rho u^\nu N_v)\langle u^\mu u^\rho\rangle$ & $O^{(3)}_{13}=\langle S\cdot u\,\,\,v\cdot u\rangle v\cdot u$\\	
\hline
\multirow{3}{*}{$u\Sigma N^2$}
& 9 & $(\Bar{N}_v v^\mu[u_\mu,\Sigma_-] N_v)$& $O^{(3)}_5=[\chi_{-},v\cdot u]$  \\	
& 10 &$(\Bar{N}_v S^\mu N_v)\langle u_\mu\Sigma_+\rangle$ & $O^{(3)}_{17}\langle S\cdot u\chi_{+}\rangle$ \\	
& 11 &$(\Bar{N}_v S^\mu u_\mu N_v)\langle \Sigma_+\rangle$ & $O^{(3)}_{17}=\langle S\cdot u\chi_{+}\rangle$ \\	
\hline
\multirow{6}{*}{$ufN^2$}
& 12 &$\varepsilon_{\mu\nu\rho\lambda}(\Bar{N}_v v^\mu N_v)\langle u^\nu f^{+\mu\nu}\rangle$ & $O^{(3)}_8=\epsilon^{\mu\nu\alpha\beta}v_{\beta}\langle\widetilde{F}_{\mu\nu}^{+}u_{\alpha}\rangle$ \\	
& 13 &$(\Bar{N}_v S^\mu [u^\nu, f^{+\mu\nu}]N_v)$ & $O^{(3)}_{21}=iS^{\mu}[\widetilde{F}_{\mu\nu}^{+},u^{\nu}]$  \\	
& 14 &$(\Bar{N}_v S^\mu N_v)\langle u^\nu \Tilde{f}^{-\mu\nu}\rangle$ & $O^{(3)}_{23}=\epsilon^{\mu\nu\alpha\beta}S_{\mu}\langle u_{\nu}F_{\alpha\beta}^{-}\rangle$ \\	
& 15 &$(\Bar{N}_v u^\mu [u^\nu, f^{-\mu\nu}]N_v)$&  $O^{(3)}_1-O^{(3)}_2=i[u_{\mu},[v\cdot D,u^{\mu}]]-i[u_{\mu},[D^{\mu},v\cdot u]]$ \\	
& 16 &$(\Bar{N}_v v_\mu S_\nu v^\rho [u^\rho, f^{+\mu\nu}]N_v)$ & $O^{(3)}_{20}=\,iS^{\mu}v^{\nu}[\widetilde{F}_{\mu\nu}^{+},v\cdot u]$ \\	
& 17 &$(\Bar{N}_v [S_\mu, S^\nu] v^\rho N_v)\langle u^\mu f_{-\nu\rho}\rangle$& $O^{(3)}_{14}+O^{(3)}_{15}=-i[S^{\mu},S^{\nu}]\langle\lbrack v\cdot D,u_{\mu}u_{\nu}\rangle+(-i)[S^{\mu},S^{\nu}]\langle u_{\mu}[D_{\nu},v\cdot u]\rangle$ \\	
\hline
\multirow{2}{*}{$DfN^2$} 
& 18 &$(\Bar{N}_v S^\mu D^\nu f_{-\mu\nu}N_v)$  & $O^{(3)}_{22}=\,S^{\mu}[D^{\nu},F_{\mu\nu}^{-}]$ \\	
& 19 &$(\Bar{N}_v v^\mu D^\nu f_{+\mu\nu}N_v)$ & $O^{(3)}_{6}=v^{\nu}[D^{\mu},\widetilde{F}_{\mu\nu}^{+}]$ \\
\hline
\multirow{2}{*}{$\Sigma N^2$}
& 20 &$D^\mu(\Bar{N}_v S_\mu \Sigma_- N_v)$ & $O^{(3)}_{18}=i [S\cdot D,\chi_{-}]$ \\	
& 21 &$D^\mu(\Bar{N}_v S_\mu N_v)\langle\Sigma_-\rangle$ & $O^{(3)}_{19}=i [S\cdot D,\langle\chi_{-}\rangle]$ \\	
\hline
\caption{Comparison between our results and Ref.~\cite{Fettes:2000gb} at $p^3$ order. The second column only provides the label $i$ for the operator $\mathcal{O}^{++}_i$ defined in Tab.~\ref{Dim-3}. }
\label{com p3}
    \end{longtable}
\end{center}

\subsection*{$\mathcal{O}(p^4)$}
\begin{center}
    \centering
    \begin{longtable}{|c|c|c|c|}
\hline
Type & $\mathcal{O}^{++(4)}_i$ & NR-form & Ref.~\cite{Fettes:2000gb} \\
\hline
\multirow{5}{*}{$D^2u^2N^2$}
& 1 &$(\Bar{N}_v N_v)\langle D_\mu u_\nu D^\mu u^\nu\rangle$ &  $O^{(4)}_{14}=\left\langle h_\mu\nu h^\mu \nu \right\rangle $ \\
& 2 &$(\Bar{N}_v [S^\mu, S^\nu][D^\sigma u_\mu,D_\sigma u_\nu] N_v)$ &  $O^{(4)}_{17}=[S^\mu ,S^\nu ]\left[ h_{\lambda \mu} ,h_{\;\;\nu}^\lambda \right] $ \\
& 3 &$(\Bar{N}_v v^\mu v^\nu N_v)\langle D_\mu u_\lambda D_\nu u^\lambda\rangle$ &  $O^{(4)}_{15}=v^\mu v^\nu \left\langle h_{\lambda \mu} h_{\;\;\nu} ^\lambda \right\rangle $ \\
& 4 &$(\Bar{N}_v [S^\mu, S^\nu] v^\alpha v^\beta [D_\mu u_\alpha,D_\nu u_\beta] N_v)$ & $O^{(4)}_{18}=[S^\mu ,S^\nu ]v^\lambda v^\rho \left[h_\lambda \mu ,h_{\nu \rho} \right] $ \\
& 5 &$(\Bar{N}_v v^\mu v^\nu v^\rho v^\lambda N_v)\langle D_\mu u_\nu D_\rho u_\lambda\rangle$ & $O^{(4)}_{13}==iv^\lambda v^\mu \left\langle h_{\lambda \mu}\left[ v\cdot u,S\cdot u\right] \right\rangle $\\
\hline
\multirow{4}{*}{$Du^3N^2$}
& 6 & $(\Bar{N}_v [v^\mu ,S^\nu] N_v)\langle D^\lambda u_\mu  [u_\lambda,u_\nu]\rangle$&  $O^{(4)}_{11}-O^{(4)}_{12}=iS^\mu \left\langle h_{\lambda \mu} \left[ u^\lambda ,v\cdot u\right] \right\rangle -iv^\mu\left\langle h_{\lambda \mu} \left[ u^\lambda ,S\cdot u\right]\right\rangle $ \\
& 7 &$(\Bar{N}_v v^\mu S^\nu N_v)\langle D^\lambda u_\mu  [u_\lambda,u_\nu]\rangle$ &  $O^{(4)}_{12}=iv^\mu\left\langle h_\lambda \mu \left[ u^\lambda ,S\cdot u\right]\right\rangle $ \\
& 8 &$\epsilon^{\mu\nu\rho\lambda}(\Bar{N}_v v_\rho S_\lambda v_\mu v_\sigma u_\nu N_v)\langle u_\rho D_\lambda u^\sigma \rangle$ &  $O^{(4)}_{10}=\varepsilon_{\quad \tau}^{\mu \nu \rho} v^\tau v^\lambda \left\langle h_{\lambda \mu} u_\nu \right\rangle u_\rho$\\
& 9 &$(\Bar{N}_v [v^\mu ,S^\nu] v^\rho v^\lambda N_v)\langle D_\mu u_\rho  [u_\nu,u_\lambda]\rangle$ &  $O^{(4)}_{13}=iv^\lambda v^\mu \left\langle h_{\lambda \mu} \left[ v\cdot u,S\cdot u\right] \right\rangle $ \\
\hline
\multirow{9}{*}{$u^4 N^2$}
& 10 &$(\Bar{N}_v N_v)\langle u^\mu u_\mu\rangle\langle u^\nu u_\nu\rangle$ &  $O^{(4)}_{1}=\left\langle u\cdot u\right\rangle \left\langle u\cdot u\right\rangle $ \\
& 11 &$(\Bar{N}_v N_v)\langle u^\mu u^\nu\rangle\langle u_\mu u_\nu\rangle$ &  $O^{(4)}_{2}=\left\langle u_\mu u_\nu \right\rangle\left\langle u^\mu u^\nu \right\rangle $ \\
& 12 &$(\Bar{N}_v [S^\mu, S^\nu][u^\mu,u^\nu] N_v)\langle u^\sigma u_\sigma\rangle$ &  $O^{(4)}_{6}=[S^\mu ,S^\nu ]\left[ u_\mu ,u_\nu \right]\left\langle u\cdot u\right\rangle $ \\
& 13 &$(\Bar{N}_v [S^\mu ,S^\nu] u^\sigma N_v)\langle u_\sigma[u^\mu,u^\nu]\rangle$ &  $O^{(4)}_{8}=[S^\mu ,S^\nu ]\left\langle \left[u_\mu ,u_\nu \right] u_\lambda \right\rangle u^\lambda $ \\
& 14 &$(\Bar{N}_v [S^\mu, S^\nu] v^\rho v^\lambda [u_\mu,u_\nu]  N_v)\langle u_\rho u_\lambda \rangle$ &  $O^{(4)}_{7}=[S^\mu,S^\nu ]\left[ u_\mu ,u_\nu \right] \left\langle (v\cdot u)^2\right\rangle $ \\
& 15 &$(\Bar{N}_v v^\mu v^\nu N_v)\langle u_\mu u_\nu\rangle\langle u_\rho u^\rho\rangle$ &  $O^{(4)}_{3}=\left\langle u\cdot u\right\rangle \left\langle (v\cdot u)^2\right\rangle $ \\
& 16 &$(\Bar{N}_v v^\mu v^\nu N_v)\langle u_\mu u_\rho\rangle\langle u_\nu u^\rho\rangle$&  $O^{(4)}_{4}=\left\langle u_\lambda v\cdot u\right\rangle \left\langle u^\lambda v\cdot u\right\rangle $ \\
& 17 &$(\Bar{N}_v [S^\mu,S^\nu] v^\alpha v^\beta u_\beta N_v)\langle u_\mu[u_\nu,u_\alpha]\rangle$ &  $O^{(4)}_{9}=[S^\mu ,S^\nu ]\left\langle \left[ u_\mu ,u_\nu \right] v\cdot u\right\rangle v\cdot u$ \\
& 18 &$(\Bar{N}_v v^\mu v^\nu v^\rho v^\lambda N_v)\langle u_\mu u_\nu\rangle\langle u_\rho u_\lambda\rangle$ & $O^{(4)}_{5}=\left\langle (v\cdot u)^2\right\rangle \left\langle (v\cdot u)^2\right\rangle $ \\
\hline
\multirow{7}{*}{$Du\Sigma N^2$}
& 19 &$(\Bar{N}_v[S^\mu,v^\nu][u_\mu,D_\nu\Sigma_+] N_v)$ &  $O^{(4)}_{29}=iS^\mu \left[ \left[ D_\mu ,\widetilde\chi _+\right] ,v\cdot u\right] $ \\
& 20 &$(\Bar{N}_v N_v)\langle u^\mu D_\mu\Sigma_-\rangle$ &  $O^{(4)}_{36}=i\left\langle u_\mu \left[ D^\mu ,\widetilde\chi _-\right] \right\rangle $ \\
& 21 &$(\Bar{N}_v [S_\mu, S_\nu][u^\mu,D^\nu\Sigma_-] N_v) $ &  $O^{(4)}_{37}=i[S^\mu ,S^\nu ]\left[ u_\mu ,\left[ D_\nu ,\widetilde\chi _-\right] \right] $\\
& 22 &$(\Bar{N}_v u^\mu N_v)D_\mu\langle \Sigma_-\rangle$ &  $O^{(4)}_{33}=iu_\mu \left[ D^\mu ,\left\langle \chi _-\right\rangle \right] $ \\
& 23 &$(\Bar{N}_v S^\mu v^\nu [u_\mu,D_\nu\Sigma_+] N_v)$ &  $O^{(4)}_{28}=iS^\mu v^\nu \left[ \widetilde\chi _+,h_\mu\nu \right] $\\
& 24 &$(\Bar{N}_v v^\mu v^\nu N_v)\langle D_\nu u_\mu\Sigma_+\rangle$ & $O^{(4)}_{35}=iv^\mu v^\nu \left\langle \widetilde\chi _-h_\mu \nu \right\rangle $ \\
& 25 &$(\Bar{N}_v v^\mu v^\nu D_\nu u_\mu N_v)\langle \Sigma_+\rangle$  &  $O^{(4)}_{32}=iv^\mu v^\nu \left\langle \chi _-\right\rangle h_{\mu \nu} $ \\
\hline
\multirow{4}{*}{$u^2\Sigma N^2$} 
& 26 &$(\Bar{N}_v [S_\mu, S_\nu] N_v)\langle \Sigma_+[u^\mu,u^\nu]\rangle $ &  $O^{(4)}_{27}=[S^\mu ,S^\nu ]\left\langle \widetilde\chi _+\left[ u_\mu ,u_\nu \right] \right\rangle $ \\
& 27 &$(\Bar{N}_v  u_\mu N_v)\langle \Sigma_+ u^\mu\rangle$ &  $O^{(4)}_{25}=u_\mu\left\langle \widetilde\chi _+u^\mu \right\rangle $ \\
& 28 &$(\Bar{N}_v \Sigma_+ N_v)\langle u_\mu u^\mu\rangle$ & $O^{(4)}_{23}=\widetilde\chi _+\left\langle u\cdot u\right\rangle $ \\
& 29 &$(\Bar{N}_v [S_\mu ,v_\nu] N_v)\langle \Sigma_-[u^\mu,u^\nu]\rangle $ &  $O^{(4)}_{34}=\left\langle \widetilde\chi _-\left[ S\cdot u,v\cdot u\right] \right\rangle $\\
\hline
\multirow{6}{*}{$u^2\Sigma N^2$}& 30 &$(\Bar{N}_v  N_v)\langle u_\mu u^\mu\rangle\langle\Sigma_+\rangle$ &  $O^{(4)}_{19}=\left\langle \chi _+\right\rangle \left\langle u\cdot u\right\rangle $\\
& 31 & $(\Bar{N}_v [S^\mu, v^\nu][u_\mu,u_\nu]  N_v)\langle\Sigma_-\rangle$&$O^{(4)}_{21}=\left\langle \chi_-\right\rangle \left[ S\cdot u,v\cdot u\right] $   \\
& 32 &$(\Bar{N}_v [S^\mu, S^\nu][u^\mu,u^\nu] N_v)\langle \Sigma_+\rangle $  &  $O^{(4)}_{31}=[S^\mu ,S^\nu ]\left\langle \chi_+\right\rangle \left[ u_\mu ,u_\nu \right] $ \\
& 33 &$(\Bar{N}_v v^\mu v^\nu u_\mu  N_v)\langle\Sigma_+u_\nu\rangle$ & $O^{(4)}_{26}=v\cdot u\left\langle \widetilde\chi _+v\cdot u\right\rangle $ \\
& 34 &$(\Bar{N}_v v^\mu v^\nu \Sigma_+  N_v)\langle u_\mu u_\nu\rangle$ &  $O^{(4)}_{24}=\widetilde\chi _+\left\langle (v\cdot u)^2\right\rangle $\\
& 35 &$(\Bar{N}_v v^\mu v^\nu N_v)\langle u_\mu u_\nu\rangle\langle\Sigma_+\rangle$ &  $O^{(4)}_{20}=\left\langle \chi_+\right\rangle \left\langle (v\cdot u)^2\right\rangle $ \\
\hline
& 36 &$(\Bar{N}_v S^\mu v^\nu N_v)\langle D^\sigma f_+^{\mu\nu}u_\sigma\rangle $ &  $O^{(4)}_{71}=S^\mu v^\nu \left\langle u^\lambda \left[ D_\lambda ,\widetilde F_{\mu \nu} ^+\right] \right\rangle $\\
& 37 &$(\Bar{N}_v [S_\mu, v_\nu] N_v)\langle D_\sigma f_+^{\nu\sigma}u^\mu\rangle $ &  $O^{(4)}_{72}-O^{(4)}_{73}=v^\nu \left\langle S\cdot u \left[ D^\lambda ,\widetilde F_{\lambda \nu} ^+\right] \right\rangle-S^\nu \left\langle v\cdot u\left[ D^\lambda ,\widetilde F_{\lambda \nu}^+\right] \right\rangle$  \\
& 38 &$(\Bar{N}_v [S^\rho,v^\lambda] N_v)\langle f_+^{\nu\sigma}D_\sigma u^\mu\rangle $ &  $O^{(4)}_{67}-O^{(4)}_{68}=S^\mu v^\nu\left\langle \widetilde F_{\lambda \mu} ^+h_{\;\nu} ^\lambda\right\rangle -S^\nu v^\mu\left\langle \widetilde F_{\lambda \mu} ^+h_{\;\nu} ^\lambda\right\rangle $ \\
& 39 &$(\Bar{N}_v  N_v)\langle D_\mu  f_-^{\mu\nu} u_\nu\rangle$ &  $O^{(4)}_{85}=\left\langle u^\mu \left[ D^\nu ,F_{\mu \nu} ^-\right] \right\rangle $ \\
& 40 &$(\Bar{N}_v [S_\mu, S_\nu][D^\mu f_-^{\nu\sigma},u_\sigma ] N_v) $ &  $O^{(4)}_{87}=[S^\mu ,S^\nu ]\left[ u^\lambda ,\left[D_\lambda ,F_{\mu \nu} ^-\right] \right] $ \\
& 41 &$(\Bar{N}_v [S_\mu, S_\nu][D_\sigma f_-^{\nu\sigma},u^\mu ] N_v) $ & $O^{(4)}_{88}=[S^\mu ,S^\nu ]\left[ u_\mu , \left[ D^\lambda,F_{\lambda \nu} ^-\right] \right] $ \\
$Duf N^2$& 42 &$(\Bar{N}_v [S_\mu, S_\nu][ f_-^{\nu\sigma},D_\sigma u^\mu ] N_v) $ & $O^{(4)}_{82}=[S^\mu ,S^\nu ]\left[ F_{\lambda \mu}^-,h_{\;\;\nu} ^\lambda \right] $\\
& 43 &$(\Bar{N}_v S^\mu v^\nu  N_v)\langle f_+^{\mu\lambda} D^\lambda u_\nu\rangle$ &  $O^{(4)}_{67}=S^\mu v^\nu\left\langle \widetilde F_{\lambda \mu} ^+h_{\;\nu} ^\lambda\right\rangle $\\
& 44 &$\varepsilon_{\mu\lambda\rho\sigma}(\Bar{N}_v v^\mu v^\nu[f_+^{\rho\sigma} ,D^\lambda u_\nu]  N_v)$ &  $O^{(4)}_{70}=i\varepsilon_{\quad \;\tau} ^{\lambda \mu \nu} v^\tau v^\rho\left[ \widetilde F_{\lambda \mu} ^+,h_{\nu \rho} \right] $ \\
& 45 &$(\Bar{N}_v S^\mu v^\nu  N_v)\langle D^\lambda f_+^{\mu\lambda} u_\nu\rangle$ & $O^{(4)}_{72}=v^\nu \left\langle S\cdot u \left[ D^\lambda ,\widetilde F_{\lambda \nu} ^+\right] \right\rangle $ \\
& 46 &$(\Bar{N}_v v^\mu v^\nu  N_v)\langle f_-^{\mu\lambda} D^\lambda u_\nu\rangle$ &  $O^{(4)}_{81}=v^\mu v^\nu \left\langle  F_{\lambda \mu}^-h_{\;\;\nu} ^\lambda \right\rangle $\\
& 47 &$(\Bar{N}_v [S^\mu,S^\nu]v^\rho v^\lambda [f_-^{\mu\rho}, D_\lambda u_\nu] N_v)$ & $O^{(4)}_{83}=[S^\mu ,S^\nu ]v^\lambda v^\rho \left[ F_{\lambda \mu} ^-,h_{\nu \rho} \right] $ \\
& 48 &$(\Bar{N}_v v^\mu v^\nu  N_v)\langle D^\lambda f_-^{\mu\lambda}u_\nu\rangle$ &  $O^{(4)}_{86}=v^\nu \left\langle v\cdot u\left[ D^\lambda , F_{\lambda \nu} ^-\right] \right\rangle $ \\
& 49 &$(\Bar{N}_v [S_\mu,S_\nu] v_\alpha v_\beta N_v)\langle f_+^{\mu\nu}D^\alpha u^\beta\rangle $ &  $O^{(4)}_{69}=S^\lambda v^\mu v^\nu v^\rho \left\langle \widetilde F_{\lambda \mu}^+h_{\nu \rho} \right\rangle $\\
& 50 &$(\Bar{N}_v [S_\mu, S_\nu] v_\alpha v_\beta [ f_-^{\mu\nu},D^\alpha u^\beta]N_v) $ &  $O^{(4)}_{84}=[S^\mu ,S^\lambda ]v^\rho v^\nu \left[F_{\mu \lambda} ^-,h_{\nu \rho} \right] $ \\
\hline
\multirow{18}{*}{$u^2fN^2$}& 51 &$(\Bar{N}_v [S_\mu, S_\nu] u^\mu N_v)\langle f_+^{\nu\sigma}u_\sigma\rangle $ &  $O^{(4)}_{62}=i[S^\mu ,S^\nu ]u^\lambda\left\langle \widetilde F_{\lambda \mu} ^+u_\nu \right\rangle $\\
& 52 &$(\Bar{N}_v [S_\mu, S_\nu] f_+^{\nu\sigma} N_v)\langle u^\mu u_\sigma\rangle $ &  $O^{(4)}_{58}=i[S^\mu ,S^\nu ]\widetilde F_{\lambda \mu} ^+\left\langle u^\lambda u_\nu \right\rangle $ \\
& 53 &$(\Bar{N}_v [S_\mu, S_\nu] u_\sigma N_v)\langle f_+^{\nu\sigma}u^\mu\rangle $ &  $O^{(4)}_{63}=i[S^\mu ,S^\nu ]u_\mu\left\langle \widetilde F_{\lambda \nu} ^+u^\lambda \right\rangle $ \\
& 54 &$(\Bar{N}_v [S_\mu, S_\nu] f_+^{\mu\nu} N_v)\langle u^\sigma u_\sigma\rangle $ & $O^{(4)}_{57}=i[S^\mu ,S^\nu ]\widetilde F_{\mu \nu}^+\left\langle u\cdot u\right\rangle $ \\
& 55 &$(\Bar{N}_v [S_\mu, S_\nu] u^\sigma N_v)\langle f_+^{\mu\nu}u_\sigma\rangle $ &  $O^{(4)}_{61}=i[S^\mu ,S^\nu ]u^\lambda\left\langle \widetilde F_{\mu \nu} ^+u_\lambda \right\rangle $ \\
& 56 & $(\Bar{N}_v N_v)\langle f_+^{\mu\nu}[u_\mu,u_\nu]\rangle $&  $O^{(4)}_{55}=i\left\langle \widetilde F_{\mu \nu} ^+\left[u^\mu ,u^\nu \right] \right\rangle $ \\
& 57 & $\varepsilon_{\mu\nu\rho\lambda}(\Bar{N}_vu_\mu N_v)\langle f_-^{\rho\lambda}u_\nu\rangle $&  $O^{(4)}_{75}=\varepsilon ^{\lambda \mu \nu \rho} u_\lambda \left\langle  F_{\mu \nu} ^-u_\rho\right\rangle $ \\
& 58 &$(\Bar{N}_v [S_\mu, v_\nu] N_v)\langle f_-^{\mu\sigma} [u^\nu, u_\sigma]\rangle $ &  $O^{(4)}_{79}-O^{(4)}_{80}=iS^\mu \left\langle F_{\lambda \mu} ^-\left[ u^\lambda ,v\cdot u\right] \right\rangle-iv^\mu\left\langle  F_{\lambda \mu} ^-\left[ u^\lambda ,S\cdot u \right]\right\rangle $ \\
& 59 &$(\Bar{N}_v [S^\mu,S^\nu]v^\lambda v^\rho u_\rho N_v)\langle f_+^{\mu\lambda} u_\nu\rangle $ &  $O^{(4)}_{65}=i[S^\mu ,S^\nu]v^\lambda v\cdot u\left\langle \widetilde F_{\mu \lambda} ^+u_\nu\right\rangle $ \\
& 60 &$(\Bar{N}_v [S^\mu,S^\nu]v^\lambda v^\rho u_\nu N_v)\langle f_+^{\mu\lambda}u_\rho \rangle $ &  $O^{(4)}_{66}=i[S^\mu ,S^\nu]v^\lambda u_\nu \left\langle \widetilde F_{\mu \lambda} ^+v\cdot u\right\rangle $ \\
& 61 &$\varepsilon^{\mu\lambda\rho\sigma}(\Bar{N}_v S^\mu v^\nu f_+^{\rho\sigma}N_v)\langle u_\lambda u_\nu \rangle $ &  $O^{(4)}_{70}=i\varepsilon_{\quad \;\tau}^{\lambda \mu \nu} v^\tau v^\rho\left[ \widetilde F_{\lambda \mu} ^+,h_\nu \rho \right] $ \\
& 62 &$(\Bar{N}_v v^\mu v^\nu N_v)\langle f_+^{\mu\lambda}[u_\lambda, u_\nu]\rangle $ &  $O^{(4)}_{56}=iv^\mu\left\langle \widetilde F_{\lambda \mu} ^+\left[ u^\lambda ,v\cdot u\right] \right\rangle $ \\
& 63 &$\varepsilon^{\mu\lambda\rho\sigma}(\Bar{N}_v v^\mu v^\nu u_\nu N_v)\langle f_-^{\rho\sigma} u_\lambda\rangle $ &  $O^{(4)}_{76}=\varepsilon_{\quad \;\tau} ^{\mu \nu \rho} v^\tau v\cdot u\left\langle  F_{\mu \nu} ^-u_\rho \right\rangle $ \\
& 64 &$\varepsilon^{\mu\lambda\rho\sigma}(\Bar{N}_v v^\mu v^\nu u_\lambda N_v)\langle f_-^{\rho\sigma}u_\nu \rangle $ & $O^{(4)}_{77}=\varepsilon_{\quad \;\tau} ^{\lambda \mu \nu} v^\tau u_\lambda \left\langle  F_{\mu \nu} ^-v\cdot u\right\rangle $ \\
& 65 &$\varepsilon^{\mu\lambda\rho\sigma}(\Bar{N}_v v^\mu v^\nu f_-^{\rho\sigma}N_v)\langle u_\lambda u_\nu \rangle $  &  $O^{(4)}_{78}=\varepsilon_{\quad \;\tau} ^{\lambda \mu \nu} v^\tau  F_{\lambda \mu} ^-\left\langle u_\nu v\cdot u\right\rangle $ \\
& 66 &$(\Bar{N}_v S^\mu v^\nu N_v)\langle f_-^{\mu\lambda}[u_\lambda, u_\nu]\rangle $&  $O^{(4)}_{79}=iS^\mu \left\langle F_{\lambda \mu} ^-\left[ u^\lambda ,v\cdot u\right] \right\rangle $ \\
& 67 &$(\Bar{N}_v [S_\mu, S_\nu] v_\alpha v_\beta u^\alpha N_v)\langle f_+^{\mu\nu}u^\beta\rangle $ & $O^{(4)}_{64}=i[S^\mu,S^\nu ]v\cdot u\left\langle \widetilde F_{\mu \nu} ^+v\cdot u\right\rangle $ \\
& 68 &$(\Bar{N}_v [S_\mu, S_\nu] v_\alpha v_\beta f_+^{\mu\nu} N_v)\langle u^\alpha u^\beta\rangle $ &  $O^{(4)}_{59}=i[S^\mu,S^\nu ]\widetilde F_{\mu \nu} ^+\left\langle (v\cdot u)^2\right\rangle $ \\
\hline
\multirow{2}{*}{$f^2N^2$}
& 69 &$(\Bar{N}_v N_v)\langle f_+^{\mu\nu}f_{+\mu\nu}\rangle $ &  $O^{(4)}_{93}=\left\langle \widetilde F_{\mu \nu} ^+\widetilde F^+\mu \nu \right\rangle $\\
& 70 &$(\Bar{N}_v [S_\mu, S_\nu] f_+^{\sigma\nu}f_{+\mu\sigma}N_v) $ &  $O^{(4)}_{95}=[S^\mu ,S^\nu ]\left[ \widetilde F_{\lambda \mu} ^+,\widetilde F_{\quad \;\nu}^{+\lambda} \right] $ \\
\hline
\multirow{10}{*}{$f^2N^2$}& 71 &$(\Bar{N}_v N_v)\langle f_-^{\mu\nu}f_{-\mu\nu}\rangle $ &  $O^{(4)}_{118}=\langle  F_{\mu \nu} ^{\rm R}F^{\rm R\,\mu \nu} + F_{\mu \nu} ^{\rm L}F^{\rm L\,\mu \nu} \rangle $                                \\
& 72 &$(\Bar{N}_v [S_\mu, S_\nu] f_-^{\sigma\nu}f_{-\mu\sigma}N_v) $ &  $O^{(4)}_{97}=[S^\mu ,S^\nu ]\left[  F_{\lambda \mu}^-,F_{\quad \nu} ^-\lambda \right] $ \\
& 73 & $(\Bar{N}_v [S^\mu, v^\nu] N_v)\langle \Tilde{f}_+^{\sigma\nu}f_{-\mu\sigma}\rangle $&  $O^{(4)}_{101}-O^{(4)}_{102}=S^\mu v^\nu \left\langle F_{\lambda \mu} ^-F_{\quad\;\nu} ^{+\lambda} \right\rangle -S^\nu v^\mu \left\langle F_{\lambda \mu} ^-F_{\quad\;\nu }^{+\lambda} \right\rangle $ \\
& 74 &$\varepsilon_{\mu\nu\rho\lambda}(\Bar{N}_v [f_+^{\mu\nu},f_{-\mu\nu}]N_v) $ &  $O^{(4)}_{103}=i\varepsilon ^{\lambda \mu \nu \rho} \left[  F_{\lambda \mu} ^-,\widetilde F_{\nu \rho} ^+\right] $ \\
& 75 & $(\Bar{N}_v v^\mu v^\nu N_v)\langle f_+^{\sigma\nu}f_{+\mu\sigma}\rangle $ &  $O^{(4)}_{94}=v^\mu v^\nu \left\langle \widetilde F_{\lambda \mu}^+\widetilde F_{\quad\;\nu}^{+\lambda} \right\rangle $ \\
& 76 &$(\Bar{N}_v [S_\mu,S_\nu] v_\rho v_\lambda [f_+^{\nu\rho},f_+^{\mu\lambda}] N_v)$ &  $O^{(4)}_{96}=[S^\mu ,S^\nu]v^\lambda v^\rho \left[ \widetilde F_{\lambda \mu} ^+,\widetilde F_{\rho \nu} ^+\right] $ \\
& 77 &$(\Bar{N}_v v^\mu v^\nu N_v)\langle f_-^{\sigma\nu}f_{-\mu\sigma}\rangle $ &  $O^{(4)}_{97}=[S^\mu ,S^\nu ]\left[  F_{\lambda \mu}^-,F_{\quad \nu} ^{-\lambda} \right] $ \\
& 78 &$(\Bar{N}_v [S_\mu ,S_\nu]v_\rho v_\lambda [f_-^{\nu\rho},f_-^{\mu\lambda}] N_v)$ &  $O^{(4)}_{98}=[S^\mu ,S^\nu ]v^\lambda v^\rho \left[  F_{\lambda \mu} ^-,F_{\rho \nu} ^-\right] $\\
& 79 &$\varepsilon_{\mu\sigma\rho\lambda}(\Bar{N}_v v^\mu v^\nu [f_-^{\sigma\nu},f_+^{\rho\lambda}] N_v)$ &  $O^{(4)}_{104}=i\varepsilon _{\quad \;\tau} ^{\lambda \mu \nu} v^\tau v^\rho \left[  F_{\lambda \mu} ^-,\widetilde F_{\nu \rho} ^+\right]$ \\
& 80 & $(\Bar{N}_v S^\mu v^\nu N_v)\langle f_-^{\sigma\nu}f_{+\mu\sigma}\rangle $&  $O^{(4)}_{101}=S^\mu v^\nu \left\langle F_{\lambda \mu} ^-F_{\quad\;\nu}^{+\lambda} \right\rangle $\\
\hline
\multirow{6}{*}{$\Sigma^2 N^2$}
& 81 &$(\Bar{N}_v N_v)\langle\Sigma_+\Sigma_+\rangle $ &  $O^{(4)}_{40}=\left\langle \widetilde\chi _+\widetilde\chi _+\right\rangle $ \\
& 82 &$(\Bar{N}_v N_v)\langle\Sigma_-\Sigma_-\rangle $  &  $O^{(4)}_{40}-O^{(4)}_{115}=\left\langle \widetilde\chi _+\widetilde\chi _+\right\rangle-\langle \chi \chi^\dagger \rangle $   \\
& 83 &$(\Bar{N}_v\Sigma_+ N_v)\langle\Sigma_+\rangle $  &  $O^{(4)}_{39}=\widetilde\chi _+\left\langle \chi _+\right\rangle $ \\
& 84 &$(\Bar{N}_v\Sigma_- N_v)\langle\Sigma_-\rangle $  &  $O^{(4)}_{41}=\widetilde\chi _-\left\langle \chi _-\right\rangle $ \\
& 85 &$(\Bar{N}_v N_v)\langle\Sigma_+\rangle\langle\Sigma_+\rangle $  & $O^{(4)}_{38}=\left\langle \chi _+\right\rangle \left\langle \chi_+\right\rangle $\\
& 86 & $(\Bar{N}_v N_v)\langle\Sigma_-\rangle\langle\Sigma_-\rangle $ & $O^{(4)}_{116}=\det \chi +\det\chi ^\dagger $ \\
\hline
\multirow{6}{*}{$\Sigma f N^2$}
& 87 &$(\Bar{N}_v [S_\mu, S_\nu] N_v)\langle \Sigma+ f_+^{\mu\nu}\rangle$ &  $O^{(4)}_{108}=i[S^\mu ,S^\nu ]\left\langle \widetilde F_{\mu \nu} ^+\widetilde\chi _+\right\rangle $ \\
& 88 &$(\Bar{N}_v S_\mu v_\nu[\Sigma_+ ,f_-^{\mu\nu}]N_v)$ &  $O^{(4)}_{109}=iS^\mu v^\nu \left[  F_{\mu \nu} ^-,\widetilde\chi _+\right] $\\
& 89 &$(\Bar{N}_v S_\mu v_\nu] N_v)\langle\Sigma_- f_+^{\mu\nu}\rangle$ &  $O^{(4)}_{113}=iS^\mu v^\nu\left\langle \widetilde F_{\mu \nu} ^+\widetilde\chi _-\right\rangle $ \\
& 90 &$(\Bar{N}_v [S_\mu, S_\nu][\Sigma_- ,f_-^{\mu\nu}]N_v)$ &  $O^{(4)}_{114}=i[S^\mu ,S^\nu ]\left[  F_{\mu \nu} ^-,\widetilde\chi _-\right] $\\
& 91 &$(\Bar{N}_v [S_\mu, S_\nu] f_+^{\mu\nu}N_v)\langle\Sigma_+\rangle$ &  $O^{(4)}_{106}=i[S^\mu ,S^\nu ]\widetilde F_{\mu \nu}^+\left\langle \chi _+\right\rangle $ \\
& 92 & $(\Bar{N}_v S_\mu v_\nu f_+^{\mu\nu}N_v)\langle\Sigma_-\rangle$&  $O^{(4)}_{111}=iS^\mu v^\nu \widetilde F_{\mu \nu}^+\left\langle \chi _-\right\rangle $\\
\hline
$D^2fN^2$&93&$(\Bar{N}_v S_\mu v_\nu D^2f_-^{\mu\nu}N_v) $& $O^{(4)}_{74}=[S^\mu,S^\nu] [D^\lambda,[D_\lambda,\widetilde F_{\mu\nu}^+]] $\\
\hline
\multirow{2}{*}{$D^2\Sigma N^2$}
& 94 & $(\Bar{N}_v D^2\Sigma_+N_v) $&  $O^{(4)}_{30}=\left[ D_\mu ,\left[ D^\mu ,\widetilde\chi _+\right] \right] $ \\
& 95 &$(\Bar{N}_v N_v) D^2\langle\Sigma_+ \rangle$ & $O^{(4)}_{22}=\left[ D_\mu ,\left[ D^\mu ,\left\langle \chi_+\right\rangle \right] \right] $ \\
\hline
    \caption{Comparison between our results and Ref.~\cite{Fettes:2000gb} at $p^4$ order. The second column only provides the label $i$ for the operator $\mathcal{O}^{++}_i$ defined in Tab.~\ref{Dim-4}. }
\label{com p4}
    \end{longtable}
\end{center}
\end{landscape}

\section{Cayley-Hamilton Relations}
In this section, we list the full Cayley-Hamilton relation for the case of 4, 5 and 6 fields of the $SU(3)$ adjoint representation, in which all of the repeated fields are also considered.
\label{app: CH relations}
\subsection{4 Fields}
\paragraph{$\mathcal{A}\mathcal{B}\mathcal{C}\mathcal{D}$}

\begin{align}
    & -\lra{\mathcal{\mathcal{A}}\mathcal{\mathcal{D}}}\lra{\mathcal{\mathcal{B}}\mathcal{\mathcal{C}}} - \lra{\mathcal{\mathcal{A}}\mathcal{\mathcal{C}}}\lra{\mathcal{\mathcal{B}}\mathcal{\mathcal{D}}} - \lra{\mathcal{\mathcal{A}}\mathcal{\mathcal{B}}}\lra{\mathcal{\mathcal{C}}\mathcal{\mathcal{D}}} \notag \\
    & + \lra{\mathcal{\mathcal{A}}\mathcal{\mathcal{B}}\mathcal{\mathcal{C}}\mathcal{\mathcal{D}}} + \lra{\mathcal{\mathcal{A}}\mathcal{\mathcal{C}}\mathcal{\mathcal{B}}\mathcal{\mathcal{D}}} + \lra{\mathcal{\mathcal{A}}\mathcal{\mathcal{B}}\mathcal{\mathcal{D}}\mathcal{\mathcal{C}}} + \lra{\mathcal{\mathcal{A}}\mathcal{\mathcal{C}}\mathcal{\mathcal{D}}\mathcal{\mathcal{B}}} + \lra{\mathcal{\mathcal{A}}\mathcal{\mathcal{D}}\mathcal{\mathcal{B}}\mathcal{\mathcal{C}}} + \lra{\mathcal{\mathcal{A}}\mathcal{\mathcal{D}}\mathcal{\mathcal{C}}\mathcal{\mathcal{B}}} \notag \\
    & = 0\,,
\end{align}

\paragraph{$\mathcal{A}\mathcal{B}\mathcal{C}^2$}

\begin{equation}
    -\lra{\mathcal{A}\mathcal{B}}\lra{\mathcal{C}\mathcal{C}} - 2 \lra{\mathcal{A}\mathcal{C}}\lra{\mathcal{B}\mathcal{C}} + 2\lra{\mathcal{A}\mathcal{B}\mathcal{C}\mathcal{C}} + 2\lra{\mathcal{A}\mathcal{C}\mathcal{B}\mathcal{C}} + 2\lra{\mathcal{A}\mathcal{C}\mathcal{C}\mathcal{B}} = 0 \,,
\end{equation}

\subsection{5 Fields}

\paragraph{$\mathcal{A}\mathcal{B}\mathcal{C}\mathcal{D}\mathcal{E}$}

\begin{align}
    1) \quad & \langle \mathcal{A}\mathcal{B}\mathcal{E}\mathcal{C}\mathcal{D}\rangle +\langle \mathcal{A}\mathcal{B}\mathcal{E}\mathcal{D}\mathcal{C}\rangle +\langle \mathcal{A}\mathcal{C}\mathcal{B}\mathcal{E}\mathcal{D}\rangle +\langle \mathcal{A}\mathcal{C}\mathcal{D}\mathcal{B}\mathcal{E}\rangle +\langle \mathcal{A}\mathcal{D}\mathcal{B}\mathcal{E}\mathcal{C}\rangle +\langle \mathcal{A}\mathcal{D}\mathcal{C}\mathcal{B}\mathcal{E}\rangle  \notag \\
           = & \langle \mathcal{A}\mathcal{B}\mathcal{E}\rangle\langle \mathcal{C}\mathcal{D}\rangle +\langle \mathcal{A}\mathcal{C}\mathcal{D}\rangle\langle \mathcal{B}\mathcal{E}\rangle +\langle \mathcal{A}\mathcal{C}\rangle\langle \mathcal{B}\mathcal{E}\mathcal{D}\rangle +\langle \mathcal{A}\mathcal{D}\mathcal{C}\rangle\langle \mathcal{B}\mathcal{E}\rangle +\langle \mathcal{A}\mathcal{D}\rangle\langle \mathcal{B}\mathcal{E}\mathcal{C}\rangle \\
    2) \quad & \langle \mathcal{A}\mathcal{C}\mathcal{D}\mathcal{E}\mathcal{B}\rangle +\langle \mathcal{A}\mathcal{C}\mathcal{E}\mathcal{B}\mathcal{D}\rangle +\langle \mathcal{A}\mathcal{D}\mathcal{C}\mathcal{E}\mathcal{B}\rangle +\langle \mathcal{A}\mathcal{D}\mathcal{E}\mathcal{B}\mathcal{C}\rangle +\langle \mathcal{A}\mathcal{E}\mathcal{B}\mathcal{C}\mathcal{D}\rangle +\langle \mathcal{A}\mathcal{E}\mathcal{B}\mathcal{D}\mathcal{C}\rangle  \notag \\
           = & \langle \mathcal{A}\mathcal{C}\mathcal{D}\rangle\langle \mathcal{B}\mathcal{E}\rangle +\langle \mathcal{A}\mathcal{C}\rangle\langle \mathcal{B}\mathcal{D}\mathcal{E}\rangle +\langle \mathcal{A}\mathcal{D}\mathcal{C}\rangle\langle \mathcal{B}\mathcal{E}\rangle +\langle \mathcal{A}\mathcal{D}\rangle\langle \mathcal{B}\mathcal{C}\mathcal{E}\rangle +\langle \mathcal{A}\mathcal{E}\mathcal{B}\rangle\langle \mathcal{C}\mathcal{D}\rangle \\
    3) \quad & \langle \mathcal{A}\mathcal{B}\mathcal{C}\mathcal{D}\mathcal{E}\rangle +\langle \mathcal{A}\mathcal{B}\mathcal{D}\mathcal{E}\mathcal{C}\rangle +\langle \mathcal{A}\mathcal{C}\mathcal{B}\mathcal{D}\mathcal{E}\rangle +\langle \mathcal{A}\mathcal{C}\mathcal{D}\mathcal{E}\mathcal{B}\rangle +\langle \mathcal{A}\mathcal{D}\mathcal{E}\mathcal{B}\mathcal{C}\rangle +\langle \mathcal{A}\mathcal{D}\mathcal{E}\mathcal{C}\mathcal{B}\rangle  \notag \\
           = & \langle \mathcal{A}\mathcal{B}\mathcal{C}\rangle\langle \mathcal{D}\mathcal{E}\rangle +\langle \mathcal{A}\mathcal{B}\rangle\langle \mathcal{C}\mathcal{D}\mathcal{E}\rangle +\langle \mathcal{A}\mathcal{C}\mathcal{B}\rangle\langle \mathcal{D}\mathcal{E}\rangle +\langle \mathcal{A}\mathcal{C}\rangle\langle \mathcal{B}\mathcal{D}\mathcal{E}\rangle +\langle \mathcal{A}\mathcal{D}\mathcal{E}\rangle\langle \mathcal{B}\mathcal{C}\rangle \\
    4) \quad & \langle \mathcal{A}\mathcal{B}\mathcal{C}\mathcal{E}\mathcal{D}\rangle +\langle \mathcal{A}\mathcal{B}\mathcal{E}\mathcal{D}\mathcal{C}\rangle +\langle \mathcal{A}\mathcal{C}\mathcal{B}\mathcal{E}\mathcal{D}\rangle +\langle \mathcal{A}\mathcal{C}\mathcal{E}\mathcal{D}\mathcal{B}\rangle +\langle \mathcal{A}\mathcal{E}\mathcal{D}\mathcal{B}\mathcal{C}\rangle +\langle \mathcal{A}\mathcal{E}\mathcal{D}\mathcal{C}\mathcal{B}\rangle  \notag \\
           = & \langle \mathcal{A}\mathcal{B}\mathcal{C}\rangle\langle \mathcal{D}\mathcal{E}\rangle +\langle \mathcal{A}\mathcal{B}\rangle\langle \mathcal{C}\mathcal{E}\mathcal{D}\rangle +\langle \mathcal{A}\mathcal{C}\mathcal{B}\rangle\langle \mathcal{D}\mathcal{E}\rangle +\langle \mathcal{A}\mathcal{C}\rangle\langle \mathcal{B}\mathcal{E}\mathcal{D}\rangle +\langle \mathcal{A}\mathcal{E}\mathcal{D}\rangle\langle \mathcal{B}\mathcal{C}\rangle \\
    5) \quad & \langle \mathcal{A}\mathcal{B}\mathcal{C}\mathcal{E}\mathcal{D}\rangle +\langle \mathcal{A}\mathcal{B}\mathcal{D}\mathcal{C}\mathcal{E}\rangle +\langle \mathcal{A}\mathcal{C}\mathcal{E}\mathcal{B}\mathcal{D}\rangle +\langle \mathcal{A}\mathcal{C}\mathcal{E}\mathcal{D}\mathcal{B}\rangle +\langle \mathcal{A}\mathcal{D}\mathcal{B}\mathcal{C}\mathcal{E}\rangle +\langle \mathcal{A}\mathcal{D}\mathcal{C}\mathcal{E}\mathcal{B}\rangle  \notag \\
           = & \langle \mathcal{A}\mathcal{B}\mathcal{D}\rangle\langle \mathcal{C}\mathcal{E}\rangle +\langle \mathcal{A}\mathcal{B}\rangle\langle \mathcal{C}\mathcal{E}\mathcal{D}\rangle +\langle \mathcal{A}\mathcal{C}\mathcal{E}\rangle\langle \mathcal{B}\mathcal{D}\rangle +\langle \mathcal{A}\mathcal{D}\mathcal{B}\rangle\langle \mathcal{C}\mathcal{E}\rangle +\langle \mathcal{A}\mathcal{D}\rangle\langle \mathcal{B}\mathcal{C}\mathcal{E}\rangle \\
    6) \quad & \langle \mathcal{A}\mathcal{B}\mathcal{D}\mathcal{E}\mathcal{C}\rangle +\langle \mathcal{A}\mathcal{B}\mathcal{E}\mathcal{C}\mathcal{D}\rangle +\langle \mathcal{A}\mathcal{D}\mathcal{B}\mathcal{E}\mathcal{C}\rangle +\langle \mathcal{A}\mathcal{D}\mathcal{E}\mathcal{C}\mathcal{B}\rangle +\langle \mathcal{A}\mathcal{E}\mathcal{C}\mathcal{B}\mathcal{D}\rangle +\langle \mathcal{A}\mathcal{E}\mathcal{C}\mathcal{D}\mathcal{B}\rangle  \notag \\
           = & \langle \mathcal{A}\mathcal{B}\mathcal{D}\rangle\langle \mathcal{C}\mathcal{E}\rangle +\langle \mathcal{A}\mathcal{B}\rangle\langle \mathcal{C}\mathcal{D}\mathcal{E}\rangle +\langle \mathcal{A}\mathcal{D}\mathcal{B}\rangle\langle \mathcal{C}\mathcal{E}\rangle +\langle \mathcal{A}\mathcal{D}\rangle\langle \mathcal{B}\mathcal{E}\mathcal{C}\rangle +\langle \mathcal{A}\mathcal{E}\mathcal{C}\rangle\langle \mathcal{B}\mathcal{D}\rangle \\
    7) \quad & \langle \mathcal{A}\mathcal{B}\mathcal{C}\mathcal{D}\mathcal{E}\rangle +\langle \mathcal{A}\mathcal{B}\mathcal{E}\mathcal{C}\mathcal{D}\rangle +\langle \mathcal{A}\mathcal{C}\mathcal{D}\mathcal{B}\mathcal{E}\rangle +\langle \mathcal{A}\mathcal{C}\mathcal{D}\mathcal{E}\mathcal{B}\rangle +\langle \mathcal{A}\mathcal{E}\mathcal{B}\mathcal{C}\mathcal{D}\rangle +\langle \mathcal{A}\mathcal{E}\mathcal{C}\mathcal{D}\mathcal{B}\rangle  \notag \\
           = & \langle \mathcal{A}\mathcal{B}\mathcal{E}\rangle\langle \mathcal{C}\mathcal{D}\rangle +\langle \mathcal{A}\mathcal{B}\rangle\langle \mathcal{C}\mathcal{D}\mathcal{E}\rangle +\langle \mathcal{A}\mathcal{C}\mathcal{D}\rangle\langle \mathcal{B}\mathcal{E}\rangle +\langle \mathcal{A}\mathcal{E}\mathcal{B}\rangle\langle \mathcal{C}\mathcal{D}\rangle +\langle \mathcal{A}\mathcal{E}\rangle\langle \mathcal{B}\mathcal{C}\mathcal{D}\rangle \\
    8) \quad & \langle \mathcal{A}\mathcal{B}\mathcal{D}\mathcal{C}\mathcal{E}\rangle +\langle \mathcal{A}\mathcal{B}\mathcal{E}\mathcal{D}\mathcal{C}\rangle +\langle \mathcal{A}\mathcal{D}\mathcal{C}\mathcal{B}\mathcal{E}\rangle +\langle \mathcal{A}\mathcal{D}\mathcal{C}\mathcal{E}\mathcal{B}\rangle +\langle \mathcal{A}\mathcal{E}\mathcal{B}\mathcal{D}\mathcal{C}\rangle +\langle \mathcal{A}\mathcal{E}\mathcal{D}\mathcal{C}\mathcal{B}\rangle  \notag \\
           = & \langle \mathcal{A}\mathcal{B}\mathcal{E}\rangle\langle \mathcal{C}\mathcal{D}\rangle +\langle \mathcal{A}\mathcal{B}\rangle\langle \mathcal{C}\mathcal{E}\mathcal{D}\rangle +\langle \mathcal{A}\mathcal{D}\mathcal{C}\rangle\langle \mathcal{B}\mathcal{E}\rangle +\langle \mathcal{A}\mathcal{E}\mathcal{B}\rangle\langle \mathcal{C}\mathcal{D}\rangle +\langle \mathcal{A}\mathcal{E}\rangle\langle \mathcal{B}\mathcal{D}\mathcal{C}\rangle \\
    9) \quad & \langle \mathcal{A}\mathcal{B}\mathcal{C}\mathcal{D}\mathcal{E}\rangle +\langle \mathcal{A}\mathcal{B}\mathcal{C}\mathcal{E}\mathcal{D}\rangle +\langle \mathcal{A}\mathcal{D}\mathcal{B}\mathcal{C}\mathcal{E}\rangle +\langle \mathcal{A}\mathcal{D}\mathcal{E}\mathcal{B}\mathcal{C}\rangle +\langle \mathcal{A}\mathcal{E}\mathcal{B}\mathcal{C}\mathcal{D}\rangle +\langle \mathcal{A}\mathcal{E}\mathcal{D}\mathcal{B}\mathcal{C}\rangle  \notag \\
           = & \langle \mathcal{A}\mathcal{B}\mathcal{C}\rangle\langle \mathcal{D}\mathcal{E}\rangle +\langle \mathcal{A}\mathcal{D}\mathcal{E}\rangle\langle \mathcal{B}\mathcal{C}\rangle +\langle \mathcal{A}\mathcal{D}\rangle\langle \mathcal{B}\mathcal{C}\mathcal{E}\rangle +\langle \mathcal{A}\mathcal{E}\mathcal{D}\rangle\langle \mathcal{B}\mathcal{C}\rangle +\langle \mathcal{A}\mathcal{E}\rangle\langle \mathcal{B}\mathcal{C}\mathcal{D}\rangle \\
    10) \quad & \langle \mathcal{A}\mathcal{C}\mathcal{B}\mathcal{D}\mathcal{E}\rangle +\langle \mathcal{A}\mathcal{C}\mathcal{B}\mathcal{E}\mathcal{D}\rangle +\langle \mathcal{A}\mathcal{D}\mathcal{C}\mathcal{B}\mathcal{E}\rangle +\langle \mathcal{A}\mathcal{D}\mathcal{E}\mathcal{C}\mathcal{B}\rangle +\langle \mathcal{A}\mathcal{E}\mathcal{C}\mathcal{B}\mathcal{D}\rangle +\langle \mathcal{A}\mathcal{E}\mathcal{D}\mathcal{C}\mathcal{B}\rangle  \notag \\
            = & \langle \mathcal{A}\mathcal{C}\mathcal{B}\rangle\langle \mathcal{D}\mathcal{E}\rangle +\langle \mathcal{A}\mathcal{D}\mathcal{E}\rangle\langle \mathcal{B}\mathcal{C}\rangle +\langle \mathcal{A}\mathcal{D}\rangle\langle \mathcal{B}\mathcal{E}\mathcal{C}\rangle +\langle \mathcal{A}\mathcal{E}\mathcal{D}\rangle\langle \mathcal{B}\mathcal{C}\rangle +\langle \mathcal{A}\mathcal{E}\rangle\langle \mathcal{B}\mathcal{D}\mathcal{C}\rangle \\
    11) \quad & \langle \mathcal{A}\mathcal{B}\mathcal{D}\mathcal{C}\mathcal{E}\rangle +\langle \mathcal{A}\mathcal{B}\mathcal{D}\mathcal{E}\mathcal{C}\rangle +\langle \mathcal{A}\mathcal{C}\mathcal{B}\mathcal{D}\mathcal{E}\rangle +\langle \mathcal{A}\mathcal{C}\mathcal{E}\mathcal{B}\mathcal{D}\rangle +\langle \mathcal{A}\mathcal{E}\mathcal{B}\mathcal{D}\mathcal{C}\rangle +\langle \mathcal{A}\mathcal{E}\mathcal{C}\mathcal{B}\mathcal{D}\rangle  \notag \\
            = & \langle \mathcal{A}\mathcal{B}\mathcal{D}\rangle\langle \mathcal{C}\mathcal{E}\rangle +\langle \mathcal{A}\mathcal{C}\mathcal{E}\rangle\langle \mathcal{B}\mathcal{D}\rangle +\langle \mathcal{A}\mathcal{C}\rangle\langle \mathcal{B}\mathcal{D}\mathcal{E}\rangle +\langle \mathcal{A}\mathcal{E}\mathcal{C}\rangle\langle \mathcal{B}\mathcal{D}\rangle +\langle \mathcal{A}\mathcal{E}\rangle\langle \mathcal{B}\mathcal{D}\mathcal{C}\rangle \\
    12) \quad & \langle \mathcal{A}\mathcal{C}\mathcal{D}\mathcal{B}\mathcal{E}\rangle +\langle \mathcal{A}\mathcal{C}\mathcal{E}\mathcal{D}\mathcal{B}\rangle +\langle \mathcal{A}\mathcal{D}\mathcal{B}\mathcal{C}\mathcal{E}\rangle +\langle \mathcal{A}\mathcal{D}\mathcal{B}\mathcal{E}\mathcal{C}\rangle +\langle \mathcal{A}\mathcal{E}\mathcal{C}\mathcal{D}\mathcal{B}\rangle +\langle \mathcal{A}\mathcal{E}\mathcal{D}\mathcal{B}\mathcal{C}\rangle  \notag \\
            = & \langle \mathcal{A}\mathcal{C}\mathcal{E}\rangle\langle \mathcal{B}\mathcal{D}\rangle +\langle \mathcal{A}\mathcal{C}\rangle\langle \mathcal{B}\mathcal{E}\mathcal{D}\rangle +\langle \mathcal{A}\mathcal{D}\mathcal{B}\rangle\langle \mathcal{C}\mathcal{E}\rangle +\langle \mathcal{A}\mathcal{E}\mathcal{C}\rangle\langle \mathcal{B}\mathcal{D}\rangle +\langle \mathcal{A}\mathcal{E}\rangle\langle \mathcal{B}\mathcal{C}\mathcal{D}\rangle
\end{align}

\paragraph{$\mathcal{A}\mathcal{B}\mathcal{C}\mathcal{D}^2$}

\begin{align}
    1) \quad & \langle \mathcal{A}\mathcal{B}\mathcal{D}\mathcal{C}\mathcal{D}\rangle +\langle \mathcal{A}\mathcal{B}\mathcal{D}\mathcal{D}\mathcal{C}\rangle +\langle \mathcal{A}\mathcal{C}\mathcal{B}\mathcal{D}\mathcal{D}\rangle +\langle \mathcal{A}\mathcal{C}\mathcal{D}\mathcal{B}\mathcal{D}\rangle +\langle \mathcal{A}\mathcal{D}\mathcal{B}\mathcal{D}\mathcal{C}\rangle +\langle \mathcal{A}\mathcal{D}\mathcal{C}\mathcal{B}\mathcal{D}\rangle  \notag \\
           = & \langle \mathcal{A}\mathcal{B}\mathcal{D}\rangle\langle \mathcal{C}\mathcal{D}\rangle +\langle \mathcal{A}\mathcal{C}\mathcal{D}\rangle\langle \mathcal{B}\mathcal{D}\rangle +\langle \mathcal{A}\mathcal{C}\rangle\langle \mathcal{B}\mathcal{D}\mathcal{D}\rangle +\langle \mathcal{A}\mathcal{D}\mathcal{C}\rangle\langle \mathcal{B}\mathcal{D}\rangle +\langle \mathcal{A}\mathcal{D}\rangle\langle \mathcal{B}\mathcal{D}\mathcal{C}\rangle \\
    2) \quad & \langle \mathcal{A}\mathcal{C}\mathcal{D}\mathcal{B}\mathcal{D}\rangle +\langle \mathcal{A}\mathcal{C}\mathcal{D}\mathcal{D}\mathcal{B}\rangle +\langle \mathcal{A}\mathcal{D}\mathcal{B}\mathcal{C}\mathcal{D}\rangle +\langle \mathcal{A}\mathcal{D}\mathcal{B}\mathcal{D}\mathcal{C}\rangle +\langle \mathcal{A}\mathcal{D}\mathcal{C}\mathcal{D}\mathcal{B}\rangle +\langle \mathcal{A}\mathcal{D}\mathcal{D}\mathcal{B}\mathcal{C}\rangle  \notag \\
           = & \langle \mathcal{A}\mathcal{C}\mathcal{D}\rangle\langle \mathcal{B}\mathcal{D}\rangle +\langle \mathcal{A}\mathcal{C}\rangle\langle \mathcal{B}\mathcal{D}\mathcal{D}\rangle +\langle \mathcal{A}\mathcal{D}\mathcal{B}\rangle\langle \mathcal{C}\mathcal{D}\rangle +\langle \mathcal{A}\mathcal{D}\mathcal{C}\rangle\langle \mathcal{B}\mathcal{D}\rangle +\langle \mathcal{A}\mathcal{D}\rangle\langle \mathcal{B}\mathcal{C}\mathcal{D}\rangle \\
    3) \quad & \langle \mathcal{A}\mathcal{B}\mathcal{C}\mathcal{D}\mathcal{D}\rangle +\langle \mathcal{A}\mathcal{B}\mathcal{D}\mathcal{D}\mathcal{C}\rangle +\langle \mathcal{A}\mathcal{C}\mathcal{B}\mathcal{D}\mathcal{D}\rangle +\langle \mathcal{A}\mathcal{C}\mathcal{D}\mathcal{D}\mathcal{B}\rangle +\langle \mathcal{A}\mathcal{D}\mathcal{D}\mathcal{B}\mathcal{C}\rangle +\langle \mathcal{A}\mathcal{D}\mathcal{D}\mathcal{C}\mathcal{B}\rangle  \notag \\
           = & \langle \mathcal{A}\mathcal{B}\mathcal{C}\rangle\langle \mathcal{D}\mathcal{D}\rangle +\langle \mathcal{A}\mathcal{B}\rangle\langle \mathcal{C}\mathcal{D}\mathcal{D}\rangle +\langle \mathcal{A}\mathcal{C}\mathcal{B}\rangle\langle \mathcal{D}\mathcal{D}\rangle +\langle \mathcal{A}\mathcal{C}\rangle\langle \mathcal{B}\mathcal{D}\mathcal{D}\rangle +\langle \mathcal{A}\mathcal{D}\mathcal{D}\rangle\langle \mathcal{B}\mathcal{C}\rangle \\
    4) \quad & \langle \mathcal{A}\mathcal{B}\mathcal{C}\mathcal{D}\mathcal{D}\rangle +\langle \mathcal{A}\mathcal{B}\mathcal{D}\mathcal{C}\mathcal{D}\rangle +\langle \mathcal{A}\mathcal{C}\mathcal{D}\mathcal{B}\mathcal{D}\rangle +\langle \mathcal{A}\mathcal{C}\mathcal{D}\mathcal{D}\mathcal{B}\rangle +\langle \mathcal{A}\mathcal{D}\mathcal{B}\mathcal{C}\mathcal{D}\rangle +\langle \mathcal{A}\mathcal{D}\mathcal{C}\mathcal{D}\mathcal{B}\rangle  \notag \\
           = & \langle \mathcal{A}\mathcal{B}\mathcal{D}\rangle\langle \mathcal{C}\mathcal{D}\rangle +\langle \mathcal{A}\mathcal{B}\rangle\langle \mathcal{C}\mathcal{D}\mathcal{D}\rangle +\langle \mathcal{A}\mathcal{C}\mathcal{D}\rangle\langle \mathcal{B}\mathcal{D}\rangle +\langle \mathcal{A}\mathcal{D}\mathcal{B}\rangle\langle \mathcal{C}\mathcal{D}\rangle +\langle \mathcal{A}\mathcal{D}\rangle\langle \mathcal{B}\mathcal{C}\mathcal{D}\rangle \\
    5) \quad & \langle \mathcal{A}\mathcal{B}\mathcal{D}\mathcal{C}\mathcal{D}\rangle +\langle \mathcal{A}\mathcal{B}\mathcal{D}\mathcal{D}\mathcal{C}\rangle +\langle \mathcal{A}\mathcal{D}\mathcal{B}\mathcal{D}\mathcal{C}\rangle +\langle \mathcal{A}\mathcal{D}\mathcal{C}\mathcal{B}\mathcal{D}\rangle +\langle \mathcal{A}\mathcal{D}\mathcal{C}\mathcal{D}\mathcal{B}\rangle +\langle \mathcal{A}\mathcal{D}\mathcal{D}\mathcal{C}\mathcal{B}\rangle  \notag \\
           = & \langle \mathcal{A}\mathcal{B}\mathcal{D}\rangle\langle \mathcal{C}\mathcal{D}\rangle +\langle \mathcal{A}\mathcal{B}\rangle\langle \mathcal{C}\mathcal{D}\mathcal{D}\rangle +\langle \mathcal{A}\mathcal{D}\mathcal{B}\rangle\langle \mathcal{C}\mathcal{D}\rangle +\langle \mathcal{A}\mathcal{D}\mathcal{C}\rangle\langle \mathcal{B}\mathcal{D}\rangle +\langle \mathcal{A}\mathcal{D}\rangle\langle \mathcal{B}\mathcal{D}\mathcal{C}\rangle \\
    6) \quad & \langle \mathcal{A}\mathcal{B}\mathcal{C}\mathcal{D}\mathcal{D}\rangle +\langle \mathcal{A}\mathcal{D}\mathcal{B}\mathcal{C}\mathcal{D}\rangle +\langle \mathcal{A}\mathcal{D}\mathcal{D}\mathcal{B}\mathcal{C}\rangle  \notag \\
           = & 2\langle \mathcal{A}\mathcal{D}\mathcal{D}\rangle\langle \mathcal{B}\mathcal{C}\rangle +2\langle \mathcal{A}\mathcal{D}\rangle\langle \mathcal{B}\mathcal{C}\mathcal{D}\rangle +\langle \mathcal{A}\mathcal{B}\mathcal{C}\rangle\langle \mathcal{D}\mathcal{D}\rangle \\
    7) \quad & \langle \mathcal{A}\mathcal{C}\mathcal{B}\mathcal{D}\mathcal{D}\rangle +\langle \mathcal{A}\mathcal{D}\mathcal{C}\mathcal{B}\mathcal{D}\rangle +\langle \mathcal{A}\mathcal{D}\mathcal{D}\mathcal{C}\mathcal{B}\rangle  \notag \\
           = & 2\langle \mathcal{A}\mathcal{D}\mathcal{D}\rangle\langle \mathcal{B}\mathcal{C}\rangle +2\langle \mathcal{A}\mathcal{D}\rangle\langle \mathcal{B}\mathcal{D}\mathcal{C}\rangle +\langle \mathcal{A}\mathcal{C}\mathcal{B}\rangle\langle \mathcal{D}\mathcal{D}\rangle
\end{align}

\paragraph{$\mathcal{A}\mathcal{B}\mathcal{C}^3$}

\begin{align}
    1) \quad & \langle \mathcal{A}\mathcal{B}\mathcal{C}\mathcal{C}\mathcal{C}\rangle +\langle \mathcal{A}\mathcal{C}\mathcal{B}\mathcal{C}\mathcal{C}\rangle +\langle \mathcal{A}\mathcal{C}\mathcal{C}\mathcal{B}\mathcal{C}\rangle  \notag \\ 
    = &  2\langle \mathcal{A}\mathcal{C}\mathcal{C}\rangle\langle \mathcal{B}\mathcal{C}\rangle +2\langle \mathcal{A}\mathcal{C}\rangle\langle \mathcal{B}\mathcal{C}\mathcal{C}\rangle +\langle \mathcal{A}\mathcal{B}\mathcal{C}\rangle\langle \mathcal{C}\mathcal{C}\rangle  \\
   2) \quad & \langle \mathcal{A}\mathcal{C}\mathcal{B}\mathcal{C}\mathcal{C}\rangle +\langle \mathcal{A}\mathcal{C}\mathcal{C}\mathcal{B}\mathcal{C}\rangle +\langle \mathcal{A}\mathcal{C}\mathcal{C}\mathcal{C}\mathcal{B}\rangle  \notag \\ 
    = &  2\langle \mathcal{A}\mathcal{C}\mathcal{C}\rangle\langle \mathcal{B}\mathcal{C}\rangle +2\langle \mathcal{A}\mathcal{C}\rangle\langle \mathcal{B}\mathcal{C}\mathcal{C}\rangle +\langle \mathcal{A}\mathcal{C}\mathcal{B}\rangle\langle \mathcal{C}\mathcal{C}\rangle  \\
   3) \quad & 2\langle \mathcal{A}\mathcal{B}\mathcal{C}\mathcal{C}\mathcal{C}\rangle +2\langle \mathcal{A}\mathcal{C}\mathcal{C}\mathcal{C}\mathcal{B}\rangle +\langle \mathcal{A}\mathcal{C}\mathcal{B}\mathcal{C}\mathcal{C}\rangle +\langle \mathcal{A}\mathcal{C}\mathcal{C}\mathcal{B}\mathcal{C}\rangle  \notag \\ 
    = &  \langle \mathcal{A}\mathcal{B}\mathcal{C}\rangle\langle \mathcal{C}\mathcal{C}\rangle +\langle \mathcal{A}\mathcal{B}\rangle\langle \mathcal{C}\mathcal{C}\mathcal{C}\rangle +\langle \mathcal{A}\mathcal{C}\mathcal{B}\rangle\langle \mathcal{C}\mathcal{C}\rangle +\langle \mathcal{A}\mathcal{C}\mathcal{C}\rangle\langle \mathcal{B}\mathcal{C}\rangle +\langle \mathcal{A}\mathcal{C}\rangle\langle \mathcal{B}\mathcal{C}\mathcal{C}\rangle  \\
\end{align}

\subsection{6 Fields}

\paragraph{$\mathcal{A}\mathcal{B}\mathcal{C}\mathcal{D}\mathcal{E}\mathcal{F}$}

\begin{align}
    1) \quad & \langle \mathcal{A}\mathcal{B}\mathcal{C}\mathcal{D}\mathcal{E}\mathcal{F}\rangle +\langle \mathcal{A}\mathcal{B}\mathcal{D}\mathcal{E}\mathcal{F}\mathcal{C}\rangle +\langle \mathcal{A}\mathcal{C}\mathcal{B}\mathcal{D}\mathcal{E}\mathcal{F}\rangle +\langle \mathcal{A}\mathcal{C}\mathcal{D}\mathcal{E}\mathcal{F}\mathcal{B}\rangle +\langle \mathcal{A}\mathcal{D}\mathcal{E}\mathcal{F}\mathcal{B}\mathcal{C}\rangle +\langle \mathcal{A}\mathcal{D}\mathcal{E}\mathcal{F}\mathcal{C}\mathcal{B}\rangle  \notag \\ 
     = &  \langle \mathcal{A}\mathcal{B}\mathcal{C}\rangle\langle \mathcal{D}\mathcal{E}\mathcal{F}\rangle +\langle \mathcal{A}\mathcal{B}\rangle\langle \mathcal{C}\mathcal{D}\mathcal{E}\mathcal{F}\rangle +\langle \mathcal{A}\mathcal{C}\mathcal{B}\rangle\langle \mathcal{D}\mathcal{E}\mathcal{F}\rangle +\langle \mathcal{A}\mathcal{C}\rangle\langle \mathcal{B}\mathcal{D}\mathcal{E}\mathcal{F}\rangle +\langle \mathcal{A}\mathcal{D}\mathcal{E}\mathcal{F}\rangle\langle \mathcal{B}\mathcal{C}\rangle  \\
    2) \quad & \langle \mathcal{A}\mathcal{B}\mathcal{C}\mathcal{D}\mathcal{F}\mathcal{E}\rangle +\langle \mathcal{A}\mathcal{B}\mathcal{D}\mathcal{F}\mathcal{E}\mathcal{C}\rangle +\langle \mathcal{A}\mathcal{C}\mathcal{B}\mathcal{D}\mathcal{F}\mathcal{E}\rangle +\langle \mathcal{A}\mathcal{C}\mathcal{D}\mathcal{F}\mathcal{E}\mathcal{B}\rangle +\langle \mathcal{A}\mathcal{D}\mathcal{F}\mathcal{E}\mathcal{B}\mathcal{C}\rangle +\langle \mathcal{A}\mathcal{D}\mathcal{F}\mathcal{E}\mathcal{C}\mathcal{B}\rangle  \notag \\ 
     = &  \langle \mathcal{A}\mathcal{B}\mathcal{C}\rangle\langle \mathcal{D}\mathcal{F}\mathcal{E}\rangle +\langle \mathcal{A}\mathcal{B}\rangle\langle \mathcal{C}\mathcal{D}\mathcal{F}\mathcal{E}\rangle +\langle \mathcal{A}\mathcal{C}\mathcal{B}\rangle\langle \mathcal{D}\mathcal{F}\mathcal{E}\rangle +\langle \mathcal{A}\mathcal{C}\rangle\langle \mathcal{B}\mathcal{D}\mathcal{F}\mathcal{E}\rangle +\langle \mathcal{A}\mathcal{D}\mathcal{F}\mathcal{E}\rangle\langle \mathcal{B}\mathcal{C}\rangle  \\
    3) \quad & \langle \mathcal{A}\mathcal{B}\mathcal{C}\mathcal{E}\mathcal{D}\mathcal{F}\rangle +\langle \mathcal{A}\mathcal{B}\mathcal{E}\mathcal{D}\mathcal{F}\mathcal{C}\rangle +\langle \mathcal{A}\mathcal{C}\mathcal{B}\mathcal{E}\mathcal{D}\mathcal{F}\rangle +\langle \mathcal{A}\mathcal{C}\mathcal{E}\mathcal{D}\mathcal{F}\mathcal{B}\rangle +\langle \mathcal{A}\mathcal{E}\mathcal{D}\mathcal{F}\mathcal{B}\mathcal{C}\rangle +\langle \mathcal{A}\mathcal{E}\mathcal{D}\mathcal{F}\mathcal{C}\mathcal{B}\rangle  \notag \\ 
     = &  \langle \mathcal{A}\mathcal{B}\mathcal{C}\rangle\langle \mathcal{D}\mathcal{F}\mathcal{E}\rangle +\langle \mathcal{A}\mathcal{B}\rangle\langle \mathcal{C}\mathcal{E}\mathcal{D}\mathcal{F}\rangle +\langle \mathcal{A}\mathcal{C}\mathcal{B}\rangle\langle \mathcal{D}\mathcal{F}\mathcal{E}\rangle +\langle \mathcal{A}\mathcal{C}\rangle\langle \mathcal{B}\mathcal{E}\mathcal{D}\mathcal{F}\rangle +\langle \mathcal{A}\mathcal{E}\mathcal{D}\mathcal{F}\rangle\langle \mathcal{B}\mathcal{C}\rangle  \\
    4) \quad & \langle \mathcal{A}\mathcal{B}\mathcal{C}\mathcal{E}\mathcal{F}\mathcal{D}\rangle +\langle \mathcal{A}\mathcal{B}\mathcal{E}\mathcal{F}\mathcal{D}\mathcal{C}\rangle +\langle \mathcal{A}\mathcal{C}\mathcal{B}\mathcal{E}\mathcal{F}\mathcal{D}\rangle +\langle \mathcal{A}\mathcal{C}\mathcal{E}\mathcal{F}\mathcal{D}\mathcal{B}\rangle +\langle \mathcal{A}\mathcal{E}\mathcal{F}\mathcal{D}\mathcal{B}\mathcal{C}\rangle +\langle \mathcal{A}\mathcal{E}\mathcal{F}\mathcal{D}\mathcal{C}\mathcal{B}\rangle  \notag \\ 
     = &  \langle \mathcal{A}\mathcal{B}\mathcal{C}\rangle\langle \mathcal{D}\mathcal{E}\mathcal{F}\rangle +\langle \mathcal{A}\mathcal{B}\rangle\langle \mathcal{C}\mathcal{E}\mathcal{F}\mathcal{D}\rangle +\langle \mathcal{A}\mathcal{C}\mathcal{B}\rangle\langle \mathcal{D}\mathcal{E}\mathcal{F}\rangle +\langle \mathcal{A}\mathcal{C}\rangle\langle \mathcal{B}\mathcal{E}\mathcal{F}\mathcal{D}\rangle +\langle \mathcal{A}\mathcal{E}\mathcal{F}\mathcal{D}\rangle\langle \mathcal{B}\mathcal{C}\rangle  \\
    5) \quad & \langle \mathcal{A}\mathcal{B}\mathcal{C}\mathcal{F}\mathcal{D}\mathcal{E}\rangle +\langle \mathcal{A}\mathcal{B}\mathcal{F}\mathcal{D}\mathcal{E}\mathcal{C}\rangle +\langle \mathcal{A}\mathcal{C}\mathcal{B}\mathcal{F}\mathcal{D}\mathcal{E}\rangle +\langle \mathcal{A}\mathcal{C}\mathcal{F}\mathcal{D}\mathcal{E}\mathcal{B}\rangle +\langle \mathcal{A}\mathcal{F}\mathcal{D}\mathcal{E}\mathcal{B}\mathcal{C}\rangle +\langle \mathcal{A}\mathcal{F}\mathcal{D}\mathcal{E}\mathcal{C}\mathcal{B}\rangle  \notag \\ 
     = &  \langle \mathcal{A}\mathcal{B}\mathcal{C}\rangle\langle \mathcal{D}\mathcal{E}\mathcal{F}\rangle +\langle \mathcal{A}\mathcal{B}\rangle\langle \mathcal{C}\mathcal{F}\mathcal{D}\mathcal{E}\rangle +\langle \mathcal{A}\mathcal{C}\mathcal{B}\rangle\langle \mathcal{D}\mathcal{E}\mathcal{F}\rangle +\langle \mathcal{A}\mathcal{C}\rangle\langle \mathcal{B}\mathcal{F}\mathcal{D}\mathcal{E}\rangle +\langle \mathcal{A}\mathcal{F}\mathcal{D}\mathcal{E}\rangle\langle \mathcal{B}\mathcal{C}\rangle  \\
    6) \quad & \langle \mathcal{A}\mathcal{B}\mathcal{C}\mathcal{F}\mathcal{E}\mathcal{D}\rangle +\langle \mathcal{A}\mathcal{B}\mathcal{F}\mathcal{E}\mathcal{D}\mathcal{C}\rangle +\langle \mathcal{A}\mathcal{C}\mathcal{B}\mathcal{F}\mathcal{E}\mathcal{D}\rangle +\langle \mathcal{A}\mathcal{C}\mathcal{F}\mathcal{E}\mathcal{D}\mathcal{B}\rangle +\langle \mathcal{A}\mathcal{F}\mathcal{E}\mathcal{D}\mathcal{B}\mathcal{C}\rangle +\langle \mathcal{A}\mathcal{F}\mathcal{E}\mathcal{D}\mathcal{C}\mathcal{B}\rangle  \notag \\ 
     = &  \langle \mathcal{A}\mathcal{B}\mathcal{C}\rangle\langle \mathcal{D}\mathcal{F}\mathcal{E}\rangle +\langle \mathcal{A}\mathcal{B}\rangle\langle \mathcal{C}\mathcal{F}\mathcal{E}\mathcal{D}\rangle +\langle \mathcal{A}\mathcal{C}\mathcal{B}\rangle\langle \mathcal{D}\mathcal{F}\mathcal{E}\rangle +\langle \mathcal{A}\mathcal{C}\rangle\langle \mathcal{B}\mathcal{F}\mathcal{E}\mathcal{D}\rangle +\langle \mathcal{A}\mathcal{F}\mathcal{E}\mathcal{D}\rangle\langle \mathcal{B}\mathcal{C}\rangle  \\
    7) \quad & \langle \mathcal{A}\mathcal{B}\mathcal{C}\mathcal{E}\mathcal{F}\mathcal{D}\rangle +\langle \mathcal{A}\mathcal{B}\mathcal{D}\mathcal{C}\mathcal{E}\mathcal{F}\rangle +\langle \mathcal{A}\mathcal{C}\mathcal{E}\mathcal{F}\mathcal{B}\mathcal{D}\rangle +\langle \mathcal{A}\mathcal{C}\mathcal{E}\mathcal{F}\mathcal{D}\mathcal{B}\rangle +\langle \mathcal{A}\mathcal{D}\mathcal{B}\mathcal{C}\mathcal{E}\mathcal{F}\rangle +\langle \mathcal{A}\mathcal{D}\mathcal{C}\mathcal{E}\mathcal{F}\mathcal{B}\rangle  \notag \\ 
     = &  \langle \mathcal{A}\mathcal{B}\mathcal{D}\rangle\langle \mathcal{C}\mathcal{E}\mathcal{F}\rangle +\langle \mathcal{A}\mathcal{B}\rangle\langle \mathcal{C}\mathcal{E}\mathcal{F}\mathcal{D}\rangle +\langle \mathcal{A}\mathcal{C}\mathcal{E}\mathcal{F}\rangle\langle \mathcal{B}\mathcal{D}\rangle +\langle \mathcal{A}\mathcal{D}\mathcal{B}\rangle\langle \mathcal{C}\mathcal{E}\mathcal{F}\rangle +\langle \mathcal{A}\mathcal{D}\rangle\langle \mathcal{B}\mathcal{C}\mathcal{E}\mathcal{F}\rangle  \\
    8) \quad & \langle \mathcal{A}\mathcal{B}\mathcal{C}\mathcal{F}\mathcal{E}\mathcal{D}\rangle +\langle \mathcal{A}\mathcal{B}\mathcal{D}\mathcal{C}\mathcal{F}\mathcal{E}\rangle +\langle \mathcal{A}\mathcal{C}\mathcal{F}\mathcal{E}\mathcal{B}\mathcal{D}\rangle +\langle \mathcal{A}\mathcal{C}\mathcal{F}\mathcal{E}\mathcal{D}\mathcal{B}\rangle +\langle \mathcal{A}\mathcal{D}\mathcal{B}\mathcal{C}\mathcal{F}\mathcal{E}\rangle +\langle \mathcal{A}\mathcal{D}\mathcal{C}\mathcal{F}\mathcal{E}\mathcal{B}\rangle  \notag \\ 
     = &  \langle \mathcal{A}\mathcal{B}\mathcal{D}\rangle\langle \mathcal{C}\mathcal{F}\mathcal{E}\rangle +\langle \mathcal{A}\mathcal{B}\rangle\langle \mathcal{C}\mathcal{F}\mathcal{E}\mathcal{D}\rangle +\langle \mathcal{A}\mathcal{C}\mathcal{F}\mathcal{E}\rangle\langle \mathcal{B}\mathcal{D}\rangle +\langle \mathcal{A}\mathcal{D}\mathcal{B}\rangle\langle \mathcal{C}\mathcal{F}\mathcal{E}\rangle +\langle \mathcal{A}\mathcal{D}\rangle\langle \mathcal{B}\mathcal{C}\mathcal{F}\mathcal{E}\rangle  \\
    9) \quad & \langle \mathcal{A}\mathcal{B}\mathcal{D}\mathcal{E}\mathcal{C}\mathcal{F}\rangle +\langle \mathcal{A}\mathcal{B}\mathcal{E}\mathcal{C}\mathcal{F}\mathcal{D}\rangle +\langle \mathcal{A}\mathcal{D}\mathcal{B}\mathcal{E}\mathcal{C}\mathcal{F}\rangle +\langle \mathcal{A}\mathcal{D}\mathcal{E}\mathcal{C}\mathcal{F}\mathcal{B}\rangle +\langle \mathcal{A}\mathcal{E}\mathcal{C}\mathcal{F}\mathcal{B}\mathcal{D}\rangle +\langle \mathcal{A}\mathcal{E}\mathcal{C}\mathcal{F}\mathcal{D}\mathcal{B}\rangle  \notag \\ 
     = &  \langle \mathcal{A}\mathcal{B}\mathcal{D}\rangle\langle \mathcal{C}\mathcal{F}\mathcal{E}\rangle +\langle \mathcal{A}\mathcal{B}\rangle\langle \mathcal{C}\mathcal{F}\mathcal{D}\mathcal{E}\rangle +\langle \mathcal{A}\mathcal{D}\mathcal{B}\rangle\langle \mathcal{C}\mathcal{F}\mathcal{E}\rangle +\langle \mathcal{A}\mathcal{D}\rangle\langle \mathcal{B}\mathcal{E}\mathcal{C}\mathcal{F}\rangle +\langle \mathcal{A}\mathcal{E}\mathcal{C}\mathcal{F}\rangle\langle \mathcal{B}\mathcal{D}\rangle  \\
    10) \quad & \langle \mathcal{A}\mathcal{B}\mathcal{D}\mathcal{E}\mathcal{F}\mathcal{C}\rangle +\langle \mathcal{A}\mathcal{B}\mathcal{E}\mathcal{F}\mathcal{C}\mathcal{D}\rangle +\langle \mathcal{A}\mathcal{D}\mathcal{B}\mathcal{E}\mathcal{F}\mathcal{C}\rangle +\langle \mathcal{A}\mathcal{D}\mathcal{E}\mathcal{F}\mathcal{C}\mathcal{B}\rangle +\langle \mathcal{A}\mathcal{E}\mathcal{F}\mathcal{C}\mathcal{B}\mathcal{D}\rangle +\langle \mathcal{A}\mathcal{E}\mathcal{F}\mathcal{C}\mathcal{D}\mathcal{B}\rangle  \notag \\ 
     = &  \langle \mathcal{A}\mathcal{B}\mathcal{D}\rangle\langle \mathcal{C}\mathcal{E}\mathcal{F}\rangle +\langle \mathcal{A}\mathcal{B}\rangle\langle \mathcal{C}\mathcal{D}\mathcal{E}\mathcal{F}\rangle +\langle \mathcal{A}\mathcal{D}\mathcal{B}\rangle\langle \mathcal{C}\mathcal{E}\mathcal{F}\rangle +\langle \mathcal{A}\mathcal{D}\rangle\langle \mathcal{B}\mathcal{E}\mathcal{F}\mathcal{C}\rangle +\langle \mathcal{A}\mathcal{E}\mathcal{F}\mathcal{C}\rangle\langle \mathcal{B}\mathcal{D}\rangle  \\
    11) \quad & \langle \mathcal{A}\mathcal{B}\mathcal{D}\mathcal{F}\mathcal{C}\mathcal{E}\rangle +\langle \mathcal{A}\mathcal{B}\mathcal{F}\mathcal{C}\mathcal{E}\mathcal{D}\rangle +\langle \mathcal{A}\mathcal{D}\mathcal{B}\mathcal{F}\mathcal{C}\mathcal{E}\rangle +\langle \mathcal{A}\mathcal{D}\mathcal{F}\mathcal{C}\mathcal{E}\mathcal{B}\rangle +\langle \mathcal{A}\mathcal{F}\mathcal{C}\mathcal{E}\mathcal{B}\mathcal{D}\rangle +\langle \mathcal{A}\mathcal{F}\mathcal{C}\mathcal{E}\mathcal{D}\mathcal{B}\rangle  \notag \\ 
     = &  \langle \mathcal{A}\mathcal{B}\mathcal{D}\rangle\langle \mathcal{C}\mathcal{E}\mathcal{F}\rangle +\langle \mathcal{A}\mathcal{B}\rangle\langle \mathcal{C}\mathcal{E}\mathcal{D}\mathcal{F}\rangle +\langle \mathcal{A}\mathcal{D}\mathcal{B}\rangle\langle \mathcal{C}\mathcal{E}\mathcal{F}\rangle +\langle \mathcal{A}\mathcal{D}\rangle\langle \mathcal{B}\mathcal{F}\mathcal{C}\mathcal{E}\rangle +\langle \mathcal{A}\mathcal{F}\mathcal{C}\mathcal{E}\rangle\langle \mathcal{B}\mathcal{D}\rangle  \\
    12) \quad & \langle \mathcal{A}\mathcal{B}\mathcal{D}\mathcal{F}\mathcal{E}\mathcal{C}\rangle +\langle \mathcal{A}\mathcal{B}\mathcal{F}\mathcal{E}\mathcal{C}\mathcal{D}\rangle +\langle \mathcal{A}\mathcal{D}\mathcal{B}\mathcal{F}\mathcal{E}\mathcal{C}\rangle +\langle \mathcal{A}\mathcal{D}\mathcal{F}\mathcal{E}\mathcal{C}\mathcal{B}\rangle +\langle \mathcal{A}\mathcal{F}\mathcal{E}\mathcal{C}\mathcal{B}\mathcal{D}\rangle +\langle \mathcal{A}\mathcal{F}\mathcal{E}\mathcal{C}\mathcal{D}\mathcal{B}\rangle  \notag \\ 
     = &  \langle \mathcal{A}\mathcal{B}\mathcal{D}\rangle\langle \mathcal{C}\mathcal{F}\mathcal{E}\rangle +\langle \mathcal{A}\mathcal{B}\rangle\langle \mathcal{C}\mathcal{D}\mathcal{F}\mathcal{E}\rangle +\langle \mathcal{A}\mathcal{D}\mathcal{B}\rangle\langle \mathcal{C}\mathcal{F}\mathcal{E}\rangle +\langle \mathcal{A}\mathcal{D}\rangle\langle \mathcal{B}\mathcal{F}\mathcal{E}\mathcal{C}\rangle +\langle \mathcal{A}\mathcal{F}\mathcal{E}\mathcal{C}\rangle\langle \mathcal{B}\mathcal{D}\rangle  \\
    13) \quad & \langle \mathcal{A}\mathcal{B}\mathcal{C}\mathcal{D}\mathcal{F}\mathcal{E}\rangle +\langle \mathcal{A}\mathcal{B}\mathcal{E}\mathcal{C}\mathcal{D}\mathcal{F}\rangle +\langle \mathcal{A}\mathcal{C}\mathcal{D}\mathcal{F}\mathcal{B}\mathcal{E}\rangle +\langle \mathcal{A}\mathcal{C}\mathcal{D}\mathcal{F}\mathcal{E}\mathcal{B}\rangle +\langle \mathcal{A}\mathcal{E}\mathcal{B}\mathcal{C}\mathcal{D}\mathcal{F}\rangle +\langle \mathcal{A}\mathcal{E}\mathcal{C}\mathcal{D}\mathcal{F}\mathcal{B}\rangle  \notag \\ 
     = &  \langle \mathcal{A}\mathcal{B}\mathcal{E}\rangle\langle \mathcal{C}\mathcal{D}\mathcal{F}\rangle +\langle \mathcal{A}\mathcal{B}\rangle\langle \mathcal{C}\mathcal{D}\mathcal{F}\mathcal{E}\rangle +\langle \mathcal{A}\mathcal{C}\mathcal{D}\mathcal{F}\rangle\langle \mathcal{B}\mathcal{E}\rangle +\langle \mathcal{A}\mathcal{E}\mathcal{B}\rangle\langle \mathcal{C}\mathcal{D}\mathcal{F}\rangle +\langle \mathcal{A}\mathcal{E}\rangle\langle \mathcal{B}\mathcal{C}\mathcal{D}\mathcal{F}\rangle  \\
    14) \quad & \langle \mathcal{A}\mathcal{B}\mathcal{C}\mathcal{F}\mathcal{D}\mathcal{E}\rangle +\langle \mathcal{A}\mathcal{B}\mathcal{E}\mathcal{C}\mathcal{F}\mathcal{D}\rangle +\langle \mathcal{A}\mathcal{C}\mathcal{F}\mathcal{D}\mathcal{B}\mathcal{E}\rangle +\langle \mathcal{A}\mathcal{C}\mathcal{F}\mathcal{D}\mathcal{E}\mathcal{B}\rangle +\langle \mathcal{A}\mathcal{E}\mathcal{B}\mathcal{C}\mathcal{F}\mathcal{D}\rangle +\langle \mathcal{A}\mathcal{E}\mathcal{C}\mathcal{F}\mathcal{D}\mathcal{B}\rangle  \notag \\ 
     = &  \langle \mathcal{A}\mathcal{B}\mathcal{E}\rangle\langle \mathcal{C}\mathcal{F}\mathcal{D}\rangle +\langle \mathcal{A}\mathcal{B}\rangle\langle \mathcal{C}\mathcal{F}\mathcal{D}\mathcal{E}\rangle +\langle \mathcal{A}\mathcal{C}\mathcal{F}\mathcal{D}\rangle\langle \mathcal{B}\mathcal{E}\rangle +\langle \mathcal{A}\mathcal{E}\mathcal{B}\rangle\langle \mathcal{C}\mathcal{F}\mathcal{D}\rangle +\langle \mathcal{A}\mathcal{E}\rangle\langle \mathcal{B}\mathcal{C}\mathcal{F}\mathcal{D}\rangle  \\
    15) \quad & \langle \mathcal{A}\mathcal{B}\mathcal{D}\mathcal{C}\mathcal{F}\mathcal{E}\rangle +\langle \mathcal{A}\mathcal{B}\mathcal{E}\mathcal{D}\mathcal{C}\mathcal{F}\rangle +\langle \mathcal{A}\mathcal{D}\mathcal{C}\mathcal{F}\mathcal{B}\mathcal{E}\rangle +\langle \mathcal{A}\mathcal{D}\mathcal{C}\mathcal{F}\mathcal{E}\mathcal{B}\rangle +\langle \mathcal{A}\mathcal{E}\mathcal{B}\mathcal{D}\mathcal{C}\mathcal{F}\rangle +\langle \mathcal{A}\mathcal{E}\mathcal{D}\mathcal{C}\mathcal{F}\mathcal{B}\rangle  \notag \\ 
     = &  \langle \mathcal{A}\mathcal{B}\mathcal{E}\rangle\langle \mathcal{C}\mathcal{F}\mathcal{D}\rangle +\langle \mathcal{A}\mathcal{B}\rangle\langle \mathcal{C}\mathcal{F}\mathcal{E}\mathcal{D}\rangle +\langle \mathcal{A}\mathcal{D}\mathcal{C}\mathcal{F}\rangle\langle \mathcal{B}\mathcal{E}\rangle +\langle \mathcal{A}\mathcal{E}\mathcal{B}\rangle\langle \mathcal{C}\mathcal{F}\mathcal{D}\rangle +\langle \mathcal{A}\mathcal{E}\rangle\langle \mathcal{B}\mathcal{D}\mathcal{C}\mathcal{F}\rangle  \\
    16) \quad & \langle \mathcal{A}\mathcal{B}\mathcal{D}\mathcal{F}\mathcal{C}\mathcal{E}\rangle +\langle \mathcal{A}\mathcal{B}\mathcal{E}\mathcal{D}\mathcal{F}\mathcal{C}\rangle +\langle \mathcal{A}\mathcal{D}\mathcal{F}\mathcal{C}\mathcal{B}\mathcal{E}\rangle +\langle \mathcal{A}\mathcal{D}\mathcal{F}\mathcal{C}\mathcal{E}\mathcal{B}\rangle +\langle \mathcal{A}\mathcal{E}\mathcal{B}\mathcal{D}\mathcal{F}\mathcal{C}\rangle +\langle \mathcal{A}\mathcal{E}\mathcal{D}\mathcal{F}\mathcal{C}\mathcal{B}\rangle  \notag \\ 
     = &  \langle \mathcal{A}\mathcal{B}\mathcal{E}\rangle\langle \mathcal{C}\mathcal{D}\mathcal{F}\rangle +\langle \mathcal{A}\mathcal{B}\rangle\langle \mathcal{C}\mathcal{E}\mathcal{D}\mathcal{F}\rangle +\langle \mathcal{A}\mathcal{D}\mathcal{F}\mathcal{C}\rangle\langle \mathcal{B}\mathcal{E}\rangle +\langle \mathcal{A}\mathcal{E}\mathcal{B}\rangle\langle \mathcal{C}\mathcal{D}\mathcal{F}\rangle +\langle \mathcal{A}\mathcal{E}\rangle\langle \mathcal{B}\mathcal{D}\mathcal{F}\mathcal{C}\rangle  \\
    17) \quad & \langle \mathcal{A}\mathcal{B}\mathcal{E}\mathcal{F}\mathcal{C}\mathcal{D}\rangle +\langle \mathcal{A}\mathcal{B}\mathcal{F}\mathcal{C}\mathcal{D}\mathcal{E}\rangle +\langle \mathcal{A}\mathcal{E}\mathcal{B}\mathcal{F}\mathcal{C}\mathcal{D}\rangle +\langle \mathcal{A}\mathcal{E}\mathcal{F}\mathcal{C}\mathcal{D}\mathcal{B}\rangle +\langle \mathcal{A}\mathcal{F}\mathcal{C}\mathcal{D}\mathcal{B}\mathcal{E}\rangle +\langle \mathcal{A}\mathcal{F}\mathcal{C}\mathcal{D}\mathcal{E}\mathcal{B}\rangle  \notag \\ 
     = &  \langle \mathcal{A}\mathcal{B}\mathcal{E}\rangle\langle \mathcal{C}\mathcal{D}\mathcal{F}\rangle +\langle \mathcal{A}\mathcal{B}\rangle\langle \mathcal{C}\mathcal{D}\mathcal{E}\mathcal{F}\rangle +\langle \mathcal{A}\mathcal{E}\mathcal{B}\rangle\langle \mathcal{C}\mathcal{D}\mathcal{F}\rangle +\langle \mathcal{A}\mathcal{E}\rangle\langle \mathcal{B}\mathcal{F}\mathcal{C}\mathcal{D}\rangle +\langle \mathcal{A}\mathcal{F}\mathcal{C}\mathcal{D}\rangle\langle \mathcal{B}\mathcal{E}\rangle  
      \end{align}
 
 \begin{align}
    18) \quad & \langle \mathcal{A}\mathcal{B}\mathcal{E}\mathcal{F}\mathcal{D}\mathcal{C}\rangle +\langle \mathcal{A}\mathcal{B}\mathcal{F}\mathcal{D}\mathcal{C}\mathcal{E}\rangle +\langle \mathcal{A}\mathcal{E}\mathcal{B}\mathcal{F}\mathcal{D}\mathcal{C}\rangle +\langle \mathcal{A}\mathcal{E}\mathcal{F}\mathcal{D}\mathcal{C}\mathcal{B}\rangle +\langle \mathcal{A}\mathcal{F}\mathcal{D}\mathcal{C}\mathcal{B}\mathcal{E}\rangle +\langle \mathcal{A}\mathcal{F}\mathcal{D}\mathcal{C}\mathcal{E}\mathcal{B}\rangle  \notag \\ 
     = &  \langle \mathcal{A}\mathcal{B}\mathcal{E}\rangle\langle \mathcal{C}\mathcal{F}\mathcal{D}\rangle +\langle \mathcal{A}\mathcal{B}\rangle\langle \mathcal{C}\mathcal{E}\mathcal{F}\mathcal{D}\rangle +\langle \mathcal{A}\mathcal{E}\mathcal{B}\rangle\langle \mathcal{C}\mathcal{F}\mathcal{D}\rangle +\langle \mathcal{A}\mathcal{E}\rangle\langle \mathcal{B}\mathcal{F}\mathcal{D}\mathcal{C}\rangle +\langle \mathcal{A}\mathcal{F}\mathcal{D}\mathcal{C}\rangle\langle \mathcal{B}\mathcal{E}\rangle  \\
    19) \quad & \langle \mathcal{A}\mathcal{B}\mathcal{C}\mathcal{D}\mathcal{E}\mathcal{F}\rangle +\langle \mathcal{A}\mathcal{B}\mathcal{F}\mathcal{C}\mathcal{D}\mathcal{E}\rangle +\langle \mathcal{A}\mathcal{C}\mathcal{D}\mathcal{E}\mathcal{B}\mathcal{F}\rangle +\langle \mathcal{A}\mathcal{C}\mathcal{D}\mathcal{E}\mathcal{F}\mathcal{B}\rangle +\langle \mathcal{A}\mathcal{F}\mathcal{B}\mathcal{C}\mathcal{D}\mathcal{E}\rangle +\langle \mathcal{A}\mathcal{F}\mathcal{C}\mathcal{D}\mathcal{E}\mathcal{B}\rangle  \notag \\ 
     = &  \langle \mathcal{A}\mathcal{B}\mathcal{F}\rangle\langle \mathcal{C}\mathcal{D}\mathcal{E}\rangle +\langle \mathcal{A}\mathcal{B}\rangle\langle \mathcal{C}\mathcal{D}\mathcal{E}\mathcal{F}\rangle +\langle \mathcal{A}\mathcal{C}\mathcal{D}\mathcal{E}\rangle\langle \mathcal{B}\mathcal{F}\rangle +\langle \mathcal{A}\mathcal{F}\mathcal{B}\rangle\langle \mathcal{C}\mathcal{D}\mathcal{E}\rangle +\langle \mathcal{A}\mathcal{F}\rangle\langle \mathcal{B}\mathcal{C}\mathcal{D}\mathcal{E}\rangle  \\
    20) \quad & \langle \mathcal{A}\mathcal{B}\mathcal{C}\mathcal{E}\mathcal{D}\mathcal{F}\rangle +\langle \mathcal{A}\mathcal{B}\mathcal{F}\mathcal{C}\mathcal{E}\mathcal{D}\rangle +\langle \mathcal{A}\mathcal{C}\mathcal{E}\mathcal{D}\mathcal{B}\mathcal{F}\rangle +\langle \mathcal{A}\mathcal{C}\mathcal{E}\mathcal{D}\mathcal{F}\mathcal{B}\rangle +\langle \mathcal{A}\mathcal{F}\mathcal{B}\mathcal{C}\mathcal{E}\mathcal{D}\rangle +\langle \mathcal{A}\mathcal{F}\mathcal{C}\mathcal{E}\mathcal{D}\mathcal{B}\rangle  \notag \\ 
     = &  \langle \mathcal{A}\mathcal{B}\mathcal{F}\rangle\langle \mathcal{C}\mathcal{E}\mathcal{D}\rangle +\langle \mathcal{A}\mathcal{B}\rangle\langle \mathcal{C}\mathcal{E}\mathcal{D}\mathcal{F}\rangle +\langle \mathcal{A}\mathcal{C}\mathcal{E}\mathcal{D}\rangle\langle \mathcal{B}\mathcal{F}\rangle +\langle \mathcal{A}\mathcal{F}\mathcal{B}\rangle\langle \mathcal{C}\mathcal{E}\mathcal{D}\rangle +\langle \mathcal{A}\mathcal{F}\rangle\langle \mathcal{B}\mathcal{C}\mathcal{E}\mathcal{D}\rangle  \\
    21) \quad & \langle \mathcal{A}\mathcal{B}\mathcal{D}\mathcal{C}\mathcal{E}\mathcal{F}\rangle +\langle \mathcal{A}\mathcal{B}\mathcal{F}\mathcal{D}\mathcal{C}\mathcal{E}\rangle +\langle \mathcal{A}\mathcal{D}\mathcal{C}\mathcal{E}\mathcal{B}\mathcal{F}\rangle +\langle \mathcal{A}\mathcal{D}\mathcal{C}\mathcal{E}\mathcal{F}\mathcal{B}\rangle +\langle \mathcal{A}\mathcal{F}\mathcal{B}\mathcal{D}\mathcal{C}\mathcal{E}\rangle +\langle \mathcal{A}\mathcal{F}\mathcal{D}\mathcal{C}\mathcal{E}\mathcal{B}\rangle  \notag \\ 
     = &  \langle \mathcal{A}\mathcal{B}\mathcal{F}\rangle\langle \mathcal{C}\mathcal{E}\mathcal{D}\rangle +\langle \mathcal{A}\mathcal{B}\rangle\langle \mathcal{C}\mathcal{E}\mathcal{F}\mathcal{D}\rangle +\langle \mathcal{A}\mathcal{D}\mathcal{C}\mathcal{E}\rangle\langle \mathcal{B}\mathcal{F}\rangle +\langle \mathcal{A}\mathcal{F}\mathcal{B}\rangle\langle \mathcal{C}\mathcal{E}\mathcal{D}\rangle +\langle \mathcal{A}\mathcal{F}\rangle\langle \mathcal{B}\mathcal{D}\mathcal{C}\mathcal{E}\rangle  \\
    22) \quad & \langle \mathcal{A}\mathcal{B}\mathcal{D}\mathcal{E}\mathcal{C}\mathcal{F}\rangle +\langle \mathcal{A}\mathcal{B}\mathcal{F}\mathcal{D}\mathcal{E}\mathcal{C}\rangle +\langle \mathcal{A}\mathcal{D}\mathcal{E}\mathcal{C}\mathcal{B}\mathcal{F}\rangle +\langle \mathcal{A}\mathcal{D}\mathcal{E}\mathcal{C}\mathcal{F}\mathcal{B}\rangle +\langle \mathcal{A}\mathcal{F}\mathcal{B}\mathcal{D}\mathcal{E}\mathcal{C}\rangle +\langle \mathcal{A}\mathcal{F}\mathcal{D}\mathcal{E}\mathcal{C}\mathcal{B}\rangle  \notag \\ 
     = &  \langle \mathcal{A}\mathcal{B}\mathcal{F}\rangle\langle \mathcal{C}\mathcal{D}\mathcal{E}\rangle +\langle \mathcal{A}\mathcal{B}\rangle\langle \mathcal{C}\mathcal{F}\mathcal{D}\mathcal{E}\rangle +\langle \mathcal{A}\mathcal{D}\mathcal{E}\mathcal{C}\rangle\langle \mathcal{B}\mathcal{F}\rangle +\langle \mathcal{A}\mathcal{F}\mathcal{B}\rangle\langle \mathcal{C}\mathcal{D}\mathcal{E}\rangle +\langle \mathcal{A}\mathcal{F}\rangle\langle \mathcal{B}\mathcal{D}\mathcal{E}\mathcal{C}\rangle  \\
    23) \quad & \langle \mathcal{A}\mathcal{B}\mathcal{E}\mathcal{C}\mathcal{D}\mathcal{F}\rangle +\langle \mathcal{A}\mathcal{B}\mathcal{F}\mathcal{E}\mathcal{C}\mathcal{D}\rangle +\langle \mathcal{A}\mathcal{E}\mathcal{C}\mathcal{D}\mathcal{B}\mathcal{F}\rangle +\langle \mathcal{A}\mathcal{E}\mathcal{C}\mathcal{D}\mathcal{F}\mathcal{B}\rangle +\langle \mathcal{A}\mathcal{F}\mathcal{B}\mathcal{E}\mathcal{C}\mathcal{D}\rangle +\langle \mathcal{A}\mathcal{F}\mathcal{E}\mathcal{C}\mathcal{D}\mathcal{B}\rangle  \notag \\ 
     = &  \langle \mathcal{A}\mathcal{B}\mathcal{F}\rangle\langle \mathcal{C}\mathcal{D}\mathcal{E}\rangle +\langle \mathcal{A}\mathcal{B}\rangle\langle \mathcal{C}\mathcal{D}\mathcal{F}\mathcal{E}\rangle +\langle \mathcal{A}\mathcal{E}\mathcal{C}\mathcal{D}\rangle\langle \mathcal{B}\mathcal{F}\rangle +\langle \mathcal{A}\mathcal{F}\mathcal{B}\rangle\langle \mathcal{C}\mathcal{D}\mathcal{E}\rangle +\langle \mathcal{A}\mathcal{F}\rangle\langle \mathcal{B}\mathcal{E}\mathcal{C}\mathcal{D}\rangle  \\
    24) \quad & \langle \mathcal{A}\mathcal{B}\mathcal{E}\mathcal{D}\mathcal{C}\mathcal{F}\rangle +\langle \mathcal{A}\mathcal{B}\mathcal{F}\mathcal{E}\mathcal{D}\mathcal{C}\rangle +\langle \mathcal{A}\mathcal{E}\mathcal{D}\mathcal{C}\mathcal{B}\mathcal{F}\rangle +\langle \mathcal{A}\mathcal{E}\mathcal{D}\mathcal{C}\mathcal{F}\mathcal{B}\rangle +\langle \mathcal{A}\mathcal{F}\mathcal{B}\mathcal{E}\mathcal{D}\mathcal{C}\rangle +\langle \mathcal{A}\mathcal{F}\mathcal{E}\mathcal{D}\mathcal{C}\mathcal{B}\rangle  \notag \\ 
     = &  \langle \mathcal{A}\mathcal{B}\mathcal{F}\rangle\langle \mathcal{C}\mathcal{E}\mathcal{D}\rangle +\langle \mathcal{A}\mathcal{B}\rangle\langle \mathcal{C}\mathcal{F}\mathcal{E}\mathcal{D}\rangle +\langle \mathcal{A}\mathcal{E}\mathcal{D}\mathcal{C}\rangle\langle \mathcal{B}\mathcal{F}\rangle +\langle \mathcal{A}\mathcal{F}\mathcal{B}\rangle\langle \mathcal{C}\mathcal{E}\mathcal{D}\rangle +\langle \mathcal{A}\mathcal{F}\rangle\langle \mathcal{B}\mathcal{E}\mathcal{D}\mathcal{C}\rangle  \\
    25) \quad & \langle \mathcal{A}\mathcal{B}\mathcal{E}\mathcal{F}\mathcal{C}\mathcal{D}\rangle +\langle \mathcal{A}\mathcal{B}\mathcal{E}\mathcal{F}\mathcal{D}\mathcal{C}\rangle +\langle \mathcal{A}\mathcal{C}\mathcal{B}\mathcal{E}\mathcal{F}\mathcal{D}\rangle +\langle \mathcal{A}\mathcal{C}\mathcal{D}\mathcal{B}\mathcal{E}\mathcal{F}\rangle +\langle \mathcal{A}\mathcal{D}\mathcal{B}\mathcal{E}\mathcal{F}\mathcal{C}\rangle +\langle \mathcal{A}\mathcal{D}\mathcal{C}\mathcal{B}\mathcal{E}\mathcal{F}\rangle  \notag \\ 
     = &  \langle \mathcal{A}\mathcal{B}\mathcal{E}\mathcal{F}\rangle\langle \mathcal{C}\mathcal{D}\rangle +\langle \mathcal{A}\mathcal{C}\mathcal{D}\rangle\langle \mathcal{B}\mathcal{E}\mathcal{F}\rangle +\langle \mathcal{A}\mathcal{C}\rangle\langle \mathcal{B}\mathcal{E}\mathcal{F}\mathcal{D}\rangle +\langle \mathcal{A}\mathcal{D}\mathcal{C}\rangle\langle \mathcal{B}\mathcal{E}\mathcal{F}\rangle +\langle \mathcal{A}\mathcal{D}\rangle\langle \mathcal{B}\mathcal{E}\mathcal{F}\mathcal{C}\rangle  \\
    26) \quad & \langle \mathcal{A}\mathcal{B}\mathcal{F}\mathcal{E}\mathcal{C}\mathcal{D}\rangle +\langle \mathcal{A}\mathcal{B}\mathcal{F}\mathcal{E}\mathcal{D}\mathcal{C}\rangle +\langle \mathcal{A}\mathcal{C}\mathcal{B}\mathcal{F}\mathcal{E}\mathcal{D}\rangle +\langle \mathcal{A}\mathcal{C}\mathcal{D}\mathcal{B}\mathcal{F}\mathcal{E}\rangle +\langle \mathcal{A}\mathcal{D}\mathcal{B}\mathcal{F}\mathcal{E}\mathcal{C}\rangle +\langle \mathcal{A}\mathcal{D}\mathcal{C}\mathcal{B}\mathcal{F}\mathcal{E}\rangle  \notag \\ 
     = &  \langle \mathcal{A}\mathcal{B}\mathcal{F}\mathcal{E}\rangle\langle \mathcal{C}\mathcal{D}\rangle +\langle \mathcal{A}\mathcal{C}\mathcal{D}\rangle\langle \mathcal{B}\mathcal{F}\mathcal{E}\rangle +\langle \mathcal{A}\mathcal{C}\rangle\langle \mathcal{B}\mathcal{F}\mathcal{E}\mathcal{D}\rangle +\langle \mathcal{A}\mathcal{D}\mathcal{C}\rangle\langle \mathcal{B}\mathcal{F}\mathcal{E}\rangle +\langle \mathcal{A}\mathcal{D}\rangle\langle \mathcal{B}\mathcal{F}\mathcal{E}\mathcal{C}\rangle  \\
    27) \quad & \langle \mathcal{A}\mathcal{C}\mathcal{D}\mathcal{E}\mathcal{B}\mathcal{F}\rangle +\langle \mathcal{A}\mathcal{C}\mathcal{E}\mathcal{B}\mathcal{F}\mathcal{D}\rangle +\langle \mathcal{A}\mathcal{D}\mathcal{C}\mathcal{E}\mathcal{B}\mathcal{F}\rangle +\langle \mathcal{A}\mathcal{D}\mathcal{E}\mathcal{B}\mathcal{F}\mathcal{C}\rangle +\langle \mathcal{A}\mathcal{E}\mathcal{B}\mathcal{F}\mathcal{C}\mathcal{D}\rangle +\langle \mathcal{A}\mathcal{E}\mathcal{B}\mathcal{F}\mathcal{D}\mathcal{C}\rangle  \notag \\ 
     = &  \langle \mathcal{A}\mathcal{C}\mathcal{D}\rangle\langle \mathcal{B}\mathcal{F}\mathcal{E}\rangle +\langle \mathcal{A}\mathcal{C}\rangle\langle \mathcal{B}\mathcal{F}\mathcal{D}\mathcal{E}\rangle +\langle \mathcal{A}\mathcal{D}\mathcal{C}\rangle\langle \mathcal{B}\mathcal{F}\mathcal{E}\rangle +\langle \mathcal{A}\mathcal{D}\rangle\langle \mathcal{B}\mathcal{F}\mathcal{C}\mathcal{E}\rangle +\langle \mathcal{A}\mathcal{E}\mathcal{B}\mathcal{F}\rangle\langle \mathcal{C}\mathcal{D}\rangle  \\
    28) \quad & \langle \mathcal{A}\mathcal{C}\mathcal{D}\mathcal{E}\mathcal{F}\mathcal{B}\rangle +\langle \mathcal{A}\mathcal{C}\mathcal{E}\mathcal{F}\mathcal{B}\mathcal{D}\rangle +\langle \mathcal{A}\mathcal{D}\mathcal{C}\mathcal{E}\mathcal{F}\mathcal{B}\rangle +\langle \mathcal{A}\mathcal{D}\mathcal{E}\mathcal{F}\mathcal{B}\mathcal{C}\rangle +\langle \mathcal{A}\mathcal{E}\mathcal{F}\mathcal{B}\mathcal{C}\mathcal{D}\rangle +\langle \mathcal{A}\mathcal{E}\mathcal{F}\mathcal{B}\mathcal{D}\mathcal{C}\rangle  \notag \\ 
     = &  \langle \mathcal{A}\mathcal{C}\mathcal{D}\rangle\langle \mathcal{B}\mathcal{E}\mathcal{F}\rangle +\langle \mathcal{A}\mathcal{C}\rangle\langle \mathcal{B}\mathcal{D}\mathcal{E}\mathcal{F}\rangle +\langle \mathcal{A}\mathcal{D}\mathcal{C}\rangle\langle \mathcal{B}\mathcal{E}\mathcal{F}\rangle +\langle \mathcal{A}\mathcal{D}\rangle\langle \mathcal{B}\mathcal{C}\mathcal{E}\mathcal{F}\rangle +\langle \mathcal{A}\mathcal{E}\mathcal{F}\mathcal{B}\rangle\langle \mathcal{C}\mathcal{D}\rangle  \\
    29) \quad & \langle \mathcal{A}\mathcal{C}\mathcal{D}\mathcal{F}\mathcal{B}\mathcal{E}\rangle +\langle \mathcal{A}\mathcal{C}\mathcal{F}\mathcal{B}\mathcal{E}\mathcal{D}\rangle +\langle \mathcal{A}\mathcal{D}\mathcal{C}\mathcal{F}\mathcal{B}\mathcal{E}\rangle +\langle \mathcal{A}\mathcal{D}\mathcal{F}\mathcal{B}\mathcal{E}\mathcal{C}\rangle +\langle \mathcal{A}\mathcal{F}\mathcal{B}\mathcal{E}\mathcal{C}\mathcal{D}\rangle +\langle \mathcal{A}\mathcal{F}\mathcal{B}\mathcal{E}\mathcal{D}\mathcal{C}\rangle  \notag \\ 
     = &  \langle \mathcal{A}\mathcal{C}\mathcal{D}\rangle\langle \mathcal{B}\mathcal{E}\mathcal{F}\rangle +\langle \mathcal{A}\mathcal{C}\rangle\langle \mathcal{B}\mathcal{E}\mathcal{D}\mathcal{F}\rangle +\langle \mathcal{A}\mathcal{D}\mathcal{C}\rangle\langle \mathcal{B}\mathcal{E}\mathcal{F}\rangle +\langle \mathcal{A}\mathcal{D}\rangle\langle \mathcal{B}\mathcal{E}\mathcal{C}\mathcal{F}\rangle +\langle \mathcal{A}\mathcal{F}\mathcal{B}\mathcal{E}\rangle\langle \mathcal{C}\mathcal{D}\rangle  \\
    30) \quad & \langle \mathcal{A}\mathcal{C}\mathcal{D}\mathcal{F}\mathcal{E}\mathcal{B}\rangle +\langle \mathcal{A}\mathcal{C}\mathcal{F}\mathcal{E}\mathcal{B}\mathcal{D}\rangle +\langle \mathcal{A}\mathcal{D}\mathcal{C}\mathcal{F}\mathcal{E}\mathcal{B}\rangle +\langle \mathcal{A}\mathcal{D}\mathcal{F}\mathcal{E}\mathcal{B}\mathcal{C}\rangle +\langle \mathcal{A}\mathcal{F}\mathcal{E}\mathcal{B}\mathcal{C}\mathcal{D}\rangle +\langle \mathcal{A}\mathcal{F}\mathcal{E}\mathcal{B}\mathcal{D}\mathcal{C}\rangle  \notag \\ 
     = &  \langle \mathcal{A}\mathcal{C}\mathcal{D}\rangle\langle \mathcal{B}\mathcal{F}\mathcal{E}\rangle +\langle \mathcal{A}\mathcal{C}\rangle\langle \mathcal{B}\mathcal{D}\mathcal{F}\mathcal{E}\rangle +\langle \mathcal{A}\mathcal{D}\mathcal{C}\rangle\langle \mathcal{B}\mathcal{F}\mathcal{E}\rangle +\langle \mathcal{A}\mathcal{D}\rangle\langle \mathcal{B}\mathcal{C}\mathcal{F}\mathcal{E}\rangle +\langle \mathcal{A}\mathcal{F}\mathcal{E}\mathcal{B}\rangle\langle \mathcal{C}\mathcal{D}\rangle  \\
    31) \quad & \langle \mathcal{A}\mathcal{B}\mathcal{D}\mathcal{F}\mathcal{C}\mathcal{E}\rangle +\langle \mathcal{A}\mathcal{B}\mathcal{D}\mathcal{F}\mathcal{E}\mathcal{C}\rangle +\langle \mathcal{A}\mathcal{C}\mathcal{B}\mathcal{D}\mathcal{F}\mathcal{E}\rangle +\langle \mathcal{A}\mathcal{C}\mathcal{E}\mathcal{B}\mathcal{D}\mathcal{F}\rangle +\langle \mathcal{A}\mathcal{E}\mathcal{B}\mathcal{D}\mathcal{F}\mathcal{C}\rangle +\langle \mathcal{A}\mathcal{E}\mathcal{C}\mathcal{B}\mathcal{D}\mathcal{F}\rangle  \notag \\ 
     = &  \langle \mathcal{A}\mathcal{B}\mathcal{D}\mathcal{F}\rangle\langle \mathcal{C}\mathcal{E}\rangle +\langle \mathcal{A}\mathcal{C}\mathcal{E}\rangle\langle \mathcal{B}\mathcal{D}\mathcal{F}\rangle +\langle \mathcal{A}\mathcal{C}\rangle\langle \mathcal{B}\mathcal{D}\mathcal{F}\mathcal{E}\rangle +\langle \mathcal{A}\mathcal{E}\mathcal{C}\rangle\langle \mathcal{B}\mathcal{D}\mathcal{F}\rangle +\langle \mathcal{A}\mathcal{E}\rangle\langle \mathcal{B}\mathcal{D}\mathcal{F}\mathcal{C}\rangle  \\
    32) \quad & \langle \mathcal{A}\mathcal{B}\mathcal{F}\mathcal{D}\mathcal{C}\mathcal{E}\rangle +\langle \mathcal{A}\mathcal{B}\mathcal{F}\mathcal{D}\mathcal{E}\mathcal{C}\rangle +\langle \mathcal{A}\mathcal{C}\mathcal{B}\mathcal{F}\mathcal{D}\mathcal{E}\rangle +\langle \mathcal{A}\mathcal{C}\mathcal{E}\mathcal{B}\mathcal{F}\mathcal{D}\rangle +\langle \mathcal{A}\mathcal{E}\mathcal{B}\mathcal{F}\mathcal{D}\mathcal{C}\rangle +\langle \mathcal{A}\mathcal{E}\mathcal{C}\mathcal{B}\mathcal{F}\mathcal{D}\rangle  \notag \\ 
     = &  \langle \mathcal{A}\mathcal{B}\mathcal{F}\mathcal{D}\rangle\langle \mathcal{C}\mathcal{E}\rangle +\langle \mathcal{A}\mathcal{C}\mathcal{E}\rangle\langle \mathcal{B}\mathcal{F}\mathcal{D}\rangle +\langle \mathcal{A}\mathcal{C}\rangle\langle \mathcal{B}\mathcal{F}\mathcal{D}\mathcal{E}\rangle +\langle \mathcal{A}\mathcal{E}\mathcal{C}\rangle\langle \mathcal{B}\mathcal{F}\mathcal{D}\rangle +\langle \mathcal{A}\mathcal{E}\rangle\langle \mathcal{B}\mathcal{F}\mathcal{D}\mathcal{C}\rangle  \\
    33) \quad & \langle \mathcal{A}\mathcal{C}\mathcal{D}\mathcal{B}\mathcal{F}\mathcal{E}\rangle +\langle \mathcal{A}\mathcal{C}\mathcal{E}\mathcal{D}\mathcal{B}\mathcal{F}\rangle +\langle \mathcal{A}\mathcal{D}\mathcal{B}\mathcal{F}\mathcal{C}\mathcal{E}\rangle +\langle \mathcal{A}\mathcal{D}\mathcal{B}\mathcal{F}\mathcal{E}\mathcal{C}\rangle +\langle \mathcal{A}\mathcal{E}\mathcal{C}\mathcal{D}\mathcal{B}\mathcal{F}\rangle +\langle \mathcal{A}\mathcal{E}\mathcal{D}\mathcal{B}\mathcal{F}\mathcal{C}\rangle  \notag \\ 
     = &  \langle \mathcal{A}\mathcal{C}\mathcal{E}\rangle\langle \mathcal{B}\mathcal{F}\mathcal{D}\rangle +\langle \mathcal{A}\mathcal{C}\rangle\langle \mathcal{B}\mathcal{F}\mathcal{E}\mathcal{D}\rangle +\langle \mathcal{A}\mathcal{D}\mathcal{B}\mathcal{F}\rangle\langle \mathcal{C}\mathcal{E}\rangle +\langle \mathcal{A}\mathcal{E}\mathcal{C}\rangle\langle \mathcal{B}\mathcal{F}\mathcal{D}\rangle +\langle \mathcal{A}\mathcal{E}\rangle\langle \mathcal{B}\mathcal{F}\mathcal{C}\mathcal{D}\rangle  
      \end{align}
 
 \begin{align}
    34) \quad & \langle \mathcal{A}\mathcal{C}\mathcal{D}\mathcal{F}\mathcal{B}\mathcal{E}\rangle +\langle \mathcal{A}\mathcal{C}\mathcal{E}\mathcal{D}\mathcal{F}\mathcal{B}\rangle +\langle \mathcal{A}\mathcal{D}\mathcal{F}\mathcal{B}\mathcal{C}\mathcal{E}\rangle +\langle \mathcal{A}\mathcal{D}\mathcal{F}\mathcal{B}\mathcal{E}\mathcal{C}\rangle +\langle \mathcal{A}\mathcal{E}\mathcal{C}\mathcal{D}\mathcal{F}\mathcal{B}\rangle +\langle \mathcal{A}\mathcal{E}\mathcal{D}\mathcal{F}\mathcal{B}\mathcal{C}\rangle  \notag \\ 
     = &  \langle \mathcal{A}\mathcal{C}\mathcal{E}\rangle\langle \mathcal{B}\mathcal{D}\mathcal{F}\rangle +\langle \mathcal{A}\mathcal{C}\rangle\langle \mathcal{B}\mathcal{E}\mathcal{D}\mathcal{F}\rangle +\langle \mathcal{A}\mathcal{D}\mathcal{F}\mathcal{B}\rangle\langle \mathcal{C}\mathcal{E}\rangle +\langle \mathcal{A}\mathcal{E}\mathcal{C}\rangle\langle \mathcal{B}\mathcal{D}\mathcal{F}\rangle +\langle \mathcal{A}\mathcal{E}\rangle\langle \mathcal{B}\mathcal{C}\mathcal{D}\mathcal{F}\rangle  \\
    35) \quad & \langle \mathcal{A}\mathcal{C}\mathcal{E}\mathcal{F}\mathcal{B}\mathcal{D}\rangle +\langle \mathcal{A}\mathcal{C}\mathcal{F}\mathcal{B}\mathcal{D}\mathcal{E}\rangle +\langle \mathcal{A}\mathcal{E}\mathcal{C}\mathcal{F}\mathcal{B}\mathcal{D}\rangle +\langle \mathcal{A}\mathcal{E}\mathcal{F}\mathcal{B}\mathcal{D}\mathcal{C}\rangle +\langle \mathcal{A}\mathcal{F}\mathcal{B}\mathcal{D}\mathcal{C}\mathcal{E}\rangle +\langle \mathcal{A}\mathcal{F}\mathcal{B}\mathcal{D}\mathcal{E}\mathcal{C}\rangle  \notag \\ 
     = &  \langle \mathcal{A}\mathcal{C}\mathcal{E}\rangle\langle \mathcal{B}\mathcal{D}\mathcal{F}\rangle +\langle \mathcal{A}\mathcal{C}\rangle\langle \mathcal{B}\mathcal{D}\mathcal{E}\mathcal{F}\rangle +\langle \mathcal{A}\mathcal{E}\mathcal{C}\rangle\langle \mathcal{B}\mathcal{D}\mathcal{F}\rangle +\langle \mathcal{A}\mathcal{E}\rangle\langle \mathcal{B}\mathcal{D}\mathcal{C}\mathcal{F}\rangle +\langle \mathcal{A}\mathcal{F}\mathcal{B}\mathcal{D}\rangle\langle \mathcal{C}\mathcal{E}\rangle  \\
    36) \quad & \langle \mathcal{A}\mathcal{C}\mathcal{E}\mathcal{F}\mathcal{D}\mathcal{B}\rangle +\langle \mathcal{A}\mathcal{C}\mathcal{F}\mathcal{D}\mathcal{B}\mathcal{E}\rangle +\langle \mathcal{A}\mathcal{E}\mathcal{C}\mathcal{F}\mathcal{D}\mathcal{B}\rangle +\langle \mathcal{A}\mathcal{E}\mathcal{F}\mathcal{D}\mathcal{B}\mathcal{C}\rangle +\langle \mathcal{A}\mathcal{F}\mathcal{D}\mathcal{B}\mathcal{C}\mathcal{E}\rangle +\langle \mathcal{A}\mathcal{F}\mathcal{D}\mathcal{B}\mathcal{E}\mathcal{C}\rangle  \notag \\ 
     = &  \langle \mathcal{A}\mathcal{C}\mathcal{E}\rangle\langle \mathcal{B}\mathcal{F}\mathcal{D}\rangle +\langle \mathcal{A}\mathcal{C}\rangle\langle \mathcal{B}\mathcal{E}\mathcal{F}\mathcal{D}\rangle +\langle \mathcal{A}\mathcal{E}\mathcal{C}\rangle\langle \mathcal{B}\mathcal{F}\mathcal{D}\rangle +\langle \mathcal{A}\mathcal{E}\rangle\langle \mathcal{B}\mathcal{C}\mathcal{F}\mathcal{D}\rangle +\langle \mathcal{A}\mathcal{F}\mathcal{D}\mathcal{B}\rangle\langle \mathcal{C}\mathcal{E}\rangle  \\
    37) \quad & \langle \mathcal{A}\mathcal{B}\mathcal{D}\mathcal{E}\mathcal{C}\mathcal{F}\rangle +\langle \mathcal{A}\mathcal{B}\mathcal{D}\mathcal{E}\mathcal{F}\mathcal{C}\rangle +\langle \mathcal{A}\mathcal{C}\mathcal{B}\mathcal{D}\mathcal{E}\mathcal{F}\rangle +\langle \mathcal{A}\mathcal{C}\mathcal{F}\mathcal{B}\mathcal{D}\mathcal{E}\rangle +\langle \mathcal{A}\mathcal{F}\mathcal{B}\mathcal{D}\mathcal{E}\mathcal{C}\rangle +\langle \mathcal{A}\mathcal{F}\mathcal{C}\mathcal{B}\mathcal{D}\mathcal{E}\rangle  \notag \\ 
     = &  \langle \mathcal{A}\mathcal{B}\mathcal{D}\mathcal{E}\rangle\langle \mathcal{C}\mathcal{F}\rangle +\langle \mathcal{A}\mathcal{C}\mathcal{F}\rangle\langle \mathcal{B}\mathcal{D}\mathcal{E}\rangle +\langle \mathcal{A}\mathcal{C}\rangle\langle \mathcal{B}\mathcal{D}\mathcal{E}\mathcal{F}\rangle +\langle \mathcal{A}\mathcal{F}\mathcal{C}\rangle\langle \mathcal{B}\mathcal{D}\mathcal{E}\rangle +\langle \mathcal{A}\mathcal{F}\rangle\langle \mathcal{B}\mathcal{D}\mathcal{E}\mathcal{C}\rangle  \\
    38) \quad & \langle \mathcal{A}\mathcal{B}\mathcal{E}\mathcal{D}\mathcal{C}\mathcal{F}\rangle +\langle \mathcal{A}\mathcal{B}\mathcal{E}\mathcal{D}\mathcal{F}\mathcal{C}\rangle +\langle \mathcal{A}\mathcal{C}\mathcal{B}\mathcal{E}\mathcal{D}\mathcal{F}\rangle +\langle \mathcal{A}\mathcal{C}\mathcal{F}\mathcal{B}\mathcal{E}\mathcal{D}\rangle +\langle \mathcal{A}\mathcal{F}\mathcal{B}\mathcal{E}\mathcal{D}\mathcal{C}\rangle +\langle \mathcal{A}\mathcal{F}\mathcal{C}\mathcal{B}\mathcal{E}\mathcal{D}\rangle  \notag \\ 
     = &  \langle \mathcal{A}\mathcal{B}\mathcal{E}\mathcal{D}\rangle\langle \mathcal{C}\mathcal{F}\rangle +\langle \mathcal{A}\mathcal{C}\mathcal{F}\rangle\langle \mathcal{B}\mathcal{E}\mathcal{D}\rangle +\langle \mathcal{A}\mathcal{C}\rangle\langle \mathcal{B}\mathcal{E}\mathcal{D}\mathcal{F}\rangle +\langle \mathcal{A}\mathcal{F}\mathcal{C}\rangle\langle \mathcal{B}\mathcal{E}\mathcal{D}\rangle +\langle \mathcal{A}\mathcal{F}\rangle\langle \mathcal{B}\mathcal{E}\mathcal{D}\mathcal{C}\rangle  \\
    39) \quad & \langle \mathcal{A}\mathcal{C}\mathcal{D}\mathcal{B}\mathcal{E}\mathcal{F}\rangle +\langle \mathcal{A}\mathcal{C}\mathcal{F}\mathcal{D}\mathcal{B}\mathcal{E}\rangle +\langle \mathcal{A}\mathcal{D}\mathcal{B}\mathcal{E}\mathcal{C}\mathcal{F}\rangle +\langle \mathcal{A}\mathcal{D}\mathcal{B}\mathcal{E}\mathcal{F}\mathcal{C}\rangle +\langle \mathcal{A}\mathcal{F}\mathcal{C}\mathcal{D}\mathcal{B}\mathcal{E}\rangle +\langle \mathcal{A}\mathcal{F}\mathcal{D}\mathcal{B}\mathcal{E}\mathcal{C}\rangle  \notag \\ 
     = &  \langle \mathcal{A}\mathcal{C}\mathcal{F}\rangle\langle \mathcal{B}\mathcal{E}\mathcal{D}\rangle +\langle \mathcal{A}\mathcal{C}\rangle\langle \mathcal{B}\mathcal{E}\mathcal{F}\mathcal{D}\rangle +\langle \mathcal{A}\mathcal{D}\mathcal{B}\mathcal{E}\rangle\langle \mathcal{C}\mathcal{F}\rangle +\langle \mathcal{A}\mathcal{F}\mathcal{C}\rangle\langle \mathcal{B}\mathcal{E}\mathcal{D}\rangle +\langle \mathcal{A}\mathcal{F}\rangle\langle \mathcal{B}\mathcal{E}\mathcal{C}\mathcal{D}\rangle  \\
    40) \quad & \langle \mathcal{A}\mathcal{C}\mathcal{D}\mathcal{E}\mathcal{B}\mathcal{F}\rangle +\langle \mathcal{A}\mathcal{C}\mathcal{F}\mathcal{D}\mathcal{E}\mathcal{B}\rangle +\langle \mathcal{A}\mathcal{D}\mathcal{E}\mathcal{B}\mathcal{C}\mathcal{F}\rangle +\langle \mathcal{A}\mathcal{D}\mathcal{E}\mathcal{B}\mathcal{F}\mathcal{C}\rangle +\langle \mathcal{A}\mathcal{F}\mathcal{C}\mathcal{D}\mathcal{E}\mathcal{B}\rangle +\langle \mathcal{A}\mathcal{F}\mathcal{D}\mathcal{E}\mathcal{B}\mathcal{C}\rangle  \notag \\ 
     = &  \langle \mathcal{A}\mathcal{C}\mathcal{F}\rangle\langle \mathcal{B}\mathcal{D}\mathcal{E}\rangle +\langle \mathcal{A}\mathcal{C}\rangle\langle \mathcal{B}\mathcal{F}\mathcal{D}\mathcal{E}\rangle +\langle \mathcal{A}\mathcal{D}\mathcal{E}\mathcal{B}\rangle\langle \mathcal{C}\mathcal{F}\rangle +\langle \mathcal{A}\mathcal{F}\mathcal{C}\rangle\langle \mathcal{B}\mathcal{D}\mathcal{E}\rangle +\langle \mathcal{A}\mathcal{F}\rangle\langle \mathcal{B}\mathcal{C}\mathcal{D}\mathcal{E}\rangle  \\
    41) \quad & \langle \mathcal{A}\mathcal{C}\mathcal{E}\mathcal{B}\mathcal{D}\mathcal{F}\rangle +\langle \mathcal{A}\mathcal{C}\mathcal{F}\mathcal{E}\mathcal{B}\mathcal{D}\rangle +\langle \mathcal{A}\mathcal{E}\mathcal{B}\mathcal{D}\mathcal{C}\mathcal{F}\rangle +\langle \mathcal{A}\mathcal{E}\mathcal{B}\mathcal{D}\mathcal{F}\mathcal{C}\rangle +\langle \mathcal{A}\mathcal{F}\mathcal{C}\mathcal{E}\mathcal{B}\mathcal{D}\rangle +\langle \mathcal{A}\mathcal{F}\mathcal{E}\mathcal{B}\mathcal{D}\mathcal{C}\rangle  \notag \\ 
     = &  \langle \mathcal{A}\mathcal{C}\mathcal{F}\rangle\langle \mathcal{B}\mathcal{D}\mathcal{E}\rangle +\langle \mathcal{A}\mathcal{C}\rangle\langle \mathcal{B}\mathcal{D}\mathcal{F}\mathcal{E}\rangle +\langle \mathcal{A}\mathcal{E}\mathcal{B}\mathcal{D}\rangle\langle \mathcal{C}\mathcal{F}\rangle +\langle \mathcal{A}\mathcal{F}\mathcal{C}\rangle\langle \mathcal{B}\mathcal{D}\mathcal{E}\rangle +\langle \mathcal{A}\mathcal{F}\rangle\langle \mathcal{B}\mathcal{D}\mathcal{C}\mathcal{E}\rangle  \\
    42) \quad & \langle \mathcal{A}\mathcal{C}\mathcal{E}\mathcal{D}\mathcal{B}\mathcal{F}\rangle +\langle \mathcal{A}\mathcal{C}\mathcal{F}\mathcal{E}\mathcal{D}\mathcal{B}\rangle +\langle \mathcal{A}\mathcal{E}\mathcal{D}\mathcal{B}\mathcal{C}\mathcal{F}\rangle +\langle \mathcal{A}\mathcal{E}\mathcal{D}\mathcal{B}\mathcal{F}\mathcal{C}\rangle +\langle \mathcal{A}\mathcal{F}\mathcal{C}\mathcal{E}\mathcal{D}\mathcal{B}\rangle +\langle \mathcal{A}\mathcal{F}\mathcal{E}\mathcal{D}\mathcal{B}\mathcal{C}\rangle  \notag \\ 
     = &  \langle \mathcal{A}\mathcal{C}\mathcal{F}\rangle\langle \mathcal{B}\mathcal{E}\mathcal{D}\rangle +\langle \mathcal{A}\mathcal{C}\rangle\langle \mathcal{B}\mathcal{F}\mathcal{E}\mathcal{D}\rangle +\langle \mathcal{A}\mathcal{E}\mathcal{D}\mathcal{B}\rangle\langle \mathcal{C}\mathcal{F}\rangle +\langle \mathcal{A}\mathcal{F}\mathcal{C}\rangle\langle \mathcal{B}\mathcal{E}\mathcal{D}\rangle +\langle \mathcal{A}\mathcal{F}\rangle\langle \mathcal{B}\mathcal{C}\mathcal{E}\mathcal{D}\rangle  \\
    43) \quad & \langle \mathcal{A}\mathcal{B}\mathcal{C}\mathcal{F}\mathcal{D}\mathcal{E}\rangle +\langle \mathcal{A}\mathcal{B}\mathcal{C}\mathcal{F}\mathcal{E}\mathcal{D}\rangle +\langle \mathcal{A}\mathcal{D}\mathcal{B}\mathcal{C}\mathcal{F}\mathcal{E}\rangle +\langle \mathcal{A}\mathcal{D}\mathcal{E}\mathcal{B}\mathcal{C}\mathcal{F}\rangle +\langle \mathcal{A}\mathcal{E}\mathcal{B}\mathcal{C}\mathcal{F}\mathcal{D}\rangle +\langle \mathcal{A}\mathcal{E}\mathcal{D}\mathcal{B}\mathcal{C}\mathcal{F}\rangle  \notag \\ 
     = &  \langle \mathcal{A}\mathcal{B}\mathcal{C}\mathcal{F}\rangle\langle \mathcal{D}\mathcal{E}\rangle +\langle \mathcal{A}\mathcal{D}\mathcal{E}\rangle\langle \mathcal{B}\mathcal{C}\mathcal{F}\rangle +\langle \mathcal{A}\mathcal{D}\rangle\langle \mathcal{B}\mathcal{C}\mathcal{F}\mathcal{E}\rangle +\langle \mathcal{A}\mathcal{E}\mathcal{D}\rangle\langle \mathcal{B}\mathcal{C}\mathcal{F}\rangle +\langle \mathcal{A}\mathcal{E}\rangle\langle \mathcal{B}\mathcal{C}\mathcal{F}\mathcal{D}\rangle  \\
    44) \quad & \langle \mathcal{A}\mathcal{B}\mathcal{F}\mathcal{C}\mathcal{D}\mathcal{E}\rangle +\langle \mathcal{A}\mathcal{B}\mathcal{F}\mathcal{C}\mathcal{E}\mathcal{D}\rangle +\langle \mathcal{A}\mathcal{D}\mathcal{B}\mathcal{F}\mathcal{C}\mathcal{E}\rangle +\langle \mathcal{A}\mathcal{D}\mathcal{E}\mathcal{B}\mathcal{F}\mathcal{C}\rangle +\langle \mathcal{A}\mathcal{E}\mathcal{B}\mathcal{F}\mathcal{C}\mathcal{D}\rangle +\langle \mathcal{A}\mathcal{E}\mathcal{D}\mathcal{B}\mathcal{F}\mathcal{C}\rangle  \notag \\ 
     = &  \langle \mathcal{A}\mathcal{B}\mathcal{F}\mathcal{C}\rangle\langle \mathcal{D}\mathcal{E}\rangle +\langle \mathcal{A}\mathcal{D}\mathcal{E}\rangle\langle \mathcal{B}\mathcal{F}\mathcal{C}\rangle +\langle \mathcal{A}\mathcal{D}\rangle\langle \mathcal{B}\mathcal{F}\mathcal{C}\mathcal{E}\rangle +\langle \mathcal{A}\mathcal{E}\mathcal{D}\rangle\langle \mathcal{B}\mathcal{F}\mathcal{C}\rangle +\langle \mathcal{A}\mathcal{E}\rangle\langle \mathcal{B}\mathcal{F}\mathcal{C}\mathcal{D}\rangle  \\
    45) \quad & \langle \mathcal{A}\mathcal{C}\mathcal{B}\mathcal{F}\mathcal{D}\mathcal{E}\rangle +\langle \mathcal{A}\mathcal{C}\mathcal{B}\mathcal{F}\mathcal{E}\mathcal{D}\rangle +\langle \mathcal{A}\mathcal{D}\mathcal{C}\mathcal{B}\mathcal{F}\mathcal{E}\rangle +\langle \mathcal{A}\mathcal{D}\mathcal{E}\mathcal{C}\mathcal{B}\mathcal{F}\rangle +\langle \mathcal{A}\mathcal{E}\mathcal{C}\mathcal{B}\mathcal{F}\mathcal{D}\rangle +\langle \mathcal{A}\mathcal{E}\mathcal{D}\mathcal{C}\mathcal{B}\mathcal{F}\rangle  \notag \\ 
     = &  \langle \mathcal{A}\mathcal{C}\mathcal{B}\mathcal{F}\rangle\langle \mathcal{D}\mathcal{E}\rangle +\langle \mathcal{A}\mathcal{D}\mathcal{E}\rangle\langle \mathcal{B}\mathcal{F}\mathcal{C}\rangle +\langle \mathcal{A}\mathcal{D}\rangle\langle \mathcal{B}\mathcal{F}\mathcal{E}\mathcal{C}\rangle +\langle \mathcal{A}\mathcal{E}\mathcal{D}\rangle\langle \mathcal{B}\mathcal{F}\mathcal{C}\rangle +\langle \mathcal{A}\mathcal{E}\rangle\langle \mathcal{B}\mathcal{F}\mathcal{D}\mathcal{C}\rangle  \\
    46) \quad & \langle \mathcal{A}\mathcal{C}\mathcal{F}\mathcal{B}\mathcal{D}\mathcal{E}\rangle +\langle \mathcal{A}\mathcal{C}\mathcal{F}\mathcal{B}\mathcal{E}\mathcal{D}\rangle +\langle \mathcal{A}\mathcal{D}\mathcal{C}\mathcal{F}\mathcal{B}\mathcal{E}\rangle +\langle \mathcal{A}\mathcal{D}\mathcal{E}\mathcal{C}\mathcal{F}\mathcal{B}\rangle +\langle \mathcal{A}\mathcal{E}\mathcal{C}\mathcal{F}\mathcal{B}\mathcal{D}\rangle +\langle \mathcal{A}\mathcal{E}\mathcal{D}\mathcal{C}\mathcal{F}\mathcal{B}\rangle  \notag \\ 
     = &  \langle \mathcal{A}\mathcal{C}\mathcal{F}\mathcal{B}\rangle\langle \mathcal{D}\mathcal{E}\rangle +\langle \mathcal{A}\mathcal{D}\mathcal{E}\rangle\langle \mathcal{B}\mathcal{C}\mathcal{F}\rangle +\langle \mathcal{A}\mathcal{D}\rangle\langle \mathcal{B}\mathcal{E}\mathcal{C}\mathcal{F}\rangle +\langle \mathcal{A}\mathcal{E}\mathcal{D}\rangle\langle \mathcal{B}\mathcal{C}\mathcal{F}\rangle +\langle \mathcal{A}\mathcal{E}\rangle\langle \mathcal{B}\mathcal{D}\mathcal{C}\mathcal{F}\rangle  \\
    47) \quad & \langle \mathcal{A}\mathcal{D}\mathcal{E}\mathcal{F}\mathcal{B}\mathcal{C}\rangle +\langle \mathcal{A}\mathcal{D}\mathcal{F}\mathcal{B}\mathcal{C}\mathcal{E}\rangle +\langle \mathcal{A}\mathcal{E}\mathcal{D}\mathcal{F}\mathcal{B}\mathcal{C}\rangle +\langle \mathcal{A}\mathcal{E}\mathcal{F}\mathcal{B}\mathcal{C}\mathcal{D}\rangle +\langle \mathcal{A}\mathcal{F}\mathcal{B}\mathcal{C}\mathcal{D}\mathcal{E}\rangle +\langle \mathcal{A}\mathcal{F}\mathcal{B}\mathcal{C}\mathcal{E}\mathcal{D}\rangle  \notag \\ 
     = &  \langle \mathcal{A}\mathcal{D}\mathcal{E}\rangle\langle \mathcal{B}\mathcal{C}\mathcal{F}\rangle +\langle \mathcal{A}\mathcal{D}\rangle\langle \mathcal{B}\mathcal{C}\mathcal{E}\mathcal{F}\rangle +\langle \mathcal{A}\mathcal{E}\mathcal{D}\rangle\langle \mathcal{B}\mathcal{C}\mathcal{F}\rangle +\langle \mathcal{A}\mathcal{E}\rangle\langle \mathcal{B}\mathcal{C}\mathcal{D}\mathcal{F}\rangle +\langle \mathcal{A}\mathcal{F}\mathcal{B}\mathcal{C}\rangle\langle \mathcal{D}\mathcal{E}\rangle  \\
    48) \quad & \langle \mathcal{A}\mathcal{D}\mathcal{E}\mathcal{F}\mathcal{C}\mathcal{B}\rangle +\langle \mathcal{A}\mathcal{D}\mathcal{F}\mathcal{C}\mathcal{B}\mathcal{E}\rangle +\langle \mathcal{A}\mathcal{E}\mathcal{D}\mathcal{F}\mathcal{C}\mathcal{B}\rangle +\langle \mathcal{A}\mathcal{E}\mathcal{F}\mathcal{C}\mathcal{B}\mathcal{D}\rangle +\langle \mathcal{A}\mathcal{F}\mathcal{C}\mathcal{B}\mathcal{D}\mathcal{E}\rangle +\langle \mathcal{A}\mathcal{F}\mathcal{C}\mathcal{B}\mathcal{E}\mathcal{D}\rangle  \notag \\ 
     = &  \langle \mathcal{A}\mathcal{D}\mathcal{E}\rangle\langle \mathcal{B}\mathcal{F}\mathcal{C}\rangle +\langle \mathcal{A}\mathcal{D}\rangle\langle \mathcal{B}\mathcal{E}\mathcal{F}\mathcal{C}\rangle +\langle \mathcal{A}\mathcal{E}\mathcal{D}\rangle\langle \mathcal{B}\mathcal{F}\mathcal{C}\rangle +\langle \mathcal{A}\mathcal{E}\rangle\langle \mathcal{B}\mathcal{D}\mathcal{F}\mathcal{C}\rangle +\langle \mathcal{A}\mathcal{F}\mathcal{C}\mathcal{B}\rangle\langle \mathcal{D}\mathcal{E}\rangle  \\
    49) \quad & \langle \mathcal{A}\mathcal{B}\mathcal{C}\mathcal{E}\mathcal{D}\mathcal{F}\rangle +\langle \mathcal{A}\mathcal{B}\mathcal{C}\mathcal{E}\mathcal{F}\mathcal{D}\rangle +\langle \mathcal{A}\mathcal{D}\mathcal{B}\mathcal{C}\mathcal{E}\mathcal{F}\rangle +\langle \mathcal{A}\mathcal{D}\mathcal{F}\mathcal{B}\mathcal{C}\mathcal{E}\rangle +\langle \mathcal{A}\mathcal{F}\mathcal{B}\mathcal{C}\mathcal{E}\mathcal{D}\rangle +\langle \mathcal{A}\mathcal{F}\mathcal{D}\mathcal{B}\mathcal{C}\mathcal{E}\rangle  \notag \\ 
     = &  \langle \mathcal{A}\mathcal{B}\mathcal{C}\mathcal{E}\rangle\langle \mathcal{D}\mathcal{F}\rangle +\langle \mathcal{A}\mathcal{D}\mathcal{F}\rangle\langle \mathcal{B}\mathcal{C}\mathcal{E}\rangle +\langle \mathcal{A}\mathcal{D}\rangle\langle \mathcal{B}\mathcal{C}\mathcal{E}\mathcal{F}\rangle +\langle \mathcal{A}\mathcal{F}\mathcal{D}\rangle\langle \mathcal{B}\mathcal{C}\mathcal{E}\rangle +\langle \mathcal{A}\mathcal{F}\rangle\langle \mathcal{B}\mathcal{C}\mathcal{E}\mathcal{D}\rangle  \\
    50) \quad & \langle \mathcal{A}\mathcal{B}\mathcal{E}\mathcal{C}\mathcal{D}\mathcal{F}\rangle +\langle \mathcal{A}\mathcal{B}\mathcal{E}\mathcal{C}\mathcal{F}\mathcal{D}\rangle +\langle \mathcal{A}\mathcal{D}\mathcal{B}\mathcal{E}\mathcal{C}\mathcal{F}\rangle +\langle \mathcal{A}\mathcal{D}\mathcal{F}\mathcal{B}\mathcal{E}\mathcal{C}\rangle +\langle \mathcal{A}\mathcal{F}\mathcal{B}\mathcal{E}\mathcal{C}\mathcal{D}\rangle +\langle \mathcal{A}\mathcal{F}\mathcal{D}\mathcal{B}\mathcal{E}\mathcal{C}\rangle  \notag \\ 
     = &  \langle \mathcal{A}\mathcal{B}\mathcal{E}\mathcal{C}\rangle\langle \mathcal{D}\mathcal{F}\rangle +\langle \mathcal{A}\mathcal{D}\mathcal{F}\rangle\langle \mathcal{B}\mathcal{E}\mathcal{C}\rangle +\langle \mathcal{A}\mathcal{D}\rangle\langle \mathcal{B}\mathcal{E}\mathcal{C}\mathcal{F}\rangle +\langle \mathcal{A}\mathcal{F}\mathcal{D}\rangle\langle \mathcal{B}\mathcal{E}\mathcal{C}\rangle +\langle \mathcal{A}\mathcal{F}\rangle\langle \mathcal{B}\mathcal{E}\mathcal{C}\mathcal{D}\rangle 
     \end{align}
 
 \begin{align}
    51) \quad & \langle \mathcal{A}\mathcal{C}\mathcal{B}\mathcal{E}\mathcal{D}\mathcal{F}\rangle +\langle \mathcal{A}\mathcal{C}\mathcal{B}\mathcal{E}\mathcal{F}\mathcal{D}\rangle +\langle \mathcal{A}\mathcal{D}\mathcal{C}\mathcal{B}\mathcal{E}\mathcal{F}\rangle +\langle \mathcal{A}\mathcal{D}\mathcal{F}\mathcal{C}\mathcal{B}\mathcal{E}\rangle +\langle \mathcal{A}\mathcal{F}\mathcal{C}\mathcal{B}\mathcal{E}\mathcal{D}\rangle +\langle \mathcal{A}\mathcal{F}\mathcal{D}\mathcal{C}\mathcal{B}\mathcal{E}\rangle  \notag \\ 
     = &  \langle \mathcal{A}\mathcal{C}\mathcal{B}\mathcal{E}\rangle\langle \mathcal{D}\mathcal{F}\rangle +\langle \mathcal{A}\mathcal{D}\mathcal{F}\rangle\langle \mathcal{B}\mathcal{E}\mathcal{C}\rangle +\langle \mathcal{A}\mathcal{D}\rangle\langle \mathcal{B}\mathcal{E}\mathcal{F}\mathcal{C}\rangle +\langle \mathcal{A}\mathcal{F}\mathcal{D}\rangle\langle \mathcal{B}\mathcal{E}\mathcal{C}\rangle +\langle \mathcal{A}\mathcal{F}\rangle\langle \mathcal{B}\mathcal{E}\mathcal{D}\mathcal{C}\rangle  \\
    52) \quad & \langle \mathcal{A}\mathcal{C}\mathcal{E}\mathcal{B}\mathcal{D}\mathcal{F}\rangle +\langle \mathcal{A}\mathcal{C}\mathcal{E}\mathcal{B}\mathcal{F}\mathcal{D}\rangle +\langle \mathcal{A}\mathcal{D}\mathcal{C}\mathcal{E}\mathcal{B}\mathcal{F}\rangle +\langle \mathcal{A}\mathcal{D}\mathcal{F}\mathcal{C}\mathcal{E}\mathcal{B}\rangle +\langle \mathcal{A}\mathcal{F}\mathcal{C}\mathcal{E}\mathcal{B}\mathcal{D}\rangle +\langle \mathcal{A}\mathcal{F}\mathcal{D}\mathcal{C}\mathcal{E}\mathcal{B}\rangle  \notag \\ 
     = &  \langle \mathcal{A}\mathcal{C}\mathcal{E}\mathcal{B}\rangle\langle \mathcal{D}\mathcal{F}\rangle +\langle \mathcal{A}\mathcal{D}\mathcal{F}\rangle\langle \mathcal{B}\mathcal{C}\mathcal{E}\rangle +\langle \mathcal{A}\mathcal{D}\rangle\langle \mathcal{B}\mathcal{F}\mathcal{C}\mathcal{E}\rangle +\langle \mathcal{A}\mathcal{F}\mathcal{D}\rangle\langle \mathcal{B}\mathcal{C}\mathcal{E}\rangle +\langle \mathcal{A}\mathcal{F}\rangle\langle \mathcal{B}\mathcal{D}\mathcal{C}\mathcal{E}\rangle  \\
    53) \quad & \langle \mathcal{A}\mathcal{D}\mathcal{E}\mathcal{B}\mathcal{C}\mathcal{F}\rangle +\langle \mathcal{A}\mathcal{D}\mathcal{F}\mathcal{E}\mathcal{B}\mathcal{C}\rangle +\langle \mathcal{A}\mathcal{E}\mathcal{B}\mathcal{C}\mathcal{D}\mathcal{F}\rangle +\langle \mathcal{A}\mathcal{E}\mathcal{B}\mathcal{C}\mathcal{F}\mathcal{D}\rangle +\langle \mathcal{A}\mathcal{F}\mathcal{D}\mathcal{E}\mathcal{B}\mathcal{C}\rangle +\langle \mathcal{A}\mathcal{F}\mathcal{E}\mathcal{B}\mathcal{C}\mathcal{D}\rangle  \notag \\ 
     = &  \langle \mathcal{A}\mathcal{D}\mathcal{F}\rangle\langle \mathcal{B}\mathcal{C}\mathcal{E}\rangle +\langle \mathcal{A}\mathcal{D}\rangle\langle \mathcal{B}\mathcal{C}\mathcal{F}\mathcal{E}\rangle +\langle \mathcal{A}\mathcal{E}\mathcal{B}\mathcal{C}\rangle\langle \mathcal{D}\mathcal{F}\rangle +\langle \mathcal{A}\mathcal{F}\mathcal{D}\rangle\langle \mathcal{B}\mathcal{C}\mathcal{E}\rangle +\langle \mathcal{A}\mathcal{F}\rangle\langle \mathcal{B}\mathcal{C}\mathcal{D}\mathcal{E}\rangle  \\
    54) \quad & \langle \mathcal{A}\mathcal{D}\mathcal{E}\mathcal{C}\mathcal{B}\mathcal{F}\rangle +\langle \mathcal{A}\mathcal{D}\mathcal{F}\mathcal{E}\mathcal{C}\mathcal{B}\rangle +\langle \mathcal{A}\mathcal{E}\mathcal{C}\mathcal{B}\mathcal{D}\mathcal{F}\rangle +\langle \mathcal{A}\mathcal{E}\mathcal{C}\mathcal{B}\mathcal{F}\mathcal{D}\rangle +\langle \mathcal{A}\mathcal{F}\mathcal{D}\mathcal{E}\mathcal{C}\mathcal{B}\rangle +\langle \mathcal{A}\mathcal{F}\mathcal{E}\mathcal{C}\mathcal{B}\mathcal{D}\rangle  \notag \\ 
     = &  \langle \mathcal{A}\mathcal{D}\mathcal{F}\rangle\langle \mathcal{B}\mathcal{E}\mathcal{C}\rangle +\langle \mathcal{A}\mathcal{D}\rangle\langle \mathcal{B}\mathcal{F}\mathcal{E}\mathcal{C}\rangle +\langle \mathcal{A}\mathcal{E}\mathcal{C}\mathcal{B}\rangle\langle \mathcal{D}\mathcal{F}\rangle +\langle \mathcal{A}\mathcal{F}\mathcal{D}\rangle\langle \mathcal{B}\mathcal{E}\mathcal{C}\rangle +\langle \mathcal{A}\mathcal{F}\rangle\langle \mathcal{B}\mathcal{D}\mathcal{E}\mathcal{C}\rangle  \\
    55) \quad & \langle \mathcal{A}\mathcal{B}\mathcal{C}\mathcal{D}\mathcal{E}\mathcal{F}\rangle +\langle \mathcal{A}\mathcal{B}\mathcal{C}\mathcal{D}\mathcal{F}\mathcal{E}\rangle +\langle \mathcal{A}\mathcal{E}\mathcal{B}\mathcal{C}\mathcal{D}\mathcal{F}\rangle +\langle \mathcal{A}\mathcal{E}\mathcal{F}\mathcal{B}\mathcal{C}\mathcal{D}\rangle +\langle \mathcal{A}\mathcal{F}\mathcal{B}\mathcal{C}\mathcal{D}\mathcal{E}\rangle +\langle \mathcal{A}\mathcal{F}\mathcal{E}\mathcal{B}\mathcal{C}\mathcal{D}\rangle  \notag \\ 
     = &  \langle \mathcal{A}\mathcal{B}\mathcal{C}\mathcal{D}\rangle\langle \mathcal{E}\mathcal{F}\rangle +\langle \mathcal{A}\mathcal{E}\mathcal{F}\rangle\langle \mathcal{B}\mathcal{C}\mathcal{D}\rangle +\langle \mathcal{A}\mathcal{E}\rangle\langle \mathcal{B}\mathcal{C}\mathcal{D}\mathcal{F}\rangle +\langle \mathcal{A}\mathcal{F}\mathcal{E}\rangle\langle \mathcal{B}\mathcal{C}\mathcal{D}\rangle +\langle \mathcal{A}\mathcal{F}\rangle\langle \mathcal{B}\mathcal{C}\mathcal{D}\mathcal{E}\rangle  \\
    56) \quad & \langle \mathcal{A}\mathcal{B}\mathcal{D}\mathcal{C}\mathcal{E}\mathcal{F}\rangle +\langle \mathcal{A}\mathcal{B}\mathcal{D}\mathcal{C}\mathcal{F}\mathcal{E}\rangle +\langle \mathcal{A}\mathcal{E}\mathcal{B}\mathcal{D}\mathcal{C}\mathcal{F}\rangle +\langle \mathcal{A}\mathcal{E}\mathcal{F}\mathcal{B}\mathcal{D}\mathcal{C}\rangle +\langle \mathcal{A}\mathcal{F}\mathcal{B}\mathcal{D}\mathcal{C}\mathcal{E}\rangle +\langle \mathcal{A}\mathcal{F}\mathcal{E}\mathcal{B}\mathcal{D}\mathcal{C}\rangle  \notag \\ 
     = &  \langle \mathcal{A}\mathcal{B}\mathcal{D}\mathcal{C}\rangle\langle \mathcal{E}\mathcal{F}\rangle +\langle \mathcal{A}\mathcal{E}\mathcal{F}\rangle\langle \mathcal{B}\mathcal{D}\mathcal{C}\rangle +\langle \mathcal{A}\mathcal{E}\rangle\langle \mathcal{B}\mathcal{D}\mathcal{C}\mathcal{F}\rangle +\langle \mathcal{A}\mathcal{F}\mathcal{E}\rangle\langle \mathcal{B}\mathcal{D}\mathcal{C}\rangle +\langle \mathcal{A}\mathcal{F}\rangle\langle \mathcal{B}\mathcal{D}\mathcal{C}\mathcal{E}\rangle  \\
    57) \quad & \langle \mathcal{A}\mathcal{C}\mathcal{B}\mathcal{D}\mathcal{E}\mathcal{F}\rangle +\langle \mathcal{A}\mathcal{C}\mathcal{B}\mathcal{D}\mathcal{F}\mathcal{E}\rangle +\langle \mathcal{A}\mathcal{E}\mathcal{C}\mathcal{B}\mathcal{D}\mathcal{F}\rangle +\langle \mathcal{A}\mathcal{E}\mathcal{F}\mathcal{C}\mathcal{B}\mathcal{D}\rangle +\langle \mathcal{A}\mathcal{F}\mathcal{C}\mathcal{B}\mathcal{D}\mathcal{E}\rangle +\langle \mathcal{A}\mathcal{F}\mathcal{E}\mathcal{C}\mathcal{B}\mathcal{D}\rangle  \notag \\ 
     = &  \langle \mathcal{A}\mathcal{C}\mathcal{B}\mathcal{D}\rangle\langle \mathcal{E}\mathcal{F}\rangle +\langle \mathcal{A}\mathcal{E}\mathcal{F}\rangle\langle \mathcal{B}\mathcal{D}\mathcal{C}\rangle +\langle \mathcal{A}\mathcal{E}\rangle\langle \mathcal{B}\mathcal{D}\mathcal{F}\mathcal{C}\rangle +\langle \mathcal{A}\mathcal{F}\mathcal{E}\rangle\langle \mathcal{B}\mathcal{D}\mathcal{C}\rangle +\langle \mathcal{A}\mathcal{F}\rangle\langle \mathcal{B}\mathcal{D}\mathcal{E}\mathcal{C}\rangle  \\
    58) \quad & \langle \mathcal{A}\mathcal{C}\mathcal{D}\mathcal{B}\mathcal{E}\mathcal{F}\rangle +\langle \mathcal{A}\mathcal{C}\mathcal{D}\mathcal{B}\mathcal{F}\mathcal{E}\rangle +\langle \mathcal{A}\mathcal{E}\mathcal{C}\mathcal{D}\mathcal{B}\mathcal{F}\rangle +\langle \mathcal{A}\mathcal{E}\mathcal{F}\mathcal{C}\mathcal{D}\mathcal{B}\rangle +\langle \mathcal{A}\mathcal{F}\mathcal{C}\mathcal{D}\mathcal{B}\mathcal{E}\rangle +\langle \mathcal{A}\mathcal{F}\mathcal{E}\mathcal{C}\mathcal{D}\mathcal{B}\rangle  \notag \\ 
     = &  \langle \mathcal{A}\mathcal{C}\mathcal{D}\mathcal{B}\rangle\langle \mathcal{E}\mathcal{F}\rangle +\langle \mathcal{A}\mathcal{E}\mathcal{F}\rangle\langle \mathcal{B}\mathcal{C}\mathcal{D}\rangle +\langle \mathcal{A}\mathcal{E}\rangle\langle \mathcal{B}\mathcal{F}\mathcal{C}\mathcal{D}\rangle +\langle \mathcal{A}\mathcal{F}\mathcal{E}\rangle\langle \mathcal{B}\mathcal{C}\mathcal{D}\rangle +\langle \mathcal{A}\mathcal{F}\rangle\langle \mathcal{B}\mathcal{E}\mathcal{C}\mathcal{D}\rangle  \\
    59) \quad & \langle \mathcal{A}\mathcal{D}\mathcal{B}\mathcal{C}\mathcal{E}\mathcal{F}\rangle +\langle \mathcal{A}\mathcal{D}\mathcal{B}\mathcal{C}\mathcal{F}\mathcal{E}\rangle +\langle \mathcal{A}\mathcal{E}\mathcal{D}\mathcal{B}\mathcal{C}\mathcal{F}\rangle +\langle \mathcal{A}\mathcal{E}\mathcal{F}\mathcal{D}\mathcal{B}\mathcal{C}\rangle +\langle \mathcal{A}\mathcal{F}\mathcal{D}\mathcal{B}\mathcal{C}\mathcal{E}\rangle +\langle \mathcal{A}\mathcal{F}\mathcal{E}\mathcal{D}\mathcal{B}\mathcal{C}\rangle  \notag \\ 
     = &  \langle \mathcal{A}\mathcal{D}\mathcal{B}\mathcal{C}\rangle\langle \mathcal{E}\mathcal{F}\rangle +\langle \mathcal{A}\mathcal{E}\mathcal{F}\rangle\langle \mathcal{B}\mathcal{C}\mathcal{D}\rangle +\langle \mathcal{A}\mathcal{E}\rangle\langle \mathcal{B}\mathcal{C}\mathcal{F}\mathcal{D}\rangle +\langle \mathcal{A}\mathcal{F}\mathcal{E}\rangle\langle \mathcal{B}\mathcal{C}\mathcal{D}\rangle +\langle \mathcal{A}\mathcal{F}\rangle\langle \mathcal{B}\mathcal{C}\mathcal{E}\mathcal{D}\rangle  \\
    60) \quad & \langle \mathcal{A}\mathcal{D}\mathcal{C}\mathcal{B}\mathcal{E}\mathcal{F}\rangle +\langle \mathcal{A}\mathcal{D}\mathcal{C}\mathcal{B}\mathcal{F}\mathcal{E}\rangle +\langle \mathcal{A}\mathcal{E}\mathcal{D}\mathcal{C}\mathcal{B}\mathcal{F}\rangle +\langle \mathcal{A}\mathcal{E}\mathcal{F}\mathcal{D}\mathcal{C}\mathcal{B}\rangle +\langle \mathcal{A}\mathcal{F}\mathcal{D}\mathcal{C}\mathcal{B}\mathcal{E}\rangle +\langle \mathcal{A}\mathcal{F}\mathcal{E}\mathcal{D}\mathcal{C}\mathcal{B}\rangle  \notag \\ 
     = &  \langle \mathcal{A}\mathcal{D}\mathcal{C}\mathcal{B}\rangle\langle \mathcal{E}\mathcal{F}\rangle +\langle \mathcal{A}\mathcal{E}\mathcal{F}\rangle\langle \mathcal{B}\mathcal{D}\mathcal{C}\rangle +\langle \mathcal{A}\mathcal{E}\rangle\langle \mathcal{B}\mathcal{F}\mathcal{D}\mathcal{C}\rangle +\langle \mathcal{A}\mathcal{F}\mathcal{E}\rangle\langle \mathcal{B}\mathcal{D}\mathcal{C}\rangle +\langle \mathcal{A}\mathcal{F}\rangle\langle \mathcal{B}\mathcal{E}\mathcal{D}\mathcal{C}\rangle  \\
    61) \quad & \langle \mathcal{A}\mathcal{E}\mathcal{F}\mathcal{B}\mathcal{C}\mathcal{D}\rangle +\langle \mathcal{A}\mathcal{E}\mathcal{F}\mathcal{B}\mathcal{D}\mathcal{C}\rangle +\langle \mathcal{A}\mathcal{E}\mathcal{F}\mathcal{C}\mathcal{B}\mathcal{D}\rangle +\langle \mathcal{A}\mathcal{E}\mathcal{F}\mathcal{C}\mathcal{D}\mathcal{B}\rangle +\langle \mathcal{A}\mathcal{E}\mathcal{F}\mathcal{D}\mathcal{B}\mathcal{C}\rangle +\langle \mathcal{A}\mathcal{E}\mathcal{F}\mathcal{D}\mathcal{C}\mathcal{B}\rangle  \notag \\ 
     = &  \langle \mathcal{A}\mathcal{E}\mathcal{F}\mathcal{B}\rangle\langle \mathcal{C}\mathcal{D}\rangle +\langle \mathcal{A}\mathcal{E}\mathcal{F}\mathcal{C}\rangle\langle \mathcal{B}\mathcal{D}\rangle +\langle \mathcal{A}\mathcal{E}\mathcal{F}\mathcal{D}\rangle\langle \mathcal{B}\mathcal{C}\rangle +\langle \mathcal{A}\mathcal{E}\mathcal{F}\rangle\langle \mathcal{B}\mathcal{C}\mathcal{D}\rangle +\langle \mathcal{A}\mathcal{E}\mathcal{F}\rangle\langle \mathcal{B}\mathcal{D}\mathcal{C}\rangle  \\
    62) \quad & \langle \mathcal{A}\mathcal{F}\mathcal{E}\mathcal{B}\mathcal{C}\mathcal{D}\rangle +\langle \mathcal{A}\mathcal{F}\mathcal{E}\mathcal{B}\mathcal{D}\mathcal{C}\rangle +\langle \mathcal{A}\mathcal{F}\mathcal{E}\mathcal{C}\mathcal{B}\mathcal{D}\rangle +\langle \mathcal{A}\mathcal{F}\mathcal{E}\mathcal{C}\mathcal{D}\mathcal{B}\rangle +\langle \mathcal{A}\mathcal{F}\mathcal{E}\mathcal{D}\mathcal{B}\mathcal{C}\rangle +\langle \mathcal{A}\mathcal{F}\mathcal{E}\mathcal{D}\mathcal{C}\mathcal{B}\rangle  \notag \\ 
     = &  \langle \mathcal{A}\mathcal{F}\mathcal{E}\mathcal{B}\rangle\langle \mathcal{C}\mathcal{D}\rangle +\langle \mathcal{A}\mathcal{F}\mathcal{E}\mathcal{C}\rangle\langle \mathcal{B}\mathcal{D}\rangle +\langle \mathcal{A}\mathcal{F}\mathcal{E}\mathcal{D}\rangle\langle \mathcal{B}\mathcal{C}\rangle +\langle \mathcal{A}\mathcal{F}\mathcal{E}\rangle\langle \mathcal{B}\mathcal{C}\mathcal{D}\rangle +\langle \mathcal{A}\mathcal{F}\mathcal{E}\rangle\langle \mathcal{B}\mathcal{D}\mathcal{C}\rangle  \\
    63) \quad & \langle \mathcal{A}\mathcal{F}\mathcal{B}\mathcal{C}\mathcal{D}\mathcal{E}\rangle +\langle \mathcal{A}\mathcal{F}\mathcal{B}\mathcal{D}\mathcal{C}\mathcal{E}\rangle +\langle \mathcal{A}\mathcal{F}\mathcal{C}\mathcal{B}\mathcal{D}\mathcal{E}\rangle +\langle \mathcal{A}\mathcal{F}\mathcal{C}\mathcal{D}\mathcal{B}\mathcal{E}\rangle +\langle \mathcal{A}\mathcal{F}\mathcal{D}\mathcal{B}\mathcal{C}\mathcal{E}\rangle +\langle \mathcal{A}\mathcal{F}\mathcal{D}\mathcal{C}\mathcal{B}\mathcal{E}\rangle  \notag \\ 
     = &  \langle \mathcal{A}\mathcal{F}\mathcal{B}\mathcal{E}\rangle\langle \mathcal{C}\mathcal{D}\rangle +\langle \mathcal{A}\mathcal{F}\mathcal{C}\mathcal{E}\rangle\langle \mathcal{B}\mathcal{D}\rangle +\langle \mathcal{A}\mathcal{F}\mathcal{D}\mathcal{E}\rangle\langle \mathcal{B}\mathcal{C}\rangle +\langle \mathcal{A}\mathcal{F}\mathcal{E}\rangle\langle \mathcal{B}\mathcal{C}\mathcal{D}\rangle +\langle \mathcal{A}\mathcal{F}\mathcal{E}\rangle\langle \mathcal{B}\mathcal{D}\mathcal{C}\rangle  \\
    64) \quad & \langle \mathcal{A}\mathcal{E}\mathcal{B}\mathcal{C}\mathcal{D}\mathcal{F}\rangle +\langle \mathcal{A}\mathcal{E}\mathcal{B}\mathcal{D}\mathcal{C}\mathcal{F}\rangle +\langle \mathcal{A}\mathcal{E}\mathcal{C}\mathcal{B}\mathcal{D}\mathcal{F}\rangle +\langle \mathcal{A}\mathcal{E}\mathcal{C}\mathcal{D}\mathcal{B}\mathcal{F}\rangle +\langle \mathcal{A}\mathcal{E}\mathcal{D}\mathcal{B}\mathcal{C}\mathcal{F}\rangle +\langle \mathcal{A}\mathcal{E}\mathcal{D}\mathcal{C}\mathcal{B}\mathcal{F}\rangle  \notag \\ 
     = &  \langle \mathcal{A}\mathcal{E}\mathcal{B}\mathcal{F}\rangle\langle \mathcal{C}\mathcal{D}\rangle +\langle \mathcal{A}\mathcal{E}\mathcal{C}\mathcal{F}\rangle\langle \mathcal{B}\mathcal{D}\rangle +\langle \mathcal{A}\mathcal{E}\mathcal{D}\mathcal{F}\rangle\langle \mathcal{B}\mathcal{C}\rangle +\langle \mathcal{A}\mathcal{E}\mathcal{F}\rangle\langle \mathcal{B}\mathcal{C}\mathcal{D}\rangle +\langle \mathcal{A}\mathcal{E}\mathcal{F}\rangle\langle \mathcal{B}\mathcal{D}\mathcal{C}\rangle  \\
    65) \quad & \langle \mathcal{A}\mathcal{D}\mathcal{F}\mathcal{B}\mathcal{C}\mathcal{E}\rangle +\langle \mathcal{A}\mathcal{D}\mathcal{F}\mathcal{B}\mathcal{E}\mathcal{C}\rangle +\langle \mathcal{A}\mathcal{D}\mathcal{F}\mathcal{C}\mathcal{B}\mathcal{E}\rangle +\langle \mathcal{A}\mathcal{D}\mathcal{F}\mathcal{C}\mathcal{E}\mathcal{B}\rangle +\langle \mathcal{A}\mathcal{D}\mathcal{F}\mathcal{E}\mathcal{B}\mathcal{C}\rangle +\langle \mathcal{A}\mathcal{D}\mathcal{F}\mathcal{E}\mathcal{C}\mathcal{B}\rangle  \notag \\ 
     = &  \langle \mathcal{A}\mathcal{D}\mathcal{F}\mathcal{B}\rangle\langle \mathcal{C}\mathcal{E}\rangle +\langle \mathcal{A}\mathcal{D}\mathcal{F}\mathcal{C}\rangle\langle \mathcal{B}\mathcal{E}\rangle +\langle \mathcal{A}\mathcal{D}\mathcal{F}\mathcal{E}\rangle\langle \mathcal{B}\mathcal{C}\rangle +\langle \mathcal{A}\mathcal{D}\mathcal{F}\rangle\langle \mathcal{B}\mathcal{C}\mathcal{E}\rangle +\langle \mathcal{A}\mathcal{D}\mathcal{F}\rangle\langle \mathcal{B}\mathcal{E}\mathcal{C}\rangle  \\
    66) \quad & \langle \mathcal{A}\mathcal{F}\mathcal{D}\mathcal{B}\mathcal{C}\mathcal{E}\rangle +\langle \mathcal{A}\mathcal{F}\mathcal{D}\mathcal{B}\mathcal{E}\mathcal{C}\rangle +\langle \mathcal{A}\mathcal{F}\mathcal{D}\mathcal{C}\mathcal{B}\mathcal{E}\rangle +\langle \mathcal{A}\mathcal{F}\mathcal{D}\mathcal{C}\mathcal{E}\mathcal{B}\rangle +\langle \mathcal{A}\mathcal{F}\mathcal{D}\mathcal{E}\mathcal{B}\mathcal{C}\rangle +\langle \mathcal{A}\mathcal{F}\mathcal{D}\mathcal{E}\mathcal{C}\mathcal{B}\rangle  \notag \\ 
     = &  \langle \mathcal{A}\mathcal{F}\mathcal{D}\mathcal{B}\rangle\langle \mathcal{C}\mathcal{E}\rangle +\langle \mathcal{A}\mathcal{F}\mathcal{D}\mathcal{C}\rangle\langle \mathcal{B}\mathcal{E}\rangle +\langle \mathcal{A}\mathcal{F}\mathcal{D}\mathcal{E}\rangle\langle \mathcal{B}\mathcal{C}\rangle +\langle \mathcal{A}\mathcal{F}\mathcal{D}\rangle\langle \mathcal{B}\mathcal{C}\mathcal{E}\rangle +\langle \mathcal{A}\mathcal{F}\mathcal{D}\rangle\langle \mathcal{B}\mathcal{E}\mathcal{C}\rangle  \\
    67) \quad & \langle \mathcal{A}\mathcal{F}\mathcal{B}\mathcal{C}\mathcal{E}\mathcal{D}\rangle +\langle \mathcal{A}\mathcal{F}\mathcal{B}\mathcal{E}\mathcal{C}\mathcal{D}\rangle +\langle \mathcal{A}\mathcal{F}\mathcal{C}\mathcal{B}\mathcal{E}\mathcal{D}\rangle +\langle \mathcal{A}\mathcal{F}\mathcal{C}\mathcal{E}\mathcal{B}\mathcal{D}\rangle +\langle \mathcal{A}\mathcal{F}\mathcal{E}\mathcal{B}\mathcal{C}\mathcal{D}\rangle +\langle \mathcal{A}\mathcal{F}\mathcal{E}\mathcal{C}\mathcal{B}\mathcal{D}\rangle  \notag \\ 
     = &  \langle \mathcal{A}\mathcal{F}\mathcal{B}\mathcal{D}\rangle\langle \mathcal{C}\mathcal{E}\rangle +\langle \mathcal{A}\mathcal{F}\mathcal{C}\mathcal{D}\rangle\langle \mathcal{B}\mathcal{E}\rangle +\langle \mathcal{A}\mathcal{F}\mathcal{D}\rangle\langle \mathcal{B}\mathcal{C}\mathcal{E}\rangle +\langle \mathcal{A}\mathcal{F}\mathcal{D}\rangle\langle \mathcal{B}\mathcal{E}\mathcal{C}\rangle +\langle \mathcal{A}\mathcal{F}\mathcal{E}\mathcal{D}\rangle\langle \mathcal{B}\mathcal{C}\rangle 
 \end{align}
 
 \begin{align}
    68) \quad & \langle \mathcal{A}\mathcal{D}\mathcal{B}\mathcal{C}\mathcal{E}\mathcal{F}\rangle +\langle \mathcal{A}\mathcal{D}\mathcal{B}\mathcal{E}\mathcal{C}\mathcal{F}\rangle +\langle \mathcal{A}\mathcal{D}\mathcal{C}\mathcal{B}\mathcal{E}\mathcal{F}\rangle +\langle \mathcal{A}\mathcal{D}\mathcal{C}\mathcal{E}\mathcal{B}\mathcal{F}\rangle +\langle \mathcal{A}\mathcal{D}\mathcal{E}\mathcal{B}\mathcal{C}\mathcal{F}\rangle +\langle \mathcal{A}\mathcal{D}\mathcal{E}\mathcal{C}\mathcal{B}\mathcal{F}\rangle  \notag \\ 
     = &  \langle \mathcal{A}\mathcal{D}\mathcal{B}\mathcal{F}\rangle\langle \mathcal{C}\mathcal{E}\rangle +\langle \mathcal{A}\mathcal{D}\mathcal{C}\mathcal{F}\rangle\langle \mathcal{B}\mathcal{E}\rangle +\langle \mathcal{A}\mathcal{D}\mathcal{E}\mathcal{F}\rangle\langle \mathcal{B}\mathcal{C}\rangle +\langle \mathcal{A}\mathcal{D}\mathcal{F}\rangle\langle \mathcal{B}\mathcal{C}\mathcal{E}\rangle +\langle \mathcal{A}\mathcal{D}\mathcal{F}\rangle\langle \mathcal{B}\mathcal{E}\mathcal{C}\rangle  \\
    69) \quad & \langle \mathcal{A}\mathcal{D}\mathcal{E}\mathcal{B}\mathcal{C}\mathcal{F}\rangle +\langle \mathcal{A}\mathcal{D}\mathcal{E}\mathcal{B}\mathcal{F}\mathcal{C}\rangle +\langle \mathcal{A}\mathcal{D}\mathcal{E}\mathcal{C}\mathcal{B}\mathcal{F}\rangle +\langle \mathcal{A}\mathcal{D}\mathcal{E}\mathcal{C}\mathcal{F}\mathcal{B}\rangle +\langle \mathcal{A}\mathcal{D}\mathcal{E}\mathcal{F}\mathcal{B}\mathcal{C}\rangle +\langle \mathcal{A}\mathcal{D}\mathcal{E}\mathcal{F}\mathcal{C}\mathcal{B}\rangle  \notag \\ 
     = &  \langle \mathcal{A}\mathcal{D}\mathcal{E}\mathcal{B}\rangle\langle \mathcal{C}\mathcal{F}\rangle +\langle \mathcal{A}\mathcal{D}\mathcal{E}\mathcal{C}\rangle\langle \mathcal{B}\mathcal{F}\rangle +\langle \mathcal{A}\mathcal{D}\mathcal{E}\mathcal{F}\rangle\langle \mathcal{B}\mathcal{C}\rangle +\langle \mathcal{A}\mathcal{D}\mathcal{E}\rangle\langle \mathcal{B}\mathcal{C}\mathcal{F}\rangle +\langle \mathcal{A}\mathcal{D}\mathcal{E}\rangle\langle \mathcal{B}\mathcal{F}\mathcal{C}\rangle  \\
    70) \quad & \langle \mathcal{A}\mathcal{E}\mathcal{D}\mathcal{B}\mathcal{C}\mathcal{F}\rangle +\langle \mathcal{A}\mathcal{E}\mathcal{D}\mathcal{B}\mathcal{F}\mathcal{C}\rangle +\langle \mathcal{A}\mathcal{E}\mathcal{D}\mathcal{C}\mathcal{B}\mathcal{F}\rangle +\langle \mathcal{A}\mathcal{E}\mathcal{D}\mathcal{C}\mathcal{F}\mathcal{B}\rangle +\langle \mathcal{A}\mathcal{E}\mathcal{D}\mathcal{F}\mathcal{B}\mathcal{C}\rangle +\langle \mathcal{A}\mathcal{E}\mathcal{D}\mathcal{F}\mathcal{C}\mathcal{B}\rangle  \notag \\ 
     = &  \langle \mathcal{A}\mathcal{E}\mathcal{D}\mathcal{B}\rangle\langle \mathcal{C}\mathcal{F}\rangle +\langle \mathcal{A}\mathcal{E}\mathcal{D}\mathcal{C}\rangle\langle \mathcal{B}\mathcal{F}\rangle +\langle \mathcal{A}\mathcal{E}\mathcal{D}\mathcal{F}\rangle\langle \mathcal{B}\mathcal{C}\rangle +\langle \mathcal{A}\mathcal{E}\mathcal{D}\rangle\langle \mathcal{B}\mathcal{C}\mathcal{F}\rangle +\langle \mathcal{A}\mathcal{E}\mathcal{D}\rangle\langle \mathcal{B}\mathcal{F}\mathcal{C}\rangle  \\
    71) \quad & \langle \mathcal{A}\mathcal{E}\mathcal{B}\mathcal{C}\mathcal{F}\mathcal{D}\rangle +\langle \mathcal{A}\mathcal{E}\mathcal{B}\mathcal{F}\mathcal{C}\mathcal{D}\rangle +\langle \mathcal{A}\mathcal{E}\mathcal{C}\mathcal{B}\mathcal{F}\mathcal{D}\rangle +\langle \mathcal{A}\mathcal{E}\mathcal{C}\mathcal{F}\mathcal{B}\mathcal{D}\rangle +\langle \mathcal{A}\mathcal{E}\mathcal{F}\mathcal{B}\mathcal{C}\mathcal{D}\rangle +\langle \mathcal{A}\mathcal{E}\mathcal{F}\mathcal{C}\mathcal{B}\mathcal{D}\rangle  \notag \\ 
     = &  \langle \mathcal{A}\mathcal{E}\mathcal{B}\mathcal{D}\rangle\langle \mathcal{C}\mathcal{F}\rangle +\langle \mathcal{A}\mathcal{E}\mathcal{C}\mathcal{D}\rangle\langle \mathcal{B}\mathcal{F}\rangle +\langle \mathcal{A}\mathcal{E}\mathcal{D}\rangle\langle \mathcal{B}\mathcal{C}\mathcal{F}\rangle +\langle \mathcal{A}\mathcal{E}\mathcal{D}\rangle\langle \mathcal{B}\mathcal{F}\mathcal{C}\rangle +\langle \mathcal{A}\mathcal{E}\mathcal{F}\mathcal{D}\rangle\langle \mathcal{B}\mathcal{C}\rangle  \\
    72) \quad & \langle \mathcal{A}\mathcal{D}\mathcal{B}\mathcal{C}\mathcal{F}\mathcal{E}\rangle +\langle \mathcal{A}\mathcal{D}\mathcal{B}\mathcal{F}\mathcal{C}\mathcal{E}\rangle +\langle \mathcal{A}\mathcal{D}\mathcal{C}\mathcal{B}\mathcal{F}\mathcal{E}\rangle +\langle \mathcal{A}\mathcal{D}\mathcal{C}\mathcal{F}\mathcal{B}\mathcal{E}\rangle +\langle \mathcal{A}\mathcal{D}\mathcal{F}\mathcal{B}\mathcal{C}\mathcal{E}\rangle +\langle \mathcal{A}\mathcal{D}\mathcal{F}\mathcal{C}\mathcal{B}\mathcal{E}\rangle  \notag \\ 
     = &  \langle \mathcal{A}\mathcal{D}\mathcal{B}\mathcal{E}\rangle\langle \mathcal{C}\mathcal{F}\rangle +\langle \mathcal{A}\mathcal{D}\mathcal{C}\mathcal{E}\rangle\langle \mathcal{B}\mathcal{F}\rangle +\langle \mathcal{A}\mathcal{D}\mathcal{E}\rangle\langle \mathcal{B}\mathcal{C}\mathcal{F}\rangle +\langle \mathcal{A}\mathcal{D}\mathcal{E}\rangle\langle \mathcal{B}\mathcal{F}\mathcal{C}\rangle +\langle \mathcal{A}\mathcal{D}\mathcal{F}\mathcal{E}\rangle\langle \mathcal{B}\mathcal{C}\rangle  \\
    73) \quad & \langle \mathcal{A}\mathcal{C}\mathcal{F}\mathcal{B}\mathcal{D}\mathcal{E}\rangle +\langle \mathcal{A}\mathcal{C}\mathcal{F}\mathcal{B}\mathcal{E}\mathcal{D}\rangle +\langle \mathcal{A}\mathcal{C}\mathcal{F}\mathcal{D}\mathcal{B}\mathcal{E}\rangle +\langle \mathcal{A}\mathcal{C}\mathcal{F}\mathcal{D}\mathcal{E}\mathcal{B}\rangle +\langle \mathcal{A}\mathcal{C}\mathcal{F}\mathcal{E}\mathcal{B}\mathcal{D}\rangle +\langle \mathcal{A}\mathcal{C}\mathcal{F}\mathcal{E}\mathcal{D}\mathcal{B}\rangle  \notag \\ 
     = &  \langle \mathcal{A}\mathcal{C}\mathcal{F}\mathcal{B}\rangle\langle \mathcal{D}\mathcal{E}\rangle +\langle \mathcal{A}\mathcal{C}\mathcal{F}\mathcal{D}\rangle\langle \mathcal{B}\mathcal{E}\rangle +\langle \mathcal{A}\mathcal{C}\mathcal{F}\mathcal{E}\rangle\langle \mathcal{B}\mathcal{D}\rangle +\langle \mathcal{A}\mathcal{C}\mathcal{F}\rangle\langle \mathcal{B}\mathcal{D}\mathcal{E}\rangle +\langle \mathcal{A}\mathcal{C}\mathcal{F}\rangle\langle \mathcal{B}\mathcal{E}\mathcal{D}\rangle  \\
    74) \quad & \langle \mathcal{A}\mathcal{F}\mathcal{C}\mathcal{B}\mathcal{D}\mathcal{E}\rangle +\langle \mathcal{A}\mathcal{F}\mathcal{C}\mathcal{B}\mathcal{E}\mathcal{D}\rangle +\langle \mathcal{A}\mathcal{F}\mathcal{C}\mathcal{D}\mathcal{B}\mathcal{E}\rangle +\langle \mathcal{A}\mathcal{F}\mathcal{C}\mathcal{D}\mathcal{E}\mathcal{B}\rangle +\langle \mathcal{A}\mathcal{F}\mathcal{C}\mathcal{E}\mathcal{B}\mathcal{D}\rangle +\langle \mathcal{A}\mathcal{F}\mathcal{C}\mathcal{E}\mathcal{D}\mathcal{B}\rangle  \notag \\ 
     = &  \langle \mathcal{A}\mathcal{F}\mathcal{C}\mathcal{B}\rangle\langle \mathcal{D}\mathcal{E}\rangle +\langle \mathcal{A}\mathcal{F}\mathcal{C}\mathcal{D}\rangle\langle \mathcal{B}\mathcal{E}\rangle +\langle \mathcal{A}\mathcal{F}\mathcal{C}\mathcal{E}\rangle\langle \mathcal{B}\mathcal{D}\rangle +\langle \mathcal{A}\mathcal{F}\mathcal{C}\rangle\langle \mathcal{B}\mathcal{D}\mathcal{E}\rangle +\langle \mathcal{A}\mathcal{F}\mathcal{C}\rangle\langle \mathcal{B}\mathcal{E}\mathcal{D}\rangle  \\
    75) \quad & \langle \mathcal{A}\mathcal{F}\mathcal{B}\mathcal{D}\mathcal{E}\mathcal{C}\rangle +\langle \mathcal{A}\mathcal{F}\mathcal{B}\mathcal{E}\mathcal{D}\mathcal{C}\rangle +\langle \mathcal{A}\mathcal{F}\mathcal{D}\mathcal{B}\mathcal{E}\mathcal{C}\rangle +\langle \mathcal{A}\mathcal{F}\mathcal{D}\mathcal{E}\mathcal{B}\mathcal{C}\rangle +\langle \mathcal{A}\mathcal{F}\mathcal{E}\mathcal{B}\mathcal{D}\mathcal{C}\rangle +\langle \mathcal{A}\mathcal{F}\mathcal{E}\mathcal{D}\mathcal{B}\mathcal{C}\rangle  \notag \\ 
     = &  \langle \mathcal{A}\mathcal{F}\mathcal{B}\mathcal{C}\rangle\langle \mathcal{D}\mathcal{E}\rangle +\langle \mathcal{A}\mathcal{F}\mathcal{C}\rangle\langle \mathcal{B}\mathcal{D}\mathcal{E}\rangle +\langle \mathcal{A}\mathcal{F}\mathcal{C}\rangle\langle \mathcal{B}\mathcal{E}\mathcal{D}\rangle +\langle \mathcal{A}\mathcal{F}\mathcal{D}\mathcal{C}\rangle\langle \mathcal{B}\mathcal{E}\rangle +\langle \mathcal{A}\mathcal{F}\mathcal{E}\mathcal{C}\rangle\langle \mathcal{B}\mathcal{D}\rangle  \\
    76) \quad & \langle \mathcal{A}\mathcal{C}\mathcal{B}\mathcal{D}\mathcal{E}\mathcal{F}\rangle +\langle \mathcal{A}\mathcal{C}\mathcal{B}\mathcal{E}\mathcal{D}\mathcal{F}\rangle +\langle \mathcal{A}\mathcal{C}\mathcal{D}\mathcal{B}\mathcal{E}\mathcal{F}\rangle +\langle \mathcal{A}\mathcal{C}\mathcal{D}\mathcal{E}\mathcal{B}\mathcal{F}\rangle +\langle \mathcal{A}\mathcal{C}\mathcal{E}\mathcal{B}\mathcal{D}\mathcal{F}\rangle +\langle \mathcal{A}\mathcal{C}\mathcal{E}\mathcal{D}\mathcal{B}\mathcal{F}\rangle  \notag \\ 
     = &  \langle \mathcal{A}\mathcal{C}\mathcal{B}\mathcal{F}\rangle\langle \mathcal{D}\mathcal{E}\rangle +\langle \mathcal{A}\mathcal{C}\mathcal{D}\mathcal{F}\rangle\langle \mathcal{B}\mathcal{E}\rangle +\langle \mathcal{A}\mathcal{C}\mathcal{E}\mathcal{F}\rangle\langle \mathcal{B}\mathcal{D}\rangle +\langle \mathcal{A}\mathcal{C}\mathcal{F}\rangle\langle \mathcal{B}\mathcal{D}\mathcal{E}\rangle +\langle \mathcal{A}\mathcal{C}\mathcal{F}\rangle\langle \mathcal{B}\mathcal{E}\mathcal{D}\rangle  \\
    77) \quad & \langle \mathcal{A}\mathcal{C}\mathcal{E}\mathcal{B}\mathcal{D}\mathcal{F}\rangle +\langle \mathcal{A}\mathcal{C}\mathcal{E}\mathcal{B}\mathcal{F}\mathcal{D}\rangle +\langle \mathcal{A}\mathcal{C}\mathcal{E}\mathcal{D}\mathcal{B}\mathcal{F}\rangle +\langle \mathcal{A}\mathcal{C}\mathcal{E}\mathcal{D}\mathcal{F}\mathcal{B}\rangle +\langle \mathcal{A}\mathcal{C}\mathcal{E}\mathcal{F}\mathcal{B}\mathcal{D}\rangle +\langle \mathcal{A}\mathcal{C}\mathcal{E}\mathcal{F}\mathcal{D}\mathcal{B}\rangle  \notag \\ 
     = &  \langle \mathcal{A}\mathcal{C}\mathcal{E}\mathcal{B}\rangle\langle \mathcal{D}\mathcal{F}\rangle +\langle \mathcal{A}\mathcal{C}\mathcal{E}\mathcal{D}\rangle\langle \mathcal{B}\mathcal{F}\rangle +\langle \mathcal{A}\mathcal{C}\mathcal{E}\mathcal{F}\rangle\langle \mathcal{B}\mathcal{D}\rangle +\langle \mathcal{A}\mathcal{C}\mathcal{E}\rangle\langle \mathcal{B}\mathcal{D}\mathcal{F}\rangle +\langle \mathcal{A}\mathcal{C}\mathcal{E}\rangle\langle \mathcal{B}\mathcal{F}\mathcal{D}\rangle  \\
    78) \quad & \langle \mathcal{A}\mathcal{E}\mathcal{C}\mathcal{B}\mathcal{D}\mathcal{F}\rangle +\langle \mathcal{A}\mathcal{E}\mathcal{C}\mathcal{B}\mathcal{F}\mathcal{D}\rangle +\langle \mathcal{A}\mathcal{E}\mathcal{C}\mathcal{D}\mathcal{B}\mathcal{F}\rangle +\langle \mathcal{A}\mathcal{E}\mathcal{C}\mathcal{D}\mathcal{F}\mathcal{B}\rangle +\langle \mathcal{A}\mathcal{E}\mathcal{C}\mathcal{F}\mathcal{B}\mathcal{D}\rangle +\langle \mathcal{A}\mathcal{E}\mathcal{C}\mathcal{F}\mathcal{D}\mathcal{B}\rangle  \notag \\ 
     = &  \langle \mathcal{A}\mathcal{E}\mathcal{C}\mathcal{B}\rangle\langle \mathcal{D}\mathcal{F}\rangle +\langle \mathcal{A}\mathcal{E}\mathcal{C}\mathcal{D}\rangle\langle \mathcal{B}\mathcal{F}\rangle +\langle \mathcal{A}\mathcal{E}\mathcal{C}\mathcal{F}\rangle\langle \mathcal{B}\mathcal{D}\rangle +\langle \mathcal{A}\mathcal{E}\mathcal{C}\rangle\langle \mathcal{B}\mathcal{D}\mathcal{F}\rangle +\langle \mathcal{A}\mathcal{E}\mathcal{C}\rangle\langle \mathcal{B}\mathcal{F}\mathcal{D}\rangle  \\
    79) \quad & \langle \mathcal{A}\mathcal{E}\mathcal{B}\mathcal{D}\mathcal{F}\mathcal{C}\rangle +\langle \mathcal{A}\mathcal{E}\mathcal{B}\mathcal{F}\mathcal{D}\mathcal{C}\rangle +\langle \mathcal{A}\mathcal{E}\mathcal{D}\mathcal{B}\mathcal{F}\mathcal{C}\rangle +\langle \mathcal{A}\mathcal{E}\mathcal{D}\mathcal{F}\mathcal{B}\mathcal{C}\rangle +\langle \mathcal{A}\mathcal{E}\mathcal{F}\mathcal{B}\mathcal{D}\mathcal{C}\rangle +\langle \mathcal{A}\mathcal{E}\mathcal{F}\mathcal{D}\mathcal{B}\mathcal{C}\rangle  \notag \\ 
     = &  \langle \mathcal{A}\mathcal{E}\mathcal{B}\mathcal{C}\rangle\langle \mathcal{D}\mathcal{F}\rangle +\langle \mathcal{A}\mathcal{E}\mathcal{C}\rangle\langle \mathcal{B}\mathcal{D}\mathcal{F}\rangle +\langle \mathcal{A}\mathcal{E}\mathcal{C}\rangle\langle \mathcal{B}\mathcal{F}\mathcal{D}\rangle +\langle \mathcal{A}\mathcal{E}\mathcal{D}\mathcal{C}\rangle\langle \mathcal{B}\mathcal{F}\rangle +\langle \mathcal{A}\mathcal{E}\mathcal{F}\mathcal{C}\rangle\langle \mathcal{B}\mathcal{D}\rangle  \\
    80) \quad & \langle \mathcal{A}\mathcal{C}\mathcal{D}\mathcal{B}\mathcal{E}\mathcal{F}\rangle +\langle \mathcal{A}\mathcal{C}\mathcal{D}\mathcal{B}\mathcal{F}\mathcal{E}\rangle +\langle \mathcal{A}\mathcal{C}\mathcal{D}\mathcal{E}\mathcal{B}\mathcal{F}\rangle +\langle \mathcal{A}\mathcal{C}\mathcal{D}\mathcal{E}\mathcal{F}\mathcal{B}\rangle +\langle \mathcal{A}\mathcal{C}\mathcal{D}\mathcal{F}\mathcal{B}\mathcal{E}\rangle +\langle \mathcal{A}\mathcal{C}\mathcal{D}\mathcal{F}\mathcal{E}\mathcal{B}\rangle  \notag \\ 
     = &  \langle \mathcal{A}\mathcal{C}\mathcal{D}\mathcal{B}\rangle\langle \mathcal{E}\mathcal{F}\rangle +\langle \mathcal{A}\mathcal{C}\mathcal{D}\mathcal{E}\rangle\langle \mathcal{B}\mathcal{F}\rangle +\langle \mathcal{A}\mathcal{C}\mathcal{D}\mathcal{F}\rangle\langle \mathcal{B}\mathcal{E}\rangle +\langle \mathcal{A}\mathcal{C}\mathcal{D}\rangle\langle \mathcal{B}\mathcal{E}\mathcal{F}\rangle +\langle \mathcal{A}\mathcal{C}\mathcal{D}\rangle\langle \mathcal{B}\mathcal{F}\mathcal{E}\rangle  \\
    81) \quad & \langle \mathcal{A}\mathcal{D}\mathcal{B}\mathcal{E}\mathcal{F}\mathcal{C}\rangle +\langle \mathcal{A}\mathcal{D}\mathcal{B}\mathcal{F}\mathcal{E}\mathcal{C}\rangle +\langle \mathcal{A}\mathcal{D}\mathcal{E}\mathcal{B}\mathcal{F}\mathcal{C}\rangle +\langle \mathcal{A}\mathcal{D}\mathcal{E}\mathcal{F}\mathcal{B}\mathcal{C}\rangle +\langle \mathcal{A}\mathcal{D}\mathcal{F}\mathcal{B}\mathcal{E}\mathcal{C}\rangle +\langle \mathcal{A}\mathcal{D}\mathcal{F}\mathcal{E}\mathcal{B}\mathcal{C}\rangle  \notag \\ 
     = &  \langle \mathcal{A}\mathcal{D}\mathcal{B}\mathcal{C}\rangle\langle \mathcal{E}\mathcal{F}\rangle +\langle \mathcal{A}\mathcal{D}\mathcal{C}\rangle\langle \mathcal{B}\mathcal{E}\mathcal{F}\rangle +\langle \mathcal{A}\mathcal{D}\mathcal{C}\rangle\langle \mathcal{B}\mathcal{F}\mathcal{E}\rangle +\langle \mathcal{A}\mathcal{D}\mathcal{E}\mathcal{C}\rangle\langle \mathcal{B}\mathcal{F}\rangle +\langle \mathcal{A}\mathcal{D}\mathcal{F}\mathcal{C}\rangle\langle \mathcal{B}\mathcal{E}\rangle  \\
    82) \quad & \langle \mathcal{A}\mathcal{B}\mathcal{F}\mathcal{C}\mathcal{D}\mathcal{E}\rangle +\langle \mathcal{A}\mathcal{B}\mathcal{F}\mathcal{C}\mathcal{E}\mathcal{D}\rangle +\langle \mathcal{A}\mathcal{B}\mathcal{F}\mathcal{D}\mathcal{C}\mathcal{E}\rangle +\langle \mathcal{A}\mathcal{B}\mathcal{F}\mathcal{D}\mathcal{E}\mathcal{C}\rangle +\langle \mathcal{A}\mathcal{B}\mathcal{F}\mathcal{E}\mathcal{C}\mathcal{D}\rangle +\langle \mathcal{A}\mathcal{B}\mathcal{F}\mathcal{E}\mathcal{D}\mathcal{C}\rangle  \notag \\ 
     = &  \langle \mathcal{A}\mathcal{B}\mathcal{F}\mathcal{C}\rangle\langle \mathcal{D}\mathcal{E}\rangle +\langle \mathcal{A}\mathcal{B}\mathcal{F}\mathcal{D}\rangle\langle \mathcal{C}\mathcal{E}\rangle +\langle \mathcal{A}\mathcal{B}\mathcal{F}\mathcal{E}\rangle\langle \mathcal{C}\mathcal{D}\rangle +\langle \mathcal{A}\mathcal{B}\mathcal{F}\rangle\langle \mathcal{C}\mathcal{D}\mathcal{E}\rangle +\langle \mathcal{A}\mathcal{B}\mathcal{F}\rangle\langle \mathcal{C}\mathcal{E}\mathcal{D}\rangle  \\
    83) \quad & \langle \mathcal{A}\mathcal{F}\mathcal{B}\mathcal{C}\mathcal{D}\mathcal{E}\rangle +\langle \mathcal{A}\mathcal{F}\mathcal{B}\mathcal{C}\mathcal{E}\mathcal{D}\rangle +\langle \mathcal{A}\mathcal{F}\mathcal{B}\mathcal{D}\mathcal{C}\mathcal{E}\rangle +\langle \mathcal{A}\mathcal{F}\mathcal{B}\mathcal{D}\mathcal{E}\mathcal{C}\rangle +\langle \mathcal{A}\mathcal{F}\mathcal{B}\mathcal{E}\mathcal{C}\mathcal{D}\rangle +\langle \mathcal{A}\mathcal{F}\mathcal{B}\mathcal{E}\mathcal{D}\mathcal{C}\rangle  \notag \\ 
     = &  \langle \mathcal{A}\mathcal{F}\mathcal{B}\mathcal{C}\rangle\langle \mathcal{D}\mathcal{E}\rangle +\langle \mathcal{A}\mathcal{F}\mathcal{B}\mathcal{D}\rangle\langle \mathcal{C}\mathcal{E}\rangle +\langle \mathcal{A}\mathcal{F}\mathcal{B}\mathcal{E}\rangle\langle \mathcal{C}\mathcal{D}\rangle +\langle \mathcal{A}\mathcal{F}\mathcal{B}\rangle\langle \mathcal{C}\mathcal{D}\mathcal{E}\rangle +\langle \mathcal{A}\mathcal{F}\mathcal{B}\rangle\langle \mathcal{C}\mathcal{E}\mathcal{D}\rangle  \\
    84) \quad & \langle \mathcal{A}\mathcal{B}\mathcal{E}\mathcal{C}\mathcal{D}\mathcal{F}\rangle +\langle \mathcal{A}\mathcal{B}\mathcal{E}\mathcal{C}\mathcal{F}\mathcal{D}\rangle +\langle \mathcal{A}\mathcal{B}\mathcal{E}\mathcal{D}\mathcal{C}\mathcal{F}\rangle +\langle \mathcal{A}\mathcal{B}\mathcal{E}\mathcal{D}\mathcal{F}\mathcal{C}\rangle +\langle \mathcal{A}\mathcal{B}\mathcal{E}\mathcal{F}\mathcal{C}\mathcal{D}\rangle +\langle \mathcal{A}\mathcal{B}\mathcal{E}\mathcal{F}\mathcal{D}\mathcal{C}\rangle  \notag \\ 
     = &  \langle \mathcal{A}\mathcal{B}\mathcal{E}\mathcal{C}\rangle\langle \mathcal{D}\mathcal{F}\rangle +\langle \mathcal{A}\mathcal{B}\mathcal{E}\mathcal{D}\rangle\langle \mathcal{C}\mathcal{F}\rangle +\langle \mathcal{A}\mathcal{B}\mathcal{E}\mathcal{F}\rangle\langle \mathcal{C}\mathcal{D}\rangle +\langle \mathcal{A}\mathcal{B}\mathcal{E}\rangle\langle \mathcal{C}\mathcal{D}\mathcal{F}\rangle +\langle \mathcal{A}\mathcal{B}\mathcal{E}\rangle\langle \mathcal{C}\mathcal{F}\mathcal{D}\rangle  \\
 \end{align}
 
 \begin{align}
    85) \quad & \langle \mathcal{A}\mathcal{B}\mathcal{D}\mathcal{C}\mathcal{E}\mathcal{F}\rangle +\langle \mathcal{A}\mathcal{B}\mathcal{D}\mathcal{C}\mathcal{F}\mathcal{E}\rangle +\langle \mathcal{A}\mathcal{B}\mathcal{D}\mathcal{E}\mathcal{C}\mathcal{F}\rangle +\langle \mathcal{A}\mathcal{B}\mathcal{D}\mathcal{E}\mathcal{F}\mathcal{C}\rangle +\langle \mathcal{A}\mathcal{B}\mathcal{D}\mathcal{F}\mathcal{C}\mathcal{E}\rangle +\langle \mathcal{A}\mathcal{B}\mathcal{D}\mathcal{F}\mathcal{E}\mathcal{C}\rangle  \notag \\ 
     = &  \langle \mathcal{A}\mathcal{B}\mathcal{D}\mathcal{C}\rangle\langle \mathcal{E}\mathcal{F}\rangle +\langle \mathcal{A}\mathcal{B}\mathcal{D}\mathcal{E}\rangle\langle \mathcal{C}\mathcal{F}\rangle +\langle \mathcal{A}\mathcal{B}\mathcal{D}\mathcal{F}\rangle\langle \mathcal{C}\mathcal{E}\rangle +\langle \mathcal{A}\mathcal{B}\mathcal{D}\rangle\langle \mathcal{C}\mathcal{E}\mathcal{F}\rangle +\langle \mathcal{A}\mathcal{B}\mathcal{D}\rangle\langle \mathcal{C}\mathcal{F}\mathcal{E}\rangle  \\
    86) \quad & \langle \mathcal{A}\mathcal{B}\mathcal{C}\mathcal{D}\mathcal{E}\mathcal{F}\rangle +\langle \mathcal{A}\mathcal{B}\mathcal{C}\mathcal{D}\mathcal{F}\mathcal{E}\rangle +\langle \mathcal{A}\mathcal{B}\mathcal{C}\mathcal{E}\mathcal{D}\mathcal{F}\rangle +\langle \mathcal{A}\mathcal{B}\mathcal{C}\mathcal{E}\mathcal{F}\mathcal{D}\rangle +\langle \mathcal{A}\mathcal{B}\mathcal{C}\mathcal{F}\mathcal{D}\mathcal{E}\rangle +\langle \mathcal{A}\mathcal{B}\mathcal{C}\mathcal{F}\mathcal{E}\mathcal{D}\rangle  \notag \\ 
     = &  \langle \mathcal{A}\mathcal{B}\mathcal{C}\mathcal{D}\rangle\langle \mathcal{E}\mathcal{F}\rangle +\langle \mathcal{A}\mathcal{B}\mathcal{C}\mathcal{E}\rangle\langle \mathcal{D}\mathcal{F}\rangle +\langle \mathcal{A}\mathcal{B}\mathcal{C}\mathcal{F}\rangle\langle \mathcal{D}\mathcal{E}\rangle +\langle \mathcal{A}\mathcal{B}\mathcal{C}\rangle\langle \mathcal{D}\mathcal{E}\mathcal{F}\rangle +\langle \mathcal{A}\mathcal{B}\mathcal{C}\rangle\langle \mathcal{D}\mathcal{F}\mathcal{E}\rangle  \\
    87) \quad & \langle \mathcal{A}\mathcal{B}\mathcal{C}\mathcal{D}\mathcal{E}\mathcal{F}\rangle +\langle \mathcal{A}\mathcal{B}\mathcal{C}\mathcal{E}\mathcal{F}\mathcal{D}\rangle +\langle \mathcal{A}\mathcal{B}\mathcal{D}\mathcal{C}\mathcal{E}\mathcal{F}\rangle +\langle \mathcal{A}\mathcal{B}\mathcal{D}\mathcal{E}\mathcal{F}\mathcal{C}\rangle +\langle \mathcal{A}\mathcal{B}\mathcal{E}\mathcal{F}\mathcal{C}\mathcal{D}\rangle +\langle \mathcal{A}\mathcal{B}\mathcal{E}\mathcal{F}\mathcal{D}\mathcal{C}\rangle  \notag \\ 
     = &  \langle \mathcal{A}\mathcal{B}\mathcal{C}\mathcal{D}\rangle\langle \mathcal{E}\mathcal{F}\rangle +\langle \mathcal{A}\mathcal{B}\mathcal{C}\rangle\langle \mathcal{D}\mathcal{E}\mathcal{F}\rangle +\langle \mathcal{A}\mathcal{B}\mathcal{D}\mathcal{C}\rangle\langle \mathcal{E}\mathcal{F}\rangle +\langle \mathcal{A}\mathcal{B}\mathcal{D}\rangle\langle \mathcal{C}\mathcal{E}\mathcal{F}\rangle +\langle \mathcal{A}\mathcal{B}\mathcal{E}\mathcal{F}\rangle\langle \mathcal{C}\mathcal{D}\rangle +\langle \mathcal{A}\mathcal{B}\rangle\langle \mathcal{C}\mathcal{D}\mathcal{E}\mathcal{F}\rangle \notag \\
       & -\langle \mathcal{A}\mathcal{B}\rangle\langle \mathcal{C}\mathcal{D}\rangle\langle \mathcal{E}\mathcal{F}\rangle +\langle \mathcal{A}\mathcal{B}\rangle\langle \mathcal{C}\mathcal{E}\mathcal{F}\mathcal{D}\rangle  \\
    88) \quad & \langle \mathcal{A}\mathcal{B}\mathcal{C}\mathcal{D}\mathcal{F}\mathcal{E}\rangle +\langle \mathcal{A}\mathcal{B}\mathcal{C}\mathcal{F}\mathcal{E}\mathcal{D}\rangle +\langle \mathcal{A}\mathcal{B}\mathcal{D}\mathcal{C}\mathcal{F}\mathcal{E}\rangle +\langle \mathcal{A}\mathcal{B}\mathcal{D}\mathcal{F}\mathcal{E}\mathcal{C}\rangle +\langle \mathcal{A}\mathcal{B}\mathcal{F}\mathcal{E}\mathcal{C}\mathcal{D}\rangle +\langle \mathcal{A}\mathcal{B}\mathcal{F}\mathcal{E}\mathcal{D}\mathcal{C}\rangle  \notag \\ 
     = &  \langle \mathcal{A}\mathcal{B}\mathcal{C}\mathcal{D}\rangle\langle \mathcal{E}\mathcal{F}\rangle +\langle \mathcal{A}\mathcal{B}\mathcal{C}\rangle\langle \mathcal{D}\mathcal{F}\mathcal{E}\rangle +\langle \mathcal{A}\mathcal{B}\mathcal{D}\mathcal{C}\rangle\langle \mathcal{E}\mathcal{F}\rangle +\langle \mathcal{A}\mathcal{B}\mathcal{D}\rangle\langle \mathcal{C}\mathcal{F}\mathcal{E}\rangle +\langle \mathcal{A}\mathcal{B}\mathcal{F}\mathcal{E}\rangle\langle \mathcal{C}\mathcal{D}\rangle +\langle \mathcal{A}\mathcal{B}\rangle\langle \mathcal{C}\mathcal{D}\mathcal{F}\mathcal{E}\rangle \notag \\
       & -\langle \mathcal{A}\mathcal{B}\rangle\langle \mathcal{C}\mathcal{D}\rangle\langle \mathcal{E}\mathcal{F}\rangle +\langle \mathcal{A}\mathcal{B}\rangle\langle \mathcal{C}\mathcal{F}\mathcal{E}\mathcal{D}\rangle  \\
    89) \quad & \langle \mathcal{A}\mathcal{B}\mathcal{C}\mathcal{D}\mathcal{F}\mathcal{E}\rangle +\langle \mathcal{A}\mathcal{B}\mathcal{C}\mathcal{E}\mathcal{D}\mathcal{F}\rangle +\langle \mathcal{A}\mathcal{B}\mathcal{D}\mathcal{F}\mathcal{C}\mathcal{E}\rangle +\langle \mathcal{A}\mathcal{B}\mathcal{D}\mathcal{F}\mathcal{E}\mathcal{C}\rangle +\langle \mathcal{A}\mathcal{B}\mathcal{E}\mathcal{C}\mathcal{D}\mathcal{F}\rangle +\langle \mathcal{A}\mathcal{B}\mathcal{E}\mathcal{D}\mathcal{F}\mathcal{C}\rangle  \notag \\ 
     = &  \langle \mathcal{A}\mathcal{B}\mathcal{C}\mathcal{E}\rangle\langle \mathcal{D}\mathcal{F}\rangle +\langle \mathcal{A}\mathcal{B}\mathcal{C}\rangle\langle \mathcal{D}\mathcal{F}\mathcal{E}\rangle +\langle \mathcal{A}\mathcal{B}\mathcal{D}\mathcal{F}\rangle\langle \mathcal{C}\mathcal{E}\rangle +\langle \mathcal{A}\mathcal{B}\mathcal{E}\mathcal{C}\rangle\langle \mathcal{D}\mathcal{F}\rangle +\langle \mathcal{A}\mathcal{B}\mathcal{E}\rangle\langle \mathcal{C}\mathcal{D}\mathcal{F}\rangle +\langle \mathcal{A}\mathcal{B}\rangle\langle \mathcal{C}\mathcal{D}\mathcal{F}\mathcal{E}\rangle \notag \\
       & +\langle \mathcal{A}\mathcal{B}\rangle\langle \mathcal{C}\mathcal{E}\mathcal{D}\mathcal{F}\rangle -\langle \mathcal{A}\mathcal{B}\rangle\langle \mathcal{C}\mathcal{E}\rangle\langle \mathcal{D}\mathcal{F}\rangle  \\
    90) \quad & \langle \mathcal{A}\mathcal{B}\mathcal{C}\mathcal{E}\mathcal{F}\mathcal{D}\rangle +\langle \mathcal{A}\mathcal{B}\mathcal{C}\mathcal{F}\mathcal{D}\mathcal{E}\rangle +\langle \mathcal{A}\mathcal{B}\mathcal{E}\mathcal{C}\mathcal{F}\mathcal{D}\rangle +\langle \mathcal{A}\mathcal{B}\mathcal{E}\mathcal{F}\mathcal{D}\mathcal{C}\rangle +\langle \mathcal{A}\mathcal{B}\mathcal{F}\mathcal{D}\mathcal{C}\mathcal{E}\rangle +\langle \mathcal{A}\mathcal{B}\mathcal{F}\mathcal{D}\mathcal{E}\mathcal{C}\rangle  \notag \\ 
     = &  \langle \mathcal{A}\mathcal{B}\mathcal{C}\mathcal{E}\rangle\langle \mathcal{D}\mathcal{F}\rangle +\langle \mathcal{A}\mathcal{B}\mathcal{C}\rangle\langle \mathcal{D}\mathcal{E}\mathcal{F}\rangle +\langle \mathcal{A}\mathcal{B}\mathcal{E}\mathcal{C}\rangle\langle \mathcal{D}\mathcal{F}\rangle +\langle \mathcal{A}\mathcal{B}\mathcal{E}\rangle\langle \mathcal{C}\mathcal{F}\mathcal{D}\rangle +\langle \mathcal{A}\mathcal{B}\mathcal{F}\mathcal{D}\rangle\langle \mathcal{C}\mathcal{E}\rangle +\langle \mathcal{A}\mathcal{B}\rangle\langle \mathcal{C}\mathcal{E}\mathcal{F}\mathcal{D}\rangle \notag \\
       & -\langle \mathcal{A}\mathcal{B}\rangle\langle \mathcal{C}\mathcal{E}\rangle\langle \mathcal{D}\mathcal{F}\rangle +\langle \mathcal{A}\mathcal{B}\rangle\langle \mathcal{C}\mathcal{F}\mathcal{D}\mathcal{E}\rangle  \\
    91) \quad & \langle \mathcal{A}\mathcal{B}\mathcal{C}\mathcal{D}\mathcal{E}\mathcal{F}\rangle +\langle \mathcal{A}\mathcal{B}\mathcal{C}\mathcal{F}\mathcal{D}\mathcal{E}\rangle +\langle \mathcal{A}\mathcal{B}\mathcal{D}\mathcal{E}\mathcal{C}\mathcal{F}\rangle +\langle \mathcal{A}\mathcal{B}\mathcal{D}\mathcal{E}\mathcal{F}\mathcal{C}\rangle +\langle \mathcal{A}\mathcal{B}\mathcal{F}\mathcal{C}\mathcal{D}\mathcal{E}\rangle +\langle \mathcal{A}\mathcal{B}\mathcal{F}\mathcal{D}\mathcal{E}\mathcal{C}\rangle  \notag \\ 
     = &  \langle \mathcal{A}\mathcal{B}\mathcal{C}\mathcal{F}\rangle\langle \mathcal{D}\mathcal{E}\rangle +\langle \mathcal{A}\mathcal{B}\mathcal{C}\rangle\langle \mathcal{D}\mathcal{E}\mathcal{F}\rangle +\langle \mathcal{A}\mathcal{B}\mathcal{D}\mathcal{E}\rangle\langle \mathcal{C}\mathcal{F}\rangle +\langle \mathcal{A}\mathcal{B}\mathcal{F}\mathcal{C}\rangle\langle \mathcal{D}\mathcal{E}\rangle +\langle \mathcal{A}\mathcal{B}\mathcal{F}\rangle\langle \mathcal{C}\mathcal{D}\mathcal{E}\rangle +\langle \mathcal{A}\mathcal{B}\rangle\langle \mathcal{C}\mathcal{D}\mathcal{E}\mathcal{F}\rangle \notag \\
       & -\langle \mathcal{A}\mathcal{B}\rangle\langle \mathcal{C}\mathcal{F}\mathcal{D}\mathcal{E}\rangle +\langle \mathcal{A}\mathcal{B}\rangle\langle \mathcal{C}\mathcal{F}\rangle\langle \mathcal{D}\mathcal{E}\rangle  \\
    92) \quad & \langle \mathcal{A}\mathcal{B}\mathcal{C}\mathcal{E}\mathcal{D}\mathcal{F}\rangle +\langle \mathcal{A}\mathcal{B}\mathcal{C}\mathcal{F}\mathcal{E}\mathcal{D}\rangle +\langle \mathcal{A}\mathcal{B}\mathcal{E}\mathcal{D}\mathcal{C}\mathcal{F}\rangle +\langle \mathcal{A}\mathcal{B}\mathcal{E}\mathcal{D}\mathcal{F}\mathcal{C}\rangle +\langle \mathcal{A}\mathcal{B}\mathcal{F}\mathcal{C}\mathcal{E}\mathcal{D}\rangle +\langle \mathcal{A}\mathcal{B}\mathcal{F}\mathcal{E}\mathcal{D}\mathcal{C}\rangle  \notag \\ 
     = &  \langle \mathcal{A}\mathcal{B}\mathcal{C}\mathcal{F}\rangle\langle \mathcal{D}\mathcal{E}\rangle +\langle \mathcal{A}\mathcal{B}\mathcal{C}\rangle\langle \mathcal{D}\mathcal{F}\mathcal{E}\rangle +\langle \mathcal{A}\mathcal{B}\mathcal{E}\mathcal{D}\rangle\langle \mathcal{C}\mathcal{F}\rangle +\langle \mathcal{A}\mathcal{B}\mathcal{F}\mathcal{C}\rangle\langle \mathcal{D}\mathcal{E}\rangle +\langle \mathcal{A}\mathcal{B}\mathcal{F}\rangle\langle \mathcal{C}\mathcal{E}\mathcal{D}\rangle +\langle \mathcal{A}\mathcal{B}\rangle\langle \mathcal{C}\mathcal{E}\mathcal{D}\mathcal{F}\rangle \notag \\
       & +\langle \mathcal{A}\mathcal{B}\rangle\langle \mathcal{C}\mathcal{F}\mathcal{E}\mathcal{D}\rangle -\langle \mathcal{A}\mathcal{B}\rangle\langle \mathcal{C}\mathcal{F}\rangle\langle \mathcal{D}\mathcal{E}\rangle  \\
    93) \quad & \langle \mathcal{A}\mathcal{C}\mathcal{D}\mathcal{E}\mathcal{F}\mathcal{B}\rangle +\langle \mathcal{A}\mathcal{C}\mathcal{E}\mathcal{F}\mathcal{D}\mathcal{B}\rangle +\langle \mathcal{A}\mathcal{D}\mathcal{C}\mathcal{E}\mathcal{F}\mathcal{B}\rangle +\langle \mathcal{A}\mathcal{D}\mathcal{E}\mathcal{F}\mathcal{C}\mathcal{B}\rangle +\langle \mathcal{A}\mathcal{E}\mathcal{F}\mathcal{C}\mathcal{D}\mathcal{B}\rangle +\langle \mathcal{A}\mathcal{E}\mathcal{F}\mathcal{D}\mathcal{C}\mathcal{B}\rangle  \notag \\ 
     = &  \langle \mathcal{A}\mathcal{B}\rangle\langle \mathcal{C}\mathcal{D}\mathcal{E}\mathcal{F}\rangle -\langle \mathcal{A}\mathcal{B}\rangle\langle \mathcal{C}\mathcal{D}\rangle\langle \mathcal{E}\mathcal{F}\rangle +\langle \mathcal{A}\mathcal{B}\rangle\langle \mathcal{C}\mathcal{E}\mathcal{F}\mathcal{D}\rangle +\langle \mathcal{A}\mathcal{C}\mathcal{B}\rangle\langle \mathcal{D}\mathcal{E}\mathcal{F}\rangle +\langle \mathcal{A}\mathcal{C}\mathcal{D}\mathcal{B}\rangle\langle \mathcal{E}\mathcal{F}\rangle +\langle \mathcal{A}\mathcal{D}\mathcal{B}\rangle\langle \mathcal{C}\mathcal{E}\mathcal{F}\rangle \notag \\
       & +\langle \mathcal{A}\mathcal{D}\mathcal{C}\mathcal{B}\rangle\langle \mathcal{E}\mathcal{F}\rangle +\langle \mathcal{A}\mathcal{E}\mathcal{F}\mathcal{B}\rangle\langle \mathcal{C}\mathcal{D}\rangle  \\
    94) \quad & \langle \mathcal{A}\mathcal{C}\mathcal{D}\mathcal{F}\mathcal{E}\mathcal{B}\rangle +\langle \mathcal{A}\mathcal{C}\mathcal{F}\mathcal{E}\mathcal{D}\mathcal{B}\rangle +\langle \mathcal{A}\mathcal{D}\mathcal{C}\mathcal{F}\mathcal{E}\mathcal{B}\rangle +\langle \mathcal{A}\mathcal{D}\mathcal{F}\mathcal{E}\mathcal{C}\mathcal{B}\rangle +\langle \mathcal{A}\mathcal{F}\mathcal{E}\mathcal{C}\mathcal{D}\mathcal{B}\rangle +\langle \mathcal{A}\mathcal{F}\mathcal{E}\mathcal{D}\mathcal{C}\mathcal{B}\rangle  \notag \\ 
     = &  \langle \mathcal{A}\mathcal{B}\rangle\langle \mathcal{C}\mathcal{D}\mathcal{F}\mathcal{E}\rangle -\langle \mathcal{A}\mathcal{B}\rangle\langle \mathcal{C}\mathcal{D}\rangle\langle \mathcal{E}\mathcal{F}\rangle +\langle \mathcal{A}\mathcal{B}\rangle\langle \mathcal{C}\mathcal{F}\mathcal{E}\mathcal{D}\rangle +\langle \mathcal{A}\mathcal{C}\mathcal{B}\rangle\langle \mathcal{D}\mathcal{F}\mathcal{E}\rangle +\langle \mathcal{A}\mathcal{C}\mathcal{D}\mathcal{B}\rangle\langle \mathcal{E}\mathcal{F}\rangle +\langle \mathcal{A}\mathcal{D}\mathcal{B}\rangle\langle \mathcal{C}\mathcal{F}\mathcal{E}\rangle \notag \\
       & +\langle \mathcal{A}\mathcal{D}\mathcal{C}\mathcal{B}\rangle\langle \mathcal{E}\mathcal{F}\rangle +\langle \mathcal{A}\mathcal{F}\mathcal{E}\mathcal{B}\rangle\langle \mathcal{C}\mathcal{D}\rangle  \\
    95) \quad & \langle \mathcal{A}\mathcal{C}\mathcal{D}\mathcal{F}\mathcal{E}\mathcal{B}\rangle +\langle \mathcal{A}\mathcal{C}\mathcal{E}\mathcal{D}\mathcal{F}\mathcal{B}\rangle +\langle \mathcal{A}\mathcal{D}\mathcal{F}\mathcal{C}\mathcal{E}\mathcal{B}\rangle +\langle \mathcal{A}\mathcal{D}\mathcal{F}\mathcal{E}\mathcal{C}\mathcal{B}\rangle +\langle \mathcal{A}\mathcal{E}\mathcal{C}\mathcal{D}\mathcal{F}\mathcal{B}\rangle +\langle \mathcal{A}\mathcal{E}\mathcal{D}\mathcal{F}\mathcal{C}\mathcal{B}\rangle  \notag \\ 
     = &  \langle \mathcal{A}\mathcal{B}\rangle\langle \mathcal{C}\mathcal{D}\mathcal{F}\mathcal{E}\rangle +\langle \mathcal{A}\mathcal{B}\rangle\langle \mathcal{C}\mathcal{E}\mathcal{D}\mathcal{F}\rangle -\langle \mathcal{A}\mathcal{B}\rangle\langle \mathcal{C}\mathcal{E}\rangle\langle \mathcal{D}\mathcal{F}\rangle +\langle \mathcal{A}\mathcal{C}\mathcal{B}\rangle\langle \mathcal{D}\mathcal{F}\mathcal{E}\rangle +\langle \mathcal{A}\mathcal{C}\mathcal{E}\mathcal{B}\rangle\langle \mathcal{D}\mathcal{F}\rangle +\langle \mathcal{A}\mathcal{D}\mathcal{F}\mathcal{B}\rangle\langle \mathcal{C}\mathcal{E}\rangle \notag \\
       & +\langle \mathcal{A}\mathcal{E}\mathcal{B}\rangle\langle \mathcal{C}\mathcal{D}\mathcal{F}\rangle +\langle \mathcal{A}\mathcal{E}\mathcal{C}\mathcal{B}\rangle\langle \mathcal{D}\mathcal{F}\rangle  \\
    96) \quad & \langle \mathcal{A}\mathcal{C}\mathcal{E}\mathcal{F}\mathcal{D}\mathcal{B}\rangle +\langle \mathcal{A}\mathcal{C}\mathcal{F}\mathcal{D}\mathcal{E}\mathcal{B}\rangle +\langle \mathcal{A}\mathcal{E}\mathcal{C}\mathcal{F}\mathcal{D}\mathcal{B}\rangle +\langle \mathcal{A}\mathcal{E}\mathcal{F}\mathcal{D}\mathcal{C}\mathcal{B}\rangle +\langle \mathcal{A}\mathcal{F}\mathcal{D}\mathcal{C}\mathcal{E}\mathcal{B}\rangle +\langle \mathcal{A}\mathcal{F}\mathcal{D}\mathcal{E}\mathcal{C}\mathcal{B}\rangle  \notag \\ 
     = &  \langle \mathcal{A}\mathcal{B}\rangle\langle \mathcal{C}\mathcal{E}\mathcal{F}\mathcal{D}\rangle -\langle \mathcal{A}\mathcal{B}\rangle\langle \mathcal{C}\mathcal{E}\rangle\langle \mathcal{D}\mathcal{F}\rangle +\langle \mathcal{A}\mathcal{B}\rangle\langle \mathcal{C}\mathcal{F}\mathcal{D}\mathcal{E}\rangle +\langle \mathcal{A}\mathcal{C}\mathcal{B}\rangle\langle \mathcal{D}\mathcal{E}\mathcal{F}\rangle +\langle \mathcal{A}\mathcal{C}\mathcal{E}\mathcal{B}\rangle\langle \mathcal{D}\mathcal{F}\rangle +\langle \mathcal{A}\mathcal{E}\mathcal{B}\rangle\langle \mathcal{C}\mathcal{F}\mathcal{D}\rangle \notag \\
       & +\langle \mathcal{A}\mathcal{E}\mathcal{C}\mathcal{B}\rangle\langle \mathcal{D}\mathcal{F}\rangle +\langle \mathcal{A}\mathcal{F}\mathcal{D}\mathcal{B}\rangle\langle \mathcal{C}\mathcal{E}\rangle 
 \end{align}
 
 \begin{align}
    97) \quad & \langle \mathcal{A}\mathcal{C}\mathcal{D}\mathcal{E}\mathcal{F}\mathcal{B}\rangle +\langle \mathcal{A}\mathcal{C}\mathcal{F}\mathcal{D}\mathcal{E}\mathcal{B}\rangle +\langle \mathcal{A}\mathcal{D}\mathcal{E}\mathcal{C}\mathcal{F}\mathcal{B}\rangle +\langle \mathcal{A}\mathcal{D}\mathcal{E}\mathcal{F}\mathcal{C}\mathcal{B}\rangle +\langle \mathcal{A}\mathcal{F}\mathcal{C}\mathcal{D}\mathcal{E}\mathcal{B}\rangle +\langle \mathcal{A}\mathcal{F}\mathcal{D}\mathcal{E}\mathcal{C}\mathcal{B}\rangle  \notag \\ 
     = &  \langle \mathcal{A}\mathcal{B}\rangle\langle \mathcal{C}\mathcal{D}\mathcal{E}\mathcal{F}\rangle +\langle \mathcal{A}\mathcal{B}\rangle\langle \mathcal{C}\mathcal{F}\mathcal{D}\mathcal{E}\rangle -\langle \mathcal{A}\mathcal{B}\rangle\langle \mathcal{C}\mathcal{F}\rangle\langle \mathcal{D}\mathcal{E}\rangle +\langle \mathcal{A}\mathcal{C}\mathcal{B}\rangle\langle \mathcal{D}\mathcal{E}\mathcal{F}\rangle +\langle \mathcal{A}\mathcal{C}\mathcal{F}\mathcal{B}\rangle\langle \mathcal{D}\mathcal{E}\rangle +\langle \mathcal{A}\mathcal{D}\mathcal{E}\mathcal{B}\rangle\langle \mathcal{C}\mathcal{F}\rangle \notag \\
       & +\langle \mathcal{A}\mathcal{F}\mathcal{B}\rangle\langle \mathcal{C}\mathcal{D}\mathcal{E}\rangle +\langle \mathcal{A}\mathcal{F}\mathcal{C}\mathcal{B}\rangle\langle \mathcal{D}\mathcal{E}\rangle  \\
    98) \quad & \langle \mathcal{A}\mathcal{C}\mathcal{B}\mathcal{D}\mathcal{E}\mathcal{F}\rangle +\langle \mathcal{A}\mathcal{C}\mathcal{B}\mathcal{E}\mathcal{F}\mathcal{D}\rangle +\langle \mathcal{A}\mathcal{C}\mathcal{D}\mathcal{B}\mathcal{E}\mathcal{F}\rangle +\langle \mathcal{A}\mathcal{C}\mathcal{D}\mathcal{E}\mathcal{F}\mathcal{B}\rangle +\langle \mathcal{A}\mathcal{C}\mathcal{E}\mathcal{F}\mathcal{B}\mathcal{D}\rangle +\langle \mathcal{A}\mathcal{C}\mathcal{E}\mathcal{F}\mathcal{D}\mathcal{B}\rangle  \notag \\ 
     = &  \langle \mathcal{A}\mathcal{C}\mathcal{B}\mathcal{D}\rangle\langle \mathcal{E}\mathcal{F}\rangle +\langle \mathcal{A}\mathcal{C}\mathcal{B}\rangle\langle \mathcal{D}\mathcal{E}\mathcal{F}\rangle +\langle \mathcal{A}\mathcal{C}\mathcal{D}\mathcal{B}\rangle\langle \mathcal{E}\mathcal{F}\rangle +\langle \mathcal{A}\mathcal{C}\mathcal{D}\rangle\langle \mathcal{B}\mathcal{E}\mathcal{F}\rangle +\langle \mathcal{A}\mathcal{C}\mathcal{E}\mathcal{F}\rangle\langle \mathcal{B}\mathcal{D}\rangle +\langle \mathcal{A}\mathcal{C}\rangle\langle \mathcal{B}\mathcal{D}\mathcal{E}\mathcal{F}\rangle \notag \\
       & -\langle \mathcal{A}\mathcal{C}\rangle\langle \mathcal{B}\mathcal{D}\rangle\langle \mathcal{E}\mathcal{F}\rangle +\langle \mathcal{A}\mathcal{C}\rangle\langle \mathcal{B}\mathcal{E}\mathcal{F}\mathcal{D}\rangle  \\
    99) \quad & \langle \mathcal{A}\mathcal{C}\mathcal{B}\mathcal{D}\mathcal{F}\mathcal{E}\rangle +\langle \mathcal{A}\mathcal{C}\mathcal{B}\mathcal{F}\mathcal{E}\mathcal{D}\rangle +\langle \mathcal{A}\mathcal{C}\mathcal{D}\mathcal{B}\mathcal{F}\mathcal{E}\rangle +\langle \mathcal{A}\mathcal{C}\mathcal{D}\mathcal{F}\mathcal{E}\mathcal{B}\rangle +\langle \mathcal{A}\mathcal{C}\mathcal{F}\mathcal{E}\mathcal{B}\mathcal{D}\rangle +\langle \mathcal{A}\mathcal{C}\mathcal{F}\mathcal{E}\mathcal{D}\mathcal{B}\rangle  \notag \\ 
     = &  \langle \mathcal{A}\mathcal{C}\mathcal{B}\mathcal{D}\rangle\langle \mathcal{E}\mathcal{F}\rangle +\langle \mathcal{A}\mathcal{C}\mathcal{B}\rangle\langle \mathcal{D}\mathcal{F}\mathcal{E}\rangle +\langle \mathcal{A}\mathcal{C}\mathcal{D}\mathcal{B}\rangle\langle \mathcal{E}\mathcal{F}\rangle +\langle \mathcal{A}\mathcal{C}\mathcal{D}\rangle\langle \mathcal{B}\mathcal{F}\mathcal{E}\rangle +\langle \mathcal{A}\mathcal{C}\mathcal{F}\mathcal{E}\rangle\langle \mathcal{B}\mathcal{D}\rangle +\langle \mathcal{A}\mathcal{C}\rangle\langle \mathcal{B}\mathcal{D}\mathcal{F}\mathcal{E}\rangle \notag \\
       & -\langle \mathcal{A}\mathcal{C}\rangle\langle \mathcal{B}\mathcal{D}\rangle\langle \mathcal{E}\mathcal{F}\rangle +\langle \mathcal{A}\mathcal{C}\rangle\langle \mathcal{B}\mathcal{F}\mathcal{E}\mathcal{D}\rangle  \\
    100) \quad & \langle \mathcal{A}\mathcal{C}\mathcal{B}\mathcal{D}\mathcal{F}\mathcal{E}\rangle +\langle \mathcal{A}\mathcal{C}\mathcal{B}\mathcal{E}\mathcal{D}\mathcal{F}\rangle +\langle \mathcal{A}\mathcal{C}\mathcal{D}\mathcal{F}\mathcal{B}\mathcal{E}\rangle +\langle \mathcal{A}\mathcal{C}\mathcal{D}\mathcal{F}\mathcal{E}\mathcal{B}\rangle +\langle \mathcal{A}\mathcal{C}\mathcal{E}\mathcal{B}\mathcal{D}\mathcal{F}\rangle +\langle \mathcal{A}\mathcal{C}\mathcal{E}\mathcal{D}\mathcal{F}\mathcal{B}\rangle  \notag \\ 
     = &  \langle \mathcal{A}\mathcal{C}\mathcal{B}\mathcal{E}\rangle\langle \mathcal{D}\mathcal{F}\rangle +\langle \mathcal{A}\mathcal{C}\mathcal{B}\rangle\langle \mathcal{D}\mathcal{F}\mathcal{E}\rangle +\langle \mathcal{A}\mathcal{C}\mathcal{D}\mathcal{F}\rangle\langle \mathcal{B}\mathcal{E}\rangle +\langle \mathcal{A}\mathcal{C}\mathcal{E}\mathcal{B}\rangle\langle \mathcal{D}\mathcal{F}\rangle +\langle \mathcal{A}\mathcal{C}\mathcal{E}\rangle\langle \mathcal{B}\mathcal{D}\mathcal{F}\rangle +\langle \mathcal{A}\mathcal{C}\rangle\langle \mathcal{B}\mathcal{D}\mathcal{F}\mathcal{E}\rangle \notag \\
       & +\langle \mathcal{A}\mathcal{C}\rangle\langle \mathcal{B}\mathcal{E}\mathcal{D}\mathcal{F}\rangle -\langle \mathcal{A}\mathcal{C}\rangle\langle \mathcal{B}\mathcal{E}\rangle\langle \mathcal{D}\mathcal{F}\rangle  \\
    101) \quad & \langle \mathcal{A}\mathcal{C}\mathcal{B}\mathcal{E}\mathcal{F}\mathcal{D}\rangle +\langle \mathcal{A}\mathcal{C}\mathcal{B}\mathcal{F}\mathcal{D}\mathcal{E}\rangle +\langle \mathcal{A}\mathcal{C}\mathcal{E}\mathcal{B}\mathcal{F}\mathcal{D}\rangle +\langle \mathcal{A}\mathcal{C}\mathcal{E}\mathcal{F}\mathcal{D}\mathcal{B}\rangle +\langle \mathcal{A}\mathcal{C}\mathcal{F}\mathcal{D}\mathcal{B}\mathcal{E}\rangle +\langle \mathcal{A}\mathcal{C}\mathcal{F}\mathcal{D}\mathcal{E}\mathcal{B}\rangle  \notag \\ 
     = &  \langle \mathcal{A}\mathcal{C}\mathcal{B}\mathcal{E}\rangle\langle \mathcal{D}\mathcal{F}\rangle +\langle \mathcal{A}\mathcal{C}\mathcal{B}\rangle\langle \mathcal{D}\mathcal{E}\mathcal{F}\rangle +\langle \mathcal{A}\mathcal{C}\mathcal{E}\mathcal{B}\rangle\langle \mathcal{D}\mathcal{F}\rangle +\langle \mathcal{A}\mathcal{C}\mathcal{E}\rangle\langle \mathcal{B}\mathcal{F}\mathcal{D}\rangle +\langle \mathcal{A}\mathcal{C}\mathcal{F}\mathcal{D}\rangle\langle \mathcal{B}\mathcal{E}\rangle +\langle \mathcal{A}\mathcal{C}\rangle\langle \mathcal{B}\mathcal{E}\mathcal{F}\mathcal{D}\rangle \notag \\
       & -\langle \mathcal{A}\mathcal{C}\rangle\langle \mathcal{B}\mathcal{E}\rangle\langle \mathcal{D}\mathcal{F}\rangle +\langle \mathcal{A}\mathcal{C}\rangle\langle \mathcal{B}\mathcal{F}\mathcal{D}\mathcal{E}\rangle  \\
    102) \quad & \langle \mathcal{A}\mathcal{C}\mathcal{B}\mathcal{D}\mathcal{E}\mathcal{F}\rangle +\langle \mathcal{A}\mathcal{C}\mathcal{B}\mathcal{F}\mathcal{D}\mathcal{E}\rangle +\langle \mathcal{A}\mathcal{C}\mathcal{D}\mathcal{E}\mathcal{B}\mathcal{F}\rangle +\langle \mathcal{A}\mathcal{C}\mathcal{D}\mathcal{E}\mathcal{F}\mathcal{B}\rangle +\langle \mathcal{A}\mathcal{C}\mathcal{F}\mathcal{B}\mathcal{D}\mathcal{E}\rangle +\langle \mathcal{A}\mathcal{C}\mathcal{F}\mathcal{D}\mathcal{E}\mathcal{B}\rangle  \notag \\ 
     = &  \langle \mathcal{A}\mathcal{C}\mathcal{B}\mathcal{F}\rangle\langle \mathcal{D}\mathcal{E}\rangle +\langle \mathcal{A}\mathcal{C}\mathcal{B}\rangle\langle \mathcal{D}\mathcal{E}\mathcal{F}\rangle +\langle \mathcal{A}\mathcal{C}\mathcal{D}\mathcal{E}\rangle\langle \mathcal{B}\mathcal{F}\rangle +\langle \mathcal{A}\mathcal{C}\mathcal{F}\mathcal{B}\rangle\langle \mathcal{D}\mathcal{E}\rangle +\langle \mathcal{A}\mathcal{C}\mathcal{F}\rangle\langle \mathcal{B}\mathcal{D}\mathcal{E}\rangle +\langle \mathcal{A}\mathcal{C}\rangle\langle \mathcal{B}\mathcal{D}\mathcal{E}\mathcal{F}\rangle \notag \\
       & +\langle \mathcal{A}\mathcal{C}\rangle\langle \mathcal{B}\mathcal{F}\mathcal{D}\mathcal{E}\rangle -\langle \mathcal{A}\mathcal{C}\rangle\langle \mathcal{B}\mathcal{F}\rangle\langle \mathcal{D}\mathcal{E}\rangle  \\
    103) \quad & \langle \mathcal{A}\mathcal{C}\mathcal{B}\mathcal{E}\mathcal{D}\mathcal{F}\rangle +\langle \mathcal{A}\mathcal{C}\mathcal{B}\mathcal{F}\mathcal{E}\mathcal{D}\rangle +\langle \mathcal{A}\mathcal{C}\mathcal{E}\mathcal{D}\mathcal{B}\mathcal{F}\rangle +\langle \mathcal{A}\mathcal{C}\mathcal{E}\mathcal{D}\mathcal{F}\mathcal{B}\rangle +\langle \mathcal{A}\mathcal{C}\mathcal{F}\mathcal{B}\mathcal{E}\mathcal{D}\rangle +\langle \mathcal{A}\mathcal{C}\mathcal{F}\mathcal{E}\mathcal{D}\mathcal{B}\rangle  \notag \\ 
     = &  \langle \mathcal{A}\mathcal{C}\mathcal{B}\mathcal{F}\rangle\langle \mathcal{D}\mathcal{E}\rangle +\langle \mathcal{A}\mathcal{C}\mathcal{B}\rangle\langle \mathcal{D}\mathcal{F}\mathcal{E}\rangle +\langle \mathcal{A}\mathcal{C}\mathcal{E}\mathcal{D}\rangle\langle \mathcal{B}\mathcal{F}\rangle +\langle \mathcal{A}\mathcal{C}\mathcal{F}\mathcal{B}\rangle\langle \mathcal{D}\mathcal{E}\rangle +\langle \mathcal{A}\mathcal{C}\mathcal{F}\rangle\langle \mathcal{B}\mathcal{E}\mathcal{D}\rangle +\langle \mathcal{A}\mathcal{C}\rangle\langle \mathcal{B}\mathcal{E}\mathcal{D}\mathcal{F}\rangle \notag \\
       & +\langle \mathcal{A}\mathcal{C}\rangle\langle \mathcal{B}\mathcal{F}\mathcal{E}\mathcal{D}\rangle -\langle \mathcal{A}\mathcal{C}\rangle\langle \mathcal{B}\mathcal{F}\rangle\langle \mathcal{D}\mathcal{E}\rangle  \\
    104) \quad & \langle \mathcal{A}\mathcal{B}\mathcal{D}\mathcal{E}\mathcal{F}\mathcal{C}\rangle +\langle \mathcal{A}\mathcal{B}\mathcal{E}\mathcal{F}\mathcal{D}\mathcal{C}\rangle +\langle \mathcal{A}\mathcal{D}\mathcal{B}\mathcal{E}\mathcal{F}\mathcal{C}\rangle +\langle \mathcal{A}\mathcal{D}\mathcal{E}\mathcal{F}\mathcal{B}\mathcal{C}\rangle +\langle \mathcal{A}\mathcal{E}\mathcal{F}\mathcal{B}\mathcal{D}\mathcal{C}\rangle +\langle \mathcal{A}\mathcal{E}\mathcal{F}\mathcal{D}\mathcal{B}\mathcal{C}\rangle  \notag \\ 
     = &  \langle \mathcal{A}\mathcal{B}\mathcal{C}\rangle\langle \mathcal{D}\mathcal{E}\mathcal{F}\rangle +\langle \mathcal{A}\mathcal{B}\mathcal{D}\mathcal{C}\rangle\langle \mathcal{E}\mathcal{F}\rangle +\langle \mathcal{A}\mathcal{C}\rangle\langle \mathcal{B}\mathcal{D}\mathcal{E}\mathcal{F}\rangle -\langle \mathcal{A}\mathcal{C}\rangle\langle \mathcal{B}\mathcal{D}\rangle\langle \mathcal{E}\mathcal{F}\rangle +\langle \mathcal{A}\mathcal{C}\rangle\langle \mathcal{B}\mathcal{E}\mathcal{F}\mathcal{D}\rangle +\langle \mathcal{A}\mathcal{D}\mathcal{B}\mathcal{C}\rangle\langle \mathcal{E}\mathcal{F}\rangle \notag \\
       & +\langle \mathcal{A}\mathcal{D}\mathcal{C}\rangle\langle \mathcal{B}\mathcal{E}\mathcal{F}\rangle +\langle \mathcal{A}\mathcal{E}\mathcal{F}\mathcal{C}\rangle\langle \mathcal{B}\mathcal{D}\rangle  \\
    105) \quad & \langle \mathcal{A}\mathcal{B}\mathcal{D}\mathcal{F}\mathcal{E}\mathcal{C}\rangle +\langle \mathcal{A}\mathcal{B}\mathcal{F}\mathcal{E}\mathcal{D}\mathcal{C}\rangle +\langle \mathcal{A}\mathcal{D}\mathcal{B}\mathcal{F}\mathcal{E}\mathcal{C}\rangle +\langle \mathcal{A}\mathcal{D}\mathcal{F}\mathcal{E}\mathcal{B}\mathcal{C}\rangle +\langle \mathcal{A}\mathcal{F}\mathcal{E}\mathcal{B}\mathcal{D}\mathcal{C}\rangle +\langle \mathcal{A}\mathcal{F}\mathcal{E}\mathcal{D}\mathcal{B}\mathcal{C}\rangle  \notag \\ 
     = &  \langle \mathcal{A}\mathcal{B}\mathcal{C}\rangle\langle \mathcal{D}\mathcal{F}\mathcal{E}\rangle +\langle \mathcal{A}\mathcal{B}\mathcal{D}\mathcal{C}\rangle\langle \mathcal{E}\mathcal{F}\rangle +\langle \mathcal{A}\mathcal{C}\rangle\langle \mathcal{B}\mathcal{D}\mathcal{F}\mathcal{E}\rangle -\langle \mathcal{A}\mathcal{C}\rangle\langle \mathcal{B}\mathcal{D}\rangle\langle \mathcal{E}\mathcal{F}\rangle +\langle \mathcal{A}\mathcal{C}\rangle\langle \mathcal{B}\mathcal{F}\mathcal{E}\mathcal{D}\rangle +\langle \mathcal{A}\mathcal{D}\mathcal{B}\mathcal{C}\rangle\langle \mathcal{E}\mathcal{F}\rangle \notag \\
       & +\langle \mathcal{A}\mathcal{D}\mathcal{C}\rangle\langle \mathcal{B}\mathcal{F}\mathcal{E}\rangle +\langle \mathcal{A}\mathcal{F}\mathcal{E}\mathcal{C}\rangle\langle \mathcal{B}\mathcal{D}\rangle  \\
    106) \quad & \langle \mathcal{A}\mathcal{B}\mathcal{E}\mathcal{F}\mathcal{D}\mathcal{C}\rangle +\langle \mathcal{A}\mathcal{B}\mathcal{F}\mathcal{D}\mathcal{E}\mathcal{C}\rangle +\langle \mathcal{A}\mathcal{E}\mathcal{B}\mathcal{F}\mathcal{D}\mathcal{C}\rangle +\langle \mathcal{A}\mathcal{E}\mathcal{F}\mathcal{D}\mathcal{B}\mathcal{C}\rangle +\langle \mathcal{A}\mathcal{F}\mathcal{D}\mathcal{B}\mathcal{E}\mathcal{C}\rangle +\langle \mathcal{A}\mathcal{F}\mathcal{D}\mathcal{E}\mathcal{B}\mathcal{C}\rangle  \notag \\ 
     = &  \langle \mathcal{A}\mathcal{B}\mathcal{C}\rangle\langle \mathcal{D}\mathcal{E}\mathcal{F}\rangle +\langle \mathcal{A}\mathcal{B}\mathcal{E}\mathcal{C}\rangle\langle \mathcal{D}\mathcal{F}\rangle +\langle \mathcal{A}\mathcal{C}\rangle\langle \mathcal{B}\mathcal{E}\mathcal{F}\mathcal{D}\rangle -\langle \mathcal{A}\mathcal{C}\rangle\langle \mathcal{B}\mathcal{E}\rangle\langle \mathcal{D}\mathcal{F}\rangle +\langle \mathcal{A}\mathcal{C}\rangle\langle \mathcal{B}\mathcal{F}\mathcal{D}\mathcal{E}\rangle +\langle \mathcal{A}\mathcal{E}\mathcal{B}\mathcal{C}\rangle\langle \mathcal{D}\mathcal{F}\rangle \notag \\
       & +\langle \mathcal{A}\mathcal{E}\mathcal{C}\rangle\langle \mathcal{B}\mathcal{F}\mathcal{D}\rangle +\langle \mathcal{A}\mathcal{F}\mathcal{D}\mathcal{C}\rangle\langle \mathcal{B}\mathcal{E}\rangle  \\
    107) \quad & \langle \mathcal{A}\mathcal{D}\mathcal{B}\mathcal{C}\mathcal{E}\mathcal{F}\rangle +\langle \mathcal{A}\mathcal{D}\mathcal{B}\mathcal{E}\mathcal{F}\mathcal{C}\rangle +\langle \mathcal{A}\mathcal{D}\mathcal{C}\mathcal{B}\mathcal{E}\mathcal{F}\rangle +\langle \mathcal{A}\mathcal{D}\mathcal{C}\mathcal{E}\mathcal{F}\mathcal{B}\rangle +\langle \mathcal{A}\mathcal{D}\mathcal{E}\mathcal{F}\mathcal{B}\mathcal{C}\rangle +\langle \mathcal{A}\mathcal{D}\mathcal{E}\mathcal{F}\mathcal{C}\mathcal{B}\rangle  \notag \\ 
     = &  \langle \mathcal{A}\mathcal{D}\mathcal{B}\mathcal{C}\rangle\langle \mathcal{E}\mathcal{F}\rangle +\langle \mathcal{A}\mathcal{D}\mathcal{B}\rangle\langle \mathcal{C}\mathcal{E}\mathcal{F}\rangle +\langle \mathcal{A}\mathcal{D}\mathcal{C}\mathcal{B}\rangle\langle \mathcal{E}\mathcal{F}\rangle +\langle \mathcal{A}\mathcal{D}\mathcal{C}\rangle\langle \mathcal{B}\mathcal{E}\mathcal{F}\rangle +\langle \mathcal{A}\mathcal{D}\mathcal{E}\mathcal{F}\rangle\langle \mathcal{B}\mathcal{C}\rangle +\langle \mathcal{A}\mathcal{D}\rangle\langle \mathcal{B}\mathcal{C}\mathcal{E}\mathcal{F}\rangle \notag \\
       & -\langle \mathcal{A}\mathcal{D}\rangle\langle \mathcal{B}\mathcal{C}\rangle\langle \mathcal{E}\mathcal{F}\rangle +\langle \mathcal{A}\mathcal{D}\rangle\langle \mathcal{B}\mathcal{E}\mathcal{F}\mathcal{C}\rangle 
\end{align}
 
 \begin{align}
    108) \quad & \langle \mathcal{A}\mathcal{D}\mathcal{B}\mathcal{C}\mathcal{F}\mathcal{E}\rangle +\langle \mathcal{A}\mathcal{D}\mathcal{B}\mathcal{E}\mathcal{C}\mathcal{F}\rangle +\langle \mathcal{A}\mathcal{D}\mathcal{C}\mathcal{F}\mathcal{B}\mathcal{E}\rangle +\langle \mathcal{A}\mathcal{D}\mathcal{C}\mathcal{F}\mathcal{E}\mathcal{B}\rangle +\langle \mathcal{A}\mathcal{D}\mathcal{E}\mathcal{B}\mathcal{C}\mathcal{F}\rangle +\langle \mathcal{A}\mathcal{D}\mathcal{E}\mathcal{C}\mathcal{F}\mathcal{B}\rangle  \notag \\ 
     = &  \langle \mathcal{A}\mathcal{D}\mathcal{B}\mathcal{E}\rangle\langle \mathcal{C}\mathcal{F}\rangle +\langle \mathcal{A}\mathcal{D}\mathcal{B}\rangle\langle \mathcal{C}\mathcal{F}\mathcal{E}\rangle +\langle \mathcal{A}\mathcal{D}\mathcal{C}\mathcal{F}\rangle\langle \mathcal{B}\mathcal{E}\rangle +\langle \mathcal{A}\mathcal{D}\mathcal{E}\mathcal{B}\rangle\langle \mathcal{C}\mathcal{F}\rangle +\langle \mathcal{A}\mathcal{D}\mathcal{E}\rangle\langle \mathcal{B}\mathcal{C}\mathcal{F}\rangle +\langle \mathcal{A}\mathcal{D}\rangle\langle \mathcal{B}\mathcal{C}\mathcal{F}\mathcal{E}\rangle \notag \\
       & +\langle \mathcal{A}\mathcal{D}\rangle\langle \mathcal{B}\mathcal{E}\mathcal{C}\mathcal{F}\rangle -\langle \mathcal{A}\mathcal{D}\rangle\langle \mathcal{B}\mathcal{E}\rangle\langle \mathcal{C}\mathcal{F}\rangle  \\
    109) \quad & \langle \mathcal{A}\mathcal{D}\mathcal{B}\mathcal{E}\mathcal{F}\mathcal{C}\rangle +\langle \mathcal{A}\mathcal{D}\mathcal{B}\mathcal{F}\mathcal{C}\mathcal{E}\rangle +\langle \mathcal{A}\mathcal{D}\mathcal{E}\mathcal{B}\mathcal{F}\mathcal{C}\rangle +\langle \mathcal{A}\mathcal{D}\mathcal{E}\mathcal{F}\mathcal{C}\mathcal{B}\rangle +\langle \mathcal{A}\mathcal{D}\mathcal{F}\mathcal{C}\mathcal{B}\mathcal{E}\rangle +\langle \mathcal{A}\mathcal{D}\mathcal{F}\mathcal{C}\mathcal{E}\mathcal{B}\rangle  \notag \\ 
     = &  \langle \mathcal{A}\mathcal{D}\mathcal{B}\mathcal{E}\rangle\langle \mathcal{C}\mathcal{F}\rangle +\langle \mathcal{A}\mathcal{D}\mathcal{B}\rangle\langle \mathcal{C}\mathcal{E}\mathcal{F}\rangle +\langle \mathcal{A}\mathcal{D}\mathcal{E}\mathcal{B}\rangle\langle \mathcal{C}\mathcal{F}\rangle +\langle \mathcal{A}\mathcal{D}\mathcal{E}\rangle\langle \mathcal{B}\mathcal{F}\mathcal{C}\rangle +\langle \mathcal{A}\mathcal{D}\mathcal{F}\mathcal{C}\rangle\langle \mathcal{B}\mathcal{E}\rangle +\langle \mathcal{A}\mathcal{D}\rangle\langle \mathcal{B}\mathcal{E}\mathcal{F}\mathcal{C}\rangle \notag \\
       & -\langle \mathcal{A}\mathcal{D}\rangle\langle \mathcal{B}\mathcal{E}\rangle\langle \mathcal{C}\mathcal{F}\rangle +\langle \mathcal{A}\mathcal{D}\rangle\langle \mathcal{B}\mathcal{F}\mathcal{C}\mathcal{E}\rangle  \\
    110) \quad & \langle \mathcal{A}\mathcal{D}\mathcal{B}\mathcal{C}\mathcal{E}\mathcal{F}\rangle +\langle \mathcal{A}\mathcal{D}\mathcal{B}\mathcal{F}\mathcal{C}\mathcal{E}\rangle +\langle \mathcal{A}\mathcal{D}\mathcal{C}\mathcal{E}\mathcal{B}\mathcal{F}\rangle +\langle \mathcal{A}\mathcal{D}\mathcal{C}\mathcal{E}\mathcal{F}\mathcal{B}\rangle +\langle \mathcal{A}\mathcal{D}\mathcal{F}\mathcal{B}\mathcal{C}\mathcal{E}\rangle +\langle \mathcal{A}\mathcal{D}\mathcal{F}\mathcal{C}\mathcal{E}\mathcal{B}\rangle  \notag \\ 
     = &  \langle \mathcal{A}\mathcal{D}\mathcal{B}\mathcal{F}\rangle\langle \mathcal{C}\mathcal{E}\rangle +\langle \mathcal{A}\mathcal{D}\mathcal{B}\rangle\langle \mathcal{C}\mathcal{E}\mathcal{F}\rangle +\langle \mathcal{A}\mathcal{D}\mathcal{C}\mathcal{E}\rangle\langle \mathcal{B}\mathcal{F}\rangle +\langle \mathcal{A}\mathcal{D}\mathcal{F}\mathcal{B}\rangle\langle \mathcal{C}\mathcal{E}\rangle +\langle \mathcal{A}\mathcal{D}\mathcal{F}\rangle\langle \mathcal{B}\mathcal{C}\mathcal{E}\rangle +\langle \mathcal{A}\mathcal{D}\rangle\langle \mathcal{B}\mathcal{C}\mathcal{E}\mathcal{F}\rangle \notag \\
       & +\langle \mathcal{A}\mathcal{D}\rangle\langle \mathcal{B}\mathcal{F}\mathcal{C}\mathcal{E}\rangle -\langle \mathcal{A}\mathcal{D}\rangle\langle \mathcal{B}\mathcal{F}\rangle\langle \mathcal{C}\mathcal{E}\rangle  \\
    111) \quad & \langle \mathcal{A}\mathcal{D}\mathcal{B}\mathcal{E}\mathcal{C}\mathcal{F}\rangle +\langle \mathcal{A}\mathcal{D}\mathcal{B}\mathcal{F}\mathcal{E}\mathcal{C}\rangle +\langle \mathcal{A}\mathcal{D}\mathcal{E}\mathcal{C}\mathcal{B}\mathcal{F}\rangle +\langle \mathcal{A}\mathcal{D}\mathcal{E}\mathcal{C}\mathcal{F}\mathcal{B}\rangle +\langle \mathcal{A}\mathcal{D}\mathcal{F}\mathcal{B}\mathcal{E}\mathcal{C}\rangle +\langle \mathcal{A}\mathcal{D}\mathcal{F}\mathcal{E}\mathcal{C}\mathcal{B}\rangle  \notag \\ 
     = &  \langle \mathcal{A}\mathcal{D}\mathcal{B}\mathcal{F}\rangle\langle \mathcal{C}\mathcal{E}\rangle +\langle \mathcal{A}\mathcal{D}\mathcal{B}\rangle\langle \mathcal{C}\mathcal{F}\mathcal{E}\rangle +\langle \mathcal{A}\mathcal{D}\mathcal{E}\mathcal{C}\rangle\langle \mathcal{B}\mathcal{F}\rangle +\langle \mathcal{A}\mathcal{D}\mathcal{F}\mathcal{B}\rangle\langle \mathcal{C}\mathcal{E}\rangle +\langle \mathcal{A}\mathcal{D}\mathcal{F}\rangle\langle \mathcal{B}\mathcal{E}\mathcal{C}\rangle +\langle \mathcal{A}\mathcal{D}\rangle\langle \mathcal{B}\mathcal{E}\mathcal{C}\mathcal{F}\rangle \notag \\
       & +\langle \mathcal{A}\mathcal{D}\rangle\langle \mathcal{B}\mathcal{F}\mathcal{E}\mathcal{C}\rangle -\langle \mathcal{A}\mathcal{D}\rangle\langle \mathcal{B}\mathcal{F}\rangle\langle \mathcal{C}\mathcal{E}\rangle  \\
    112) \quad & \langle \mathcal{A}\mathcal{B}\mathcal{E}\mathcal{F}\mathcal{C}\mathcal{D}\rangle +\langle \mathcal{A}\mathcal{B}\mathcal{F}\mathcal{C}\mathcal{E}\mathcal{D}\rangle +\langle \mathcal{A}\mathcal{E}\mathcal{B}\mathcal{F}\mathcal{C}\mathcal{D}\rangle +\langle \mathcal{A}\mathcal{E}\mathcal{F}\mathcal{C}\mathcal{B}\mathcal{D}\rangle +\langle \mathcal{A}\mathcal{F}\mathcal{C}\mathcal{B}\mathcal{E}\mathcal{D}\rangle +\langle \mathcal{A}\mathcal{F}\mathcal{C}\mathcal{E}\mathcal{B}\mathcal{D}\rangle  \notag \\ 
     = &  \langle \mathcal{A}\mathcal{B}\mathcal{D}\rangle\langle \mathcal{C}\mathcal{E}\mathcal{F}\rangle +\langle \mathcal{A}\mathcal{B}\mathcal{E}\mathcal{D}\rangle\langle \mathcal{C}\mathcal{F}\rangle +\langle \mathcal{A}\mathcal{D}\rangle\langle \mathcal{B}\mathcal{E}\mathcal{F}\mathcal{C}\rangle -\langle \mathcal{A}\mathcal{D}\rangle\langle \mathcal{B}\mathcal{E}\rangle\langle \mathcal{C}\mathcal{F}\rangle +\langle \mathcal{A}\mathcal{D}\rangle\langle \mathcal{B}\mathcal{F}\mathcal{C}\mathcal{E}\rangle +\langle \mathcal{A}\mathcal{E}\mathcal{B}\mathcal{D}\rangle\langle \mathcal{C}\mathcal{F}\rangle \notag \\
       & +\langle \mathcal{A}\mathcal{E}\mathcal{D}\rangle\langle \mathcal{B}\mathcal{F}\mathcal{C}\rangle +\langle \mathcal{A}\mathcal{F}\mathcal{C}\mathcal{D}\rangle\langle \mathcal{B}\mathcal{E}\rangle  \\
    113) \quad & \langle \mathcal{A}\mathcal{E}\mathcal{B}\mathcal{C}\mathcal{D}\mathcal{F}\rangle +\langle \mathcal{A}\mathcal{E}\mathcal{B}\mathcal{D}\mathcal{F}\mathcal{C}\rangle +\langle \mathcal{A}\mathcal{E}\mathcal{C}\mathcal{B}\mathcal{D}\mathcal{F}\rangle +\langle \mathcal{A}\mathcal{E}\mathcal{C}\mathcal{D}\mathcal{F}\mathcal{B}\rangle +\langle \mathcal{A}\mathcal{E}\mathcal{D}\mathcal{F}\mathcal{B}\mathcal{C}\rangle +\langle \mathcal{A}\mathcal{E}\mathcal{D}\mathcal{F}\mathcal{C}\mathcal{B}\rangle  \notag \\ 
     = &  \langle \mathcal{A}\mathcal{E}\mathcal{B}\mathcal{C}\rangle\langle \mathcal{D}\mathcal{F}\rangle +\langle \mathcal{A}\mathcal{E}\mathcal{B}\rangle\langle \mathcal{C}\mathcal{D}\mathcal{F}\rangle +\langle \mathcal{A}\mathcal{E}\mathcal{C}\mathcal{B}\rangle\langle \mathcal{D}\mathcal{F}\rangle +\langle \mathcal{A}\mathcal{E}\mathcal{C}\rangle\langle \mathcal{B}\mathcal{D}\mathcal{F}\rangle +\langle \mathcal{A}\mathcal{E}\mathcal{D}\mathcal{F}\rangle\langle \mathcal{B}\mathcal{C}\rangle +\langle \mathcal{A}\mathcal{E}\rangle\langle \mathcal{B}\mathcal{C}\mathcal{D}\mathcal{F}\rangle \notag \\
       & -\langle \mathcal{A}\mathcal{E}\rangle\langle \mathcal{B}\mathcal{C}\rangle\langle \mathcal{D}\mathcal{F}\rangle +\langle \mathcal{A}\mathcal{E}\rangle\langle \mathcal{B}\mathcal{D}\mathcal{F}\mathcal{C}\rangle  \\
    114) \quad & \langle \mathcal{A}\mathcal{E}\mathcal{B}\mathcal{C}\mathcal{F}\mathcal{D}\rangle +\langle \mathcal{A}\mathcal{E}\mathcal{B}\mathcal{D}\mathcal{C}\mathcal{F}\rangle +\langle \mathcal{A}\mathcal{E}\mathcal{C}\mathcal{F}\mathcal{B}\mathcal{D}\rangle +\langle \mathcal{A}\mathcal{E}\mathcal{C}\mathcal{F}\mathcal{D}\mathcal{B}\rangle +\langle \mathcal{A}\mathcal{E}\mathcal{D}\mathcal{B}\mathcal{C}\mathcal{F}\rangle +\langle \mathcal{A}\mathcal{E}\mathcal{D}\mathcal{C}\mathcal{F}\mathcal{B}\rangle  \notag \\ 
     = &  \langle \mathcal{A}\mathcal{E}\mathcal{B}\mathcal{D}\rangle\langle \mathcal{C}\mathcal{F}\rangle +\langle \mathcal{A}\mathcal{E}\mathcal{B}\rangle\langle \mathcal{C}\mathcal{F}\mathcal{D}\rangle +\langle \mathcal{A}\mathcal{E}\mathcal{C}\mathcal{F}\rangle\langle \mathcal{B}\mathcal{D}\rangle +\langle \mathcal{A}\mathcal{E}\mathcal{D}\mathcal{B}\rangle\langle \mathcal{C}\mathcal{F}\rangle +\langle \mathcal{A}\mathcal{E}\mathcal{D}\rangle\langle \mathcal{B}\mathcal{C}\mathcal{F}\rangle +\langle \mathcal{A}\mathcal{E}\rangle\langle \mathcal{B}\mathcal{C}\mathcal{F}\mathcal{D}\rangle \notag \\
       & +\langle \mathcal{A}\mathcal{E}\rangle\langle \mathcal{B}\mathcal{D}\mathcal{C}\mathcal{F}\rangle -\langle \mathcal{A}\mathcal{E}\rangle\langle \mathcal{B}\mathcal{D}\rangle\langle \mathcal{C}\mathcal{F}\rangle  \\
    115) \quad & \langle \mathcal{A}\mathcal{E}\mathcal{B}\mathcal{C}\mathcal{D}\mathcal{F}\rangle +\langle \mathcal{A}\mathcal{E}\mathcal{B}\mathcal{F}\mathcal{C}\mathcal{D}\rangle +\langle \mathcal{A}\mathcal{E}\mathcal{C}\mathcal{D}\mathcal{B}\mathcal{F}\rangle +\langle \mathcal{A}\mathcal{E}\mathcal{C}\mathcal{D}\mathcal{F}\mathcal{B}\rangle +\langle \mathcal{A}\mathcal{E}\mathcal{F}\mathcal{B}\mathcal{C}\mathcal{D}\rangle +\langle \mathcal{A}\mathcal{E}\mathcal{F}\mathcal{C}\mathcal{D}\mathcal{B}\rangle  \notag \\ 
     = &  \langle \mathcal{A}\mathcal{E}\mathcal{B}\mathcal{F}\rangle\langle \mathcal{C}\mathcal{D}\rangle +\langle \mathcal{A}\mathcal{E}\mathcal{B}\rangle\langle \mathcal{C}\mathcal{D}\mathcal{F}\rangle +\langle \mathcal{A}\mathcal{E}\mathcal{C}\mathcal{D}\rangle\langle \mathcal{B}\mathcal{F}\rangle +\langle \mathcal{A}\mathcal{E}\mathcal{F}\mathcal{B}\rangle\langle \mathcal{C}\mathcal{D}\rangle +\langle \mathcal{A}\mathcal{E}\mathcal{F}\rangle\langle \mathcal{B}\mathcal{C}\mathcal{D}\rangle +\langle \mathcal{A}\mathcal{E}\rangle\langle \mathcal{B}\mathcal{C}\mathcal{D}\mathcal{F}\rangle \notag \\
       & +\langle \mathcal{A}\mathcal{E}\rangle\langle \mathcal{B}\mathcal{F}\mathcal{C}\mathcal{D}\rangle -\langle \mathcal{A}\mathcal{E}\rangle\langle \mathcal{B}\mathcal{F}\rangle\langle \mathcal{C}\mathcal{D}\rangle  \\
    116) \quad & \langle \mathcal{A}\mathcal{E}\mathcal{B}\mathcal{D}\mathcal{C}\mathcal{F}\rangle +\langle \mathcal{A}\mathcal{E}\mathcal{B}\mathcal{F}\mathcal{D}\mathcal{C}\rangle +\langle \mathcal{A}\mathcal{E}\mathcal{D}\mathcal{C}\mathcal{B}\mathcal{F}\rangle +\langle \mathcal{A}\mathcal{E}\mathcal{D}\mathcal{C}\mathcal{F}\mathcal{B}\rangle +\langle \mathcal{A}\mathcal{E}\mathcal{F}\mathcal{B}\mathcal{D}\mathcal{C}\rangle +\langle \mathcal{A}\mathcal{E}\mathcal{F}\mathcal{D}\mathcal{C}\mathcal{B}\rangle  \notag \\ 
     = &  \langle \mathcal{A}\mathcal{E}\mathcal{B}\mathcal{F}\rangle\langle \mathcal{C}\mathcal{D}\rangle +\langle \mathcal{A}\mathcal{E}\mathcal{B}\rangle\langle \mathcal{C}\mathcal{F}\mathcal{D}\rangle +\langle \mathcal{A}\mathcal{E}\mathcal{D}\mathcal{C}\rangle\langle \mathcal{B}\mathcal{F}\rangle +\langle \mathcal{A}\mathcal{E}\mathcal{F}\mathcal{B}\rangle\langle \mathcal{C}\mathcal{D}\rangle +\langle \mathcal{A}\mathcal{E}\mathcal{F}\rangle\langle \mathcal{B}\mathcal{D}\mathcal{C}\rangle +\langle \mathcal{A}\mathcal{E}\rangle\langle \mathcal{B}\mathcal{D}\mathcal{C}\mathcal{F}\rangle \notag \\
       & +\langle \mathcal{A}\mathcal{E}\rangle\langle \mathcal{B}\mathcal{F}\mathcal{D}\mathcal{C}\rangle -\langle \mathcal{A}\mathcal{E}\rangle\langle \mathcal{B}\mathcal{F}\rangle\langle \mathcal{C}\mathcal{D}\rangle  \\
    117) \quad & \langle \mathcal{A}\mathcal{F}\mathcal{B}\mathcal{C}\mathcal{D}\mathcal{E}\rangle +\langle \mathcal{A}\mathcal{F}\mathcal{B}\mathcal{D}\mathcal{E}\mathcal{C}\rangle +\langle \mathcal{A}\mathcal{F}\mathcal{C}\mathcal{B}\mathcal{D}\mathcal{E}\rangle +\langle \mathcal{A}\mathcal{F}\mathcal{C}\mathcal{D}\mathcal{E}\mathcal{B}\rangle +\langle \mathcal{A}\mathcal{F}\mathcal{D}\mathcal{E}\mathcal{B}\mathcal{C}\rangle +\langle \mathcal{A}\mathcal{F}\mathcal{D}\mathcal{E}\mathcal{C}\mathcal{B}\rangle  \notag \\ 
     = &  \langle \mathcal{A}\mathcal{F}\mathcal{B}\mathcal{C}\rangle\langle \mathcal{D}\mathcal{E}\rangle +\langle \mathcal{A}\mathcal{F}\mathcal{B}\rangle\langle \mathcal{C}\mathcal{D}\mathcal{E}\rangle +\langle \mathcal{A}\mathcal{F}\mathcal{C}\mathcal{B}\rangle\langle \mathcal{D}\mathcal{E}\rangle +\langle \mathcal{A}\mathcal{F}\mathcal{C}\rangle\langle \mathcal{B}\mathcal{D}\mathcal{E}\rangle +\langle \mathcal{A}\mathcal{F}\mathcal{D}\mathcal{E}\rangle\langle \mathcal{B}\mathcal{C}\rangle +\langle \mathcal{A}\mathcal{F}\rangle\langle \mathcal{B}\mathcal{C}\mathcal{D}\mathcal{E}\rangle \notag \\
       & -\langle \mathcal{A}\mathcal{F}\rangle\langle \mathcal{B}\mathcal{C}\rangle\langle \mathcal{D}\mathcal{E}\rangle +\langle \mathcal{A}\mathcal{F}\rangle\langle \mathcal{B}\mathcal{D}\mathcal{E}\mathcal{C}\rangle  \\
    118) \quad & \langle \mathcal{A}\mathcal{F}\mathcal{B}\mathcal{C}\mathcal{E}\mathcal{D}\rangle +\langle \mathcal{A}\mathcal{F}\mathcal{B}\mathcal{D}\mathcal{C}\mathcal{E}\rangle +\langle \mathcal{A}\mathcal{F}\mathcal{C}\mathcal{E}\mathcal{B}\mathcal{D}\rangle +\langle \mathcal{A}\mathcal{F}\mathcal{C}\mathcal{E}\mathcal{D}\mathcal{B}\rangle +\langle \mathcal{A}\mathcal{F}\mathcal{D}\mathcal{B}\mathcal{C}\mathcal{E}\rangle +\langle \mathcal{A}\mathcal{F}\mathcal{D}\mathcal{C}\mathcal{E}\mathcal{B}\rangle  \notag \\ 
     = &  \langle \mathcal{A}\mathcal{F}\mathcal{B}\mathcal{D}\rangle\langle \mathcal{C}\mathcal{E}\rangle +\langle \mathcal{A}\mathcal{F}\mathcal{B}\rangle\langle \mathcal{C}\mathcal{E}\mathcal{D}\rangle +\langle \mathcal{A}\mathcal{F}\mathcal{C}\mathcal{E}\rangle\langle \mathcal{B}\mathcal{D}\rangle +\langle \mathcal{A}\mathcal{F}\mathcal{D}\mathcal{B}\rangle\langle \mathcal{C}\mathcal{E}\rangle +\langle \mathcal{A}\mathcal{F}\mathcal{D}\rangle\langle \mathcal{B}\mathcal{C}\mathcal{E}\rangle +\langle \mathcal{A}\mathcal{F}\rangle\langle \mathcal{B}\mathcal{C}\mathcal{E}\mathcal{D}\rangle \notag \\
       & +\langle \mathcal{A}\mathcal{F}\rangle\langle \mathcal{B}\mathcal{D}\mathcal{C}\mathcal{E}\rangle -\langle \mathcal{A}\mathcal{F}\rangle\langle \mathcal{B}\mathcal{D}\rangle\langle \mathcal{C}\mathcal{E}\rangle  
 \end{align}
 
 \begin{align}
    119) \quad & \langle \mathcal{A}\mathcal{F}\mathcal{B}\mathcal{C}\mathcal{D}\mathcal{E}\rangle +\langle \mathcal{A}\mathcal{F}\mathcal{B}\mathcal{E}\mathcal{C}\mathcal{D}\rangle +\langle \mathcal{A}\mathcal{F}\mathcal{C}\mathcal{D}\mathcal{B}\mathcal{E}\rangle +\langle \mathcal{A}\mathcal{F}\mathcal{C}\mathcal{D}\mathcal{E}\mathcal{B}\rangle +\langle \mathcal{A}\mathcal{F}\mathcal{E}\mathcal{B}\mathcal{C}\mathcal{D}\rangle +\langle \mathcal{A}\mathcal{F}\mathcal{E}\mathcal{C}\mathcal{D}\mathcal{B}\rangle  \notag \\ 
     = &  \langle \mathcal{A}\mathcal{F}\mathcal{B}\mathcal{E}\rangle\langle \mathcal{C}\mathcal{D}\rangle +\langle \mathcal{A}\mathcal{F}\mathcal{B}\rangle\langle \mathcal{C}\mathcal{D}\mathcal{E}\rangle +\langle \mathcal{A}\mathcal{F}\mathcal{C}\mathcal{D}\rangle\langle \mathcal{B}\mathcal{E}\rangle +\langle \mathcal{A}\mathcal{F}\mathcal{E}\mathcal{B}\rangle\langle \mathcal{C}\mathcal{D}\rangle +\langle \mathcal{A}\mathcal{F}\mathcal{E}\rangle\langle \mathcal{B}\mathcal{C}\mathcal{D}\rangle +\langle \mathcal{A}\mathcal{F}\rangle\langle \mathcal{B}\mathcal{C}\mathcal{D}\mathcal{E}\rangle \notag \\
       & +\langle \mathcal{A}\mathcal{F}\rangle\langle \mathcal{B}\mathcal{E}\mathcal{C}\mathcal{D}\rangle -\langle \mathcal{A}\mathcal{F}\rangle\langle \mathcal{B}\mathcal{E}\rangle\langle \mathcal{C}\mathcal{D}\rangle  \\
    120) \quad & \langle \mathcal{A}\mathcal{B}\mathcal{C}\mathcal{D}\mathcal{E}\mathcal{F}\rangle +\langle \mathcal{A}\mathcal{B}\mathcal{C}\mathcal{E}\mathcal{F}\mathcal{D}\rangle +\langle \mathcal{A}\mathcal{D}\mathcal{B}\mathcal{C}\mathcal{E}\mathcal{F}\rangle +\langle \mathcal{A}\mathcal{D}\mathcal{E}\mathcal{F}\mathcal{B}\mathcal{C}\rangle +\langle \mathcal{A}\mathcal{E}\mathcal{F}\mathcal{B}\mathcal{C}\mathcal{D}\rangle +\langle \mathcal{A}\mathcal{E}\mathcal{F}\mathcal{D}\mathcal{B}\mathcal{C}\rangle  \notag \\ 
     = &  \langle \mathcal{A}\mathcal{B}\mathcal{C}\mathcal{D}\rangle\langle \mathcal{E}\mathcal{F}\rangle +\langle \mathcal{A}\mathcal{B}\mathcal{C}\rangle\langle \mathcal{D}\mathcal{E}\mathcal{F}\rangle +\langle \mathcal{A}\mathcal{D}\mathcal{B}\mathcal{C}\rangle\langle \mathcal{E}\mathcal{F}\rangle +\langle \mathcal{A}\mathcal{D}\mathcal{E}\mathcal{F}\rangle\langle \mathcal{B}\mathcal{C}\rangle +\langle \mathcal{A}\mathcal{D}\rangle\langle \mathcal{B}\mathcal{C}\mathcal{E}\mathcal{F}\rangle -\langle \mathcal{A}\mathcal{D}\rangle\langle \mathcal{B}\mathcal{C}\rangle\langle \mathcal{E}\mathcal{F}\rangle \notag \\
       & +\langle \mathcal{A}\mathcal{E}\mathcal{F}\mathcal{D}\rangle\langle \mathcal{B}\mathcal{C}\rangle +\langle \mathcal{A}\mathcal{E}\mathcal{F}\rangle\langle \mathcal{B}\mathcal{C}\mathcal{D}\rangle  
 \end{align}

\paragraph{$\mathcal{A}\mathcal{B}\mathcal{C}\mathcal{D}\mathcal{E}^2$}

\begin{align}
    1) \quad & \langle \mathcal{A}\mathcal{B}\mathcal{C}\mathcal{D}\mathcal{E}\mathcal{E}\rangle +\langle \mathcal{A}\mathcal{B}\mathcal{D}\mathcal{E}\mathcal{E}\mathcal{C}\rangle +\langle \mathcal{A}\mathcal{C}\mathcal{B}\mathcal{D}\mathcal{E}\mathcal{E}\rangle +\langle \mathcal{A}\mathcal{C}\mathcal{D}\mathcal{E}\mathcal{E}\mathcal{B}\rangle +\langle \mathcal{A}\mathcal{D}\mathcal{E}\mathcal{E}\mathcal{B}\mathcal{C}\rangle +\langle \mathcal{A}\mathcal{D}\mathcal{E}\mathcal{E}\mathcal{C}\mathcal{B}\rangle  \notag \\ 
    = &  \langle \mathcal{A}\mathcal{B}\mathcal{C}\rangle\langle \mathcal{D}\mathcal{E}\mathcal{E}\rangle +\langle \mathcal{A}\mathcal{B}\rangle\langle \mathcal{C}\mathcal{D}\mathcal{E}\mathcal{E}\rangle +\langle \mathcal{A}\mathcal{C}\mathcal{B}\rangle\langle \mathcal{D}\mathcal{E}\mathcal{E}\rangle +\langle \mathcal{A}\mathcal{C}\rangle\langle \mathcal{B}\mathcal{D}\mathcal{E}\mathcal{E}\rangle +\langle \mathcal{A}\mathcal{D}\mathcal{E}\mathcal{E}\rangle\langle \mathcal{B}\mathcal{C}\rangle  \\
   2) \quad & \langle \mathcal{A}\mathcal{B}\mathcal{C}\mathcal{E}\mathcal{D}\mathcal{E}\rangle +\langle \mathcal{A}\mathcal{B}\mathcal{E}\mathcal{D}\mathcal{E}\mathcal{C}\rangle +\langle \mathcal{A}\mathcal{C}\mathcal{B}\mathcal{E}\mathcal{D}\mathcal{E}\rangle +\langle \mathcal{A}\mathcal{C}\mathcal{E}\mathcal{D}\mathcal{E}\mathcal{B}\rangle +\langle \mathcal{A}\mathcal{E}\mathcal{D}\mathcal{E}\mathcal{B}\mathcal{C}\rangle +\langle \mathcal{A}\mathcal{E}\mathcal{D}\mathcal{E}\mathcal{C}\mathcal{B}\rangle  \notag \\ 
    = &  \langle \mathcal{A}\mathcal{B}\mathcal{C}\rangle\langle \mathcal{D}\mathcal{E}\mathcal{E}\rangle +\langle \mathcal{A}\mathcal{B}\rangle\langle \mathcal{C}\mathcal{E}\mathcal{D}\mathcal{E}\rangle +\langle \mathcal{A}\mathcal{C}\mathcal{B}\rangle\langle \mathcal{D}\mathcal{E}\mathcal{E}\rangle +\langle \mathcal{A}\mathcal{C}\rangle\langle \mathcal{B}\mathcal{E}\mathcal{D}\mathcal{E}\rangle +\langle \mathcal{A}\mathcal{E}\mathcal{D}\mathcal{E}\rangle\langle \mathcal{B}\mathcal{C}\rangle  \\
   3) \quad & \langle \mathcal{A}\mathcal{B}\mathcal{C}\mathcal{E}\mathcal{E}\mathcal{D}\rangle +\langle \mathcal{A}\mathcal{B}\mathcal{E}\mathcal{E}\mathcal{D}\mathcal{C}\rangle +\langle \mathcal{A}\mathcal{C}\mathcal{B}\mathcal{E}\mathcal{E}\mathcal{D}\rangle +\langle \mathcal{A}\mathcal{C}\mathcal{E}\mathcal{E}\mathcal{D}\mathcal{B}\rangle +\langle \mathcal{A}\mathcal{E}\mathcal{E}\mathcal{D}\mathcal{B}\mathcal{C}\rangle +\langle \mathcal{A}\mathcal{E}\mathcal{E}\mathcal{D}\mathcal{C}\mathcal{B}\rangle  \notag \\ 
    = &  \langle \mathcal{A}\mathcal{B}\mathcal{C}\rangle\langle \mathcal{D}\mathcal{E}\mathcal{E}\rangle +\langle \mathcal{A}\mathcal{B}\rangle\langle \mathcal{C}\mathcal{E}\mathcal{E}\mathcal{D}\rangle +\langle \mathcal{A}\mathcal{C}\mathcal{B}\rangle\langle \mathcal{D}\mathcal{E}\mathcal{E}\rangle +\langle \mathcal{A}\mathcal{C}\rangle\langle \mathcal{B}\mathcal{E}\mathcal{E}\mathcal{D}\rangle +\langle \mathcal{A}\mathcal{E}\mathcal{E}\mathcal{D}\rangle\langle \mathcal{B}\mathcal{C}\rangle  \\
   4) \quad & \langle \mathcal{A}\mathcal{B}\mathcal{C}\mathcal{E}\mathcal{E}\mathcal{D}\rangle +\langle \mathcal{A}\mathcal{B}\mathcal{D}\mathcal{C}\mathcal{E}\mathcal{E}\rangle +\langle \mathcal{A}\mathcal{C}\mathcal{E}\mathcal{E}\mathcal{B}\mathcal{D}\rangle +\langle \mathcal{A}\mathcal{C}\mathcal{E}\mathcal{E}\mathcal{D}\mathcal{B}\rangle +\langle \mathcal{A}\mathcal{D}\mathcal{B}\mathcal{C}\mathcal{E}\mathcal{E}\rangle +\langle \mathcal{A}\mathcal{D}\mathcal{C}\mathcal{E}\mathcal{E}\mathcal{B}\rangle  \notag \\ 
    = &  \langle \mathcal{A}\mathcal{B}\mathcal{D}\rangle\langle \mathcal{C}\mathcal{E}\mathcal{E}\rangle +\langle \mathcal{A}\mathcal{B}\rangle\langle \mathcal{C}\mathcal{E}\mathcal{E}\mathcal{D}\rangle +\langle \mathcal{A}\mathcal{C}\mathcal{E}\mathcal{E}\rangle\langle \mathcal{B}\mathcal{D}\rangle +\langle \mathcal{A}\mathcal{D}\mathcal{B}\rangle\langle \mathcal{C}\mathcal{E}\mathcal{E}\rangle +\langle \mathcal{A}\mathcal{D}\rangle\langle \mathcal{B}\mathcal{C}\mathcal{E}\mathcal{E}\rangle  \\
   5) \quad & \langle \mathcal{A}\mathcal{B}\mathcal{D}\mathcal{E}\mathcal{C}\mathcal{E}\rangle +\langle \mathcal{A}\mathcal{B}\mathcal{E}\mathcal{C}\mathcal{E}\mathcal{D}\rangle +\langle \mathcal{A}\mathcal{D}\mathcal{B}\mathcal{E}\mathcal{C}\mathcal{E}\rangle +\langle \mathcal{A}\mathcal{D}\mathcal{E}\mathcal{C}\mathcal{E}\mathcal{B}\rangle +\langle \mathcal{A}\mathcal{E}\mathcal{C}\mathcal{E}\mathcal{B}\mathcal{D}\rangle +\langle \mathcal{A}\mathcal{E}\mathcal{C}\mathcal{E}\mathcal{D}\mathcal{B}\rangle  \notag \\ 
    = &  \langle \mathcal{A}\mathcal{B}\mathcal{D}\rangle\langle \mathcal{C}\mathcal{E}\mathcal{E}\rangle +\langle \mathcal{A}\mathcal{B}\rangle\langle \mathcal{C}\mathcal{E}\mathcal{D}\mathcal{E}\rangle +\langle \mathcal{A}\mathcal{D}\mathcal{B}\rangle\langle \mathcal{C}\mathcal{E}\mathcal{E}\rangle +\langle \mathcal{A}\mathcal{D}\rangle\langle \mathcal{B}\mathcal{E}\mathcal{C}\mathcal{E}\rangle +\langle \mathcal{A}\mathcal{E}\mathcal{C}\mathcal{E}\rangle\langle \mathcal{B}\mathcal{D}\rangle  \\
   6) \quad & \langle \mathcal{A}\mathcal{B}\mathcal{D}\mathcal{E}\mathcal{E}\mathcal{C}\rangle +\langle \mathcal{A}\mathcal{B}\mathcal{E}\mathcal{E}\mathcal{C}\mathcal{D}\rangle +\langle \mathcal{A}\mathcal{D}\mathcal{B}\mathcal{E}\mathcal{E}\mathcal{C}\rangle +\langle \mathcal{A}\mathcal{D}\mathcal{E}\mathcal{E}\mathcal{C}\mathcal{B}\rangle +\langle \mathcal{A}\mathcal{E}\mathcal{E}\mathcal{C}\mathcal{B}\mathcal{D}\rangle +\langle \mathcal{A}\mathcal{E}\mathcal{E}\mathcal{C}\mathcal{D}\mathcal{B}\rangle  \notag \\ 
    = &  \langle \mathcal{A}\mathcal{B}\mathcal{D}\rangle\langle \mathcal{C}\mathcal{E}\mathcal{E}\rangle +\langle \mathcal{A}\mathcal{B}\rangle\langle \mathcal{C}\mathcal{D}\mathcal{E}\mathcal{E}\rangle +\langle \mathcal{A}\mathcal{D}\mathcal{B}\rangle\langle \mathcal{C}\mathcal{E}\mathcal{E}\rangle +\langle \mathcal{A}\mathcal{D}\rangle\langle \mathcal{B}\mathcal{E}\mathcal{E}\mathcal{C}\rangle +\langle \mathcal{A}\mathcal{E}\mathcal{E}\mathcal{C}\rangle\langle \mathcal{B}\mathcal{D}\rangle  \\
   7) \quad & \langle \mathcal{A}\mathcal{B}\mathcal{C}\mathcal{D}\mathcal{E}\mathcal{E}\rangle +\langle \mathcal{A}\mathcal{B}\mathcal{E}\mathcal{C}\mathcal{D}\mathcal{E}\rangle +\langle \mathcal{A}\mathcal{C}\mathcal{D}\mathcal{E}\mathcal{B}\mathcal{E}\rangle +\langle \mathcal{A}\mathcal{C}\mathcal{D}\mathcal{E}\mathcal{E}\mathcal{B}\rangle +\langle \mathcal{A}\mathcal{E}\mathcal{B}\mathcal{C}\mathcal{D}\mathcal{E}\rangle +\langle \mathcal{A}\mathcal{E}\mathcal{C}\mathcal{D}\mathcal{E}\mathcal{B}\rangle  \notag \\ 
    = &  \langle \mathcal{A}\mathcal{B}\mathcal{E}\rangle\langle \mathcal{C}\mathcal{D}\mathcal{E}\rangle +\langle \mathcal{A}\mathcal{B}\rangle\langle \mathcal{C}\mathcal{D}\mathcal{E}\mathcal{E}\rangle +\langle \mathcal{A}\mathcal{C}\mathcal{D}\mathcal{E}\rangle\langle \mathcal{B}\mathcal{E}\rangle +\langle \mathcal{A}\mathcal{E}\mathcal{B}\rangle\langle \mathcal{C}\mathcal{D}\mathcal{E}\rangle +\langle \mathcal{A}\mathcal{E}\rangle\langle \mathcal{B}\mathcal{C}\mathcal{D}\mathcal{E}\rangle  \\
   8) \quad & \langle \mathcal{A}\mathcal{B}\mathcal{C}\mathcal{E}\mathcal{D}\mathcal{E}\rangle +\langle \mathcal{A}\mathcal{B}\mathcal{E}\mathcal{C}\mathcal{E}\mathcal{D}\rangle +\langle \mathcal{A}\mathcal{C}\mathcal{E}\mathcal{D}\mathcal{B}\mathcal{E}\rangle +\langle \mathcal{A}\mathcal{C}\mathcal{E}\mathcal{D}\mathcal{E}\mathcal{B}\rangle +\langle \mathcal{A}\mathcal{E}\mathcal{B}\mathcal{C}\mathcal{E}\mathcal{D}\rangle +\langle \mathcal{A}\mathcal{E}\mathcal{C}\mathcal{E}\mathcal{D}\mathcal{B}\rangle  \notag \\ 
    = &  \langle \mathcal{A}\mathcal{B}\mathcal{E}\rangle\langle \mathcal{C}\mathcal{E}\mathcal{D}\rangle +\langle \mathcal{A}\mathcal{B}\rangle\langle \mathcal{C}\mathcal{E}\mathcal{D}\mathcal{E}\rangle +\langle \mathcal{A}\mathcal{C}\mathcal{E}\mathcal{D}\rangle\langle \mathcal{B}\mathcal{E}\rangle +\langle \mathcal{A}\mathcal{E}\mathcal{B}\rangle\langle \mathcal{C}\mathcal{E}\mathcal{D}\rangle +\langle \mathcal{A}\mathcal{E}\rangle\langle \mathcal{B}\mathcal{C}\mathcal{E}\mathcal{D}\rangle  \\
   9) \quad & \langle \mathcal{A}\mathcal{B}\mathcal{D}\mathcal{C}\mathcal{E}\mathcal{E}\rangle +\langle \mathcal{A}\mathcal{B}\mathcal{E}\mathcal{D}\mathcal{C}\mathcal{E}\rangle +\langle \mathcal{A}\mathcal{D}\mathcal{C}\mathcal{E}\mathcal{B}\mathcal{E}\rangle +\langle \mathcal{A}\mathcal{D}\mathcal{C}\mathcal{E}\mathcal{E}\mathcal{B}\rangle +\langle \mathcal{A}\mathcal{E}\mathcal{B}\mathcal{D}\mathcal{C}\mathcal{E}\rangle +\langle \mathcal{A}\mathcal{E}\mathcal{D}\mathcal{C}\mathcal{E}\mathcal{B}\rangle  \notag \\ 
    = &  \langle \mathcal{A}\mathcal{B}\mathcal{E}\rangle\langle \mathcal{C}\mathcal{E}\mathcal{D}\rangle +\langle \mathcal{A}\mathcal{B}\rangle\langle \mathcal{C}\mathcal{E}\mathcal{E}\mathcal{D}\rangle +\langle \mathcal{A}\mathcal{D}\mathcal{C}\mathcal{E}\rangle\langle \mathcal{B}\mathcal{E}\rangle +\langle \mathcal{A}\mathcal{E}\mathcal{B}\rangle\langle \mathcal{C}\mathcal{E}\mathcal{D}\rangle +\langle \mathcal{A}\mathcal{E}\rangle\langle \mathcal{B}\mathcal{D}\mathcal{C}\mathcal{E}\rangle  \\
   10) \quad & \langle \mathcal{A}\mathcal{B}\mathcal{D}\mathcal{E}\mathcal{C}\mathcal{E}\rangle +\langle \mathcal{A}\mathcal{B}\mathcal{E}\mathcal{D}\mathcal{E}\mathcal{C}\rangle +\langle \mathcal{A}\mathcal{D}\mathcal{E}\mathcal{C}\mathcal{B}\mathcal{E}\rangle +\langle \mathcal{A}\mathcal{D}\mathcal{E}\mathcal{C}\mathcal{E}\mathcal{B}\rangle +\langle \mathcal{A}\mathcal{E}\mathcal{B}\mathcal{D}\mathcal{E}\mathcal{C}\rangle +\langle \mathcal{A}\mathcal{E}\mathcal{D}\mathcal{E}\mathcal{C}\mathcal{B}\rangle  \notag \\ 
    = &  \langle \mathcal{A}\mathcal{B}\mathcal{E}\rangle\langle \mathcal{C}\mathcal{D}\mathcal{E}\rangle +\langle \mathcal{A}\mathcal{B}\rangle\langle \mathcal{C}\mathcal{E}\mathcal{D}\mathcal{E}\rangle +\langle \mathcal{A}\mathcal{D}\mathcal{E}\mathcal{C}\rangle\langle \mathcal{B}\mathcal{E}\rangle +\langle \mathcal{A}\mathcal{E}\mathcal{B}\rangle\langle \mathcal{C}\mathcal{D}\mathcal{E}\rangle +\langle \mathcal{A}\mathcal{E}\rangle\langle \mathcal{B}\mathcal{D}\mathcal{E}\mathcal{C}\rangle  \\
   11) \quad & \langle \mathcal{A}\mathcal{B}\mathcal{E}\mathcal{C}\mathcal{D}\mathcal{E}\rangle +\langle \mathcal{A}\mathcal{B}\mathcal{E}\mathcal{E}\mathcal{C}\mathcal{D}\rangle +\langle \mathcal{A}\mathcal{E}\mathcal{B}\mathcal{E}\mathcal{C}\mathcal{D}\rangle +\langle \mathcal{A}\mathcal{E}\mathcal{C}\mathcal{D}\mathcal{B}\mathcal{E}\rangle +\langle \mathcal{A}\mathcal{E}\mathcal{C}\mathcal{D}\mathcal{E}\mathcal{B}\rangle +\langle \mathcal{A}\mathcal{E}\mathcal{E}\mathcal{C}\mathcal{D}\mathcal{B}\rangle  \notag \\ 
    = &  \langle \mathcal{A}\mathcal{B}\mathcal{E}\rangle\langle \mathcal{C}\mathcal{D}\mathcal{E}\rangle +\langle \mathcal{A}\mathcal{B}\rangle\langle \mathcal{C}\mathcal{D}\mathcal{E}\mathcal{E}\rangle +\langle \mathcal{A}\mathcal{E}\mathcal{B}\rangle\langle \mathcal{C}\mathcal{D}\mathcal{E}\rangle +\langle \mathcal{A}\mathcal{E}\mathcal{C}\mathcal{D}\rangle\langle \mathcal{B}\mathcal{E}\rangle +\langle \mathcal{A}\mathcal{E}\rangle\langle \mathcal{B}\mathcal{E}\mathcal{C}\mathcal{D}\rangle  \\
   12) \quad & \langle \mathcal{A}\mathcal{B}\mathcal{E}\mathcal{D}\mathcal{C}\mathcal{E}\rangle +\langle \mathcal{A}\mathcal{B}\mathcal{E}\mathcal{E}\mathcal{D}\mathcal{C}\rangle +\langle \mathcal{A}\mathcal{E}\mathcal{B}\mathcal{E}\mathcal{D}\mathcal{C}\rangle +\langle \mathcal{A}\mathcal{E}\mathcal{D}\mathcal{C}\mathcal{B}\mathcal{E}\rangle +\langle \mathcal{A}\mathcal{E}\mathcal{D}\mathcal{C}\mathcal{E}\mathcal{B}\rangle +\langle \mathcal{A}\mathcal{E}\mathcal{E}\mathcal{D}\mathcal{C}\mathcal{B}\rangle  \notag \\ 
    = &  \langle \mathcal{A}\mathcal{B}\mathcal{E}\rangle\langle \mathcal{C}\mathcal{E}\mathcal{D}\rangle +\langle \mathcal{A}\mathcal{B}\rangle\langle \mathcal{C}\mathcal{E}\mathcal{E}\mathcal{D}\rangle +\langle \mathcal{A}\mathcal{E}\mathcal{B}\rangle\langle \mathcal{C}\mathcal{E}\mathcal{D}\rangle +\langle \mathcal{A}\mathcal{E}\mathcal{D}\mathcal{C}\rangle\langle \mathcal{B}\mathcal{E}\rangle +\langle \mathcal{A}\mathcal{E}\rangle\langle \mathcal{B}\mathcal{E}\mathcal{D}\mathcal{C}\rangle  \\
   13) \quad & \langle \mathcal{A}\mathcal{B}\mathcal{E}\mathcal{E}\mathcal{C}\mathcal{D}\rangle +\langle \mathcal{A}\mathcal{B}\mathcal{E}\mathcal{E}\mathcal{D}\mathcal{C}\rangle +\langle \mathcal{A}\mathcal{C}\mathcal{B}\mathcal{E}\mathcal{E}\mathcal{D}\rangle +\langle \mathcal{A}\mathcal{C}\mathcal{D}\mathcal{B}\mathcal{E}\mathcal{E}\rangle +\langle \mathcal{A}\mathcal{D}\mathcal{B}\mathcal{E}\mathcal{E}\mathcal{C}\rangle +\langle \mathcal{A}\mathcal{D}\mathcal{C}\mathcal{B}\mathcal{E}\mathcal{E}\rangle  \notag \\ 
    = &  \langle \mathcal{A}\mathcal{B}\mathcal{E}\mathcal{E}\rangle\langle \mathcal{C}\mathcal{D}\rangle +\langle \mathcal{A}\mathcal{C}\mathcal{D}\rangle\langle \mathcal{B}\mathcal{E}\mathcal{E}\rangle +\langle \mathcal{A}\mathcal{C}\rangle\langle \mathcal{B}\mathcal{E}\mathcal{E}\mathcal{D}\rangle +\langle \mathcal{A}\mathcal{D}\mathcal{C}\rangle\langle \mathcal{B}\mathcal{E}\mathcal{E}\rangle +\langle \mathcal{A}\mathcal{D}\rangle\langle \mathcal{B}\mathcal{E}\mathcal{E}\mathcal{C}\rangle  \\
   14) \quad & \langle \mathcal{A}\mathcal{C}\mathcal{D}\mathcal{E}\mathcal{B}\mathcal{E}\rangle +\langle \mathcal{A}\mathcal{C}\mathcal{E}\mathcal{B}\mathcal{E}\mathcal{D}\rangle +\langle \mathcal{A}\mathcal{D}\mathcal{C}\mathcal{E}\mathcal{B}\mathcal{E}\rangle +\langle \mathcal{A}\mathcal{D}\mathcal{E}\mathcal{B}\mathcal{E}\mathcal{C}\rangle +\langle \mathcal{A}\mathcal{E}\mathcal{B}\mathcal{E}\mathcal{C}\mathcal{D}\rangle +\langle \mathcal{A}\mathcal{E}\mathcal{B}\mathcal{E}\mathcal{D}\mathcal{C}\rangle  \notag \\ 
    = &  \langle \mathcal{A}\mathcal{C}\mathcal{D}\rangle\langle \mathcal{B}\mathcal{E}\mathcal{E}\rangle +\langle \mathcal{A}\mathcal{C}\rangle\langle \mathcal{B}\mathcal{E}\mathcal{D}\mathcal{E}\rangle +\langle \mathcal{A}\mathcal{D}\mathcal{C}\rangle\langle \mathcal{B}\mathcal{E}\mathcal{E}\rangle +\langle \mathcal{A}\mathcal{D}\rangle\langle \mathcal{B}\mathcal{E}\mathcal{C}\mathcal{E}\rangle +\langle \mathcal{A}\mathcal{E}\mathcal{B}\mathcal{E}\rangle\langle \mathcal{C}\mathcal{D}\rangle  \\
   15) \quad & \langle \mathcal{A}\mathcal{C}\mathcal{D}\mathcal{E}\mathcal{E}\mathcal{B}\rangle +\langle \mathcal{A}\mathcal{C}\mathcal{E}\mathcal{E}\mathcal{B}\mathcal{D}\rangle +\langle \mathcal{A}\mathcal{D}\mathcal{C}\mathcal{E}\mathcal{E}\mathcal{B}\rangle +\langle \mathcal{A}\mathcal{D}\mathcal{E}\mathcal{E}\mathcal{B}\mathcal{C}\rangle +\langle \mathcal{A}\mathcal{E}\mathcal{E}\mathcal{B}\mathcal{C}\mathcal{D}\rangle +\langle \mathcal{A}\mathcal{E}\mathcal{E}\mathcal{B}\mathcal{D}\mathcal{C}\rangle  \notag \\ 
    = &  \langle \mathcal{A}\mathcal{C}\mathcal{D}\rangle\langle \mathcal{B}\mathcal{E}\mathcal{E}\rangle +\langle \mathcal{A}\mathcal{C}\rangle\langle \mathcal{B}\mathcal{D}\mathcal{E}\mathcal{E}\rangle +\langle \mathcal{A}\mathcal{D}\mathcal{C}\rangle\langle \mathcal{B}\mathcal{E}\mathcal{E}\rangle +\langle \mathcal{A}\mathcal{D}\rangle\langle \mathcal{B}\mathcal{C}\mathcal{E}\mathcal{E}\rangle +\langle \mathcal{A}\mathcal{E}\mathcal{E}\mathcal{B}\rangle\langle \mathcal{C}\mathcal{D}\rangle  \\
   16) \quad & \langle \mathcal{A}\mathcal{B}\mathcal{D}\mathcal{E}\mathcal{C}\mathcal{E}\rangle +\langle \mathcal{A}\mathcal{B}\mathcal{D}\mathcal{E}\mathcal{E}\mathcal{C}\rangle +\langle \mathcal{A}\mathcal{C}\mathcal{B}\mathcal{D}\mathcal{E}\mathcal{E}\rangle +\langle \mathcal{A}\mathcal{C}\mathcal{E}\mathcal{B}\mathcal{D}\mathcal{E}\rangle +\langle \mathcal{A}\mathcal{E}\mathcal{B}\mathcal{D}\mathcal{E}\mathcal{C}\rangle +\langle \mathcal{A}\mathcal{E}\mathcal{C}\mathcal{B}\mathcal{D}\mathcal{E}\rangle  \notag \\ 
    = &  \langle \mathcal{A}\mathcal{B}\mathcal{D}\mathcal{E}\rangle\langle \mathcal{C}\mathcal{E}\rangle +\langle \mathcal{A}\mathcal{C}\mathcal{E}\rangle\langle \mathcal{B}\mathcal{D}\mathcal{E}\rangle +\langle \mathcal{A}\mathcal{C}\rangle\langle \mathcal{B}\mathcal{D}\mathcal{E}\mathcal{E}\rangle +\langle \mathcal{A}\mathcal{E}\mathcal{C}\rangle\langle \mathcal{B}\mathcal{D}\mathcal{E}\rangle +\langle \mathcal{A}\mathcal{E}\rangle\langle \mathcal{B}\mathcal{D}\mathcal{E}\mathcal{C}\rangle  \\
   17) \quad & \langle \mathcal{A}\mathcal{B}\mathcal{E}\mathcal{D}\mathcal{C}\mathcal{E}\rangle +\langle \mathcal{A}\mathcal{B}\mathcal{E}\mathcal{D}\mathcal{E}\mathcal{C}\rangle +\langle \mathcal{A}\mathcal{C}\mathcal{B}\mathcal{E}\mathcal{D}\mathcal{E}\rangle +\langle \mathcal{A}\mathcal{C}\mathcal{E}\mathcal{B}\mathcal{E}\mathcal{D}\rangle +\langle \mathcal{A}\mathcal{E}\mathcal{B}\mathcal{E}\mathcal{D}\mathcal{C}\rangle +\langle \mathcal{A}\mathcal{E}\mathcal{C}\mathcal{B}\mathcal{E}\mathcal{D}\rangle  \notag \\ 
    = &  \langle \mathcal{A}\mathcal{B}\mathcal{E}\mathcal{D}\rangle\langle \mathcal{C}\mathcal{E}\rangle +\langle \mathcal{A}\mathcal{C}\mathcal{E}\rangle\langle \mathcal{B}\mathcal{E}\mathcal{D}\rangle +\langle \mathcal{A}\mathcal{C}\rangle\langle \mathcal{B}\mathcal{E}\mathcal{D}\mathcal{E}\rangle +\langle \mathcal{A}\mathcal{E}\mathcal{C}\rangle\langle \mathcal{B}\mathcal{E}\mathcal{D}\rangle +\langle \mathcal{A}\mathcal{E}\rangle\langle \mathcal{B}\mathcal{E}\mathcal{D}\mathcal{C}\rangle
 \end{align}
 
 \begin{align}
   18) \quad & \langle \mathcal{A}\mathcal{C}\mathcal{D}\mathcal{B}\mathcal{E}\mathcal{E}\rangle +\langle \mathcal{A}\mathcal{C}\mathcal{E}\mathcal{D}\mathcal{B}\mathcal{E}\rangle +\langle \mathcal{A}\mathcal{D}\mathcal{B}\mathcal{E}\mathcal{C}\mathcal{E}\rangle +\langle \mathcal{A}\mathcal{D}\mathcal{B}\mathcal{E}\mathcal{E}\mathcal{C}\rangle +\langle \mathcal{A}\mathcal{E}\mathcal{C}\mathcal{D}\mathcal{B}\mathcal{E}\rangle +\langle \mathcal{A}\mathcal{E}\mathcal{D}\mathcal{B}\mathcal{E}\mathcal{C}\rangle  \notag \\ 
    = &  \langle \mathcal{A}\mathcal{C}\mathcal{E}\rangle\langle \mathcal{B}\mathcal{E}\mathcal{D}\rangle +\langle \mathcal{A}\mathcal{C}\rangle\langle \mathcal{B}\mathcal{E}\mathcal{E}\mathcal{D}\rangle +\langle \mathcal{A}\mathcal{D}\mathcal{B}\mathcal{E}\rangle\langle \mathcal{C}\mathcal{E}\rangle +\langle \mathcal{A}\mathcal{E}\mathcal{C}\rangle\langle \mathcal{B}\mathcal{E}\mathcal{D}\rangle +\langle \mathcal{A}\mathcal{E}\rangle\langle \mathcal{B}\mathcal{E}\mathcal{C}\mathcal{D}\rangle  \\
   19) \quad & \langle \mathcal{A}\mathcal{C}\mathcal{D}\mathcal{E}\mathcal{B}\mathcal{E}\rangle +\langle \mathcal{A}\mathcal{C}\mathcal{E}\mathcal{D}\mathcal{E}\mathcal{B}\rangle +\langle \mathcal{A}\mathcal{D}\mathcal{E}\mathcal{B}\mathcal{C}\mathcal{E}\rangle +\langle \mathcal{A}\mathcal{D}\mathcal{E}\mathcal{B}\mathcal{E}\mathcal{C}\rangle +\langle \mathcal{A}\mathcal{E}\mathcal{C}\mathcal{D}\mathcal{E}\mathcal{B}\rangle +\langle \mathcal{A}\mathcal{E}\mathcal{D}\mathcal{E}\mathcal{B}\mathcal{C}\rangle  \notag \\ 
    = &  \langle \mathcal{A}\mathcal{C}\mathcal{E}\rangle\langle \mathcal{B}\mathcal{D}\mathcal{E}\rangle +\langle \mathcal{A}\mathcal{C}\rangle\langle \mathcal{B}\mathcal{E}\mathcal{D}\mathcal{E}\rangle +\langle \mathcal{A}\mathcal{D}\mathcal{E}\mathcal{B}\rangle\langle \mathcal{C}\mathcal{E}\rangle +\langle \mathcal{A}\mathcal{E}\mathcal{C}\rangle\langle \mathcal{B}\mathcal{D}\mathcal{E}\rangle +\langle \mathcal{A}\mathcal{E}\rangle\langle \mathcal{B}\mathcal{C}\mathcal{D}\mathcal{E}\rangle  \\
   20) \quad & \langle \mathcal{A}\mathcal{C}\mathcal{E}\mathcal{B}\mathcal{D}\mathcal{E}\rangle +\langle \mathcal{A}\mathcal{C}\mathcal{E}\mathcal{E}\mathcal{B}\mathcal{D}\rangle +\langle \mathcal{A}\mathcal{E}\mathcal{B}\mathcal{D}\mathcal{C}\mathcal{E}\rangle +\langle \mathcal{A}\mathcal{E}\mathcal{B}\mathcal{D}\mathcal{E}\mathcal{C}\rangle +\langle \mathcal{A}\mathcal{E}\mathcal{C}\mathcal{E}\mathcal{B}\mathcal{D}\rangle +\langle \mathcal{A}\mathcal{E}\mathcal{E}\mathcal{B}\mathcal{D}\mathcal{C}\rangle  \notag \\ 
    = &  \langle \mathcal{A}\mathcal{C}\mathcal{E}\rangle\langle \mathcal{B}\mathcal{D}\mathcal{E}\rangle +\langle \mathcal{A}\mathcal{C}\rangle\langle \mathcal{B}\mathcal{D}\mathcal{E}\mathcal{E}\rangle +\langle \mathcal{A}\mathcal{E}\mathcal{B}\mathcal{D}\rangle\langle \mathcal{C}\mathcal{E}\rangle +\langle \mathcal{A}\mathcal{E}\mathcal{C}\rangle\langle \mathcal{B}\mathcal{D}\mathcal{E}\rangle +\langle \mathcal{A}\mathcal{E}\rangle\langle \mathcal{B}\mathcal{D}\mathcal{C}\mathcal{E}\rangle  \\
    21) \quad & \langle \mathcal{A}\mathcal{C}\mathcal{E}\mathcal{D}\mathcal{B}\mathcal{E}\rangle +\langle \mathcal{A}\mathcal{C}\mathcal{E}\mathcal{E}\mathcal{D}\mathcal{B}\rangle +\langle \mathcal{A}\mathcal{E}\mathcal{C}\mathcal{E}\mathcal{D}\mathcal{B}\rangle +\langle \mathcal{A}\mathcal{E}\mathcal{D}\mathcal{B}\mathcal{C}\mathcal{E}\rangle +\langle \mathcal{A}\mathcal{E}\mathcal{D}\mathcal{B}\mathcal{E}\mathcal{C}\rangle +\langle \mathcal{A}\mathcal{E}\mathcal{E}\mathcal{D}\mathcal{B}\mathcal{C}\rangle  \notag \\ 
    = &  \langle \mathcal{A}\mathcal{C}\mathcal{E}\rangle\langle \mathcal{B}\mathcal{E}\mathcal{D}\rangle +\langle \mathcal{A}\mathcal{C}\rangle\langle \mathcal{B}\mathcal{E}\mathcal{E}\mathcal{D}\rangle +\langle \mathcal{A}\mathcal{E}\mathcal{C}\rangle\langle \mathcal{B}\mathcal{E}\mathcal{D}\rangle +\langle \mathcal{A}\mathcal{E}\mathcal{D}\mathcal{B}\rangle\langle \mathcal{C}\mathcal{E}\rangle +\langle \mathcal{A}\mathcal{E}\rangle\langle \mathcal{B}\mathcal{C}\mathcal{E}\mathcal{D}\rangle  \\
   22) \quad & \langle \mathcal{A}\mathcal{B}\mathcal{C}\mathcal{E}\mathcal{D}\mathcal{E}\rangle +\langle \mathcal{A}\mathcal{B}\mathcal{C}\mathcal{E}\mathcal{E}\mathcal{D}\rangle +\langle \mathcal{A}\mathcal{D}\mathcal{B}\mathcal{C}\mathcal{E}\mathcal{E}\rangle +\langle \mathcal{A}\mathcal{D}\mathcal{E}\mathcal{B}\mathcal{C}\mathcal{E}\rangle +\langle \mathcal{A}\mathcal{E}\mathcal{B}\mathcal{C}\mathcal{E}\mathcal{D}\rangle +\langle \mathcal{A}\mathcal{E}\mathcal{D}\mathcal{B}\mathcal{C}\mathcal{E}\rangle  \notag \\ 
    = &  \langle \mathcal{A}\mathcal{B}\mathcal{C}\mathcal{E}\rangle\langle \mathcal{D}\mathcal{E}\rangle +\langle \mathcal{A}\mathcal{D}\mathcal{E}\rangle\langle \mathcal{B}\mathcal{C}\mathcal{E}\rangle +\langle \mathcal{A}\mathcal{D}\rangle\langle \mathcal{B}\mathcal{C}\mathcal{E}\mathcal{E}\rangle +\langle \mathcal{A}\mathcal{E}\mathcal{D}\rangle\langle \mathcal{B}\mathcal{C}\mathcal{E}\rangle +\langle \mathcal{A}\mathcal{E}\rangle\langle \mathcal{B}\mathcal{C}\mathcal{E}\mathcal{D}\rangle  \\
   23) \quad & \langle \mathcal{A}\mathcal{B}\mathcal{E}\mathcal{C}\mathcal{D}\mathcal{E}\rangle +\langle \mathcal{A}\mathcal{B}\mathcal{E}\mathcal{C}\mathcal{E}\mathcal{D}\rangle +\langle \mathcal{A}\mathcal{D}\mathcal{B}\mathcal{E}\mathcal{C}\mathcal{E}\rangle +\langle \mathcal{A}\mathcal{D}\mathcal{E}\mathcal{B}\mathcal{E}\mathcal{C}\rangle +\langle \mathcal{A}\mathcal{E}\mathcal{B}\mathcal{E}\mathcal{C}\mathcal{D}\rangle +\langle \mathcal{A}\mathcal{E}\mathcal{D}\mathcal{B}\mathcal{E}\mathcal{C}\rangle  \notag \\ 
    = &  \langle \mathcal{A}\mathcal{B}\mathcal{E}\mathcal{C}\rangle\langle \mathcal{D}\mathcal{E}\rangle +\langle \mathcal{A}\mathcal{D}\mathcal{E}\rangle\langle \mathcal{B}\mathcal{E}\mathcal{C}\rangle +\langle \mathcal{A}\mathcal{D}\rangle\langle \mathcal{B}\mathcal{E}\mathcal{C}\mathcal{E}\rangle +\langle \mathcal{A}\mathcal{E}\mathcal{D}\rangle\langle \mathcal{B}\mathcal{E}\mathcal{C}\rangle +\langle \mathcal{A}\mathcal{E}\rangle\langle \mathcal{B}\mathcal{E}\mathcal{C}\mathcal{D}\rangle  \\
   24) \quad & \langle \mathcal{A}\mathcal{C}\mathcal{B}\mathcal{E}\mathcal{D}\mathcal{E}\rangle +\langle \mathcal{A}\mathcal{C}\mathcal{B}\mathcal{E}\mathcal{E}\mathcal{D}\rangle +\langle \mathcal{A}\mathcal{D}\mathcal{C}\mathcal{B}\mathcal{E}\mathcal{E}\rangle +\langle \mathcal{A}\mathcal{D}\mathcal{E}\mathcal{C}\mathcal{B}\mathcal{E}\rangle +\langle \mathcal{A}\mathcal{E}\mathcal{C}\mathcal{B}\mathcal{E}\mathcal{D}\rangle +\langle \mathcal{A}\mathcal{E}\mathcal{D}\mathcal{C}\mathcal{B}\mathcal{E}\rangle  \notag \\ 
    = &  \langle \mathcal{A}\mathcal{C}\mathcal{B}\mathcal{E}\rangle\langle \mathcal{D}\mathcal{E}\rangle +\langle \mathcal{A}\mathcal{D}\mathcal{E}\rangle\langle \mathcal{B}\mathcal{E}\mathcal{C}\rangle +\langle \mathcal{A}\mathcal{D}\rangle\langle \mathcal{B}\mathcal{E}\mathcal{E}\mathcal{C}\rangle +\langle \mathcal{A}\mathcal{E}\mathcal{D}\rangle\langle \mathcal{B}\mathcal{E}\mathcal{C}\rangle +\langle \mathcal{A}\mathcal{E}\rangle\langle \mathcal{B}\mathcal{E}\mathcal{D}\mathcal{C}\rangle  \\
   25) \quad & \langle \mathcal{A}\mathcal{C}\mathcal{E}\mathcal{B}\mathcal{D}\mathcal{E}\rangle +\langle \mathcal{A}\mathcal{C}\mathcal{E}\mathcal{B}\mathcal{E}\mathcal{D}\rangle +\langle \mathcal{A}\mathcal{D}\mathcal{C}\mathcal{E}\mathcal{B}\mathcal{E}\rangle +\langle \mathcal{A}\mathcal{D}\mathcal{E}\mathcal{C}\mathcal{E}\mathcal{B}\rangle +\langle \mathcal{A}\mathcal{E}\mathcal{C}\mathcal{E}\mathcal{B}\mathcal{D}\rangle +\langle \mathcal{A}\mathcal{E}\mathcal{D}\mathcal{C}\mathcal{E}\mathcal{B}\rangle  \notag \\ 
    = &  \langle \mathcal{A}\mathcal{C}\mathcal{E}\mathcal{B}\rangle\langle \mathcal{D}\mathcal{E}\rangle +\langle \mathcal{A}\mathcal{D}\mathcal{E}\rangle\langle \mathcal{B}\mathcal{C}\mathcal{E}\rangle +\langle \mathcal{A}\mathcal{D}\rangle\langle \mathcal{B}\mathcal{E}\mathcal{C}\mathcal{E}\rangle +\langle \mathcal{A}\mathcal{E}\mathcal{D}\rangle\langle \mathcal{B}\mathcal{C}\mathcal{E}\rangle +\langle \mathcal{A}\mathcal{E}\rangle\langle \mathcal{B}\mathcal{D}\mathcal{C}\mathcal{E}\rangle  \\
   26) \quad & \langle \mathcal{A}\mathcal{D}\mathcal{E}\mathcal{B}\mathcal{C}\mathcal{E}\rangle +\langle \mathcal{A}\mathcal{D}\mathcal{E}\mathcal{E}\mathcal{B}\mathcal{C}\rangle +\langle \mathcal{A}\mathcal{E}\mathcal{B}\mathcal{C}\mathcal{D}\mathcal{E}\rangle +\langle \mathcal{A}\mathcal{E}\mathcal{B}\mathcal{C}\mathcal{E}\mathcal{D}\rangle +\langle \mathcal{A}\mathcal{E}\mathcal{D}\mathcal{E}\mathcal{B}\mathcal{C}\rangle +\langle \mathcal{A}\mathcal{E}\mathcal{E}\mathcal{B}\mathcal{C}\mathcal{D}\rangle  \notag \\ 
    = &  \langle \mathcal{A}\mathcal{D}\mathcal{E}\rangle\langle \mathcal{B}\mathcal{C}\mathcal{E}\rangle +\langle \mathcal{A}\mathcal{D}\rangle\langle \mathcal{B}\mathcal{C}\mathcal{E}\mathcal{E}\rangle +\langle \mathcal{A}\mathcal{E}\mathcal{B}\mathcal{C}\rangle\langle \mathcal{D}\mathcal{E}\rangle +\langle \mathcal{A}\mathcal{E}\mathcal{D}\rangle\langle \mathcal{B}\mathcal{C}\mathcal{E}\rangle +\langle \mathcal{A}\mathcal{E}\rangle\langle \mathcal{B}\mathcal{C}\mathcal{D}\mathcal{E}\rangle  \\
   27) \quad & \langle \mathcal{A}\mathcal{D}\mathcal{E}\mathcal{C}\mathcal{B}\mathcal{E}\rangle +\langle \mathcal{A}\mathcal{D}\mathcal{E}\mathcal{E}\mathcal{C}\mathcal{B}\rangle +\langle \mathcal{A}\mathcal{E}\mathcal{C}\mathcal{B}\mathcal{D}\mathcal{E}\rangle +\langle \mathcal{A}\mathcal{E}\mathcal{C}\mathcal{B}\mathcal{E}\mathcal{D}\rangle +\langle \mathcal{A}\mathcal{E}\mathcal{D}\mathcal{E}\mathcal{C}\mathcal{B}\rangle +\langle \mathcal{A}\mathcal{E}\mathcal{E}\mathcal{C}\mathcal{B}\mathcal{D}\rangle  \notag \\ 
    = &  \langle \mathcal{A}\mathcal{D}\mathcal{E}\rangle\langle \mathcal{B}\mathcal{E}\mathcal{C}\rangle +\langle \mathcal{A}\mathcal{D}\rangle\langle \mathcal{B}\mathcal{E}\mathcal{E}\mathcal{C}\rangle +\langle \mathcal{A}\mathcal{E}\mathcal{C}\mathcal{B}\rangle\langle \mathcal{D}\mathcal{E}\rangle +\langle \mathcal{A}\mathcal{E}\mathcal{D}\rangle\langle \mathcal{B}\mathcal{E}\mathcal{C}\rangle +\langle \mathcal{A}\mathcal{E}\rangle\langle \mathcal{B}\mathcal{D}\mathcal{E}\mathcal{C}\rangle  \\
   28) \quad & \langle \mathcal{A}\mathcal{B}\mathcal{C}\mathcal{D}\mathcal{E}\mathcal{E}\rangle +\langle \mathcal{A}\mathcal{E}\mathcal{B}\mathcal{C}\mathcal{D}\mathcal{E}\rangle +\langle \mathcal{A}\mathcal{E}\mathcal{E}\mathcal{B}\mathcal{C}\mathcal{D}\rangle  \notag \\ 
    = &  \langle \mathcal{A}\mathcal{B}\mathcal{C}\mathcal{D}\rangle\langle \mathcal{E}\mathcal{E}\rangle +\langle \mathcal{A}\mathcal{E}\mathcal{E}\rangle\langle \mathcal{B}\mathcal{C}\mathcal{D}\rangle +\langle \mathcal{A}\mathcal{E}\rangle\langle \mathcal{B}\mathcal{C}\mathcal{D}\mathcal{E}\rangle  \\
   29) \quad & \langle \mathcal{A}\mathcal{B}\mathcal{D}\mathcal{C}\mathcal{E}\mathcal{E}\rangle +\langle \mathcal{A}\mathcal{E}\mathcal{B}\mathcal{D}\mathcal{C}\mathcal{E}\rangle +\langle \mathcal{A}\mathcal{E}\mathcal{E}\mathcal{B}\mathcal{D}\mathcal{C}\rangle  \notag \\ 
    = &  \langle \mathcal{A}\mathcal{B}\mathcal{D}\mathcal{C}\rangle\langle \mathcal{E}\mathcal{E}\rangle +\langle \mathcal{A}\mathcal{E}\mathcal{E}\rangle\langle \mathcal{B}\mathcal{D}\mathcal{C}\rangle +\langle \mathcal{A}\mathcal{E}\rangle\langle \mathcal{B}\mathcal{D}\mathcal{C}\mathcal{E}\rangle  \\
   30) \quad & \langle \mathcal{A}\mathcal{C}\mathcal{B}\mathcal{D}\mathcal{E}\mathcal{E}\rangle +\langle \mathcal{A}\mathcal{E}\mathcal{C}\mathcal{B}\mathcal{D}\mathcal{E}\rangle +\langle \mathcal{A}\mathcal{E}\mathcal{E}\mathcal{C}\mathcal{B}\mathcal{D}\rangle  \notag \\ 
    = &  \langle \mathcal{A}\mathcal{C}\mathcal{B}\mathcal{D}\rangle\langle \mathcal{E}\mathcal{E}\rangle +\langle \mathcal{A}\mathcal{E}\mathcal{E}\rangle\langle \mathcal{B}\mathcal{D}\mathcal{C}\rangle +\langle \mathcal{A}\mathcal{E}\rangle\langle \mathcal{B}\mathcal{D}\mathcal{E}\mathcal{C}\rangle  \\
   31) \quad & \langle \mathcal{A}\mathcal{C}\mathcal{D}\mathcal{B}\mathcal{E}\mathcal{E}\rangle +\langle \mathcal{A}\mathcal{E}\mathcal{C}\mathcal{D}\mathcal{B}\mathcal{E}\rangle +\langle \mathcal{A}\mathcal{E}\mathcal{E}\mathcal{C}\mathcal{D}\mathcal{B}\rangle  \notag \\ 
    = &  \langle \mathcal{A}\mathcal{C}\mathcal{D}\mathcal{B}\rangle\langle \mathcal{E}\mathcal{E}\rangle +\langle \mathcal{A}\mathcal{E}\mathcal{E}\rangle\langle \mathcal{B}\mathcal{C}\mathcal{D}\rangle +\langle \mathcal{A}\mathcal{E}\rangle\langle \mathcal{B}\mathcal{E}\mathcal{C}\mathcal{D}\rangle  \\
   32) \quad & \langle \mathcal{A}\mathcal{D}\mathcal{B}\mathcal{C}\mathcal{E}\mathcal{E}\rangle +\langle \mathcal{A}\mathcal{E}\mathcal{D}\mathcal{B}\mathcal{C}\mathcal{E}\rangle +\langle \mathcal{A}\mathcal{E}\mathcal{E}\mathcal{D}\mathcal{B}\mathcal{C}\rangle  \notag \\ 
    = &  \langle \mathcal{A}\mathcal{D}\mathcal{B}\mathcal{C}\rangle\langle \mathcal{E}\mathcal{E}\rangle +\langle \mathcal{A}\mathcal{E}\mathcal{E}\rangle\langle \mathcal{B}\mathcal{C}\mathcal{D}\rangle +\langle \mathcal{A}\mathcal{E}\rangle\langle \mathcal{B}\mathcal{C}\mathcal{E}\mathcal{D}\rangle  \\
   33) \quad & \langle \mathcal{A}\mathcal{D}\mathcal{C}\mathcal{B}\mathcal{E}\mathcal{E}\rangle +\langle \mathcal{A}\mathcal{E}\mathcal{D}\mathcal{C}\mathcal{B}\mathcal{E}\rangle +\langle \mathcal{A}\mathcal{E}\mathcal{E}\mathcal{D}\mathcal{C}\mathcal{B}\rangle  \notag \\ 
    = &  \langle \mathcal{A}\mathcal{D}\mathcal{C}\mathcal{B}\rangle\langle \mathcal{E}\mathcal{E}\rangle +\langle \mathcal{A}\mathcal{E}\mathcal{E}\rangle\langle \mathcal{B}\mathcal{D}\mathcal{C}\rangle +\langle \mathcal{A}\mathcal{E}\rangle\langle \mathcal{B}\mathcal{E}\mathcal{D}\mathcal{C}\rangle  \\
   34) \quad & \langle \mathcal{A}\mathcal{E}\mathcal{E}\mathcal{B}\mathcal{C}\mathcal{D}\rangle +\langle \mathcal{A}\mathcal{E}\mathcal{E}\mathcal{B}\mathcal{D}\mathcal{C}\rangle +\langle \mathcal{A}\mathcal{E}\mathcal{E}\mathcal{C}\mathcal{B}\mathcal{D}\rangle +\langle \mathcal{A}\mathcal{E}\mathcal{E}\mathcal{C}\mathcal{D}\mathcal{B}\rangle +\langle \mathcal{A}\mathcal{E}\mathcal{E}\mathcal{D}\mathcal{B}\mathcal{C}\rangle +\langle \mathcal{A}\mathcal{E}\mathcal{E}\mathcal{D}\mathcal{C}\mathcal{B}\rangle  \notag \\ 
    = &  \langle \mathcal{A}\mathcal{E}\mathcal{E}\mathcal{B}\rangle\langle \mathcal{C}\mathcal{D}\rangle +\langle \mathcal{A}\mathcal{E}\mathcal{E}\mathcal{C}\rangle\langle \mathcal{B}\mathcal{D}\rangle +\langle \mathcal{A}\mathcal{E}\mathcal{E}\mathcal{D}\rangle\langle \mathcal{B}\mathcal{C}\rangle +\langle \mathcal{A}\mathcal{E}\mathcal{E}\rangle\langle \mathcal{B}\mathcal{C}\mathcal{D}\rangle +\langle \mathcal{A}\mathcal{E}\mathcal{E}\rangle\langle \mathcal{B}\mathcal{D}\mathcal{C}\rangle
 \end{align}
 
 \begin{align}
   35) \quad & \langle \mathcal{A}\mathcal{E}\mathcal{B}\mathcal{C}\mathcal{D}\mathcal{E}\rangle +\langle \mathcal{A}\mathcal{E}\mathcal{B}\mathcal{D}\mathcal{C}\mathcal{E}\rangle +\langle \mathcal{A}\mathcal{E}\mathcal{C}\mathcal{B}\mathcal{D}\mathcal{E}\rangle +\langle \mathcal{A}\mathcal{E}\mathcal{C}\mathcal{D}\mathcal{B}\mathcal{E}\rangle +\langle \mathcal{A}\mathcal{E}\mathcal{D}\mathcal{B}\mathcal{C}\mathcal{E}\rangle +\langle \mathcal{A}\mathcal{E}\mathcal{D}\mathcal{C}\mathcal{B}\mathcal{E}\rangle  \notag \\ 
    = &  \langle \mathcal{A}\mathcal{E}\mathcal{B}\mathcal{E}\rangle\langle \mathcal{C}\mathcal{D}\rangle +\langle \mathcal{A}\mathcal{E}\mathcal{C}\mathcal{E}\rangle\langle \mathcal{B}\mathcal{D}\rangle +\langle \mathcal{A}\mathcal{E}\mathcal{D}\mathcal{E}\rangle\langle \mathcal{B}\mathcal{C}\rangle +\langle \mathcal{A}\mathcal{E}\mathcal{E}\rangle\langle \mathcal{B}\mathcal{C}\mathcal{D}\rangle +\langle \mathcal{A}\mathcal{E}\mathcal{E}\rangle\langle \mathcal{B}\mathcal{D}\mathcal{C}\rangle  \\
   36) \quad & \langle \mathcal{A}\mathcal{D}\mathcal{E}\mathcal{B}\mathcal{C}\mathcal{E}\rangle +\langle \mathcal{A}\mathcal{D}\mathcal{E}\mathcal{B}\mathcal{E}\mathcal{C}\rangle +\langle \mathcal{A}\mathcal{D}\mathcal{E}\mathcal{C}\mathcal{B}\mathcal{E}\rangle +\langle \mathcal{A}\mathcal{D}\mathcal{E}\mathcal{C}\mathcal{E}\mathcal{B}\rangle +\langle \mathcal{A}\mathcal{D}\mathcal{E}\mathcal{E}\mathcal{B}\mathcal{C}\rangle +\langle \mathcal{A}\mathcal{D}\mathcal{E}\mathcal{E}\mathcal{C}\mathcal{B}\rangle  \notag \\ 
    = &  \langle \mathcal{A}\mathcal{D}\mathcal{E}\mathcal{B}\rangle\langle \mathcal{C}\mathcal{E}\rangle +\langle \mathcal{A}\mathcal{D}\mathcal{E}\mathcal{C}\rangle\langle \mathcal{B}\mathcal{E}\rangle +\langle \mathcal{A}\mathcal{D}\mathcal{E}\mathcal{E}\rangle\langle \mathcal{B}\mathcal{C}\rangle +\langle \mathcal{A}\mathcal{D}\mathcal{E}\rangle\langle \mathcal{B}\mathcal{C}\mathcal{E}\rangle +\langle \mathcal{A}\mathcal{D}\mathcal{E}\rangle\langle \mathcal{B}\mathcal{E}\mathcal{C}\rangle  \\
   37) \quad & \langle \mathcal{A}\mathcal{E}\mathcal{D}\mathcal{B}\mathcal{C}\mathcal{E}\rangle +\langle \mathcal{A}\mathcal{E}\mathcal{D}\mathcal{B}\mathcal{E}\mathcal{C}\rangle +\langle \mathcal{A}\mathcal{E}\mathcal{D}\mathcal{C}\mathcal{B}\mathcal{E}\rangle +\langle \mathcal{A}\mathcal{E}\mathcal{D}\mathcal{C}\mathcal{E}\mathcal{B}\rangle +\langle \mathcal{A}\mathcal{E}\mathcal{D}\mathcal{E}\mathcal{B}\mathcal{C}\rangle +\langle \mathcal{A}\mathcal{E}\mathcal{D}\mathcal{E}\mathcal{C}\mathcal{B}\rangle  \notag \\ 
    = &  \langle \mathcal{A}\mathcal{E}\mathcal{D}\mathcal{B}\rangle\langle \mathcal{C}\mathcal{E}\rangle +\langle \mathcal{A}\mathcal{E}\mathcal{D}\mathcal{C}\rangle\langle \mathcal{B}\mathcal{E}\rangle +\langle \mathcal{A}\mathcal{E}\mathcal{D}\mathcal{E}\rangle\langle \mathcal{B}\mathcal{C}\rangle +\langle \mathcal{A}\mathcal{E}\mathcal{D}\rangle\langle \mathcal{B}\mathcal{C}\mathcal{E}\rangle +\langle \mathcal{A}\mathcal{E}\mathcal{D}\rangle\langle \mathcal{B}\mathcal{E}\mathcal{C}\rangle  \\
   38) \quad & \langle \mathcal{A}\mathcal{E}\mathcal{B}\mathcal{C}\mathcal{E}\mathcal{D}\rangle +\langle \mathcal{A}\mathcal{E}\mathcal{B}\mathcal{E}\mathcal{C}\mathcal{D}\rangle +\langle \mathcal{A}\mathcal{E}\mathcal{C}\mathcal{B}\mathcal{E}\mathcal{D}\rangle +\langle \mathcal{A}\mathcal{E}\mathcal{C}\mathcal{E}\mathcal{B}\mathcal{D}\rangle +\langle \mathcal{A}\mathcal{E}\mathcal{E}\mathcal{B}\mathcal{C}\mathcal{D}\rangle +\langle \mathcal{A}\mathcal{E}\mathcal{E}\mathcal{C}\mathcal{B}\mathcal{D}\rangle  \notag \\ 
    = &  \langle \mathcal{A}\mathcal{E}\mathcal{B}\mathcal{D}\rangle\langle \mathcal{C}\mathcal{E}\rangle +\langle \mathcal{A}\mathcal{E}\mathcal{C}\mathcal{D}\rangle\langle \mathcal{B}\mathcal{E}\rangle +\langle \mathcal{A}\mathcal{E}\mathcal{D}\rangle\langle \mathcal{B}\mathcal{C}\mathcal{E}\rangle +\langle \mathcal{A}\mathcal{E}\mathcal{D}\rangle\langle \mathcal{B}\mathcal{E}\mathcal{C}\rangle +\langle \mathcal{A}\mathcal{E}\mathcal{E}\mathcal{D}\rangle\langle \mathcal{B}\mathcal{C}\rangle  \\
   39) \quad & \langle \mathcal{A}\mathcal{D}\mathcal{B}\mathcal{C}\mathcal{E}\mathcal{E}\rangle +\langle \mathcal{A}\mathcal{D}\mathcal{B}\mathcal{E}\mathcal{C}\mathcal{E}\rangle +\langle \mathcal{A}\mathcal{D}\mathcal{C}\mathcal{B}\mathcal{E}\mathcal{E}\rangle +\langle \mathcal{A}\mathcal{D}\mathcal{C}\mathcal{E}\mathcal{B}\mathcal{E}\rangle +\langle \mathcal{A}\mathcal{D}\mathcal{E}\mathcal{B}\mathcal{C}\mathcal{E}\rangle +\langle \mathcal{A}\mathcal{D}\mathcal{E}\mathcal{C}\mathcal{B}\mathcal{E}\rangle  \notag \\ 
    = &  \langle \mathcal{A}\mathcal{D}\mathcal{B}\mathcal{E}\rangle\langle \mathcal{C}\mathcal{E}\rangle +\langle \mathcal{A}\mathcal{D}\mathcal{C}\mathcal{E}\rangle\langle \mathcal{B}\mathcal{E}\rangle +\langle \mathcal{A}\mathcal{D}\mathcal{E}\mathcal{E}\rangle\langle \mathcal{B}\mathcal{C}\rangle +\langle \mathcal{A}\mathcal{D}\mathcal{E}\rangle\langle \mathcal{B}\mathcal{C}\mathcal{E}\rangle +\langle \mathcal{A}\mathcal{D}\mathcal{E}\rangle\langle \mathcal{B}\mathcal{E}\mathcal{C}\rangle  \\
   40) \quad & \langle \mathcal{A}\mathcal{C}\mathcal{E}\mathcal{B}\mathcal{D}\mathcal{E}\rangle +\langle \mathcal{A}\mathcal{C}\mathcal{E}\mathcal{B}\mathcal{E}\mathcal{D}\rangle +\langle \mathcal{A}\mathcal{C}\mathcal{E}\mathcal{D}\mathcal{B}\mathcal{E}\rangle +\langle \mathcal{A}\mathcal{C}\mathcal{E}\mathcal{D}\mathcal{E}\mathcal{B}\rangle +\langle \mathcal{A}\mathcal{C}\mathcal{E}\mathcal{E}\mathcal{B}\mathcal{D}\rangle +\langle \mathcal{A}\mathcal{C}\mathcal{E}\mathcal{E}\mathcal{D}\mathcal{B}\rangle  \notag \\ 
    = &  \langle \mathcal{A}\mathcal{C}\mathcal{E}\mathcal{B}\rangle\langle \mathcal{D}\mathcal{E}\rangle +\langle \mathcal{A}\mathcal{C}\mathcal{E}\mathcal{D}\rangle\langle \mathcal{B}\mathcal{E}\rangle +\langle \mathcal{A}\mathcal{C}\mathcal{E}\mathcal{E}\rangle\langle \mathcal{B}\mathcal{D}\rangle +\langle \mathcal{A}\mathcal{C}\mathcal{E}\rangle\langle \mathcal{B}\mathcal{D}\mathcal{E}\rangle +\langle \mathcal{A}\mathcal{C}\mathcal{E}\rangle\langle \mathcal{B}\mathcal{E}\mathcal{D}\rangle  \\
    41) \quad & \langle \mathcal{A}\mathcal{E}\mathcal{C}\mathcal{B}\mathcal{D}\mathcal{E}\rangle +\langle \mathcal{A}\mathcal{E}\mathcal{C}\mathcal{B}\mathcal{E}\mathcal{D}\rangle +\langle \mathcal{A}\mathcal{E}\mathcal{C}\mathcal{D}\mathcal{B}\mathcal{E}\rangle +\langle \mathcal{A}\mathcal{E}\mathcal{C}\mathcal{D}\mathcal{E}\mathcal{B}\rangle +\langle \mathcal{A}\mathcal{E}\mathcal{C}\mathcal{E}\mathcal{B}\mathcal{D}\rangle +\langle \mathcal{A}\mathcal{E}\mathcal{C}\mathcal{E}\mathcal{D}\mathcal{B}\rangle  \notag \\ 
    = &  \langle \mathcal{A}\mathcal{E}\mathcal{C}\mathcal{B}\rangle\langle \mathcal{D}\mathcal{E}\rangle +\langle \mathcal{A}\mathcal{E}\mathcal{C}\mathcal{D}\rangle\langle \mathcal{B}\mathcal{E}\rangle +\langle \mathcal{A}\mathcal{E}\mathcal{C}\mathcal{E}\rangle\langle \mathcal{B}\mathcal{D}\rangle +\langle \mathcal{A}\mathcal{E}\mathcal{C}\rangle\langle \mathcal{B}\mathcal{D}\mathcal{E}\rangle +\langle \mathcal{A}\mathcal{E}\mathcal{C}\rangle\langle \mathcal{B}\mathcal{E}\mathcal{D}\rangle  \\
   42) \quad & \langle \mathcal{A}\mathcal{E}\mathcal{B}\mathcal{D}\mathcal{E}\mathcal{C}\rangle +\langle \mathcal{A}\mathcal{E}\mathcal{B}\mathcal{E}\mathcal{D}\mathcal{C}\rangle +\langle \mathcal{A}\mathcal{E}\mathcal{D}\mathcal{B}\mathcal{E}\mathcal{C}\rangle +\langle \mathcal{A}\mathcal{E}\mathcal{D}\mathcal{E}\mathcal{B}\mathcal{C}\rangle +\langle \mathcal{A}\mathcal{E}\mathcal{E}\mathcal{B}\mathcal{D}\mathcal{C}\rangle +\langle \mathcal{A}\mathcal{E}\mathcal{E}\mathcal{D}\mathcal{B}\mathcal{C}\rangle  \notag \\ 
    = &  \langle \mathcal{A}\mathcal{E}\mathcal{B}\mathcal{C}\rangle\langle \mathcal{D}\mathcal{E}\rangle +\langle \mathcal{A}\mathcal{E}\mathcal{C}\rangle\langle \mathcal{B}\mathcal{D}\mathcal{E}\rangle +\langle \mathcal{A}\mathcal{E}\mathcal{C}\rangle\langle \mathcal{B}\mathcal{E}\mathcal{D}\rangle +\langle \mathcal{A}\mathcal{E}\mathcal{D}\mathcal{C}\rangle\langle \mathcal{B}\mathcal{E}\rangle +\langle \mathcal{A}\mathcal{E}\mathcal{E}\mathcal{C}\rangle\langle \mathcal{B}\mathcal{D}\rangle  \\
   43) \quad & \langle \mathcal{A}\mathcal{C}\mathcal{B}\mathcal{D}\mathcal{E}\mathcal{E}\rangle +\langle \mathcal{A}\mathcal{C}\mathcal{B}\mathcal{E}\mathcal{D}\mathcal{E}\rangle +\langle \mathcal{A}\mathcal{C}\mathcal{D}\mathcal{B}\mathcal{E}\mathcal{E}\rangle +\langle \mathcal{A}\mathcal{C}\mathcal{D}\mathcal{E}\mathcal{B}\mathcal{E}\rangle +\langle \mathcal{A}\mathcal{C}\mathcal{E}\mathcal{B}\mathcal{D}\mathcal{E}\rangle +\langle \mathcal{A}\mathcal{C}\mathcal{E}\mathcal{D}\mathcal{B}\mathcal{E}\rangle  \notag \\ 
    = &  \langle \mathcal{A}\mathcal{C}\mathcal{B}\mathcal{E}\rangle\langle \mathcal{D}\mathcal{E}\rangle +\langle \mathcal{A}\mathcal{C}\mathcal{D}\mathcal{E}\rangle\langle \mathcal{B}\mathcal{E}\rangle +\langle \mathcal{A}\mathcal{C}\mathcal{E}\mathcal{E}\rangle\langle \mathcal{B}\mathcal{D}\rangle +\langle \mathcal{A}\mathcal{C}\mathcal{E}\rangle\langle \mathcal{B}\mathcal{D}\mathcal{E}\rangle +\langle \mathcal{A}\mathcal{C}\mathcal{E}\rangle\langle \mathcal{B}\mathcal{E}\mathcal{D}\rangle  \\
   44) \quad & \langle \mathcal{A}\mathcal{C}\mathcal{D}\mathcal{B}\mathcal{E}\mathcal{E}\rangle +\langle \mathcal{A}\mathcal{C}\mathcal{D}\mathcal{E}\mathcal{B}\mathcal{E}\rangle +\langle \mathcal{A}\mathcal{C}\mathcal{D}\mathcal{E}\mathcal{E}\mathcal{B}\rangle  \notag \\ 
    = &  \langle \mathcal{A}\mathcal{C}\mathcal{D}\mathcal{B}\rangle\langle \mathcal{E}\mathcal{E}\rangle +\langle \mathcal{A}\mathcal{C}\mathcal{D}\mathcal{E}\rangle\langle \mathcal{B}\mathcal{E}\rangle +\langle \mathcal{A}\mathcal{C}\mathcal{D}\rangle\langle \mathcal{B}\mathcal{E}\mathcal{E}\rangle  \\
   45) \quad & \langle \mathcal{A}\mathcal{D}\mathcal{B}\mathcal{E}\mathcal{E}\mathcal{C}\rangle +\langle \mathcal{A}\mathcal{D}\mathcal{E}\mathcal{B}\mathcal{E}\mathcal{C}\rangle +\langle \mathcal{A}\mathcal{D}\mathcal{E}\mathcal{E}\mathcal{B}\mathcal{C}\rangle  \notag \\ 
    = &  \langle \mathcal{A}\mathcal{D}\mathcal{B}\mathcal{C}\rangle\langle \mathcal{E}\mathcal{E}\rangle +\langle \mathcal{A}\mathcal{D}\mathcal{C}\rangle\langle \mathcal{B}\mathcal{E}\mathcal{E}\rangle +\langle \mathcal{A}\mathcal{D}\mathcal{E}\mathcal{C}\rangle\langle \mathcal{B}\mathcal{E}\rangle  \\
   46) \quad & \langle \mathcal{A}\mathcal{B}\mathcal{E}\mathcal{C}\mathcal{D}\mathcal{E}\rangle +\langle \mathcal{A}\mathcal{B}\mathcal{E}\mathcal{C}\mathcal{E}\mathcal{D}\rangle +\langle \mathcal{A}\mathcal{B}\mathcal{E}\mathcal{D}\mathcal{C}\mathcal{E}\rangle +\langle \mathcal{A}\mathcal{B}\mathcal{E}\mathcal{D}\mathcal{E}\mathcal{C}\rangle +\langle \mathcal{A}\mathcal{B}\mathcal{E}\mathcal{E}\mathcal{C}\mathcal{D}\rangle +\langle \mathcal{A}\mathcal{B}\mathcal{E}\mathcal{E}\mathcal{D}\mathcal{C}\rangle  \notag \\ 
    = &  \langle \mathcal{A}\mathcal{B}\mathcal{E}\mathcal{C}\rangle\langle \mathcal{D}\mathcal{E}\rangle +\langle \mathcal{A}\mathcal{B}\mathcal{E}\mathcal{D}\rangle\langle \mathcal{C}\mathcal{E}\rangle +\langle \mathcal{A}\mathcal{B}\mathcal{E}\mathcal{E}\rangle\langle \mathcal{C}\mathcal{D}\rangle +\langle \mathcal{A}\mathcal{B}\mathcal{E}\rangle\langle \mathcal{C}\mathcal{D}\mathcal{E}\rangle +\langle \mathcal{A}\mathcal{B}\mathcal{E}\rangle\langle \mathcal{C}\mathcal{E}\mathcal{D}\rangle  \\
   47) \quad & \langle \mathcal{A}\mathcal{E}\mathcal{B}\mathcal{C}\mathcal{D}\mathcal{E}\rangle +\langle \mathcal{A}\mathcal{E}\mathcal{B}\mathcal{C}\mathcal{E}\mathcal{D}\rangle +\langle \mathcal{A}\mathcal{E}\mathcal{B}\mathcal{D}\mathcal{C}\mathcal{E}\rangle +\langle \mathcal{A}\mathcal{E}\mathcal{B}\mathcal{D}\mathcal{E}\mathcal{C}\rangle +\langle \mathcal{A}\mathcal{E}\mathcal{B}\mathcal{E}\mathcal{C}\mathcal{D}\rangle +\langle \mathcal{A}\mathcal{E}\mathcal{B}\mathcal{E}\mathcal{D}\mathcal{C}\rangle  \notag \\ 
    = &  \langle \mathcal{A}\mathcal{E}\mathcal{B}\mathcal{C}\rangle\langle \mathcal{D}\mathcal{E}\rangle +\langle \mathcal{A}\mathcal{E}\mathcal{B}\mathcal{D}\rangle\langle \mathcal{C}\mathcal{E}\rangle +\langle \mathcal{A}\mathcal{E}\mathcal{B}\mathcal{E}\rangle\langle \mathcal{C}\mathcal{D}\rangle +\langle \mathcal{A}\mathcal{E}\mathcal{B}\rangle\langle \mathcal{C}\mathcal{D}\mathcal{E}\rangle +\langle \mathcal{A}\mathcal{E}\mathcal{B}\rangle\langle \mathcal{C}\mathcal{E}\mathcal{D}\rangle  \\
   48) \quad & \langle \mathcal{A}\mathcal{B}\mathcal{D}\mathcal{C}\mathcal{E}\mathcal{E}\rangle +\langle \mathcal{A}\mathcal{B}\mathcal{D}\mathcal{E}\mathcal{C}\mathcal{E}\rangle +\langle \mathcal{A}\mathcal{B}\mathcal{D}\mathcal{E}\mathcal{E}\mathcal{C}\rangle  \notag \\ 
    = &  \langle \mathcal{A}\mathcal{B}\mathcal{D}\mathcal{C}\rangle\langle \mathcal{E}\mathcal{E}\rangle +\langle \mathcal{A}\mathcal{B}\mathcal{D}\mathcal{E}\rangle\langle \mathcal{C}\mathcal{E}\rangle +\langle \mathcal{A}\mathcal{B}\mathcal{D}\rangle\langle \mathcal{C}\mathcal{E}\mathcal{E}\rangle  \\
   49) \quad & \langle \mathcal{A}\mathcal{B}\mathcal{C}\mathcal{D}\mathcal{E}\mathcal{E}\rangle +\langle \mathcal{A}\mathcal{B}\mathcal{C}\mathcal{E}\mathcal{D}\mathcal{E}\rangle +\langle \mathcal{A}\mathcal{B}\mathcal{C}\mathcal{E}\mathcal{E}\mathcal{D}\rangle  \notag \\ 
    = &  \langle \mathcal{A}\mathcal{B}\mathcal{C}\mathcal{D}\rangle\langle \mathcal{E}\mathcal{E}\rangle +\langle \mathcal{A}\mathcal{B}\mathcal{C}\mathcal{E}\rangle\langle \mathcal{D}\mathcal{E}\rangle +\langle \mathcal{A}\mathcal{B}\mathcal{C}\rangle\langle \mathcal{D}\mathcal{E}\mathcal{E}\rangle  \\
    50) \quad & \langle \mathcal{A}\mathcal{B}\mathcal{C}\mathcal{D}\mathcal{E}\mathcal{E}\rangle +\langle \mathcal{A}\mathcal{B}\mathcal{C}\mathcal{E}\mathcal{E}\mathcal{D}\rangle +\langle \mathcal{A}\mathcal{B}\mathcal{D}\mathcal{C}\mathcal{E}\mathcal{E}\rangle +\langle \mathcal{A}\mathcal{B}\mathcal{D}\mathcal{E}\mathcal{E}\mathcal{C}\rangle +\langle \mathcal{A}\mathcal{B}\mathcal{E}\mathcal{E}\mathcal{C}\mathcal{D}\rangle +\langle \mathcal{A}\mathcal{B}\mathcal{E}\mathcal{E}\mathcal{D}\mathcal{C}\rangle  \notag \\ 
    = &  \langle \mathcal{A}\mathcal{B}\mathcal{C}\mathcal{D}\rangle\langle \mathcal{E}\mathcal{E}\rangle +\langle \mathcal{A}\mathcal{B}\mathcal{C}\rangle\langle \mathcal{D}\mathcal{E}\mathcal{E}\rangle +\langle \mathcal{A}\mathcal{B}\mathcal{D}\mathcal{C}\rangle\langle \mathcal{E}\mathcal{E}\rangle +\langle \mathcal{A}\mathcal{B}\mathcal{D}\rangle\langle \mathcal{C}\mathcal{E}\mathcal{E}\rangle +\langle \mathcal{A}\mathcal{B}\mathcal{E}\mathcal{E}\rangle\langle \mathcal{C}\mathcal{D}\rangle +\langle \mathcal{A}\mathcal{B}\rangle\langle \mathcal{C}\mathcal{D}\mathcal{E}\mathcal{E}\rangle \notag \\
      & +\langle \mathcal{A}\mathcal{B}\rangle\langle \mathcal{C}\mathcal{D}\rangle\langle \mathcal{E}\mathcal{E}\rangle +\langle \mathcal{A}\mathcal{B}\rangle\langle \mathcal{C}\mathcal{E}\mathcal{E}\mathcal{D}\rangle  \\
  \end{align}
  
  \begin{align}
   51) \quad & \langle \mathcal{A}\mathcal{B}\mathcal{C}\mathcal{D}\mathcal{E}\mathcal{E}\rangle +\langle \mathcal{A}\mathcal{B}\mathcal{C}\mathcal{E}\mathcal{D}\mathcal{E}\rangle +\langle \mathcal{A}\mathcal{B}\mathcal{D}\mathcal{E}\mathcal{C}\mathcal{E}\rangle +\langle \mathcal{A}\mathcal{B}\mathcal{D}\mathcal{E}\mathcal{E}\mathcal{C}\rangle +\langle \mathcal{A}\mathcal{B}\mathcal{E}\mathcal{C}\mathcal{D}\mathcal{E}\rangle +\langle \mathcal{A}\mathcal{B}\mathcal{E}\mathcal{D}\mathcal{E}\mathcal{C}\rangle  \notag \\ 
    = &  \langle \mathcal{A}\mathcal{B}\mathcal{C}\mathcal{E}\rangle\langle \mathcal{D}\mathcal{E}\rangle +\langle \mathcal{A}\mathcal{B}\mathcal{C}\rangle\langle \mathcal{D}\mathcal{E}\mathcal{E}\rangle +\langle \mathcal{A}\mathcal{B}\mathcal{D}\mathcal{E}\rangle\langle \mathcal{C}\mathcal{E}\rangle +\langle \mathcal{A}\mathcal{B}\mathcal{E}\mathcal{C}\rangle\langle \mathcal{D}\mathcal{E}\rangle \notag \\
      & +\langle \mathcal{A}\mathcal{B}\mathcal{E}\rangle\langle \mathcal{C}\mathcal{D}\mathcal{E}\rangle +\langle \mathcal{A}\mathcal{B}\rangle\langle \mathcal{C}\mathcal{D}\mathcal{E}\mathcal{E}\rangle +\langle \mathcal{A}\mathcal{B}\rangle\langle \mathcal{C}\mathcal{E}\mathcal{D}\mathcal{E}\rangle +\langle \mathcal{A}\mathcal{B}\rangle\langle \mathcal{C}\mathcal{E}\rangle\langle \mathcal{D}\mathcal{E}\rangle  \\
   52) \quad & \langle \mathcal{A}\mathcal{B}\mathcal{C}\mathcal{E}\mathcal{D}\mathcal{E}\rangle +\langle \mathcal{A}\mathcal{B}\mathcal{C}\mathcal{E}\mathcal{E}\mathcal{D}\rangle +\langle \mathcal{A}\mathcal{B}\mathcal{E}\mathcal{C}\mathcal{E}\mathcal{D}\rangle +\langle \mathcal{A}\mathcal{B}\mathcal{E}\mathcal{D}\mathcal{C}\mathcal{E}\rangle +\langle \mathcal{A}\mathcal{B}\mathcal{E}\mathcal{D}\mathcal{E}\mathcal{C}\rangle +\langle \mathcal{A}\mathcal{B}\mathcal{E}\mathcal{E}\mathcal{D}\mathcal{C}\rangle  \notag \\ 
    = &  \langle \mathcal{A}\mathcal{B}\mathcal{C}\mathcal{E}\rangle\langle \mathcal{D}\mathcal{E}\rangle +\langle \mathcal{A}\mathcal{B}\mathcal{C}\rangle\langle \mathcal{D}\mathcal{E}\mathcal{E}\rangle +\langle \mathcal{A}\mathcal{B}\mathcal{E}\mathcal{C}\rangle\langle \mathcal{D}\mathcal{E}\rangle +\langle \mathcal{A}\mathcal{B}\mathcal{E}\mathcal{D}\rangle\langle \mathcal{C}\mathcal{E}\rangle +\langle \mathcal{A}\mathcal{B}\mathcal{E}\rangle\langle \mathcal{C}\mathcal{E}\mathcal{D}\rangle +\langle \mathcal{A}\mathcal{B}\rangle\langle \mathcal{C}\mathcal{E}\mathcal{D}\mathcal{E}\rangle \notag \\
      & +\langle \mathcal{A}\mathcal{B}\rangle\langle \mathcal{C}\mathcal{E}\mathcal{E}\mathcal{D}\rangle +\langle \mathcal{A}\mathcal{B}\rangle\langle \mathcal{C}\mathcal{E}\rangle\langle \mathcal{D}\mathcal{E}\rangle  \\
   53) \quad & \langle \mathcal{A}\mathcal{C}\mathcal{D}\mathcal{E}\mathcal{E}\mathcal{B}\rangle +\langle \mathcal{A}\mathcal{C}\mathcal{E}\mathcal{E}\mathcal{D}\mathcal{B}\rangle +\langle \mathcal{A}\mathcal{D}\mathcal{C}\mathcal{E}\mathcal{E}\mathcal{B}\rangle +\langle \mathcal{A}\mathcal{D}\mathcal{E}\mathcal{E}\mathcal{C}\mathcal{B}\rangle +\langle \mathcal{A}\mathcal{E}\mathcal{E}\mathcal{C}\mathcal{D}\mathcal{B}\rangle +\langle \mathcal{A}\mathcal{E}\mathcal{E}\mathcal{D}\mathcal{C}\mathcal{B}\rangle  \notag \\ 
    = &  \langle \mathcal{A}\mathcal{B}\rangle\langle \mathcal{C}\mathcal{D}\mathcal{E}\mathcal{E}\rangle +\langle \mathcal{A}\mathcal{B}\rangle\langle \mathcal{C}\mathcal{D}\rangle\langle \mathcal{E}\mathcal{E}\rangle +\langle \mathcal{A}\mathcal{B}\rangle\langle \mathcal{C}\mathcal{E}\mathcal{E}\mathcal{D}\rangle +\langle \mathcal{A}\mathcal{C}\mathcal{B}\rangle\langle \mathcal{D}\mathcal{E}\mathcal{E}\rangle +\langle \mathcal{A}\mathcal{C}\mathcal{D}\mathcal{B}\rangle\langle \mathcal{E}\mathcal{E}\rangle +\langle \mathcal{A}\mathcal{D}\mathcal{B}\rangle\langle \mathcal{C}\mathcal{E}\mathcal{E}\rangle \notag \\
      & +\langle \mathcal{A}\mathcal{D}\mathcal{C}\mathcal{B}\rangle\langle \mathcal{E}\mathcal{E}\rangle +\langle \mathcal{A}\mathcal{E}\mathcal{E}\mathcal{B}\rangle\langle \mathcal{C}\mathcal{D}\rangle  \\
   54) \quad & \langle \mathcal{A}\mathcal{C}\mathcal{D}\mathcal{E}\mathcal{E}\mathcal{B}\rangle +\langle \mathcal{A}\mathcal{C}\mathcal{E}\mathcal{D}\mathcal{E}\mathcal{B}\rangle +\langle \mathcal{A}\mathcal{D}\mathcal{E}\mathcal{C}\mathcal{E}\mathcal{B}\rangle +\langle \mathcal{A}\mathcal{D}\mathcal{E}\mathcal{E}\mathcal{C}\mathcal{B}\rangle +\langle \mathcal{A}\mathcal{E}\mathcal{C}\mathcal{D}\mathcal{E}\mathcal{B}\rangle +\langle \mathcal{A}\mathcal{E}\mathcal{D}\mathcal{E}\mathcal{C}\mathcal{B}\rangle  \notag \\ 
    = &  \langle \mathcal{A}\mathcal{B}\rangle\langle \mathcal{C}\mathcal{D}\mathcal{E}\mathcal{E}\rangle +\langle \mathcal{A}\mathcal{B}\rangle\langle \mathcal{C}\mathcal{E}\mathcal{D}\mathcal{E}\rangle +\langle \mathcal{A}\mathcal{B}\rangle\langle \mathcal{C}\mathcal{E}\rangle\langle \mathcal{D}\mathcal{E}\rangle +\langle \mathcal{A}\mathcal{C}\mathcal{B}\rangle\langle \mathcal{D}\mathcal{E}\mathcal{E}\rangle +\langle \mathcal{A}\mathcal{C}\mathcal{E}\mathcal{B}\rangle\langle \mathcal{D}\mathcal{E}\rangle +\langle \mathcal{A}\mathcal{D}\mathcal{E}\mathcal{B}\rangle\langle \mathcal{C}\mathcal{E}\rangle \notag \\
      & +\langle \mathcal{A}\mathcal{E}\mathcal{B}\rangle\langle \mathcal{C}\mathcal{D}\mathcal{E}\rangle +\langle \mathcal{A}\mathcal{E}\mathcal{C}\mathcal{B}\rangle\langle \mathcal{D}\mathcal{E}\rangle  \\
   55) \quad & \langle \mathcal{A}\mathcal{C}\mathcal{B}\mathcal{D}\mathcal{E}\mathcal{E}\rangle +\langle \mathcal{A}\mathcal{C}\mathcal{B}\mathcal{E}\mathcal{E}\mathcal{D}\rangle +\langle \mathcal{A}\mathcal{C}\mathcal{D}\mathcal{B}\mathcal{E}\mathcal{E}\rangle +\langle \mathcal{A}\mathcal{C}\mathcal{D}\mathcal{E}\mathcal{E}\mathcal{B}\rangle +\langle \mathcal{A}\mathcal{C}\mathcal{E}\mathcal{E}\mathcal{B}\mathcal{D}\rangle +\langle \mathcal{A}\mathcal{C}\mathcal{E}\mathcal{E}\mathcal{D}\mathcal{B}\rangle  \notag \\ 
    = &  \langle \mathcal{A}\mathcal{C}\mathcal{B}\mathcal{D}\rangle\langle \mathcal{E}\mathcal{E}\rangle +\langle \mathcal{A}\mathcal{C}\mathcal{B}\rangle\langle \mathcal{D}\mathcal{E}\mathcal{E}\rangle +\langle \mathcal{A}\mathcal{C}\mathcal{D}\mathcal{B}\rangle\langle \mathcal{E}\mathcal{E}\rangle +\langle \mathcal{A}\mathcal{C}\mathcal{D}\rangle\langle \mathcal{B}\mathcal{E}\mathcal{E}\rangle +\langle \mathcal{A}\mathcal{C}\mathcal{E}\mathcal{E}\rangle\langle \mathcal{B}\mathcal{D}\rangle +\langle \mathcal{A}\mathcal{C}\rangle\langle \mathcal{B}\mathcal{D}\mathcal{E}\mathcal{E}\rangle \notag \\
      & +\langle \mathcal{A}\mathcal{C}\rangle\langle \mathcal{B}\mathcal{D}\rangle\langle \mathcal{E}\mathcal{E}\rangle +\langle \mathcal{A}\mathcal{C}\rangle\langle \mathcal{B}\mathcal{E}\mathcal{E}\mathcal{D}\rangle  \\
   56) \quad & \langle \mathcal{A}\mathcal{C}\mathcal{B}\mathcal{D}\mathcal{E}\mathcal{E}\rangle +\langle \mathcal{A}\mathcal{C}\mathcal{B}\mathcal{E}\mathcal{D}\mathcal{E}\rangle +\langle \mathcal{A}\mathcal{C}\mathcal{D}\mathcal{E}\mathcal{B}\mathcal{E}\rangle +\langle \mathcal{A}\mathcal{C}\mathcal{D}\mathcal{E}\mathcal{E}\mathcal{B}\rangle +\langle \mathcal{A}\mathcal{C}\mathcal{E}\mathcal{B}\mathcal{D}\mathcal{E}\rangle +\langle \mathcal{A}\mathcal{C}\mathcal{E}\mathcal{D}\mathcal{E}\mathcal{B}\rangle  \notag \\ 
    = &  \langle \mathcal{A}\mathcal{C}\mathcal{B}\mathcal{E}\rangle\langle \mathcal{D}\mathcal{E}\rangle +\langle \mathcal{A}\mathcal{C}\mathcal{B}\rangle\langle \mathcal{D}\mathcal{E}\mathcal{E}\rangle +\langle \mathcal{A}\mathcal{C}\mathcal{D}\mathcal{E}\rangle\langle \mathcal{B}\mathcal{E}\rangle +\langle \mathcal{A}\mathcal{C}\mathcal{E}\mathcal{B}\rangle\langle \mathcal{D}\mathcal{E}\rangle +\langle \mathcal{A}\mathcal{C}\mathcal{E}\rangle\langle \mathcal{B}\mathcal{D}\mathcal{E}\rangle +\langle \mathcal{A}\mathcal{C}\rangle\langle \mathcal{B}\mathcal{D}\mathcal{E}\mathcal{E}\rangle \notag \\
      & +\langle \mathcal{A}\mathcal{C}\rangle\langle \mathcal{B}\mathcal{E}\mathcal{D}\mathcal{E}\rangle +\langle \mathcal{A}\mathcal{C}\rangle\langle \mathcal{B}\mathcal{E}\rangle\langle \mathcal{D}\mathcal{E}\rangle  \\
   57) \quad & \langle \mathcal{A}\mathcal{C}\mathcal{B}\mathcal{E}\mathcal{D}\mathcal{E}\rangle +\langle \mathcal{A}\mathcal{C}\mathcal{B}\mathcal{E}\mathcal{E}\mathcal{D}\rangle +\langle \mathcal{A}\mathcal{C}\mathcal{E}\mathcal{B}\mathcal{E}\mathcal{D}\rangle +\langle \mathcal{A}\mathcal{C}\mathcal{E}\mathcal{D}\mathcal{B}\mathcal{E}\rangle +\langle \mathcal{A}\mathcal{C}\mathcal{E}\mathcal{D}\mathcal{E}\mathcal{B}\rangle +\langle \mathcal{A}\mathcal{C}\mathcal{E}\mathcal{E}\mathcal{D}\mathcal{B}\rangle  \notag \\ 
    = &  \langle \mathcal{A}\mathcal{C}\mathcal{B}\mathcal{E}\rangle\langle \mathcal{D}\mathcal{E}\rangle +\langle \mathcal{A}\mathcal{C}\mathcal{B}\rangle\langle \mathcal{D}\mathcal{E}\mathcal{E}\rangle +\langle \mathcal{A}\mathcal{C}\mathcal{E}\mathcal{B}\rangle\langle \mathcal{D}\mathcal{E}\rangle +\langle \mathcal{A}\mathcal{C}\mathcal{E}\mathcal{D}\rangle\langle \mathcal{B}\mathcal{E}\rangle +\langle \mathcal{A}\mathcal{C}\mathcal{E}\rangle\langle \mathcal{B}\mathcal{E}\mathcal{D}\rangle +\langle \mathcal{A}\mathcal{C}\rangle\langle \mathcal{B}\mathcal{E}\mathcal{D}\mathcal{E}\rangle \notag \\
      & +\langle \mathcal{A}\mathcal{C}\rangle\langle \mathcal{B}\mathcal{E}\mathcal{E}\mathcal{D}\rangle +\langle \mathcal{A}\mathcal{C}\rangle\langle \mathcal{B}\mathcal{E}\rangle\langle \mathcal{D}\mathcal{E}\rangle  \\
   58) \quad & \langle \mathcal{A}\mathcal{B}\mathcal{D}\mathcal{E}\mathcal{E}\mathcal{C}\rangle +\langle \mathcal{A}\mathcal{B}\mathcal{E}\mathcal{E}\mathcal{D}\mathcal{C}\rangle +\langle \mathcal{A}\mathcal{D}\mathcal{B}\mathcal{E}\mathcal{E}\mathcal{C}\rangle +\langle \mathcal{A}\mathcal{D}\mathcal{E}\mathcal{E}\mathcal{B}\mathcal{C}\rangle +\langle \mathcal{A}\mathcal{E}\mathcal{E}\mathcal{B}\mathcal{D}\mathcal{C}\rangle +\langle \mathcal{A}\mathcal{E}\mathcal{E}\mathcal{D}\mathcal{B}\mathcal{C}\rangle  \notag \\ 
    = &  \langle \mathcal{A}\mathcal{B}\mathcal{C}\rangle\langle \mathcal{D}\mathcal{E}\mathcal{E}\rangle +\langle \mathcal{A}\mathcal{B}\mathcal{D}\mathcal{C}\rangle\langle \mathcal{E}\mathcal{E}\rangle +\langle \mathcal{A}\mathcal{C}\rangle\langle \mathcal{B}\mathcal{D}\mathcal{E}\mathcal{E}\rangle +\langle \mathcal{A}\mathcal{C}\rangle\langle \mathcal{B}\mathcal{D}\rangle\langle \mathcal{E}\mathcal{E}\rangle +\langle \mathcal{A}\mathcal{C}\rangle\langle \mathcal{B}\mathcal{E}\mathcal{E}\mathcal{D}\rangle +\langle \mathcal{A}\mathcal{D}\mathcal{B}\mathcal{C}\rangle\langle \mathcal{E}\mathcal{E}\rangle \notag \\
      & +\langle \mathcal{A}\mathcal{D}\mathcal{C}\rangle\langle \mathcal{B}\mathcal{E}\mathcal{E}\rangle +\langle \mathcal{A}\mathcal{E}\mathcal{E}\mathcal{C}\rangle\langle \mathcal{B}\mathcal{D}\rangle  \\
   59) \quad & \langle \mathcal{A}\mathcal{D}\mathcal{B}\mathcal{C}\mathcal{E}\mathcal{E}\rangle +\langle \mathcal{A}\mathcal{D}\mathcal{B}\mathcal{E}\mathcal{E}\mathcal{C}\rangle +\langle \mathcal{A}\mathcal{D}\mathcal{C}\mathcal{B}\mathcal{E}\mathcal{E}\rangle +\langle \mathcal{A}\mathcal{D}\mathcal{C}\mathcal{E}\mathcal{E}\mathcal{B}\rangle +\langle \mathcal{A}\mathcal{D}\mathcal{E}\mathcal{E}\mathcal{B}\mathcal{C}\rangle +\langle \mathcal{A}\mathcal{D}\mathcal{E}\mathcal{E}\mathcal{C}\mathcal{B}\rangle  \notag \\ 
    = &  \langle \mathcal{A}\mathcal{D}\mathcal{B}\mathcal{C}\rangle\langle \mathcal{E}\mathcal{E}\rangle +\langle \mathcal{A}\mathcal{D}\mathcal{B}\rangle\langle \mathcal{C}\mathcal{E}\mathcal{E}\rangle +\langle \mathcal{A}\mathcal{D}\mathcal{C}\mathcal{B}\rangle\langle \mathcal{E}\mathcal{E}\rangle +\langle \mathcal{A}\mathcal{D}\mathcal{C}\rangle\langle \mathcal{B}\mathcal{E}\mathcal{E}\rangle +\langle \mathcal{A}\mathcal{D}\mathcal{E}\mathcal{E}\rangle\langle \mathcal{B}\mathcal{C}\rangle +\langle \mathcal{A}\mathcal{D}\rangle\langle \mathcal{B}\mathcal{C}\mathcal{E}\mathcal{E}\rangle \notag \\ 
      & +\langle \mathcal{A}\mathcal{D}\rangle\langle \mathcal{B}\mathcal{C}\rangle\langle \mathcal{E}\mathcal{E}\rangle +\langle \mathcal{A}\mathcal{D}\rangle\langle \mathcal{B}\mathcal{E}\mathcal{E}\mathcal{C}\rangle  \\
   60) \quad & \langle \mathcal{A}\mathcal{D}\mathcal{B}\mathcal{C}\mathcal{E}\mathcal{E}\rangle +\langle \mathcal{A}\mathcal{D}\mathcal{B}\mathcal{E}\mathcal{C}\mathcal{E}\rangle +\langle \mathcal{A}\mathcal{D}\mathcal{C}\mathcal{E}\mathcal{B}\mathcal{E}\rangle +\langle \mathcal{A}\mathcal{D}\mathcal{C}\mathcal{E}\mathcal{E}\mathcal{B}\rangle +\langle \mathcal{A}\mathcal{D}\mathcal{E}\mathcal{B}\mathcal{C}\mathcal{E}\rangle +\langle \mathcal{A}\mathcal{D}\mathcal{E}\mathcal{C}\mathcal{E}\mathcal{B}\rangle  \notag \\ 
    = &  \langle \mathcal{A}\mathcal{D}\mathcal{B}\mathcal{E}\rangle\langle \mathcal{C}\mathcal{E}\rangle +\langle \mathcal{A}\mathcal{D}\mathcal{B}\rangle\langle \mathcal{C}\mathcal{E}\mathcal{E}\rangle +\langle \mathcal{A}\mathcal{D}\mathcal{C}\mathcal{E}\rangle\langle \mathcal{B}\mathcal{E}\rangle +\langle \mathcal{A}\mathcal{D}\mathcal{E}\mathcal{B}\rangle\langle \mathcal{C}\mathcal{E}\rangle +\langle \mathcal{A}\mathcal{D}\mathcal{E}\rangle\langle \mathcal{B}\mathcal{C}\mathcal{E}\rangle +\langle \mathcal{A}\mathcal{D}\rangle\langle \mathcal{B}\mathcal{C}\mathcal{E}\mathcal{E}\rangle \notag \\
      & +\langle \mathcal{A}\mathcal{D}\rangle\langle \mathcal{B}\mathcal{E}\mathcal{C}\mathcal{E}\rangle +\langle \mathcal{A}\mathcal{D}\rangle\langle \mathcal{B}\mathcal{E}\rangle\langle \mathcal{C}\mathcal{E}\rangle  \\
   61) \quad & \langle \mathcal{A}\mathcal{D}\mathcal{B}\mathcal{E}\mathcal{C}\mathcal{E}\rangle +\langle \mathcal{A}\mathcal{D}\mathcal{B}\mathcal{E}\mathcal{E}\mathcal{C}\rangle +\langle \mathcal{A}\mathcal{D}\mathcal{E}\mathcal{B}\mathcal{E}\mathcal{C}\rangle +\langle \mathcal{A}\mathcal{D}\mathcal{E}\mathcal{C}\mathcal{B}\mathcal{E}\rangle +\langle \mathcal{A}\mathcal{D}\mathcal{E}\mathcal{C}\mathcal{E}\mathcal{B}\rangle +\langle \mathcal{A}\mathcal{D}\mathcal{E}\mathcal{E}\mathcal{C}\mathcal{B}\rangle  \notag \\ 
    = &  \langle \mathcal{A}\mathcal{D}\mathcal{B}\mathcal{E}\rangle\langle \mathcal{C}\mathcal{E}\rangle +\langle \mathcal{A}\mathcal{D}\mathcal{B}\rangle\langle \mathcal{C}\mathcal{E}\mathcal{E}\rangle +\langle \mathcal{A}\mathcal{D}\mathcal{E}\mathcal{B}\rangle\langle \mathcal{C}\mathcal{E}\rangle +\langle \mathcal{A}\mathcal{D}\mathcal{E}\mathcal{C}\rangle\langle \mathcal{B}\mathcal{E}\rangle +\langle \mathcal{A}\mathcal{D}\mathcal{E}\rangle\langle \mathcal{B}\mathcal{E}\mathcal{C}\rangle +\langle \mathcal{A}\mathcal{D}\rangle\langle \mathcal{B}\mathcal{E}\mathcal{C}\mathcal{E}\rangle \notag \\
      & +\langle \mathcal{A}\mathcal{D}\rangle\langle \mathcal{B}\mathcal{E}\mathcal{E}\mathcal{C}\rangle +\langle \mathcal{A}\mathcal{D}\rangle\langle \mathcal{B}\mathcal{E}\rangle\langle \mathcal{C}\mathcal{E}\rangle 
  \end{align}
  
  \begin{align}
   62) \quad & \langle \mathcal{A}\mathcal{E}\mathcal{B}\mathcal{C}\mathcal{D}\mathcal{E}\rangle +\langle \mathcal{A}\mathcal{E}\mathcal{B}\mathcal{D}\mathcal{E}\mathcal{C}\rangle +\langle \mathcal{A}\mathcal{E}\mathcal{C}\mathcal{B}\mathcal{D}\mathcal{E}\rangle +\langle \mathcal{A}\mathcal{E}\mathcal{C}\mathcal{D}\mathcal{E}\mathcal{B}\rangle +\langle \mathcal{A}\mathcal{E}\mathcal{D}\mathcal{E}\mathcal{B}\mathcal{C}\rangle +\langle \mathcal{A}\mathcal{E}\mathcal{D}\mathcal{E}\mathcal{C}\mathcal{B}\rangle  \notag \\ 
    = &  \langle \mathcal{A}\mathcal{E}\mathcal{B}\mathcal{C}\rangle\langle \mathcal{D}\mathcal{E}\rangle +\langle \mathcal{A}\mathcal{E}\mathcal{B}\rangle\langle \mathcal{C}\mathcal{D}\mathcal{E}\rangle +\langle \mathcal{A}\mathcal{E}\mathcal{C}\mathcal{B}\rangle\langle \mathcal{D}\mathcal{E}\rangle +\langle \mathcal{A}\mathcal{E}\mathcal{C}\rangle\langle \mathcal{B}\mathcal{D}\mathcal{E}\rangle +\langle \mathcal{A}\mathcal{E}\mathcal{D}\mathcal{E}\rangle\langle \mathcal{B}\mathcal{C}\rangle +\langle \mathcal{A}\mathcal{E}\rangle\langle \mathcal{B}\mathcal{C}\mathcal{D}\mathcal{E}\rangle \notag \\
      & +\langle \mathcal{A}\mathcal{E}\rangle\langle \mathcal{B}\mathcal{C}\rangle\langle \mathcal{D}\mathcal{E}\rangle +\langle \mathcal{A}\mathcal{E}\rangle\langle \mathcal{B}\mathcal{D}\mathcal{E}\mathcal{C}\rangle  \\
   63) \quad & \langle \mathcal{A}\mathcal{E}\mathcal{B}\mathcal{C}\mathcal{E}\mathcal{D}\rangle +\langle \mathcal{A}\mathcal{E}\mathcal{B}\mathcal{D}\mathcal{C}\mathcal{E}\rangle +\langle \mathcal{A}\mathcal{E}\mathcal{C}\mathcal{E}\mathcal{B}\mathcal{D}\rangle +\langle \mathcal{A}\mathcal{E}\mathcal{C}\mathcal{E}\mathcal{D}\mathcal{B}\rangle +\langle \mathcal{A}\mathcal{E}\mathcal{D}\mathcal{B}\mathcal{C}\mathcal{E}\rangle +\langle \mathcal{A}\mathcal{E}\mathcal{D}\mathcal{C}\mathcal{E}\mathcal{B}\rangle  \notag \\ 
    = &  \langle \mathcal{A}\mathcal{E}\mathcal{B}\mathcal{D}\rangle\langle \mathcal{C}\mathcal{E}\rangle +\langle \mathcal{A}\mathcal{E}\mathcal{B}\rangle\langle \mathcal{C}\mathcal{E}\mathcal{D}\rangle +\langle \mathcal{A}\mathcal{E}\mathcal{C}\mathcal{E}\rangle\langle \mathcal{B}\mathcal{D}\rangle +\langle \mathcal{A}\mathcal{E}\mathcal{D}\mathcal{B}\rangle\langle \mathcal{C}\mathcal{E}\rangle +\langle \mathcal{A}\mathcal{E}\mathcal{D}\rangle\langle \mathcal{B}\mathcal{C}\mathcal{E}\rangle +\langle \mathcal{A}\mathcal{E}\rangle\langle \mathcal{B}\mathcal{C}\mathcal{E}\mathcal{D}\rangle \notag \\
      & +\langle \mathcal{A}\mathcal{E}\rangle\langle \mathcal{B}\mathcal{D}\mathcal{C}\mathcal{E}\rangle +\langle \mathcal{A}\mathcal{E}\rangle\langle \mathcal{B}\mathcal{D}\rangle\langle \mathcal{C}\mathcal{E}\rangle  \\
   64) \quad & \langle \mathcal{A}\mathcal{E}\mathcal{B}\mathcal{C}\mathcal{D}\mathcal{E}\rangle +\langle \mathcal{A}\mathcal{E}\mathcal{B}\mathcal{E}\mathcal{C}\mathcal{D}\rangle +\langle \mathcal{A}\mathcal{E}\mathcal{C}\mathcal{D}\mathcal{B}\mathcal{E}\rangle +\langle \mathcal{A}\mathcal{E}\mathcal{C}\mathcal{D}\mathcal{E}\mathcal{B}\rangle +\langle \mathcal{A}\mathcal{E}\mathcal{E}\mathcal{B}\mathcal{C}\mathcal{D}\rangle +\langle \mathcal{A}\mathcal{E}\mathcal{E}\mathcal{C}\mathcal{D}\mathcal{B}\rangle  \notag \\ 
    = &  \langle \mathcal{A}\mathcal{E}\mathcal{B}\mathcal{E}\rangle\langle \mathcal{C}\mathcal{D}\rangle +\langle \mathcal{A}\mathcal{E}\mathcal{B}\rangle\langle \mathcal{C}\mathcal{D}\mathcal{E}\rangle +\langle \mathcal{A}\mathcal{E}\mathcal{C}\mathcal{D}\rangle\langle \mathcal{B}\mathcal{E}\rangle +\langle \mathcal{A}\mathcal{E}\mathcal{E}\mathcal{B}\rangle\langle \mathcal{C}\mathcal{D}\rangle +\langle \mathcal{A}\mathcal{E}\mathcal{E}\rangle\langle \mathcal{B}\mathcal{C}\mathcal{D}\rangle +\langle \mathcal{A}\mathcal{E}\rangle\langle \mathcal{B}\mathcal{C}\mathcal{D}\mathcal{E}\rangle \notag \\
      & +\langle \mathcal{A}\mathcal{E}\rangle\langle \mathcal{B}\mathcal{E}\mathcal{C}\mathcal{D}\rangle +\langle \mathcal{A}\mathcal{E}\rangle\langle \mathcal{B}\mathcal{E}\rangle\langle \mathcal{C}\mathcal{D}\rangle  
\end{align}

\paragraph{$\mathcal{A}\mathcal{B}\mathcal{C}\mathcal{D}^3$}

\begin{align}
    1) \quad & \langle \mathcal{A}\mathcal{B}\mathcal{C}\mathcal{D}\mathcal{D}\mathcal{D}\rangle +\langle \mathcal{A}\mathcal{B}\mathcal{D}\mathcal{D}\mathcal{D}\mathcal{C}\rangle +\langle \mathcal{A}\mathcal{C}\mathcal{B}\mathcal{D}\mathcal{D}\mathcal{D}\rangle +\langle \mathcal{A}\mathcal{C}\mathcal{D}\mathcal{D}\mathcal{D}\mathcal{B}\rangle +\langle \mathcal{A}\mathcal{D}\mathcal{D}\mathcal{D}\mathcal{B}\mathcal{C}\rangle +\langle \mathcal{A}\mathcal{D}\mathcal{D}\mathcal{D}\mathcal{C}\mathcal{B}\rangle  \notag \\ 
    = &  \langle \mathcal{A}\mathcal{B}\mathcal{C}\rangle\langle \mathcal{D}\mathcal{D}\mathcal{D}\rangle +\langle \mathcal{A}\mathcal{B}\rangle\langle \mathcal{C}\mathcal{D}\mathcal{D}\mathcal{D}\rangle +\langle \mathcal{A}\mathcal{C}\mathcal{B}\rangle\langle \mathcal{D}\mathcal{D}\mathcal{D}\rangle +\langle \mathcal{A}\mathcal{C}\rangle\langle \mathcal{B}\mathcal{D}\mathcal{D}\mathcal{D}\rangle +\langle \mathcal{A}\mathcal{D}\mathcal{D}\mathcal{D}\rangle\langle \mathcal{B}\mathcal{C}\rangle  \\
   2) \quad & \langle \mathcal{A}\mathcal{B}\mathcal{C}\mathcal{D}\mathcal{D}\mathcal{D}\rangle +\langle \mathcal{A}\mathcal{B}\mathcal{D}\mathcal{C}\mathcal{D}\mathcal{D}\rangle +\langle \mathcal{A}\mathcal{C}\mathcal{D}\mathcal{D}\mathcal{B}\mathcal{D}\rangle +\langle \mathcal{A}\mathcal{C}\mathcal{D}\mathcal{D}\mathcal{D}\mathcal{B}\rangle +\langle \mathcal{A}\mathcal{D}\mathcal{B}\mathcal{C}\mathcal{D}\mathcal{D}\rangle +\langle \mathcal{A}\mathcal{D}\mathcal{C}\mathcal{D}\mathcal{D}\mathcal{B}\rangle  \notag \\ 
    = &  \langle \mathcal{A}\mathcal{B}\mathcal{D}\rangle\langle \mathcal{C}\mathcal{D}\mathcal{D}\rangle +\langle \mathcal{A}\mathcal{B}\rangle\langle \mathcal{C}\mathcal{D}\mathcal{D}\mathcal{D}\rangle +\langle \mathcal{A}\mathcal{C}\mathcal{D}\mathcal{D}\rangle\langle \mathcal{B}\mathcal{D}\rangle +\langle \mathcal{A}\mathcal{D}\mathcal{B}\rangle\langle \mathcal{C}\mathcal{D}\mathcal{D}\rangle +\langle \mathcal{A}\mathcal{D}\rangle\langle \mathcal{B}\mathcal{C}\mathcal{D}\mathcal{D}\rangle  \\
   3) \quad & \langle \mathcal{A}\mathcal{B}\mathcal{D}\mathcal{C}\mathcal{D}\mathcal{D}\rangle +\langle \mathcal{A}\mathcal{B}\mathcal{D}\mathcal{D}\mathcal{C}\mathcal{D}\rangle +\langle \mathcal{A}\mathcal{D}\mathcal{B}\mathcal{D}\mathcal{C}\mathcal{D}\rangle +\langle \mathcal{A}\mathcal{D}\mathcal{C}\mathcal{D}\mathcal{B}\mathcal{D}\rangle +\langle \mathcal{A}\mathcal{D}\mathcal{C}\mathcal{D}\mathcal{D}\mathcal{B}\rangle +\langle \mathcal{A}\mathcal{D}\mathcal{D}\mathcal{C}\mathcal{D}\mathcal{B}\rangle  \notag \\ 
    = &  \langle \mathcal{A}\mathcal{B}\mathcal{D}\rangle\langle \mathcal{C}\mathcal{D}\mathcal{D}\rangle +\langle \mathcal{A}\mathcal{B}\rangle\langle \mathcal{C}\mathcal{D}\mathcal{D}\mathcal{D}\rangle +\langle \mathcal{A}\mathcal{D}\mathcal{B}\rangle\langle \mathcal{C}\mathcal{D}\mathcal{D}\rangle +\langle \mathcal{A}\mathcal{D}\mathcal{C}\mathcal{D}\rangle\langle \mathcal{B}\mathcal{D}\rangle +\langle \mathcal{A}\mathcal{D}\rangle\langle \mathcal{B}\mathcal{D}\mathcal{C}\mathcal{D}\rangle  \\
   4) \quad & \langle \mathcal{A}\mathcal{B}\mathcal{D}\mathcal{D}\mathcal{C}\mathcal{D}\rangle +\langle \mathcal{A}\mathcal{B}\mathcal{D}\mathcal{D}\mathcal{D}\mathcal{C}\rangle +\langle \mathcal{A}\mathcal{D}\mathcal{B}\mathcal{D}\mathcal{D}\mathcal{C}\rangle +\langle \mathcal{A}\mathcal{D}\mathcal{D}\mathcal{C}\mathcal{B}\mathcal{D}\rangle +\langle \mathcal{A}\mathcal{D}\mathcal{D}\mathcal{C}\mathcal{D}\mathcal{B}\rangle +\langle \mathcal{A}\mathcal{D}\mathcal{D}\mathcal{D}\mathcal{C}\mathcal{B}\rangle  \notag \\ 
    = &  \langle \mathcal{A}\mathcal{B}\mathcal{D}\rangle\langle \mathcal{C}\mathcal{D}\mathcal{D}\rangle +\langle \mathcal{A}\mathcal{B}\rangle\langle \mathcal{C}\mathcal{D}\mathcal{D}\mathcal{D}\rangle +\langle \mathcal{A}\mathcal{D}\mathcal{B}\rangle\langle \mathcal{C}\mathcal{D}\mathcal{D}\rangle +\langle \mathcal{A}\mathcal{D}\mathcal{D}\mathcal{C}\rangle\langle \mathcal{B}\mathcal{D}\rangle +\langle \mathcal{A}\mathcal{D}\rangle\langle \mathcal{B}\mathcal{D}\mathcal{D}\mathcal{C}\rangle  \\
   5) \quad & \langle \mathcal{A}\mathcal{B}\mathcal{D}\mathcal{D}\mathcal{C}\mathcal{D}\rangle +\langle \mathcal{A}\mathcal{B}\mathcal{D}\mathcal{D}\mathcal{D}\mathcal{C}\rangle +\langle \mathcal{A}\mathcal{C}\mathcal{B}\mathcal{D}\mathcal{D}\mathcal{D}\rangle +\langle \mathcal{A}\mathcal{C}\mathcal{D}\mathcal{B}\mathcal{D}\mathcal{D}\rangle +\langle \mathcal{A}\mathcal{D}\mathcal{B}\mathcal{D}\mathcal{D}\mathcal{C}\rangle +\langle \mathcal{A}\mathcal{D}\mathcal{C}\mathcal{B}\mathcal{D}\mathcal{D}\rangle  \notag \\ 
    = &  \langle \mathcal{A}\mathcal{B}\mathcal{D}\mathcal{D}\rangle\langle \mathcal{C}\mathcal{D}\rangle +\langle \mathcal{A}\mathcal{C}\mathcal{D}\rangle\langle \mathcal{B}\mathcal{D}\mathcal{D}\rangle +\langle \mathcal{A}\mathcal{C}\rangle\langle \mathcal{B}\mathcal{D}\mathcal{D}\mathcal{D}\rangle +\langle \mathcal{A}\mathcal{D}\mathcal{C}\rangle\langle \mathcal{B}\mathcal{D}\mathcal{D}\rangle +\langle \mathcal{A}\mathcal{D}\rangle\langle \mathcal{B}\mathcal{D}\mathcal{D}\mathcal{C}\rangle  \\
   6) \quad & \langle \mathcal{A}\mathcal{C}\mathcal{D}\mathcal{B}\mathcal{D}\mathcal{D}\rangle +\langle \mathcal{A}\mathcal{C}\mathcal{D}\mathcal{D}\mathcal{B}\mathcal{D}\rangle +\langle \mathcal{A}\mathcal{D}\mathcal{B}\mathcal{D}\mathcal{C}\mathcal{D}\rangle +\langle \mathcal{A}\mathcal{D}\mathcal{B}\mathcal{D}\mathcal{D}\mathcal{C}\rangle +\langle \mathcal{A}\mathcal{D}\mathcal{C}\mathcal{D}\mathcal{B}\mathcal{D}\rangle +\langle \mathcal{A}\mathcal{D}\mathcal{D}\mathcal{B}\mathcal{D}\mathcal{C}\rangle  \notag \\ 
    = &  \langle \mathcal{A}\mathcal{C}\mathcal{D}\rangle\langle \mathcal{B}\mathcal{D}\mathcal{D}\rangle +\langle \mathcal{A}\mathcal{C}\rangle\langle \mathcal{B}\mathcal{D}\mathcal{D}\mathcal{D}\rangle +\langle \mathcal{A}\mathcal{D}\mathcal{B}\mathcal{D}\rangle\langle \mathcal{C}\mathcal{D}\rangle +\langle \mathcal{A}\mathcal{D}\mathcal{C}\rangle\langle \mathcal{B}\mathcal{D}\mathcal{D}\rangle +\langle \mathcal{A}\mathcal{D}\rangle\langle \mathcal{B}\mathcal{D}\mathcal{C}\mathcal{D}\rangle  \\
   7) \quad & \langle \mathcal{A}\mathcal{C}\mathcal{D}\mathcal{D}\mathcal{B}\mathcal{D}\rangle +\langle \mathcal{A}\mathcal{C}\mathcal{D}\mathcal{D}\mathcal{D}\mathcal{B}\rangle +\langle \mathcal{A}\mathcal{D}\mathcal{C}\mathcal{D}\mathcal{D}\mathcal{B}\rangle +\langle \mathcal{A}\mathcal{D}\mathcal{D}\mathcal{B}\mathcal{C}\mathcal{D}\rangle +\langle \mathcal{A}\mathcal{D}\mathcal{D}\mathcal{B}\mathcal{D}\mathcal{C}\rangle +\langle \mathcal{A}\mathcal{D}\mathcal{D}\mathcal{D}\mathcal{B}\mathcal{C}\rangle  \notag \\ 
    = &  \langle \mathcal{A}\mathcal{C}\mathcal{D}\rangle\langle \mathcal{B}\mathcal{D}\mathcal{D}\rangle +\langle \mathcal{A}\mathcal{C}\rangle\langle \mathcal{B}\mathcal{D}\mathcal{D}\mathcal{D}\rangle +\langle \mathcal{A}\mathcal{D}\mathcal{C}\rangle\langle \mathcal{B}\mathcal{D}\mathcal{D}\rangle +\langle \mathcal{A}\mathcal{D}\mathcal{D}\mathcal{B}\rangle\langle \mathcal{C}\mathcal{D}\rangle +\langle \mathcal{A}\mathcal{D}\rangle\langle \mathcal{B}\mathcal{C}\mathcal{D}\mathcal{D}\rangle  \\
   8) \quad & \langle \mathcal{A}\mathcal{B}\mathcal{C}\mathcal{D}\mathcal{D}\mathcal{D}\rangle +\langle \mathcal{A}\mathcal{D}\mathcal{B}\mathcal{C}\mathcal{D}\mathcal{D}\rangle +\langle \mathcal{A}\mathcal{D}\mathcal{D}\mathcal{B}\mathcal{C}\mathcal{D}\rangle  \notag \\ 
    = &  \langle \mathcal{A}\mathcal{B}\mathcal{C}\mathcal{D}\rangle\langle \mathcal{D}\mathcal{D}\rangle +\langle \mathcal{A}\mathcal{D}\mathcal{D}\rangle\langle \mathcal{B}\mathcal{C}\mathcal{D}\rangle +\langle \mathcal{A}\mathcal{D}\rangle\langle \mathcal{B}\mathcal{C}\mathcal{D}\mathcal{D}\rangle  \\
   9) \quad & \langle \mathcal{A}\mathcal{B}\mathcal{D}\mathcal{C}\mathcal{D}\mathcal{D}\rangle +\langle \mathcal{A}\mathcal{D}\mathcal{B}\mathcal{D}\mathcal{C}\mathcal{D}\rangle +\langle \mathcal{A}\mathcal{D}\mathcal{D}\mathcal{B}\mathcal{D}\mathcal{C}\rangle  \notag \\ 
    = &  \langle \mathcal{A}\mathcal{B}\mathcal{D}\mathcal{C}\rangle\langle \mathcal{D}\mathcal{D}\rangle +\langle \mathcal{A}\mathcal{D}\mathcal{D}\rangle\langle \mathcal{B}\mathcal{D}\mathcal{C}\rangle +\langle \mathcal{A}\mathcal{D}\rangle\langle \mathcal{B}\mathcal{D}\mathcal{C}\mathcal{D}\rangle  \\
   10) \quad & \langle \mathcal{A}\mathcal{C}\mathcal{B}\mathcal{D}\mathcal{D}\mathcal{D}\rangle +\langle \mathcal{A}\mathcal{D}\mathcal{C}\mathcal{B}\mathcal{D}\mathcal{D}\rangle +\langle \mathcal{A}\mathcal{D}\mathcal{D}\mathcal{C}\mathcal{B}\mathcal{D}\rangle  \notag \\ 
    = &  \langle \mathcal{A}\mathcal{C}\mathcal{B}\mathcal{D}\rangle\langle \mathcal{D}\mathcal{D}\rangle +\langle \mathcal{A}\mathcal{D}\mathcal{D}\rangle\langle \mathcal{B}\mathcal{D}\mathcal{C}\rangle +\langle \mathcal{A}\mathcal{D}\rangle\langle \mathcal{B}\mathcal{D}\mathcal{D}\mathcal{C}\rangle  \\
   11) \quad & \langle \mathcal{A}\mathcal{C}\mathcal{D}\mathcal{B}\mathcal{D}\mathcal{D}\rangle +\langle \mathcal{A}\mathcal{D}\mathcal{C}\mathcal{D}\mathcal{B}\mathcal{D}\rangle +\langle \mathcal{A}\mathcal{D}\mathcal{D}\mathcal{C}\mathcal{D}\mathcal{B}\rangle  \notag \\ 
    = &  \langle \mathcal{A}\mathcal{C}\mathcal{D}\mathcal{B}\rangle\langle \mathcal{D}\mathcal{D}\rangle +\langle \mathcal{A}\mathcal{D}\mathcal{D}\rangle\langle \mathcal{B}\mathcal{C}\mathcal{D}\rangle +\langle \mathcal{A}\mathcal{D}\rangle\langle \mathcal{B}\mathcal{D}\mathcal{C}\mathcal{D}\rangle  \\
   12) \quad & \langle \mathcal{A}\mathcal{D}\mathcal{B}\mathcal{C}\mathcal{D}\mathcal{D}\rangle +\langle \mathcal{A}\mathcal{D}\mathcal{D}\mathcal{B}\mathcal{C}\mathcal{D}\rangle +\langle \mathcal{A}\mathcal{D}\mathcal{D}\mathcal{D}\mathcal{B}\mathcal{C}\rangle  \notag \\ 
    = &  \langle \mathcal{A}\mathcal{D}\mathcal{B}\mathcal{C}\rangle\langle \mathcal{D}\mathcal{D}\rangle +\langle \mathcal{A}\mathcal{D}\mathcal{D}\rangle\langle \mathcal{B}\mathcal{C}\mathcal{D}\rangle +\langle \mathcal{A}\mathcal{D}\rangle\langle \mathcal{B}\mathcal{C}\mathcal{D}\mathcal{D}\rangle  \\
   13) \quad & \langle \mathcal{A}\mathcal{D}\mathcal{C}\mathcal{B}\mathcal{D}\mathcal{D}\rangle +\langle \mathcal{A}\mathcal{D}\mathcal{D}\mathcal{C}\mathcal{B}\mathcal{D}\rangle +\langle \mathcal{A}\mathcal{D}\mathcal{D}\mathcal{D}\mathcal{C}\mathcal{B}\rangle  \notag \\ 
    = &  \langle \mathcal{A}\mathcal{D}\mathcal{C}\mathcal{B}\rangle\langle \mathcal{D}\mathcal{D}\rangle +\langle \mathcal{A}\mathcal{D}\mathcal{D}\rangle\langle \mathcal{B}\mathcal{D}\mathcal{C}\rangle +\langle \mathcal{A}\mathcal{D}\rangle\langle \mathcal{B}\mathcal{D}\mathcal{D}\mathcal{C}\rangle  \\
   14) \quad & \langle \mathcal{A}\mathcal{D}\mathcal{D}\mathcal{B}\mathcal{C}\mathcal{D}\rangle +\langle \mathcal{A}\mathcal{D}\mathcal{D}\mathcal{B}\mathcal{D}\mathcal{C}\rangle +\langle \mathcal{A}\mathcal{D}\mathcal{D}\mathcal{C}\mathcal{B}\mathcal{D}\rangle +\langle \mathcal{A}\mathcal{D}\mathcal{D}\mathcal{C}\mathcal{D}\mathcal{B}\rangle +\langle \mathcal{A}\mathcal{D}\mathcal{D}\mathcal{D}\mathcal{B}\mathcal{C}\rangle +\langle \mathcal{A}\mathcal{D}\mathcal{D}\mathcal{D}\mathcal{C}\mathcal{B}\rangle  \notag \\ 
    = &  \langle \mathcal{A}\mathcal{D}\mathcal{D}\mathcal{B}\rangle\langle \mathcal{C}\mathcal{D}\rangle +\langle \mathcal{A}\mathcal{D}\mathcal{D}\mathcal{C}\rangle\langle \mathcal{B}\mathcal{D}\rangle +\langle \mathcal{A}\mathcal{D}\mathcal{D}\mathcal{D}\rangle\langle \mathcal{B}\mathcal{C}\rangle +\langle \mathcal{A}\mathcal{D}\mathcal{D}\rangle\langle \mathcal{B}\mathcal{C}\mathcal{D}\rangle +\langle \mathcal{A}\mathcal{D}\mathcal{D}\rangle\langle \mathcal{B}\mathcal{D}\mathcal{C}\rangle  \\
   15) \quad & \langle \mathcal{A}\mathcal{D}\mathcal{B}\mathcal{C}\mathcal{D}\mathcal{D}\rangle +\langle \mathcal{A}\mathcal{D}\mathcal{B}\mathcal{D}\mathcal{C}\mathcal{D}\rangle +\langle \mathcal{A}\mathcal{D}\mathcal{C}\mathcal{B}\mathcal{D}\mathcal{D}\rangle +\langle \mathcal{A}\mathcal{D}\mathcal{C}\mathcal{D}\mathcal{B}\mathcal{D}\rangle +\langle \mathcal{A}\mathcal{D}\mathcal{D}\mathcal{B}\mathcal{C}\mathcal{D}\rangle +\langle \mathcal{A}\mathcal{D}\mathcal{D}\mathcal{C}\mathcal{B}\mathcal{D}\rangle  \notag \\ 
    = &  \langle \mathcal{A}\mathcal{D}\mathcal{B}\mathcal{D}\rangle\langle \mathcal{C}\mathcal{D}\rangle +\langle \mathcal{A}\mathcal{D}\mathcal{C}\mathcal{D}\rangle\langle \mathcal{B}\mathcal{D}\rangle +\langle \mathcal{A}\mathcal{D}\mathcal{D}\mathcal{D}\rangle\langle \mathcal{B}\mathcal{C}\rangle +\langle \mathcal{A}\mathcal{D}\mathcal{D}\rangle\langle \mathcal{B}\mathcal{C}\mathcal{D}\rangle +\langle \mathcal{A}\mathcal{D}\mathcal{D}\rangle\langle \mathcal{B}\mathcal{D}\mathcal{C}\rangle  \\
   16) \quad & \langle \mathcal{A}\mathcal{C}\mathcal{D}\mathcal{B}\mathcal{D}\mathcal{D}\rangle +\langle \mathcal{A}\mathcal{C}\mathcal{D}\mathcal{D}\mathcal{B}\mathcal{D}\rangle +\langle \mathcal{A}\mathcal{C}\mathcal{D}\mathcal{D}\mathcal{D}\mathcal{B}\rangle  \notag \\ 
    = &  \langle \mathcal{A}\mathcal{C}\mathcal{D}\mathcal{B}\rangle\langle \mathcal{D}\mathcal{D}\rangle +\langle \mathcal{A}\mathcal{C}\mathcal{D}\mathcal{D}\rangle\langle \mathcal{B}\mathcal{D}\rangle +\langle \mathcal{A}\mathcal{C}\mathcal{D}\rangle\langle \mathcal{B}\mathcal{D}\mathcal{D}\rangle  \\
   17) \quad & \langle \mathcal{A}\mathcal{D}\mathcal{C}\mathcal{B}\mathcal{D}\mathcal{D}\rangle +\langle \mathcal{A}\mathcal{D}\mathcal{C}\mathcal{D}\mathcal{B}\mathcal{D}\rangle +\langle \mathcal{A}\mathcal{D}\mathcal{C}\mathcal{D}\mathcal{D}\mathcal{B}\rangle  \notag \\ 
    = &  \langle \mathcal{A}\mathcal{D}\mathcal{C}\mathcal{B}\rangle\langle \mathcal{D}\mathcal{D}\rangle +\langle \mathcal{A}\mathcal{D}\mathcal{C}\mathcal{D}\rangle\langle \mathcal{B}\mathcal{D}\rangle +\langle \mathcal{A}\mathcal{D}\mathcal{C}\rangle\langle \mathcal{B}\mathcal{D}\mathcal{D}\rangle 
  \end{align}
  
  \begin{align}
   18) \quad & \langle \mathcal{A}\mathcal{D}\mathcal{B}\mathcal{D}\mathcal{D}\mathcal{C}\rangle +\langle \mathcal{A}\mathcal{D}\mathcal{D}\mathcal{B}\mathcal{D}\mathcal{C}\rangle +\langle \mathcal{A}\mathcal{D}\mathcal{D}\mathcal{D}\mathcal{B}\mathcal{C}\rangle  \notag \\ 
    = &  \langle \mathcal{A}\mathcal{D}\mathcal{B}\mathcal{C}\rangle\langle \mathcal{D}\mathcal{D}\rangle +\langle \mathcal{A}\mathcal{D}\mathcal{C}\rangle\langle \mathcal{B}\mathcal{D}\mathcal{D}\rangle +\langle \mathcal{A}\mathcal{D}\mathcal{D}\mathcal{C}\rangle\langle \mathcal{B}\mathcal{D}\rangle  \\
   19) \quad & \langle \mathcal{A}\mathcal{B}\mathcal{D}\mathcal{C}\mathcal{D}\mathcal{D}\rangle +\langle \mathcal{A}\mathcal{B}\mathcal{D}\mathcal{D}\mathcal{C}\mathcal{D}\rangle +\langle \mathcal{A}\mathcal{B}\mathcal{D}\mathcal{D}\mathcal{D}\mathcal{C}\rangle  \notag \\ 
    = &  \langle \mathcal{A}\mathcal{B}\mathcal{D}\mathcal{C}\rangle\langle \mathcal{D}\mathcal{D}\rangle +\langle \mathcal{A}\mathcal{B}\mathcal{D}\mathcal{D}\rangle\langle \mathcal{C}\mathcal{D}\rangle +\langle \mathcal{A}\mathcal{B}\mathcal{D}\rangle\langle \mathcal{C}\mathcal{D}\mathcal{D}\rangle  \\
   20) \quad & \langle \mathcal{A}\mathcal{D}\mathcal{B}\mathcal{C}\mathcal{D}\mathcal{D}\rangle +\langle \mathcal{A}\mathcal{D}\mathcal{B}\mathcal{D}\mathcal{C}\mathcal{D}\rangle +\langle \mathcal{A}\mathcal{D}\mathcal{B}\mathcal{D}\mathcal{D}\mathcal{C}\rangle  \notag \\ 
    = &  \langle \mathcal{A}\mathcal{D}\mathcal{B}\mathcal{C}\rangle\langle \mathcal{D}\mathcal{D}\rangle +\langle \mathcal{A}\mathcal{D}\mathcal{B}\mathcal{D}\rangle\langle \mathcal{C}\mathcal{D}\rangle +\langle \mathcal{A}\mathcal{D}\mathcal{B}\rangle\langle \mathcal{C}\mathcal{D}\mathcal{D}\rangle  \\
   21) \quad & \langle \mathcal{A}\mathcal{B}\mathcal{C}\mathcal{D}\mathcal{D}\mathcal{D} \rangle \notag \\ 
    = &  \langle \mathcal{A}\mathcal{B}\mathcal{C}\mathcal{D}\rangle\langle \mathcal{D}\mathcal{D}\rangle +\langle \mathcal{A}\mathcal{B}\mathcal{C}\rangle\langle \mathcal{D}\mathcal{D}\mathcal{D}\rangle  \\
    22) \quad & \langle \mathcal{A}\mathcal{B}\mathcal{C}\mathcal{D}\mathcal{D}\mathcal{D}\rangle +\langle \mathcal{A}\mathcal{B}\mathcal{D}\mathcal{C}\mathcal{D}\mathcal{D}\rangle +\langle \mathcal{A}\mathcal{B}\mathcal{D}\mathcal{D}\mathcal{C}\mathcal{D}\rangle +\langle \mathcal{A}\mathcal{B}\mathcal{D}\mathcal{D}\mathcal{D}\mathcal{C}\rangle  \notag \\ 
    = &  \langle \mathcal{A}\mathcal{B}\mathcal{C}\mathcal{D}\rangle\langle \mathcal{D}\mathcal{D}\rangle +\langle \mathcal{A}\mathcal{B}\mathcal{C}\rangle\langle \mathcal{D}\mathcal{D}\mathcal{D}\rangle +\langle \mathcal{A}\mathcal{B}\mathcal{D}\mathcal{C}\rangle\langle \mathcal{D}\mathcal{D}\rangle +\langle \mathcal{A}\mathcal{B}\mathcal{D}\mathcal{D}\rangle\langle \mathcal{C}\mathcal{D}\rangle +\langle \mathcal{A}\mathcal{B}\mathcal{D}\rangle\langle \mathcal{C}\mathcal{D}\mathcal{D}\rangle +\langle \mathcal{A}\mathcal{B}\rangle\langle \mathcal{C}\mathcal{D}\mathcal{D}\mathcal{D}\rangle \notag \\
      & +\langle \mathcal{A}\mathcal{B}\rangle\langle \mathcal{C}\mathcal{D}\rangle\langle \mathcal{D}\mathcal{D}\rangle  \\
   23) \quad & \langle \mathcal{A}\mathcal{C}\mathcal{B}\mathcal{D}\mathcal{D}\mathcal{D}\rangle +\langle \mathcal{A}\mathcal{C}\mathcal{D}\mathcal{B}\mathcal{D}\mathcal{D}\rangle +\langle \mathcal{A}\mathcal{C}\mathcal{D}\mathcal{D}\mathcal{B}\mathcal{D}\rangle +\langle \mathcal{A}\mathcal{C}\mathcal{D}\mathcal{D}\mathcal{D}\mathcal{B}\rangle  \notag \\ 
    = &  \langle \mathcal{A}\mathcal{C}\mathcal{B}\mathcal{D}\rangle\langle \mathcal{D}\mathcal{D}\rangle +\langle \mathcal{A}\mathcal{C}\mathcal{B}\rangle\langle \mathcal{D}\mathcal{D}\mathcal{D}\rangle +\langle \mathcal{A}\mathcal{C}\mathcal{D}\mathcal{B}\rangle\langle \mathcal{D}\mathcal{D}\rangle +\langle \mathcal{A}\mathcal{C}\mathcal{D}\mathcal{D}\rangle\langle \mathcal{B}\mathcal{D}\rangle +\langle \mathcal{A}\mathcal{C}\mathcal{D}\rangle\langle \mathcal{B}\mathcal{D}\mathcal{D}\rangle +\langle \mathcal{A}\mathcal{C}\rangle\langle \mathcal{B}\mathcal{D}\mathcal{D}\mathcal{D}\rangle \notag \\
      & +\langle \mathcal{A}\mathcal{C}\rangle\langle \mathcal{B}\mathcal{D}\rangle\langle \mathcal{D}\mathcal{D}\rangle  \\
   24) \quad & \langle \mathcal{A}\mathcal{D}\mathcal{B}\mathcal{C}\mathcal{D}\mathcal{D}\rangle +\langle \mathcal{A}\mathcal{D}\mathcal{B}\mathcal{D}\mathcal{D}\mathcal{C}\rangle +\langle \mathcal{A}\mathcal{D}\mathcal{C}\mathcal{B}\mathcal{D}\mathcal{D}\rangle +\langle \mathcal{A}\mathcal{D}\mathcal{C}\mathcal{D}\mathcal{D}\mathcal{B}\rangle +\langle \mathcal{A}\mathcal{D}\mathcal{D}\mathcal{D}\mathcal{B}\mathcal{C}\rangle +\langle \mathcal{A}\mathcal{D}\mathcal{D}\mathcal{D}\mathcal{C}\mathcal{B}\rangle  \notag \\ 
    = &  \langle \mathcal{A}\mathcal{D}\mathcal{B}\mathcal{C}\rangle\langle \mathcal{D}\mathcal{D}\rangle +\langle \mathcal{A}\mathcal{D}\mathcal{B}\rangle\langle \mathcal{C}\mathcal{D}\mathcal{D}\rangle +\langle \mathcal{A}\mathcal{D}\mathcal{C}\mathcal{B}\rangle\langle \mathcal{D}\mathcal{D}\rangle +\langle \mathcal{A}\mathcal{D}\mathcal{C}\rangle\langle \mathcal{B}\mathcal{D}\mathcal{D}\rangle +\langle \mathcal{A}\mathcal{D}\mathcal{D}\mathcal{D}\rangle\langle \mathcal{B}\mathcal{C}\rangle +\langle \mathcal{A}\mathcal{D}\rangle\langle \mathcal{B}\mathcal{C}\mathcal{D}\mathcal{D}\rangle \notag \\
      & +\langle \mathcal{A}\mathcal{D}\rangle\langle \mathcal{B}\mathcal{C}\rangle\langle \mathcal{D}\mathcal{D}\rangle +\langle \mathcal{A}\mathcal{D}\rangle\langle \mathcal{B}\mathcal{D}\mathcal{D}\mathcal{C}\rangle  \\
   25) \quad & \langle \mathcal{A}\mathcal{D}\mathcal{B}\mathcal{C}\mathcal{D}\mathcal{D}\rangle +\langle \mathcal{A}\mathcal{D}\mathcal{B}\mathcal{D}\mathcal{C}\mathcal{D}\rangle +\langle \mathcal{A}\mathcal{D}\mathcal{C}\mathcal{D}\mathcal{B}\mathcal{D}\rangle +\langle \mathcal{A}\mathcal{D}\mathcal{C}\mathcal{D}\mathcal{D}\mathcal{B}\rangle +\langle \mathcal{A}\mathcal{D}\mathcal{D}\mathcal{B}\mathcal{C}\mathcal{D}\rangle +\langle \mathcal{A}\mathcal{D}\mathcal{D}\mathcal{C}\mathcal{D}\mathcal{B}\rangle  \notag \\ 
    = &  \langle \mathcal{A}\mathcal{D}\mathcal{B}\mathcal{D}\rangle\langle \mathcal{C}\mathcal{D}\rangle +\langle \mathcal{A}\mathcal{D}\mathcal{B}\rangle\langle \mathcal{C}\mathcal{D}\mathcal{D}\rangle +\langle \mathcal{A}\mathcal{D}\mathcal{C}\mathcal{D}\rangle\langle \mathcal{B}\mathcal{D}\rangle +\langle \mathcal{A}\mathcal{D}\mathcal{D}\mathcal{B}\rangle\langle \mathcal{C}\mathcal{D}\rangle +\langle \mathcal{A}\mathcal{D}\mathcal{D}\rangle\langle \mathcal{B}\mathcal{C}\mathcal{D}\rangle +\langle \mathcal{A}\mathcal{D}\rangle\langle \mathcal{B}\mathcal{C}\mathcal{D}\mathcal{D}\rangle \notag \\
      & +\langle \mathcal{A}\mathcal{D}\rangle\langle \mathcal{B}\mathcal{D}\mathcal{C}\mathcal{D}\rangle +\langle \mathcal{A}\mathcal{D}\rangle\langle \mathcal{B}\mathcal{D}\rangle\langle \mathcal{C}\mathcal{D}\rangle  
  \end{align}

\paragraph{$\mathcal{A}\mathcal{B}\mathcal{\mathcal{C}}^4$}

\begin{align}
    1) \quad & \langle \mathcal{A}\mathcal{B}\mathcal{C}\mathcal{C}\mathcal{C}\mathcal{C}\rangle +\langle \mathcal{A}\mathcal{C}\mathcal{B}\mathcal{C}\mathcal{C}\mathcal{C}\rangle +\langle \mathcal{A}\mathcal{C}\mathcal{C}\mathcal{C}\mathcal{B}\mathcal{C}\rangle +\langle \mathcal{A}\mathcal{C}\mathcal{C}\mathcal{C}\mathcal{C}\mathcal{B}\rangle  \notag \\ 
    = &  \langle \mathcal{A}\mathcal{B}\mathcal{C}\rangle\langle \mathcal{C}\mathcal{C}\mathcal{C}\rangle +\langle \mathcal{A}\mathcal{B}\rangle\langle \mathcal{C}\mathcal{C}\mathcal{C}\mathcal{C}\rangle +\langle \mathcal{A}\mathcal{C}\mathcal{B}\rangle\langle \mathcal{C}\mathcal{C}\mathcal{C}\rangle +\langle \mathcal{A}\mathcal{C}\mathcal{C}\mathcal{C}\rangle\langle \mathcal{B}\mathcal{C}\rangle +\langle \mathcal{A}\mathcal{C}\rangle\langle \mathcal{B}\mathcal{C}\mathcal{C}\mathcal{C}\rangle  \\
   2) \quad & \langle \mathcal{A}\mathcal{B}\mathcal{C}\mathcal{C}\mathcal{C}\mathcal{C}\rangle +\langle \mathcal{A}\mathcal{C}\mathcal{B}\mathcal{C}\mathcal{C}\mathcal{C}\rangle +\langle \mathcal{A}\mathcal{C}\mathcal{C}\mathcal{B}\mathcal{C}\mathcal{C}\rangle  \notag \\ 
    = &  \langle \mathcal{A}\mathcal{B}\mathcal{C}\mathcal{C}\rangle\langle \mathcal{C}\mathcal{C}\rangle +\langle \mathcal{A}\mathcal{C}\mathcal{C}\rangle\langle \mathcal{B}\mathcal{C}\mathcal{C}\rangle +\langle \mathcal{A}\mathcal{C}\rangle\langle \mathcal{B}\mathcal{C}\mathcal{C}\mathcal{C}\rangle  \\
   3) \quad & \langle \mathcal{A}\mathcal{C}\mathcal{B}\mathcal{C}\mathcal{C}\mathcal{C}\rangle +\langle \mathcal{A}\mathcal{C}\mathcal{C}\mathcal{B}\mathcal{C}\mathcal{C}\rangle +\langle \mathcal{A}\mathcal{C}\mathcal{C}\mathcal{C}\mathcal{B}\mathcal{C}\rangle  \notag \\ 
    = &  \langle \mathcal{A}\mathcal{C}\mathcal{B}\mathcal{C}\rangle\langle \mathcal{C}\mathcal{C}\rangle +\langle \mathcal{A}\mathcal{C}\mathcal{C}\rangle\langle \mathcal{B}\mathcal{C}\mathcal{C}\rangle +\langle \mathcal{A}\mathcal{C}\rangle\langle \mathcal{B}\mathcal{C}\mathcal{C}\mathcal{C}\rangle  \\
   4) \quad & \langle \mathcal{A}\mathcal{C}\mathcal{C}\mathcal{B}\mathcal{C}\mathcal{C}\rangle +\langle \mathcal{A}\mathcal{C}\mathcal{C}\mathcal{C}\mathcal{B}\mathcal{C}\rangle +\langle \mathcal{A}\mathcal{C}\mathcal{C}\mathcal{C}\mathcal{C}\mathcal{B}\rangle  \notag \\ 
    = &  \langle \mathcal{A}\mathcal{C}\mathcal{C}\mathcal{B}\rangle\langle \mathcal{C}\mathcal{C}\rangle +\langle \mathcal{A}\mathcal{C}\mathcal{C}\rangle\langle \mathcal{B}\mathcal{C}\mathcal{C}\rangle +\langle \mathcal{A}\mathcal{C}\rangle\langle \mathcal{B}\mathcal{C}\mathcal{C}\mathcal{C}\rangle  \\
   5) \quad & \langle \mathcal{A}\mathcal{C}\mathcal{C}\mathcal{B}\mathcal{C}\mathcal{C}\rangle +\langle \mathcal{A}\mathcal{C}\mathcal{C}\mathcal{C}\mathcal{B}\mathcal{C}\rangle +\langle \mathcal{A}\mathcal{C}\mathcal{C}\mathcal{C}\mathcal{C}\mathcal{B}\rangle  \notag \\ 
    = &  \langle \mathcal{A}\mathcal{C}\mathcal{C}\mathcal{B}\rangle\langle \mathcal{C}\mathcal{C}\rangle +\langle \mathcal{A}\mathcal{C}\mathcal{C}\mathcal{C}\rangle\langle \mathcal{B}\mathcal{C}\rangle +\langle \mathcal{A}\mathcal{C}\mathcal{C}\rangle\langle \mathcal{B}\mathcal{C}\mathcal{C}\rangle  \\
   6) \quad & \langle \mathcal{A}\mathcal{B}\mathcal{C}\mathcal{C}\mathcal{C}\mathcal{C}\rangle \notag \\ 
    = &  \langle \mathcal{A}\mathcal{B}\mathcal{C}\mathcal{C}\rangle\langle \mathcal{C}\mathcal{C}\rangle +\langle \mathcal{A}\mathcal{B}\mathcal{C}\rangle\langle \mathcal{C}\mathcal{C}\mathcal{C}\rangle  \\
   7) \quad & \langle \mathcal{A}\mathcal{C}\mathcal{B}\mathcal{C}\mathcal{C}\mathcal{C}\rangle \notag \\ 
    = &  \langle \mathcal{A}\mathcal{C}\mathcal{B}\mathcal{C}\rangle\langle \mathcal{C}\mathcal{C}\rangle +\langle \mathcal{A}\mathcal{C}\mathcal{B}\rangle\langle \mathcal{C}\mathcal{C}\mathcal{C}\rangle  \\
   8) \quad & \langle \mathcal{A}\mathcal{B}\mathcal{C}\mathcal{C}\mathcal{C}\mathcal{C}\rangle \notag \\ 
    = &  \langle \mathcal{A}\mathcal{B}\mathcal{C}\mathcal{C}\rangle\langle \mathcal{C}\mathcal{C}\rangle +\langle \mathcal{A}\mathcal{B}\mathcal{C}\rangle\langle \mathcal{C}\mathcal{C}\mathcal{C}\rangle +\langle \mathcal{A}\mathcal{B}\rangle\langle \mathcal{C}\mathcal{C}\mathcal{C}\mathcal{C}\rangle +\langle \mathcal{A}\mathcal{B}\rangle\langle \mathcal{C}\mathcal{C}\rangle^2  \\
   9) \quad & \langle \mathcal{A}\mathcal{C}\mathcal{B}\mathcal{C}\mathcal{C}\mathcal{C}\rangle +\langle \mathcal{A}\mathcal{C}\mathcal{C}\mathcal{B}\mathcal{C}\mathcal{C}\rangle +\langle \mathcal{A}\mathcal{C}\mathcal{C}\mathcal{C}\mathcal{B}\mathcal{C}\rangle +\langle \mathcal{A}\mathcal{C}\mathcal{C}\mathcal{C}\mathcal{C}\mathcal{B}\rangle  \notag \\ 
    = &  \langle \mathcal{A}\mathcal{C}\mathcal{B}\mathcal{C}\rangle\langle \mathcal{C}\mathcal{C}\rangle +\langle \mathcal{A}\mathcal{C}\mathcal{B}\rangle\langle \mathcal{C}\mathcal{C}\mathcal{C}\rangle +\langle \mathcal{A}\mathcal{C}\mathcal{C}\mathcal{B}\rangle\langle \mathcal{C}\mathcal{C}\rangle +\langle \mathcal{A}\mathcal{C}\mathcal{C}\mathcal{C}\rangle\langle \mathcal{B}\mathcal{C}\rangle +\langle \mathcal{A}\mathcal{C}\mathcal{C}\rangle\langle \mathcal{B}\mathcal{C}\mathcal{C}\rangle \notag \\
      & +\langle \mathcal{A}\mathcal{C}\rangle\langle \mathcal{B}\mathcal{C}\mathcal{C}\mathcal{C}\rangle +\langle \mathcal{A}\mathcal{C}\rangle\langle \mathcal{B}\mathcal{C}\rangle\langle \mathcal{C}\mathcal{C}\rangle  
  \end{align}


\section{Complete $SU(2)$ Operator Set for $SU(2)$}
\label{app:SU2}
\subsection{Chiral Dim-2 Operator List}

\setlength\tabcolsep{0pt} 
\begin{center}
\begin{longtable}{|c|c|c||c||c||c|}
\hline
& & $C+P+$ & $C+P-$ & $C-P+$ & $C-P-$\\
\hline  
Type & $i$ & $\mathcal{O}^{++}{}^{(2)}_i$ & $\mathcal{O}^{+-}{}^{(2)}_i$ & $\mathcal{O}^{-+}{}^{(2)}_i$ & $\mathcal{O}^{--}{}^{(2)}_i$ \\
\hline

\multirow{3}{*}{$u^2N^2$}&1&$(\Bar{N} N)\langle u_\mu u^\mu\rangle$&$(\Bar{N}\gamma^5\sigma_{\mu\nu}[u^\mu,u^\nu] N)$&&$(\Bar{N}\gamma^5\gamma^\mu\lrpartial^\nu N)\langle u_\mu u_\nu\rangle$\\
        &2&$(\Bar{N}\sigma^{\mu\nu}[u_\mu,u_\nu] N)$&&&\\
        &3&$(\Bar{N}\gamma^\mu\lrpartial^\nu N)\langle u_\mu u_\nu\rangle$&&&\\
\hline       
        \multirow{2}{*}
        {$\Sigma N^2$}&4&$(\Bar{N}\Sigma_+ N)$&$(\Bar{N}\Sigma_- N)$&&\\
        &5&$(\Bar{N} N)\langle\Sigma_+\rangle$&$(\Bar{N} N)\langle\Sigma_-\rangle$&&\\
\hline
{$fN^2$} & 6 & $(\Bar{N}\sigma^{\mu\nu} f_{+\mu\nu} N)$ & $(\Bar{N}\sigma^{\mu\nu}\Tilde{f}_{+\mu\nu} N)$ & $(\Bar{N}\sigma^{\mu\nu}\Tilde{f}_{-\mu\nu} N)$ & $(\Bar{N}\sigma^{\mu\nu} f_{\mu\nu} N)$ \\
\hline
        \caption{Chiral dim-2 operator list for $SU(2)$.}
        \label{Dim-2}
\end{longtable}
\end{center}

\subsection{Chiral Dim-3 Operator List}
\setlength\tabcolsep{0pt} 
\fontsize{8pt}{10pt}\selectfont
\begin{center}
\begin{longtable}{|c|c|c||c||c||c|}
\hline
& & $C+P+$ & $C+P-$ & $C-P+$ & $C-P-$\\
\hline  
Type & $i$ & $\mathcal{O}^{++}{}^{(3)}_i$ & $\mathcal{O}^{+-}{}^{(3)}_i$ & $\mathcal{O}^{-+}{}^{(3)}_i$ & $\mathcal{O}^{--}{}^{(3)}_i$ \\
\hline

\multirow{3}{*}{$Du^2N^2$} & $1$ & $(\Bar{N}\gamma^\mu [D^\nu u_\mu,u_\nu] N)$ 
 &$(\Bar{N} \gamma^5\sigma^{\mu\nu}\lrpartial^\lambda N)\langle u_\mu D_\lambda u_\nu\rangle$ 
 & 
 & $(\Bar{N}\gamma^5\gamma^\mu [D^\nu u_\mu,u_\nu] N)$ \\
 & $2$ & $(\Bar{N}\sigma_{\mu\nu}\lrpartial^\lambda N)\langle u^\mu D_\lambda u^\nu\rangle$ 
 & 
 & 
 & $(\bar{N}\gamma^5\gamma^\mu\lrpartial^\nu\lrpartial^\rho [D_\mu u_\nu,u_\lambda]N)$\\
 & $3$ & $(\bar{N}\gamma^\mu\lrpartial^\nu\lrpartial^\rho [D_\mu u_\nu,u_\rho]N)$ 
 & 
 &
 &  \\
\hline

\multirow{5}{*}{$u^3N^2$} & $4$ & $\varepsilon^{\mu\nu\rho\lambda}(\Bar{N}\gamma_\mu N)\langle [u_\nu,u_\rho] u_\lambda \rangle$ 
 & $(\Bar{N} \sigma_{\mu\nu}\lrpartial^\rho u^\mu N)\langle u_\rho u^\nu\rangle$
 &
 & $\varepsilon^{\mu\nu\rho\lambda}(\Bar{N}\gamma^5\gamma_\mu N)\langle [u_\nu,u_\rho] u_\lambda \rangle$ \\
 & $5$ & $(\Bar{N}\gamma^5\gamma^\nu u^\mu N)\langle u_\mu u_\nu \rangle $ 
 & 
 &
 & $(\Bar{N}\gamma^\nu u^\mu N)\langle u_\mu u_\nu \rangle $ \\
 & $6$ & $(\Bar{N}\gamma^5\gamma^\nu u_\nu N)\langle u^\mu u_\mu \rangle $ & 
 &
 & $(\Bar{N}\gamma^\nu u_\nu N)\langle u^\mu u_\mu \rangle $ \\
\end{longtable}
\end{center}
\begin{landscape}
\subsection{Chiral Dim-4 Operator List}
\setlength\tabcolsep{0pt} 
\fontsize{8pt}{10pt}\selectfont
\begin{center}
\begin{longtable}{|c|c|c||c||c||c|}
\hline
& & $C+P+$ & $C+P-$ & $C-P+$ & $C-P-$\\
\hline  
Type & $i$ & $\mathcal{O}^{++}{}^{(4)}_i$ & $\mathcal{O}^{+-}{}^{(4)}_i$ & $\mathcal{O}^{-+}{}^{(4)}_i$ & $\mathcal{O}^{--}{}^{(4)}_i$ \\
\hline
\multirow{5}{*}{$D^2u^2 N^2$}
& 1& $\bin{}\lra{D_\mu u_\nu D^\mu u^\nu}$ & $\bin{\gamma^5\sigma^{\mu\nu}[D_\lambda u_\mu,D^\lambda u_\nu]}$& & $\bin{\gamma^5\gamma^\mu\lrpartial^\nu}\lra{D_\mu u_\lambda D^\lambda  u_\nu}$\\
& 2& $\bin{\sigma^{\mu\nu}[D_\lambda u_\mu,D^\lambda u_\nu]}$ & $\bin{\gamma^5\sigma_{\mu\nu}\lrpartial_\lambda\lrpartial_\rho [D^\nu u^\rho,D^\lambda u^\mu]}$ & &$\bin{\gamma^5\gamma_\mu\lrpartial_\lambda\lrpartial_\nu\lrpartial_\rho }\lra{D^\lambda u^\mu D^\nu u^\rho}$ \\
& 3& $\bin{\gamma^\mu\lrpartial^\nu}\lra{D_\mu u_\lambda D^\lambda  u_\nu}$ & & &  \\
& 4& $\bin{\sigma_{\mu\nu}\lrpartial_\lambda\lrpartial_\rho [D^\nu u^\rho,D^\lambda u^\mu]}$ &  & & \\
&5 & $\bin{\gamma_\mu\lrpartial_\lambda\lrpartial_\nu\lrpartial_\rho }\lra{D^\lambda u^\mu D^\nu u^\rho}$ & & &  \\
\hline
\multirow{6}{*}{$Du^3 N^2$}
& 6& $\bin{\gamma^5\sigma^{\mu\nu}}\lra{D_\lambda u_\mu [u^\lambda,u_\nu]}$ & $\bin{\sigma^{\mu\nu}}\lra{D_\lambda u_\mu [u^\lambda,u_\nu]}$ & $\bin{\gamma^5\sigma_{\mu\nu}u_\lambda}\lra{D^\lambda u_\mu u_\nu}$ & $\bin{\sigma^{\mu\nu}u_\lambda}\lra{D^\lambda u_\mu u_\nu}$ \\
\end{longtable}
\end{center}
\end{landscape}

\begin{landscape}
\subsection{Chiral Dim-5 Operator List}
\setlength\tabcolsep{-1.5pt} 
\fontsize{6.4pt}{8pt}\selectfont
\begin{center}
    \begin{longtable}{|c|c|c|c|c|c|}
    \hline
        &&$C+P+$&$C+P-$&$C-P+$&$C-P-$\\
        \hline
         Type&$i$&${\mathcal{O}^{++}}_i^{(5)}$&${\mathcal{O}^{+-}}_i^{(5)}$&${\mathcal{O}^{-+}}_i^{(5)}$&${\mathcal{O}^{--}}_i^{(5)}$\\
         \hline
         \multirow{2}{*}{$D^3fN^2$}&1&$(\Bar{N}\gamma^\mu (D^\nu D^2f_{+\mu\nu})N)$&$(\Bar{N}\gamma^\mu (D^\nu D^2\tilde f_{+\mu\nu})N)$&$(\Bar{N}\gamma^5\gamma^\mu (D^\nu D^2\tilde f_{+\mu\nu})N)$&$(\Bar{N}\gamma^5\gamma^\mu (D^\nu D^2f_{+\mu\nu})N)$\\
         &2&$(\Bar{N}\gamma^5\gamma^\mu (D^\nu D^2f_{-\mu\nu})N)$&$(\Bar{N}\gamma^5\gamma^\mu (D^\nu D^2\Tilde{f}_{-\mu\nu})N)$&$(\Bar{N}\gamma^\mu (D^\nu D^2\tilde f_{-\mu\nu})N)$&$(\Bar{N}\gamma^\mu (D^\nu D^2f_{-\mu\nu})N)$\\
         \hline
         \multirow{22}{*}{$Df\Sigma N^2$}&3&$(\Bar{N}\gamma^\mu N)\langle f_{+\mu\nu}D^\nu \Sigma_+\rangle$&$(\Bar{N}\gamma^\mu N)\langle \Tilde{f}_{+\mu\nu}D^\nu \Sigma_+\rangle$&$(\Bar{N}\gamma^5\gamma^\mu N)\langle \Tilde{f}_{+\mu\nu}D^\nu \Sigma_+\rangle$&$(\Bar{N}\gamma^5\gamma^\mu N)\langle f_{+\mu\nu}D^\nu \Sigma_+\rangle$\\
         &4&$(\Bar{N}\gamma^5\gamma^\mu [\Tilde{f}_{+\mu\nu},D^\nu \Sigma_+]N)$&$(\Bar{N}\gamma^5\gamma^\mu [f_{+\mu\nu},D^\nu \Sigma_+]N)$&$(\Bar{N}\gamma^\mu [f_{+\mu\nu},D^\nu \Sigma_+]N)$&$(\Bar{N}\gamma^\mu [\Tilde{f}_{+\mu\nu},D^\nu \Sigma_+]N)$\\
         &5&$(\Bar{N}\gamma^\mu N)\langle D^\nu f_{+\mu\nu} \Sigma_+\rangle$&$(\Bar{N}\gamma^5\gamma^\mu  [D^\nu f_{+\mu\nu}, \Sigma_+]N)$&$(\Bar{N}\gamma^\mu [D^\nu f_{+\mu\nu}, \Sigma_+]N)$&$(\Bar{N}\gamma^5\gamma^\mu N)\langle D^\nu f_{+\mu\nu} \Sigma_+\rangle$\\
         &6&$(\Bar{N}\gamma^5\gamma^\mu [f_{+\mu\nu},D^\nu \Sigma_-]N)$&$(\Bar{N}\gamma^5\gamma^\mu [\Tilde{f}_{+\mu\nu},D^\nu \Sigma_-]N)$&$(\Bar{N}\gamma^\mu [\Tilde{f}_{+\mu\nu},D^\nu \Sigma_-]N)$&$(\Bar{N}\gamma^\mu [f_{+\mu\nu},D^\nu \Sigma_-]N)$\\
\end{longtable}
\end{center}
\end{landscape}

\section{Complete $SU(3)$ Operator Set for $SU(3)$}
\label{app:SU3}
\subsection{Chiral Dim-2 Operator List}

\setlength\tabcolsep{0pt} 
\fontsize{8pt}{10pt}\selectfont
\begin{center}
    \begin{longtable}{|c|c|c|c|c|c|}
    \hline
        &&$C+P+$&$C+P-$&$C-P+$&$C-P-$\\
        \hline
         Type&$i$&${\mathcal{O}^{++}}_i^{(2)}$&${\mathcal{O}^{+-}}_i^{(2)}$&${\mathcal{O}^{-+}}_i^{(2)}$&${\mathcal{O}^{--}}_i^{(2)}$\\
         \hline
         \multirow{2}{*}
        {$fB^2$}
        &1&$\langle\Bar{B}\sigma_{\mu\nu} B f_+^{\mu\nu}\rangle$&$\langle\Bar{B}\sigma^{\rho\lambda} B f_+^{\mu\nu}\rangle\epsilon_{\mu\nu\rho\lambda}$&$\langle\Bar{B}\sigma^{mu\nu} B f_-^{\rho\lambda}\rangle\epsilon_{\mu\nu\rho\lambda}$&$\langle\Bar{B}\sigma_{\mu\nu} B f_-^{\mu\nu}\rangle$\\
        &2&$\langle\Bar{B}\sigma_{\mu\nu} f_+^{\mu\nu} B\rangle$&$\langle\Bar{B}\sigma^{\rho\lambda} f_+^{\mu\nu} B\rangle\epsilon_{\mu\nu\rho\lambda}$&$\langle\Bar{B}\sigma^{\mu\nu} f_-^{\rho\lambda} B\rangle\epsilon_{\mu\nu\rho\lambda}$&$\langle\Bar{B}\sigma_{\mu\nu} f_-^{\mu\nu} B\rangle$\\
         \hline
         \multirow{11}{*}
        {$u^2B^2$}
        &3&$\langle\Bar{B} B\rangle\langle u_\mu u^\mu\rangle$&$\langle\Bar{B}\sigma^{\mu\nu} u^\rho\rangle\langle u^\lambda B\rangle$&$\langle\Bar{B}\sigma^{\mu\nu} u_\mu B u_\nu\rangle$&$\langle\Bar{B}\sigma^{\mu\nu} u^\rho B u^\lambda\rangle\epsilon_{\mu\nu\rho\lambda}$\\
        &4&$\langle\Bar{B} u_\mu\rangle\langle u^\mu B\rangle$&$\langle\Bar{B}\sigma^{\mu\nu} B[u^\rho,u^\lambda]\rangle\epsilon_{\mu\nu\rho\lambda}$&&$\langle\Bar{B} \gamma^5\gamma^\mu\lrpartial^\nu B\rangle\langle u_\mu u_\nu\rangle$\\
        &5&$\langle\Bar{B}\sigma^{\mu\nu} u_\mu\rangle\langle u_\nu B\rangle$&$\langle\Bar{B}\sigma^{\mu\nu}[u^\rho,u^\lambda] B\rangle\epsilon_{\mu\nu\rho\lambda}$&&$\langle\Bar{B}\gamma^5 \gamma^\mu\lrpartial^\nu u_\mu\rangle\langle  u_\nu B\rangle$\\
        &6&$\langle\Bar{B} B u_\mu u^\mu\rangle$&&&$\langle\Bar{B} \gamma^5\gamma^\mu\lrpartial^\nu B u_\mu u_\nu\rangle$\\
\end{longtable}
\end{center}

\subsection{Chiral Dim-3 Operator List}

\setlength\tabcolsep{0pt} 
\fontsize{8pt}{10pt}\selectfont
\begin{center}
    \begin{longtable}{|c|c|c|c|c|c|}
    \hline
        &&$C+P+$&$C+P-$&$C-P+$&$C-P-$\\
        \hline
         Type&$i$&${\mathcal{O}^{++}}_i^{(3)}$&${\mathcal{O}^{+-}}_i^{(3)}$&${\mathcal{O}^{-+}}_i^{(3)}$&${\mathcal{O}^{--}}_i^{(3)}$\\
         \hline
         \multirow{4}{*}
        {$DfB^2$}
        &1&$\langle\Bar{B}\gamma^5\gamma_\nu D_\mu f_-^{\mu\nu} B\rangle$&&&$\langle\Bar{B}\gamma_\nu D_\mu f_-^{\mu\nu} B\rangle$\\
        &2&$\langle\Bar{B}\gamma^5\gamma_\nu B D_\mu f_-^{\mu\nu}\rangle$&&&$\langle\Bar{B}\gamma_\nu B D_\mu f_-^{\mu\nu}\rangle$\\
        &3&$\langle\Bar{B}\gamma_\nu D_\mu f_+^{\mu\nu} B\rangle$&&&$\langle\Bar{B}\gamma^5\gamma_\nu D_\mu f_+^{\mu\nu} B\rangle$\\
        &4&$\langle\Bar{B}\gamma_\nu B D_\mu f_+^{\mu\nu}\rangle$&&&$\langle\Bar{B}\gamma^5\gamma_\nu B D_\mu f_+^{\mu\nu}\rangle$\\
         \hline
        {$ufB^2$}
        &5
        &$\langle\Bar{B}\gamma^5\gamma_\mu B\rangle\langle\Tilde{f}_-^{\mu\nu} u_\nu\rangle$
        &$\langle\Bar{B}\gamma^5\gamma_\mu B\rangle\langle f_-^{\mu\nu} u_\nu\rangle$
        &$\langle\Bar{B}\gamma_\mu B\rangle\langle f_-^{\mu\nu} u_\nu\rangle$
        &$\langle\Bar{B}\gamma_\mu B\rangle\langle\Tilde{f}_-^{\mu\nu} u_\nu\rangle$\\
        &6
        &$\langle\Bar{B}\gamma_\mu f_-^{\mu\nu}\rangle\langle u_\nu B\rangle-h.c.$
        &$\langle\Bar{B}\gamma_\mu\Tilde{f}_-^{\mu\nu}\rangle\langle u_\nu B\rangle-h.c.$
        &$\langle\Bar{B}\gamma^5\gamma_\mu\Tilde{f}_-^{\mu\nu}\rangle\langle u_\nu B\rangle-h.c.$
        &$\langle\Bar{B}\gamma^5\gamma_\mu f_-^{\mu\nu}\rangle\langle u_\nu B\rangle-h.c.$\\
\end{longtable} 
    
\end{center}
\begin{landscape}
\subsection{Chiral Dim-4 Operator List}

\setlength\tabcolsep{0pt} 
\fontsize{8pt}{10pt}\selectfont
\begin{center}
    \begin{longtable}{|c|c|c|c|c|c|}
    \hline
        &&$C+P+$&$C+P-$&$C-P+$&$C-P-$\\
        \hline
         Type&$i$&${\mathcal{O}^{++}}_i^{(4)}$&${\mathcal{O}^{+-}}_i^{(4)}$&${\mathcal{O}^{-+}}_i^{(4)}$&${\mathcal{O}^{--}}_i^{(4)}$\\
        \hline
        \multirow{2}{*}{$D^2fB^2$}&1&$\langle\Bar{B}\sigma^{\mu\nu}BD_\mu D^\lambda f_{+\nu\lambda}\rangle$&$\langle\Bar{B}\sigma^{\mu\nu}BD_\mu D^\lambda f_{+\nu\lambda}\rangle$&$\langle\Bar{B}\sigma^{\mu\nu}BD_\mu D^\lambda f_{-\nu\lambda}\rangle$&$\langle\Bar{B}\sigma^{\mu\nu}BD_\mu D^\lambda f_{-\nu\lambda}\rangle$\\
        &2&$\langle\Bar{B}\sigma^{\mu\nu}D_\mu D^\lambda f_{+\nu\lambda}B\rangle$&$\langle\Bar{B}\sigma^{\mu\nu}D_\mu D^\lambda f_{+\nu\lambda}B\rangle$&$\langle\Bar{B}\sigma^{\mu\nu}D_\mu D^\lambda f_{-\nu\lambda}B\rangle$&$\langle\Bar{B}\sigma^{\mu\nu}D_\mu D^\lambda f_{-\nu\lambda}B\rangle$\\
        \hline
        \multirow{3}{*}{$D^2\Sigma B^2$}&3&$\langle \Bar{B}B\rangle D^2\langle\Sigma_+\rangle$&$\langle \Bar{B}B\rangle D^2\langle
\Sigma_+\rangle$&&\\
        &4&$\langle\Bar{B}BD^2\Sigma_+\rangle$&$\langle\Bar{B}BD^2\Sigma_+\rangle$&&\\
        &5&$\langle\Bar{B}D^2\Sigma_+B\rangle$&$\langle\Bar{B}D^2\Sigma_+B\rangle$&&\\
        \hline
        {$f^2B^2$}&6&$\langle f_-^{\mu\nu}f_{-\mu\nu}\rangle\langle\Bar{B}B\rangle$&$\langle \Bar{B}\Tilde{f}_-^{\mu\nu}\rangle\langle f_{-\mu\nu}B\rangle$&$\langle\Bar{B}\sigma^{\mu\nu} f_{-\mu}^\lambda Bf_{-\nu\lambda}\rangle$&$\langle\Bar{B}\sigma^{\mu\nu} \Tilde{f}_{-\mu}^\lambda Bf_{-\nu\lambda}\rangle$\\
\end{longtable}
\end{center}
\end{landscape}

\end{document}